\documentclass[11pt]{article}

\usepackage{amsmath,amssymb}
\usepackage{comment}
\usepackage{ulem}

\usepackage{times}
\usepackage{bm}
\usepackage{natbib}
\usepackage{algorithm}
\usepackage{amssymb}
\usepackage{mathrsfs}
\usepackage{graphicx}
\usepackage[flushleft]{threeparttable}
\usepackage{enumitem} 
\usepackage{caption}
\usepackage{subcaption}
\usepackage[font={small,it}]{caption}
\usepackage[utf8]{inputenc} 
\usepackage[T1]{fontenc}    
\usepackage{url}           
\usepackage{booktabs}      
\usepackage{amsfonts}      
\usepackage{nicefrac}       
\usepackage{microtype}   

\usepackage{amsmath,amssymb,amsfonts}
\usepackage{graphicx}
\usepackage{accents}

\newcommand{\R}{\mathbb{R}}
\newcommand{\E}{\mathbb{E}}

\newcommand{\y}{\mathbf{y}}
\newcommand{\X}{\mathbf{X}}
\newcommand{\Z}{\mathbf{Z}}
\newcommand{\x}{\mathbf{x}}

\newcommand{\W}{\mathbf{W}}

\newcommand{\M}{\mathbf{M}}
\newcommand{\A}{\mathbf{A}}

\newcommand{\bmS}{\mathbf{S}}

\newcommand{\probP}{\mathbb{P}}
\newcommand{\bmSigma}{\mathbf{{\bSigma}}}

\newcommand{\bmeps}{\boldsymbol{\bepsilon}}

\DeclareMathOperator\erf{erf}

\newcommand{\sg}{\mbox{sg}}
\newcommand{\barbeta}{\bar{\beta}}

\usepackage{amsthm}
\newtheorem{theorem}{Theorem}
\newtheorem{proposition}{Proposition}
\newtheorem{lemma}{Lemma}
\newtheorem{corollary}{Corollary}
\newtheorem{definition}{Definition}
\newtheorem{condition}{Condition}
\newtheorem{remark}{Remark}
\newtheorem{remark.s}{Remark S\ignorespaces}

\makeatletter

\newcommand{\Rmnum}[1]{\expandafter\@slowromancap\romannumeral #1@}
\makeatother

\usepackage{xcolor}

\usepackage{star}

\usepackage[colorlinks,
linkcolor=blue,
anchorcolor=blue,
citecolor=blue
]{hyperref}

\providecommand{\bnorm}[1]{\left\|#1\right\|}

\newcommand*{\supp}{\mathrm{supp}}
\newcommand*{\var}{\mathrm{var}}

\def\T{{ \mathrm{\scriptscriptstyle T} }}

\def\##1\#{\begin{align}#1\end{align}}
\def\$#1\${\begin{align*}#1\end{align*}}

\newcommand {\vecc}{\textnormal {vec}}
\newcommand {\cov}{\textnormal {cov}}
\newcommand {\rank}{\textnormal {rank}}
\newcommand {\soft}{\textnormal {soft}}

\newcommand {\RW}{\textnormal {RW}}
\newcommand {\RA}{\textnormal {R}}
\newcommand {\LA}{\textnormal {L}}
\newcommand {\LW}{\textnormal {LW}}
\def\T{{ \mathrm{\scriptscriptstyle T} }} 
\newcommand {\erfc}{\textnormal {erfc}}
\newcommand{\wrt}{{\rm w.r.t.}}

\newcommand{\as}{{\rm a.s.}}
\newcommand{\qas}{{\quad \rm a.s.}}
\newcommand{\qprob}{{\quad \rm in~probability}}

\newcommand{\Rom}[1]{\text{\uppercase\expandafter{\romannumeral #1\relax}}}

\usepackage{txfonts}

\usepackage{geometry}

\def\T{{ \mathrm{\scriptscriptstyle T} }}

\newcommand{\nn}{\nonumber}
\def\bxz{\color{black}}

\def\BS{\color{black}}

\usepackage{tocloft}
\makeatletter
\renewcommand{\numberline}[1]{%
  \@cftbsnum #1\@cftasnum~\@cftasnumb%
}
\makeatother

\usepackage{minitoc}

\begin{document}

\title{\Large
The exact risks of reference panel-based regularized estimators
} 
 \author{Buxin Su\thanks{Department of Mathematics, University of Pennsylvania; Email: \texttt{subuxin@sas.upenn.edu}.} \and Qiang Sun\thanks{Department of Statistical Sciences, University of Toronto; Email: \texttt{qiang.sun@utoronto.ca}.} \and Xiaochen Yang\thanks{Department of Statistics, Purdue University; Email: \texttt{yang1641@purdue.edu}. } \and  Bingxin Zhao\thanks{Department of Statistics and Data Science, University of Pennsylvania; Email: \texttt{bxzhao@upenn.edu}. 
   Authors are listed in alphabetical order.}
 }    

\maketitle
\date{}


\abstract{

Reference panel-based estimators have become widely used in genetic prediction of complex traits due to their ability to address data privacy concerns and reduce computational and communication costs. These estimators estimate the covariance matrix of predictors using an external reference panel, instead of relying solely on the original training data. In this paper, we investigate the performance of reference panel-based $L_1$ and $L_2$ regularized estimators within a unified framework based on approximate message passing (AMP). We uncover several key factors that influence the accuracy of reference panel-based estimators, including the sample sizes of the training data and reference panels, the signal-to-noise ratio, the underlying sparsity of the signal, and the covariance matrix among predictors. Our findings reveal that, even when the sample size of the reference panel matches that of the training data, reference panel-based estimators tend to exhibit lower accuracy compared to traditional regularized estimators. Furthermore, we observe that this performance gap widens as the amount of training data increases, highlighting the importance of constructing large-scale reference panels to mitigate this issue. To support our theoretical analysis, we develop a novel non-separable matrix AMP framework capable of handling the complexities introduced by a general covariance matrix and the additional randomness associated with a reference panel. We validate our theoretical results through extensive simulation studies and real data analyses using the UK Biobank database. \\

}


\section{Introduction}\label{sec1}


The utilization of genetic data in the prediction of complex traits and diseases holds the potential to facilitate early detection and prevention of clinical outcomes, thereby advancing the field of precision medicine \citep{torkamani2018personal}. Genome-wide association studies (GWAS) 
examine millions of common genetic variants to estimate their effects on the phenotypes of interest \citep{uffelmann2021genome}.  Recent GWAS have analyzed millions of samples  across {various} phenotypes, creating unprecedented opportunities for predicting genetic risk in new individuals \citep{martin2019clinical}. Over the past two decades, numerous genetic risk prediction models have been developed, including those mentioned in \cite{marquez2021incorporating,ge2019polygenic,vilhjalmsson2015modeling,mak2017polygenic,hu2017leveraging,yang2020accurate,lloyd2019improved,pattee2020penalized,song2020leveraging}. We refer readers  to \cite{ma2021genetic} for a recent statistical review of these models. These models have demonstrated success in predicting a wide range of phenotypes, leading to thousands of publications each year in the fields of genomics and clinical research \citep{wray2021basic}.


Consider a generic training dataset $(\X, \y_x)$ obtained from genome-wide association studies (GWAS), which consists of data from $n_x$ individuals, such as half a million subjects from the UK Biobank (UKB) study \citep{sudlow2015uk}. Here, $\X \in \RR^{n_x \times p}$ represents the genetic variants, which can include millions of single nucleotide polymorphisms (SNPs), and $\y_x \in \RR^{n_x}$ represents the corresponding continuous phenotype of interest, such as blood pressure, for the respective $n_x$ samples. The goal of genetic prediction is to predict the same phenotype for a testing dataset $\Sbb$, which comprises $n_s$ new subjects and the same set of $p$ genetic variants. In practice, the sharing and access to individual-level data $(\X, \y_x)$ can be challenging due to privacy restrictions and the costs associated with data communication. Consequently, the scientific community has adopted the standard practice of sharing GWAS marginal summary-level data, such as $\hat{\bbeta}_S = \X^{\T}\y_x$ \citep{hayhurst2022community}. As a result, the majority of genetic data prediction methods rely on the widely available GWAS summary statistics $\hat{\bbeta}_S$ \citep{ma2021genetic}.


When constructing models based on GWAS marginal summary-level data $\hat{\bbeta}_{S}$, adjusting for the linkage disequilibrium (LD) pattern among  genetic variants is crucial for accurate genetic data prediction. The LD pattern can be characterized by the covariance structure $\bmSigma$ of the $p$ genetic variants \citep{pasaniuc2017dissecting}, which is consistent within a specific population (e.g., European population) but may vary across different populations.   To estimate the LD pattern, we often use an external genetic dataset $\W$, known as a reference panel dataset, from a {cohort} that matches the reference population. This reference panel dataset is typically independent of the original training GWAS data $(\X, \y_x)$. For example, many methods, such as those by \cite{ge2019polygenic, vilhjalmsson2015modeling, mak2017polygenic, hu2017leveraging, wang2022sparse}, utilize the 1000 Genomes data \citep{10002015global} to estimate LD patterns. Other reference panels, such as the UK10K dataset \citep{uk10k2015uk10k}, TOPMed dataset \citep{taliun2021sequencing}, and the testing dataset $\Sbb$ \citep{lloyd2019improved, yang2020accurate}, are also frequently employed for this purpose.


Despite the increasing popularity of combining summary statistics and reference panels in genetic data prediction, there is still limited understanding of their statistical properties. This paper aims to investigate reference panel-based estimators within a unified high-dimensional prediction framework that encompasses a wide range of complex traits with varying genetic architectures and causal variants \citep{timpson2018genetic}. Specifically, we focus on the reference panel-based $L_1$ and $L_2$ regularized estimators, which are widely used in genetic risk prediction methods. The reference panel-based $L_1$ regularized estimator \citep{li2022estimation} can be formulated as follows: 
\begin{equation} \label{eq:ref_lasso_arg_min}
\hat{\bbeta}_{\LW}(\lambda) = \arg\min_{\bbeta \in \R^p} \frac{1}{2 n_w} \bbeta^{\T} \W^{\T} \W \bbeta - \frac{1}{n_x} \bbeta^{\T} \X^{\T} \y_x + \frac{\lambda}{\sqrt{p}} \| \bbeta \|_1,
\end{equation}
whereas the reference panel-based $L_2$ regularized estimator \citep{zhao2022block} is given by:
\begin{equation} \label{eq:ref_ridge_formula}
\hat{\bbeta}_{\RW}(\lambda) = \frac{1}{n_x} (\W^{\T} \W/n_w + \lambda \bI_{p})^{-1}\X^{\T}\y_x. 
\end{equation}
In comparison to the traditional $L_1$ and $L_2$ regularized estimators $\hat{\bbeta}_{\textnormal {L}}(\lambda)$ and $\hat{\bbeta}_{\textnormal{R}}(\lambda)$ \citep{tibshirani1996regression,hoerl1970ridge},  the key distinction lies in the estimation of the covariance matrix. The reference panel estimators employ $\W^{\T} \W/n_w$ instead of $\X^{\T} \X/n_x$. This distinction naturally raises the following questions
\begin{quote}
{\it Do $\hat{\bbeta}_{\LW}(\lambda)$ and $\hat{\bbeta}_{\textnormal {L}}(\lambda)$  have the same prediction performance? If not, will the use of the reference panel have similar influences on the $L_1$ and $L_2$ regularized estimators?
}
\end{quote}


To quantitatively assess the impact of the reference panel on $L_1$ regularized estimators that rely on GWAS summary statistics, it is important to compare the performance of $\hat{\bbeta}_{\LW}(\lambda)$ and $\hat{\bbeta}_{\textnormal {L}}(\lambda)$.  Examples of such estimators include lassosum \citep{mak2017polygenic} and its variants \citep{hahn2022smoothed,prive2022identifying}. As the reference panel approach is privacy-friendly and computationally efficient, it may trade some prediction accuracy for convenience.
While the comparison between $L_2$ estimators $\hat{\bbeta}_{\textnormal {R}}(\lambda)$ and $\hat{\bbeta}_{\RW}(\lambda)$  has been explored using random matrix theory (RMT) techniques \citep{zhao2022block}, analyzing $L_1$ estimators using RMT appears infeasible due to the lack of closed-form expressions. To investigate the asymptotic behaviors of $L_1$ estimators, this paper employs approximate message passing (AMP) techniques to derive new theoretical results for $\hat{\bbeta}_{\LW}(\lambda)$. 
In the asymptotic limit, the iterative process of AMP algorithms exhibits a simple distributional state, known as the state evolution of AMP \citep{donoho2009message}. AMP results with an i.i.d. Gaussian matrix $\X$ have been established in previous works, such as \cite{bolthausen2014iterative} and \cite{bayati2011dynamics}, laying the foundation for analyzing more generalized AMP algorithms. Subsequent studies by  \cite{bayati2011lasso}, \cite{rangan2011generalized}, \cite{javanmard2013state}, \cite{bu2019algorithmic}, \cite{berthier2020state}, \cite{gerbelot2021graph}, \cite{wang2020bridge}, and \cite{huang2022lasso}, have demonstrated the applicability of AMP techniques to various scenarios. However, none of these works can be directly used to characterize the performance of $\hat{\bbeta}_{\LW}(\lambda)$.


In this paper, we propose a novel matrix AMP framework for non-separable functions, which enables us to analyze reference panel-based estimators. This framework addresses the challenges arising from a general covariance structure $\bmSigma$ and the additional randomness introduced by a reference panel. By leveraging this framework, we provide a unified analysis of both reference panel-based $L_1$ and $L_2$ regularized estimators. Our theoretical results offer valuable insights into reference panel-based estimators and provide practical guidance for real-world genetic data applications. Specifically, we address the following questions:
\begin{quote}
{\it Can the prediction accuracy of $\hat{\bbeta}_{\LW}(\lambda)$ and $\hat{\bbeta}_{\RW}(\lambda)$ be improved by using better reference panels? If so, which specific complex traits and datasets, characterized by what particular characteristics, would benefit the most from the availability of better reference panels?}
\end{quote}
Our contributions in this paper are as follows. First, we develop AMP results for $\hat{\bbeta}_{\LW}(\lambda)$ and demonstrate that it exhibits different out-of-sample prediction behavior compared to $\hat{\bbeta}_{\textnormal {L}}(\lambda)$ in high-dimensional settings. We evaluate key factors that influence the performance of $\hat{\bbeta}_{\LW}(\lambda)$, including the sample size of the training GWAS data $n_x$, the sample size of the reference panel $n_w$, the dimension of predictors $p$, the overall signal-to-noise ratio (i.e., the heritability in genetics), the underlying signal sparsity, and the LD covariance matrix $\bmSigma$. As reference panel-based $L_1$ and $L_2$ regularized estimators are widely used in various genomic data types \citep{hu2019statistical}, our AMP results provide insights into a variety of high-dimensional prediction problems \citep{wang2020methods}. Second, we find that $\hat{\bbeta}_{\LW}(\lambda)$ is likely to be less accurate than $\hat{\bbeta}_{\textnormal {L}}(\lambda)$, similar to the relationship observed between $\hat{\bbeta}_{\RW}(\lambda)$ and $\hat{\bbeta}_{\textnormal {L}}(\lambda)$. The limited sample size of commonly used reference panels may lead to poorer performance of $\hat{\bbeta}_{\LW}(\lambda)$, and the performance gap between $\hat{\bbeta}_{\LW}(\lambda)$ and $\hat{\bbeta}_{\textnormal {L}}(\lambda)$ may widen as more GWAS training samples are acquired \citep{zhou2022global}. Consequently, constructing larger reference panels can potentially improve prediction performance using genetic data resources. Last, we use AMP to re-derive the exact risks of $L_2$ estimators, which are previously derived using RMT, demonstrating possible wider applicability of AMP, because  RMT can not be used to derive the exact risks for $L_1$ estimtors.   In summary, our results shed light on the trade-offs associated with using reference panels and provide recommendations for the development of genetic data prediction infrastructure. A summary of our main results, together with results from existing literature,  is presented in Table \ref{tab:summary}.

\paragraph{Paper overview}
The rest of the paper proceeds as follows. In Section~\ref{sec2}, we introduce the model setup and the estimators under consideration. Section~\ref{sec: i.i.d. REF-LASSO} presents the AMP results for $\hat{\bbeta}_{\LW}(\lambda)$ in the case of $\bmSigma={\bI_p}$, and compares its performance with that of $\hat{\bbeta}_{\textnormal {L}}(\lambda)$.
Section~\ref{sec: L_1 estimator} introduces the non-separable matrix AMP framework and provides the results for $\hat{\bbeta}_{\LW}(\lambda)$ with a general $\bmSigma$.
In Section~\ref{sec: L_2 estimator}, we analyze the performance of $\hat{\bbeta}_{\RW}(\lambda)$ within our AMP framework, and compare it with $\hat{\bbeta}_{\textnormal {R}}(\lambda)$. Numerical experiments based on real data from the UKB study are presented in Section~\ref{sec: numerical experiments}. In Section~\ref{sec: discussion}, we discuss potential future research directions. Appendix collects proof to the main results and technical lemmas.

\begin{table}
\caption{{\bxz {\bf Summary of our main results and their connections with the existing literature.} 
In this paper, we provide the theoretical results for $\hat{\bbeta}_{\LW}(\lambda)$ and $\hat{\bbeta}_{\RW}(\lambda)$ using AMP. We consider both general $\bmSigma$ (Theorems \ref{thm: i.i.d. ridge mse + R2} and \ref{thm: ridge mse + R2}) and the special case ${\bSigma}={\bI_p}$ (Theorems \ref{thm: i.i.d. lasso mse + R2} and Theorem \ref{thm: general lasso mse + R2}). } } \label{tab:summary}
{%
\begin{center}
\begin{tabular}{ c c c }
 & ${\bSigma}={\bI_p}$ & General ${\bSigma}$\\
 \hline
 $\hat{\bbeta}_{\textnormal {L}}(\lambda)$ & \cite{bayati2011lasso} & \cite{huang2022lasso} \\
 \hline
$\hat{\bbeta}_{\LW}(\lambda)$ & Theorem \ref{thm: i.i.d. lasso mse + R2} & Theorem \ref{thm: general lasso mse + R2}\\ 
\hline
$\hat{\bbeta}_{\textnormal {R}}(\lambda)$ & \cite{ledoit2011eigenvectors} &  \cite{dobriban2018high}\\
\hline
$\hat{\bbeta}_{\RW}(\lambda)$ & 
\begin{tabular}{@{}c@{}}Theorem \ref{thm: i.i.d. ridge mse + R2}(AMP) and \\ \cite{zhao2022block} (RMT) \end{tabular}   & 
\begin{tabular}{@{}c@{}}Theorem \ref{thm: ridge mse + R2} (AMP) and \\ \cite{zhao2022block} (RMT) \end{tabular}\\
\end{tabular}
\end{center}}
\end{table}

\paragraph{Notation}

We define $\RR$ and $\RR_{+}$ as the sets of all real numbers and positive real numbers, respectively. For any positive natural number $n \in \NN_{>0}$, we define $[n]$ to be the set of $\{1, 2, \cdots ,n\}$. For any function $f: \RR \times \RR \mapsto \RR$ and vectors $\vb, \ub \in \RR^{p}$, we define the $i$-th coordinate of $f(\ub, \vb) \in \RR^{p}$ as {$f(\ub, \vb)_{i} := f(\ub_i, \vb_i).$} We denote $\supp(f)$ {as} the support of function $f$, and $f'$ as the derivative of function $f$. 
We say that $\cF: \RR^{p} \mapsto \RR^{p}$ is separable if $\cF (x_1, x_2, \cdots, x_p)_{i} = F_{i}(x_i)$ for some function $F_i: \RR \mapsto \RR$. 
For any $p\in \NN_{>0}$, a function $\phi: \RR^p \mapsto \RR$ is called pseudo-Lipschitz if there exists a constant $L$ such that for any $x, y \in \RR^p$, {\bxz we have}
\begin{equation}
    |\phi(x) - \phi(y)| \leq L \left(1 + {\bnorm{x}_2} + {\bnorm{y}_2} \right) {\bnorm{x - y}_2},
\end{equation}
$L$ is then called the pseudo-Lipschitz constant of $\phi$. Note this definition is the same as the one introduced by \cite{berthier2020state}. Moreover, a function $\Phi: \RR^{p \times q} \mapsto \RR$ is said to be pseudo-Lipschitz if $\Phi(\vecc(\cdot)): \RR^{pq} \mapsto \RR$ is pseudo-Lipschitz function, where $\vecc(\cdot)$ the vectorization of the $p \times q$ matrix.
A sequence (in $p$) of pseudo-Lipschitz functions $\{\phi^p\}_{p \in \NN_{>0}}$ is called uniformly pseudo-Lipschitz  if we have $L_p < \infty$ for each $p$ and $\limsup_{p \to \infty} L_p < \infty$, where $L_p$ is the pseudo-Lipschitz constant of $\phi^p$. 
Note that the input and output dimensions of each $\phi^p$ may
depend on $p$. We call any $L \geq \limsup_{p \to \infty} L_p$ a pseudo-Lipschitz constant of the sequence.

For two sequences of random variables $\{X^n\}_{n \in \NN}$ and $\{Y^n\}_{n \in \NN}$, we write $X^n \overset{P}{\approx} Y^n$ when their difference converges in probability to $0$, that is, $X^n - Y^n \overset{P}{\to} 0$. 
Unless explicitly stated otherwise, all convergences are considered as convergence in probability.
For any sequence of vectors $\{\ub^{t}\}_{t \in \NN}$ or vector-valued function $\{\cF^{t}\}_{t \in \NN}$,  we denote the $t$-th element of the sequence as $\ub^{t}$ or $\cF^{t}$.  Additionally, we use the notation $\ub^{t}_{i}$ and $\cF^{t}_{i}$ to represent the $i$-th coordinate of $\ub^{n}$ and $\cF^{t}$.
For any random variable $X$, we denote $\PP_{X}$ to be the push-forward measure of $X$. That is, for any measurable set $B$, {we have $\PP_{X}(B) = \PP (X^{-1}(B)).$} We say $X_{n}$ converges weakly to random variable $X$ if the push-forward measure $\PP_{X_n}$ converges to $\PP_{X}$ in the sense of weak convergence of measures. 
For any vectors $\bv, \bu \in \RR^p$, we denote the standard inner product by 
$\left\langle \bv, \bu \right\rangle := \sum_{i=1}^{p} \bv_{i} \bu_{i}.$ Moreover, $\|\bv\|_{r}$ denotes the $\ell_{r}$ norm of $\bv$.
For any matrix $\Ab$, $\|\Ab\|_{\textnormal{op}}$ represents the operator norm of $\Ab$, $\|\Ab\|_{\textnormal{F}}$ denotes the Frobenius norm, and $\|\Ab\|_{\tr}$ denotes the trace norm. 
In addition, we define the kernel of $\Ab \in \RR^{n \times p}$ to be 
\begin{align*}
    \ker(\Ab) = \{\xb \in \RR^{p}: \Ab \xb = 0\}. 
\end{align*}

Let $\Kb = (K_{s,r})_{1 \leq s,r \leq t}$ be a $t \times t$ covariance matrix, we  write $(\Zb^1, . . . ,\Zb^t) \sim N(0, \Kb 
\otimes {\bI_p})$ to indicate that $(\Zb^1, \cdots ,\Zb^t)$ is a collection of centered, jointly Gaussian random vectors in $\RR^p$, with
covariances $E[\Zb^s(\Zb^r)^{\T}] = K_{s,r} \otimes {\bI_p}$, for $1 \leq s, r \leq t$. 
{Here $\otimes$ denotes the Kronecker product.}
For any sequence of matrices $\Kb_{i,j} \in \RR^{q \times q}$, 
we define $\Kb^{(t)} = [\Kb_{i,j}]_{1 \leq i,j \leq t}$ to be a $\RR^{qt \times qt}$ matrix as follows: 
\begin{align*}
    \Kb^{(t)} = \begin{bmatrix}
    \Kb_{1,1} & \cdots & \Kb_{1, t}\\
     & \cdots & \\
     \Kb_{t,1} & \cdots & \Kb_{t, t}\\
\end{bmatrix}
\end{align*}
{For a} matrix {$\Xb \in \R^{p \times p}$} sampled from the Gaussian orthogonal ensemble $\GOE(p)$,  {we have} $\Xb = \Gb + \Gb^{\T}$ for $\Gb \in \RR^{p \times p}$ with i.i.d. entries $\Gb_{ij} \sim N(0, 1/(2p))$.


\section{Reference panel-based estimators}\label{sec2}

This section introduces the model, the underlying assumptions, our estimators, and two associated risk measures.


\subsection{The model}\label{sec2.1}

In this section, we specify  the data generation model for the training data $(\X, \y_x)$ and test data $(\Sbb, \y_s)$, where $\X \in \R^{n_x \times p}$, $\y_x \in \R^{n_x}$, $\Sbb \in \R^{n_s \times p}$, and $\y_s \in \R^{n_s}$.
The linear model relating the phenotype and genetic data can be expressed as follows:
\begin{align}\label{linear model}
\begin{split}
\y_x = \X \bbeta_0 + \bmeps_x \quad \text{and} \quad \y_s = \Sbb \bbeta_0 + \bmeps_s.
\end{split}
\end{align}
Here, the coefficient vector $\bbeta_0 \in \R^p$, as well as the noise vectors $\bmeps_x \in \R^{n_x}$ and $\bmeps_s \in \R^{n_s}$, are random variables.
They are mutually independent and also independent of $\X$ and $\Sbb$.
In the context of genetic data predictions, individual-level genotypes in the training data are often not publicly accessible due to privacy and transmission concerns. Consequently, neither $\X$ nor the sample covariance $\X^{\T} \X$ are available.
In such scenarios, an external reference panel $\W \in \R^{n_w \times p}$ is commonly used as a substitute for $\X$, with $\W^{\T} \W$ serving the role of $\X^{\T} \X$.  We assume that the training data $(\X, \y_x)$, test data $(\Sbb, \y_s)$, and reference panel $\W$ are all independent of one another.


To establish the AMP framework, we adopt the following conditions, which are commonly assumed in the  literature \citep{huang2022lasso, bayati2011lasso}. Condition \ref{cond1-np-ratio} indicates that our results are derived under a high-dimensional regime where the sample size and feature dimensionality are proportional.

\begin{condition} \label{cond1-np-ratio} 
The sample sizes $n_x, n_s, n_w \to \infty$ while the dimensionality $p \to \infty$ as well, such that the aspect ratios $p/n_x \to \gamma_x >0$, $p/n_s \to \gamma_s >0$, and $p/n_w \to\gamma_w > 0$. 
\end{condition}


The following condition is for the coefficient vector $\bbeta_0$. 

\begin{condition} [Distribution assumption for $\bbeta_0$] \label{cond2-mp}
We assume that $\bbeta_0$ follows a distribution with mean $\mathbf{0}$ and covariance $\bmSigma_{\bbeta_0}/p$, where $\bmSigma_{\bbeta_0} \in \R^{p \times p}$ is diagonal and has uniformly bounded eigenvalues. 
For $i=m+1, \cdots, p$, $(\bmSigma_{\bbeta_0})_{ii} = 0$, and for $i=1, \cdots, m$, $(\bmSigma_{\bbeta_0})_{ii} \neq 0$. 
Thus, there are $m$ nonzero signals in $\bbeta_0$, and the sparsity $m/p \to \kappa \in (0,1]$ as $m, p \to \infty$. 
There exists $\sigma_{\bbeta}^2>0$ such that
\begin{align*}
    \kappa \cdot \sigma_{\bbeta}^2 = p^{-1} \cdot \sum_{i=1}^p (\bmSigma_{\bbeta_0})_{ii}.
\end{align*}
Furthermore, as $n_x, p \to \infty$ with $p/n_x \to \gamma_x >0$, the distribution of the sum of entries of $\bbeta_0$, denoted as $\sum_{j=1}^{p} \bbeta_{0,j}$, converges weakly to a probability measure $\probP_{\barbeta}$ on $\R$ with a bounded second moment. 
We also assume that $\sum_{j=1}^p (\bbeta_{0,j})^2 \to \E_{\probP_{\barbeta}}(\bar{\beta}^2)$.
\end{condition}

The following condition imposes similar constraints on the noise vectors ${\bepsilon_x}$ and ${\bmeps_s}$.
\begin{condition}[Noise assumption] \label{cond3-dmm-def1} 
As $n_x, n_s, p \to \infty$ with $p/n_x \to \gamma_x >0$ and $p/n_s \to \gamma_s >0$, the empirical distribution of the entries of $\bmeps_x$ and $\bmeps_s$ respectively converges weakly to probability measures $\probP_{\bar{\bepsilon}_x}$ and $\probP_{\bar{\bepsilon}_s}$ on $\R$ with bounded second moments. 
Additionally, we assume that $n_x^{-1} \sum_{i=1}^{n_x} (\bmeps_{x,i})^2 \to \sigma_{{\bepsilon}_{x}}^2 < \infty$ and $n_s^{-1} \sum_{i=1}^{n_s} (\bmeps_{s,i})^2 \to \sigma_{{\bepsilon}_{s}}^2 < \infty$.
\end{condition}

The next condition assumes that each row of the datasets $\Xb$, $\Sbb$, and $\Wb$ follows a Gaussian distribution.  The assumption of normality is commonly used in the derivation of AMP theoretical results.  Simulation and real data analyses in Section~\ref{sec: numerical experiments} demonstrate that the Gaussian distribution assumption could be relaxed for GWAS applications.
\begin{condition} [Distribution of genetic data] \label{cond-distn-data} 
Let $\X = \X_0 \bmSigma^{1/2} \in \R^{n_x \times p}$, $\Sbb = \Sbb_0 \bmSigma^{1/2} \in \R^{n_s \times p}$, and $\W = \W_0 \bmSigma^{1/2} \in \R^{n_w \times p}$, where the entries of $\X_0 \in \R^{n_x \times p}$, $\Sbb_0 \in \R^{n_s \times p}$, and $\W_0 \in \R^{n_w \times p}$ are i.i.d. $N(0,1)$.
Furthermore, we assume $\bmSigma \in \R^{p \times p}$ to be positive definite, with uniformly bounded eigenvalues. That is, there exist absolute constants $c$ and $C$ such that $0 < c \leq \lambda_{\min}(\bmSigma) \leq \lambda_{\max}(\bmSigma) \leq C < \infty$.
\end{condition}

{We define heritability as follows, representing the proportion of phenotypic variance attributable to genetic data \citep{yang2017concepts}.
Intuitively, a larger heritability implies a higher signal-to-noise ratio.}

\begin{definition}[Heritability]
Conditional on $\bbeta_0$, the heritability $h_x^2$ of the training data $(\X, \y_x)$ is defined as 
\begin{align}
    \begin{split}
    h_x^2 = \lim_{p \to \infty} \frac{\var(\X \bbeta_0)}{\var(\y_x)} = \lim_{p \to \infty} \frac{\bbeta_0^{\T} \X^{\T} \X \bbeta_0}{\bbeta_0^{\T} \X^{\T} \X \bbeta_0 + \bmeps_x^{\T} \bmeps_x}.
    \end{split}
\end{align}
Similarly, the heritability $h_s^2$ of the test data $(\Sbb, \y_s)$ is defined as 
\begin{align}
\begin{split}
h_s^2 
= \lim_{p \to \infty} \frac{\var(\Sbb \bbeta_0)}{\var(\y_s)} = \lim_{p \to \infty} \frac{\bbeta_0^{\T} \Sbb^{\T} \Sbb \bbeta_0}{\bbeta_0^{\T} \Sbb^{\T} \Sbb \bbeta_0 + \bmeps_s^{\T} \bmeps_s}.
\end{split}
\end{align}
{Thus, we have $h_x^2$ and $h_s^2 \in [0, 1]$.
}
\end{definition}

\begin{remark}
    As $p \to \infty$ with $p/n_x \to \gamma_x >0$, $h_x^2$ can be asymptotically represented as 
\begin{align}
\begin{split}
h_x^2 
= \lim_{p \to \infty} \frac{\bbeta_0^{\T} \bmSigma \bbeta_0}{\bbeta_0^{\T} \bmSigma \bbeta_0 + \sigma_{{\bepsilon}_x}^2}
= \lim_{p \to \infty} \frac{ \| \bbeta_0 \|_{\bmSigma}^2 }{\| \bbeta_0 \|_{\bmSigma}^2 + \sigma_{{\bepsilon}_x}^2},
\end{split}
\end{align}
where $\| \x \|_{\bmSigma}^2 := \xb^{\T} \bmSigma \x$. Note that this implies $\sigma_{{\bepsilon}_x}^2 = \| \bbeta_0 \|_{\bmSigma}^2  \cdot (1-h_x^2)/h^2_x$. 
{\bxz Similarly, we have $\sigma_{{\bepsilon}_s}^2 = \| \bbeta_0 \|_{\bmSigma}^2  \cdot (1-h_s^2)/h_s^2$. More details can be found in the proof of Theorem \ref{thm: general lasso mse + R2}.} 
\end{remark}

The following three conditions introduce assumptions on the general covariance matrix $\bmSigma$, which are necessary for our proof. 
Condition \ref{cond: exchange limit and derivative} allows for the exchange of the limit and derivative operations with respect to the parameters $a$ and $b$. 

\begin{condition} \label{cond: exchange limit and derivative}
{For any $a, b\in \RR_{+}$ and $v\in \RR^{p}$, the sequence of functions}
\begin{equation}
    \cE^{(p)}(a,b) := p^{-1} \cdot \EE \min_{v \in \RR^{p}} \left\{\frac{1}{2} \bnorm{v - \bbeta_0 - \sqrt{a} {\bSigma}^{-1/2} \Zb}_{{\bSigma}}^2 + b \bnorm{v}_1 \right\}
\end{equation}
admits a differentiable limit $\cE(a,b)$ on $\RR_+ \times \RR_+$ with 
$$
\lim_{p \to \infty} \frac{\partial \cE^{(p)}(a ,b)}{\partial a} \to \frac{\partial \cE(a ,b)}{\partial a}\quad \mbox{and} \quad \lim_{p \to \infty} \frac{\partial \cE^{(p)}(a ,b)}{\partial b} \to \frac{\partial \cE(a ,b)}{\partial b},
$$
 where $\cE(a ,b) = \lim_{p \to \infty} \cE^{(p)}(a ,b)$, and $\Zb \sim N(0, \bI_{p})$  is independent of $\bbeta_0$.
\end{condition}

The following condition ensures the existence of the state evolution in the AMP algorithm.
\begin{condition} \label{cond: limit exist}
    For any $a_1, b_1, a_2, b_2 \in \RR_{+}$ and any $2 \times 2$ positive definite matrix $\bm{M}$, the limit 
    \begin{equation*}
        \lim_{p \to \infty} p^{-1} \cdot \left\langle \hat{\bbeta}_1^{(p)},\hat{\bbeta}_2^{(p)} \right\rangle
    \end{equation*}
    exists and is finite, 
    where $\langle \cdot, \cdot \rangle$ is the standard dot product. Here 
    \begin{equation*}
    \begin{split}
        \hat{\bbeta}_1^{(p)} =& \arg\min_{\bbeta \in \RR^{p}} \left\{\frac{1}{2} \|\bbeta - \bbeta_0 - \sqrt{a_1} {\bSigma}^{-1/2} \Zb_1\|_{\bSigma}^2 + b_1 \|\bbeta\|_1 \right\}\quad \mbox{and} \\
        \hat{\bbeta}_2^{(p)} =& \arg\min_{\bbeta \in \RR^{p}} \left\{\frac{1}{2} \|\bbeta - \bbeta_0 - \sqrt{a_2} {\bSigma}^{-1/2} \Zb_2\|_{\bSigma}^2 + b_2 \|\bbeta\|_1 \right\},
    \end{split}
    \end{equation*}
    where $(\Zb_1, \Zb_2) \sim N(0, \bm{M} \otimes \bI_{p})$ is independent of $\bbeta_0$. 
\end{condition}

Similar to Condition \ref{cond: exchange limit and derivative}, the following condition allows for the exchange of the limit between the dimension $p$ and the parameter $b$.
\begin{condition} \label{cond: exchange limit}
    The following two limits both exist and enable the exchange of limits between the dimension $p$ and the parameter $b$: 
    \begin{align*}
        &\lim_{b \to 0^+} \lim_{p \to \infty} p^{-1} \cdot \bnorm{ \hat{\bbeta}^{(p)}}_{2}^{2} = \lim_{p \to \infty} \lim_{b \to 0^+} p^{-1} \cdot \bnorm{ \hat{\bbeta}^{(p)}}_{2}^{2} \quad \mbox{and}\\
        &\lim_{b \to \infty} \lim_{p \to \infty} p^{-1} \cdot \bnorm{ \hat{\bbeta}^{(p)}}_{2}^{2} = \lim_{p \to \infty} \lim_{b \to \infty} p^{-1} \cdot \bnorm{ \hat{\bbeta}^{(p)}}_{2}^{2},
    \end{align*}
    where 
    \begin{equation*}
    \begin{split}
        \hat{\bbeta}^{(p)} =& \arg\min_{\bbeta \in \RR^{p}} \left\{\frac{1}{2} \|\bbeta - \bbeta_0 - \sqrt{a} {\bSigma}^{-1/2} \Zb\|_{\bSigma}^2 + b \|\bbeta\|_1 \right\},
    \end{split}
    \end{equation*}
    and $\Zb \sim N(0, \bI_{p})$ is independent of $\bbeta_0$. 
\end{condition}

Conditions \ref{cond1-np-ratio} - \ref{cond-distn-data} are assumptions commonly used in previous AMP literature on regularized estimators. Conditions \ref{cond: exchange limit and derivative} - \ref{cond: exchange limit} impose restrictions on $\bmSigma$ that allow for the exchange of limits or differentiation, which are useful in proving the existence of a fixed-point solution for the state evolution. In the special case where $\bmSigma = \bI_{p}$, Conditions \ref{cond: exchange limit and derivative} - \ref{cond: exchange limit} are trivially satisfied.

A sequence of instances
\$
\{\bbeta_0(p), {\bepsilon}_x(n_x), {\bepsilon}_s(n_s), {\bSigma}(p), \X(n_x, p), \Sbb(n_s, p),\W(n_w, p)\}_{(p, n_x, n_s, n_w) \in \NN^4},
\$
indexed by $p, n_x, n_s$ and $n_w$, is referred to as a {\it converging sequence} if $\bbeta_0(p) \in \RR^{p}$, ${\bepsilon}_x(n_x) \in \RR^{n_x}$, ${\bepsilon}_s(n_s) \in \RR^{n_s}$, ${\bSigma}(p) \in \RR^{p \times p}$, $\X(n_x, p) \in \RR^{n_x \times p}$, $\Sbb(n_s, p) \in \RR^{n_s \times p}$, and $\W(n_w, p) \in \RR^{n_w \times p}$ with $n_x = n_x(p)$, $n_s = n_s(p)$ and $n_w = n_w(p)$,  and such that $p/n_x \to \gamma_x > 0$,  $p/n_s \to \gamma_s > 0$, and $p/n_w \to \gamma_w> 0$. This sequence is also assumed to satisfy all the aforementioned conditions. Moreover, a sequence of instances 
\$
\{\bbeta_0(p), {\bepsilon}_x(n_x), {\bepsilon}_s(n_s), {\bSigma}(p), \Xb (n_x, p), \Sbb(n_s, p)\}_{(p, n_x, n_s) \in \NN^3}
\$ 
is called a {\it non-reference panel converging sequence} if $\bbeta_0(p), {\bepsilon}_x(n_x), {\bepsilon}_s(n_s), {\bSigma}(p), \Xb(n_x, p)$ and $\Sbb(n_s, p)$ satisfy all the conditions mentioned above without any requirements on $\Wb(n_w, p)$ or $n_w$.

To facilitate easier reading, we will typically omit the indexes $p, n_x, n_s$, and $n_w$ in later sections.


\subsection{Estimators and risk measures}\label{sec2.2}
The traditional $L_1$ regularized estimator can be formulated as
\begin{align} \label{eqn:-lasso-est with normalization}
\begin{split}
\hat{\bbeta}_{\textnormal{L}}(\lambda) = \arg\min_{\bbeta \in \R^p} \frac{1}{2 n_x} \bbeta^{\T} \X^{\T} \X \bbeta -  \frac{1}{n_x} \bbeta^{\T} \X^{\T} \y_x + \frac{\lambda}{\sqrt{p}} \| \bbeta \|_1.
\end{split}
\end{align}
{\bxz Similar to  \cite{li2022estimation}, when $\X^{\T} \X$ is unavailable and is substituted with $\W^{\T} \W$, we can define the reference panel-based $L_1$ regularized estimator as}
\begin{align} \label{eqn:ref-panel-lasso-est without normalization}
\begin{split}
\hat{\bbeta}_{\LW}(\lambda) = \arg\min_{\bbeta \in \R^p} \frac{1}{2 n_w} \bbeta^{\T} \W^{\T} \W \bbeta - \frac{1}{n_x} \bbeta^{\T} \X^{\T} \y_x + \frac{\lambda}{\sqrt{p}} \| \bbeta \|_1.
\end{split}
\end{align}
Furthermore, let $\hat{\bbeta}_{\textnormal{R}}(\lambda)$ and $\hat{\bbeta}_{\textnormal{RW}}(\lambda)$ be the traditional and reference panel-based $L_2$ regularized estimators, respectively. We have 
\begin{equation} \label{eqn:ref-panel-ridge-est without normalization}
\hat{\bbeta}_{\RA}(\lambda) = \frac{1}{n_x} (\X^{\T} \X/n_x + \lambda \bI_{p})^{-1}\X^{\T}\y_x
\quad \mbox{and} \quad
\hat{\bbeta}_{\RW}(\lambda) = \frac{1}{n_x} (\W^{\T} \W/n_w + \lambda \bI_{p})^{-1}\X^{\T}\y_x.
\end{equation}

{In this paper, we analyze two risk measures: the mean {squared} error ($\MSE$) and the out-of-sample $R$-square ($R^2$).} 
Let $\hat{\bbeta} \in \R^p$ be a {\bxz generic estimator of $\bbeta_0$; 
for example, $\hat{\bbeta}_{\LW}(\lambda)$ or $\hat{\bbeta}_{\textnormal{L}}(\lambda)$). 
Consider a test point $\xb_{\textnormal{new}} \sim N(0, \bmSigma)$, independent of training data.  We define the following MSE, which agrees with the out-of-sample prediction risk by \cite{hastie2022surprises} and \cite{chen2023sketched}: 
\begin{align}
\begin{split}
    R := \EE \left[ \left(x_{\textnormal{new}}^{\T} \hat{\bbeta} - x_{\textnormal{new}}^{\T} {\bbeta_0} \right)^2 \bigg| \Xb, \bbeta_0 \right] = \EE \left[ \left\|\hat{\bbeta} - \bbeta_0 \right\|_{\bmSigma}^2 \bigg| \Xb, \bbeta_0 \right]. 
\end{split}
\end{align}
We emphasize that the expectation is conditional on $\bbeta_0$ and $\Xb$ and taken with respect to $\bmeps_{x}$ and $\xb_{\textnormal{new}}$. 
In the special case of $\bmSigma = \bI_p$, {we have} 
$\| \hat{\bbeta} - \bbeta_0 \|_{\bmSigma}^2 = \| \hat{\bbeta} - \bbeta_0 \|_{2}^2$  \citep{bayati2011lasso}. 
{In addition, the predictor of $\y_s$ in the test data $(\Sbb, \y_s)$ is given by $\hat{\bmS}_{\Sbb}(\hat{\bbeta}) = \Sbb \hat{\bbeta}$. 
Conditioning on $\bbeta_0$, the out-of-sample $R^2$ is defined as $A^2(\hat{\bbeta})$, where}  
\begin{align}\label{R2 def}
    \begin{split}
        A^2(\hat{\bbeta}) = \frac{\big(\y_s^{\T} \hat{\bmS}_{\Sbb}(\hat{\bbeta})\big)^2}{ \| \y_s \|_2^2 \cdot \| \hat{\bmS}_{\Sbb}(\hat{\bbeta}) \|_2^2 }.
    \end{split}
\end{align}
{The out-of-sample $R^2$ is commonly used to  measure the  prediction accuracy in genetic data prediction \citep{ma2021genetic}. 
For instance, a recent study finds that the prediction accuracy of genetic data for educational attainment can reach up to $16\%$ \citep{okbay2022polygenic}
}

\section{\texorpdfstring{$L_1$}{TEXT} regularized estimators with isotropic features }\label{sec: i.i.d. REF-LASSO}

\subsection{Asymptotic results}\label{sec3.1}
{In this section, we present the asymptotic $\MSE$ and out-of-sample $R^2$ of $\hat{\bbeta}_{\LW}(\lambda)$ under the assumption that $\bmSigma = \bI_{p}$. We compare these results with those for $\hat{\bbeta}_{\textnormal{L}}(\lambda)$ and provide extensive numerical illustrations to gain further insights into our theoretical findings.}

We begin by introducing the soft-threshold operator,  $\eta_{\soft}(v, \theta): \RR \times \RR \mapsto \RR$, as 
\begin{align} \label{eta_soft}
\begin{split}
    \eta_{\soft}(v,\theta) := \sign(v) (|v| - \theta)_{+} = \arg\min_{w \in \RR} \left(\frac{1}{2} \bnorm{w - v}_{2}^2 + \theta \bnorm{w}_1
    \right).
\end{split}
\end{align}
The following theorem characterizes {\bxz the $\MSE$ and out-of-sample $R^2$ of $\hat{\bbeta}_{\LW}(\lambda)$} when $\bmSigma = \bI_{p}$. 
\begin{theorem} \label{thm: i.i.d. lasso mse + R2}
Let $\{\bbeta_0, \bepsilon_x,\bmeps_s, \bSigma, \bX,  \Sbb, \bW\}$ be a converging sequence of instances with $\PP\left({\bbeta_0}(p) \neq 0 \right) > 0$. Each row of $\Xb$, $\Sbb$, and 
$\Wb$ 
is i.i.d. Gaussian with mean ${\bm 0}$ and covariance $\bI_p$.
In probability, the MSE {\bxz of $\hat{\bbeta}_{\LW}(\lambda)$} is
\begin{equation} \label{eqn: i.i.d. l_1 ref mse}
    {\bxz R_{\LW}(\lambda)} = \lim_{p \to \infty}  \bnorm{\hat{\bbeta}_{\LW}(\lambda) - \bbeta_0}_2^2 = \EE \bigg[\eta_{\soft}\big\{\tilde{\tau}_{*} z+(1 + \tilde{b}_*)\bar{\beta}, \lambda (1 + \tilde{b}_*)\big\} - \bar{\beta} \bigg]^2
\end{equation}
and the out-of-sample $R^2$ {\bxz of $\hat{\bbeta}_{\LW}(\lambda)$} is 
\begin{equation} \label{eqn: i.i.d. l_1 ref R2}
    A^2_{\LW}(\lambda) = \frac{h_{s}^2}{\kappa \sigma_{\beta}^2} \cdot \frac{ \EE \left[ \left \langle  \bar{\beta}, \eta_{\soft}\{\tilde{\tau}_* z +  (1 + \tilde{b}_*)\bar{\beta}, \lambda (1 + \tilde{b}_*)\} \right \rangle \right]^2}{\EE \left[ \eta_{\soft}\{\tilde{\tau}_* z +  (1 + \tilde{b}_*)\bar{\beta}, \lambda (1 + \tilde{b}_*)\} \right]^2}.
\end{equation}
Here $z \sim N(0,1)$ and is independent of $\barbeta \sim \PP_{\barbeta}$, and $(\tilde{\tau}_*^2, \tilde{b}_*)$ satisfy the fixed point equations
\begin{align} \label{eqn: i.i.d. l1 state evo}
\begin{split}
    &\tilde{\tau}^2_* = 
    \gamma_x (1 + \tilde{b}_*)^2 \cdot \frac{\kappa \sigma_{\beta}^2}{h_{x}^2} + \gamma_w \EE \eta_{\soft}^2\{\tilde{\tau}_* z +  (1+\tilde{b}_{*})\barbeta, \lambda (1 + \tilde{b}_*)\} \quad \mbox{and} \\
    &(1 + \tilde{b}_{*})^{-1} = 1 - \gamma_w \EE \eta'_{\soft} \{\tilde{\tau}_* z +  (1 + \tilde{b}_*)\barbeta, \lambda (1 + \tilde{b}_*)\},
\end{split}
\end{align}
where $\eta_{\soft}'(v, \theta)$ {denotes} the derivative of $\eta_{\soft}(v, \theta)$ with respect to $v$. 
\end{theorem}

Proposition~\ref{prop: i.i.d. non-ref lasso mse + R2} is taken from \cite{bayati2011lasso} and  provides a summary of the $\MSE$ and out-of-sample $R^2$ for $\hat{\bbeta}_{\textnormal{L}}(\lambda)$ under the  model  described in Section~\ref{sec2.1}. 

\begin{proposition}[Theorem 1.5 in \cite{bayati2011lasso}]
\label{prop: i.i.d. non-ref lasso mse + R2}
Let $\{\bbeta_0, \bepsilon_x,\bmeps_s, \bSigma, \bX,  \Sbb\}$ be a non-reference panel converging sequence of instances with $\PP\left({\bbeta_0}(p) \neq 0 \right) > 0$. 
Each row of $\Xb$ and $\Sbb$
is i.i.d. Gaussian with mean ${\bm 0}$ and covariance $\bI_p$. Almost surely, 
the $\MSE$ of $\hat{\bbeta}_{\textnormal{L}}(\lambda)$ is
\begin{equation}\label{eqn: i.i.d. l_1 mse}
\begin{split}
    R_{\textnormal {L}}(\lambda) = \lim_{p \to \infty}  \bnorm{\hat{\bbeta}_{\textnormal {L}}(\lambda) - \bbeta_0}_2^2 &= \EE \left[ \eta_{\soft}\{\bar{\tau}_* z+ \bar{\beta}, \lambda (1 + \bar{b}_*)\} -  \barbeta \right]^2
\end{split}
\end{equation}
{\bxz and the {out-of-sample} $R^2$ of $\hat{\bbeta}_{\textnormal{L}}(\lambda)$ is }
\begin{equation*}
\begin{split}
    A^2_{\textnormal {L}}(\lambda) = \frac{h_{s}^2}{\kappa \sigma_{\beta}^2} \cdot \frac{ \EE \left[\left \langle \barbeta, \eta_{\soft}\{\bar{\tau}_* z +  \bar{\beta}, \lambda (1 + \bar{b}_*) \} \right \rangle \right]^2}{\EE \left[ \eta_{\soft}\{\bar{\tau}_* z +  \bar{\beta}, \lambda (1 + \bar{b}_*)\} \right]^2 }.
\end{split}
\end{equation*}
{\bxz Here $z \sim N(0,1)$ and is independent of $\barbeta \sim \PP_{\barbeta}$, and
 $(\bar{\tau}_*^2, \bar{b}_*)$ satisfy the fixed point equations} 
\begin{align} \label{eqn: lasso state evolution}
\begin{split}
    &  \bar{\tau}_*^2 = \gamma_x \cdot \frac{\kappa \sigma_{\beta}^2}{h_{x}^2} + \gamma_x \EE  \eta^2_{\soft} \big\{ \bar{\tau}_* z+ \bar{\beta}, \lambda (1 + \bar{b}_*) \big\} - 2 \gamma_x \EE \left[ \left \langle \bar\bbeta, \eta_{\soft}\{\bar{\tau}_* z +  \barbeta, \lambda (1 + \bar{b}_*)\} \right \rangle \right]
    \\
    &\mbox{and} \quad(1 + \bar{b}_*)^{-1} = 1 - \gamma_x \EE \eta_{\soft}' \big\{\bar{\tau}_* z+ \bar{\beta}, \lambda (1 + \bar{b}_*)\big\}. 
\end{split}
\end{align}
\end{proposition}

{\bxz As $p \to \infty$, the asymptotic MSE of the $p$-dimensional vector $\hat{\bbeta}_{\LW}(\lambda)$ with respect to 
$\bbeta_0$ will converge to the MSE of the one-dimensional random variable $\eta_{\soft} \big\{\tilde{\tau}_{*} z+ (1 + \tilde{b}_*)\bar{\beta}, \lambda (1 + \tilde{b}_*) \big\}$ with respect to 
$\barbeta$, assuming that the empirical distribution of $\sum_{j=1}^p (\bbeta_{0,j})^2$ converges weakly to that of $\barbeta$. 
The term $\eta_{\soft} \big\{\tilde{\tau}_{*} z+ (1 + \tilde{b}_*)\bar{\beta}, \lambda (1 + \tilde{b}_*) \big\}$ represents the output of a soft-threshold operator of a random variable centered around $(1 + b_*) \bar\beta$ with Gaussian noise having variance $\tilde{\tau}_{*}^2$. 
The function $\eta_{\soft}$ uses the parameter $\lambda (1 + \tilde{b}_*)$  to  control the sparsity of the estimator, penalizing small coefficients to be exactly zero.
When comparing with the results of $\hat{\bbeta}_{\textnormal{L}}(\lambda)$ from Equation~(\ref{eqn: i.i.d. l_1 mse}), an immediate distinction is that the random variable within the soft-threshold operator is centered around $\bar\beta$, rather than $(1 + b_*) \bar\beta$ as observed in $\hat{\bbeta}_{\LW}(\lambda)$. 
Such differences clearly demonstrate the contrasting prediction performance of  $\hat{\bbeta}_{\LW}(\lambda)$ and $\hat{\bbeta}_{\textnormal{L}}(\lambda)$.

Analyzing the special case $\bmSigma = \bI_{p}$ allows us to numerically evaluate the theoretical results of the $\MSE$ and out-of-sample $R^2$, based on a specific distribution function of $\bbeta_0$. Similar to previous literature (e.g., \cite{gerbelot2020asymptotic}), we assume a Bernoulli-Gaussian density distribution for $\bbeta_0$ and conduct numerical comparisons of the asymptotic results between $\hat{\bbeta}_{\textnormal{L}}(\lambda)$ and $\hat{\bbeta}_{\LW}(\lambda)$. Specifically, entries of $\bbeta_0$ are i.i.d. random variables following the density function below:
\begin{align} \label{true_beta0_bernoulli_Gaussian}
    f(\bbeta_{0_j})=(1-\kappa)\delta(\bbeta_{0_j})+\kappa\exp(-\bbeta_{0_j}^2/2)/\sqrt{2\pi},\ \forall j \in [p],
\end{align}
where $\delta$ denotes the Dirac measure function. 
}

\begin{figure}[!t] 
\includegraphics[page=1,width=0.85\linewidth]{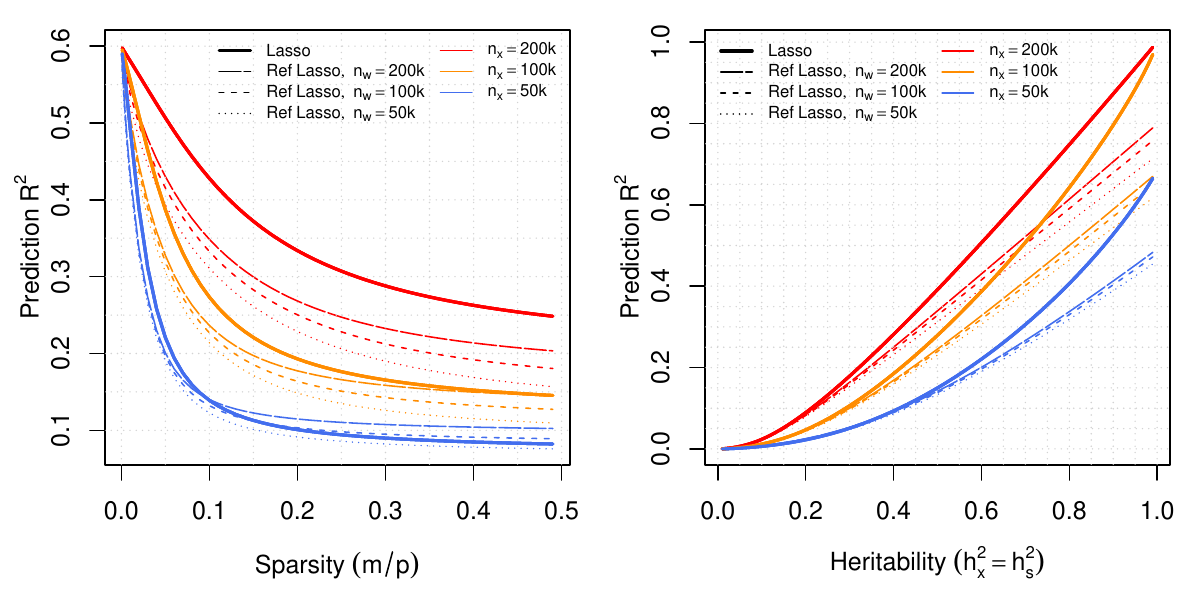}
\centering
\caption{
\textbf{Comparing the theoretical out-of-sample $R^2$ of $\hat{\bbeta}_{\textnormal{L}}(\lambda)$ and $\hat{\bbeta}_{\LW}(\lambda)$ as sparsity and heritability vary.
} 
The out-of-sample $R^2$ of $\hat{\bbeta}_{\textnormal{L}}(\lambda)$ (`Lasso') and $\hat{\bbeta}_{\LW}(\lambda)$  (`Ref Lasso') is calculated according to Proposition~\ref{prop: i.i.d. non-ref lasso mse + R2} and Theorem~\ref{thm: i.i.d. lasso mse + R2}, respectively. 
Here we set $\bmSigma = \bI_{p}$, 
 $p~=~$461,488, $n_x~=~$50,000, 100,000, 200,000, and $n_w~=~$50,000, 100,000, 200,000. Entries of $\bbeta_0$ are i.i.d. random variables following the Bernoulli-Gaussian distribution. 
 For each level of sparsity or heritability, we compute $A^2_{\LW}(\lambda)$ and $A^2_{\textnormal{L}}(\lambda)$ for various values of $\lambda$ and present the results with the respective best-performing $\lambda$ in the figure.
 {\textbf {Left:}} Heritability $h_x^2=h_s^2=0.6$, and sparsity $m/p$ varies from 0.001 to 0.49.  {\textbf {Right:}} Sparsity $m/p=0.05$, and heritability $h_x^2=h_s^2$ varies from 0.01 to 0.99. 
}
\label{main_fig_1}
\end{figure}

\begin{figure}[!ht] 
\includegraphics[page=1,width=0.85\linewidth]{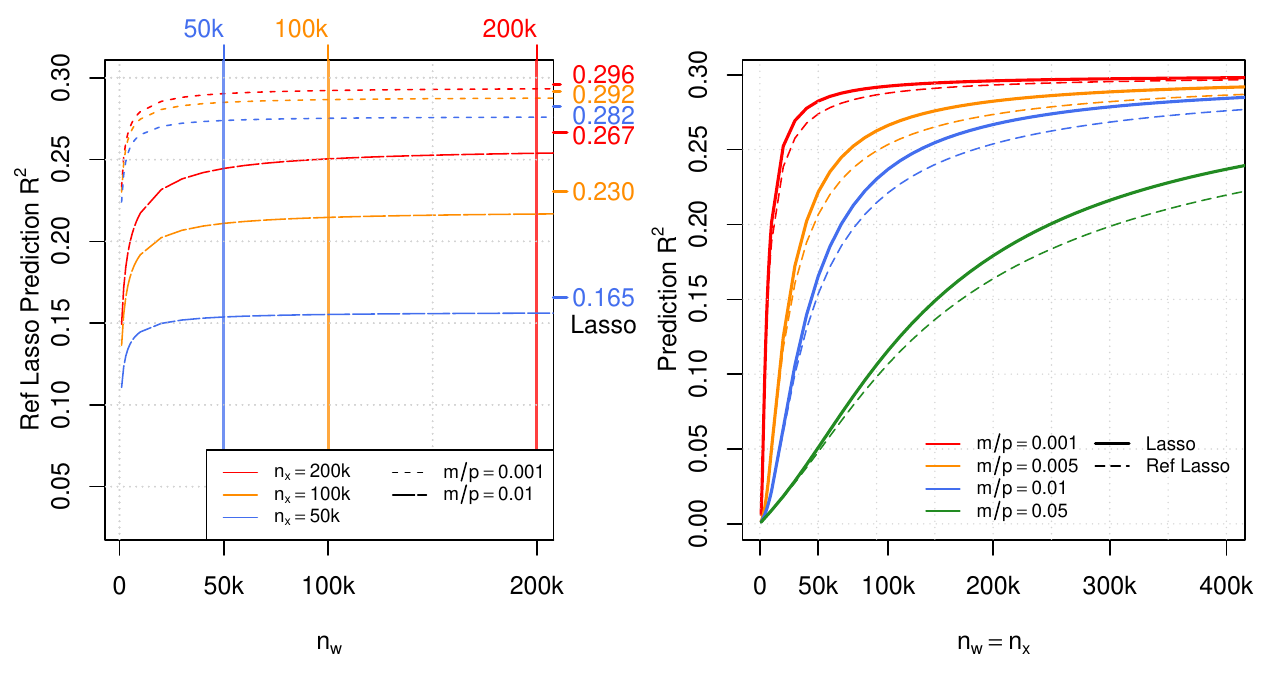}
\centering
\caption{
\textbf{Comparing the theoretical out-of-sample $R^2$ of $\hat{\bbeta}_{\textnormal{L}}(\lambda)$ and $\hat{\bbeta}_{\LW}(\lambda)$ as $n_w$ and $n_x$ vary.} 
The out-of-sample $R^2$ of $\hat{\bbeta}_{\textnormal{L}}(\lambda)$ (`Lasso') and $\hat{\bbeta}_{\LW}(\lambda)$ (`Ref Lasso') is calculated according to Proposition~\ref{prop: i.i.d. non-ref lasso mse + R2} and Theorem~\ref{thm: i.i.d. lasso mse + R2}, respectively. 
Here we set $\bmSigma = \bI_{p}$, 
 $p~=~$461,488, and heritability $h_x^2=h_s^2=0.3$. Entries of  $\bbeta_0$ are i.i.d. random variables following the Bernoulli-Gaussian distribution. 
{\textbf {Left:}} $n_x~=~$50,000, 100,000, 200,000, sparsity $m/p=0.001,0.01$, and reference panel size $n_w$ varies from 1000 to 200,000. Vertical lines colored by blue, orange and red are at $n_w~=~$50,000, 100,000, 200,000, whose intersections with the respective same-colored dashed curves represent the prediction accuracy of $\hat{\bbeta}_{\LW}(\lambda)$  when $n_w=n_x$. Decimal numbers on the right side, in blue, orange, and red, represent the prediction accuracy of $\hat{\bbeta}_{\textnormal{L}}(\lambda)$ under $n_x~=~$50,000, 100,000, 200,000, respectively, with the shorter ticks corresponding to $m/p=0.001$, and longer ticks corresponding to $m/p=0.01$. {\textbf {Right:}} Sparsity $m/p=0.001,0.005,0.01,0.05$, $n_x$ is equal to $n_w$, and they vary together from $1,000$ to $400,000$. 
}
\label{main_fig_2}
\end{figure}

{
{\bxz Figure~\ref{main_fig_1} 
 displays the theoretical out-of-sample $R^2$
of $\hat{\bbeta}_{\textnormal{L}}(\lambda)$ and $\hat{\bbeta}_{\LW}(\lambda)$ across different sample sizes $n_x$ and $n_w$ as the model sparsity and heritability change.
As expected, the prediction accuracy of both $\hat{\bbeta}_{\textnormal{L}}(\lambda)$ and $\hat{\bbeta}_{\LW}(\lambda)$  decrease as sparsity increases.
In general, the performance of $\hat{\bbeta}_{\textnormal{L}}(\lambda)$ is better to that of $\hat{\bbeta}_{\LW}(\lambda)$, except in cases where the sparsity level is high and the training data sample size is small. However, it is worth noting that in such scenarios, both estimators have low prediction accuracy, making them less practically relevant.
The gap between $A_{\LA}^2(\lambda)$ and $A_{\LW}^2(\lambda)$ tends to be larger as the training data sample size increases. Moreover, 
across a wide range of sparsity levels, we consistently observe that the difference among the $A_{\text{LW}}^2(\lambda)$ curves with the same $n_x$ but different $n_w$ (indicated by the same color in the figure) is smaller than the difference among the $A_{\text{LW}}^2(\lambda)$ curves with different $n_x$ but the same $n_w$ (indicated by the same dash type). These results suggest that the training data sample size has a larger impact on $A_{\text{LW}}^2(\lambda)$ compared to the size of the reference panel.  

Furthermore, we find that the difference between $A_{\LA}^2(\lambda)$ and $A_{\LW}^2(\lambda)$ increases as the heritability increases. As the heritability approaches $1$, $A_{\LA}^2(\lambda)$ can be very close to $1$, indicating a strong predictive power, while $A_{\LW}^2(\lambda)$ is always noticeably smaller than $1$. This suggests that the $\hat{\bbeta}_{\textnormal{L}}(\lambda)$ tends to have higher prediction accuracy  compared to $\hat{\bbeta}_{\LW}(\lambda)$ for highly heritable traits.
In addition, we observe that both $A_{\LA}^2(\lambda)$ and $A_{\LW}^2(\lambda)$ may increase non-linearly as the heritability increases, indicating a complex dependence on $h_x^2$ of the fixed points in Proposition~\ref{prop: i.i.d. non-ref lasso mse + R2} and Theorem~\ref{thm: i.i.d. lasso mse + R2}. 

To further demonstrate the impact of sample sizes $n_w$ and $n_x$ on the relative performance of $\hat{\bbeta}_{\textnormal{L}}(\lambda)$ and $\hat{\bbeta}_{\LW}(\lambda)$, we illustrate the theoretical out-of-sample $R^2$ for various sample sizes at different levels of heritability and sparsity in Figure~\ref{main_fig_2}. 
We observe that $A_{\LW}^2(\lambda)$ shows a rapid initial increase with $n_w$ when $n_w$ is small (for example, less than $20,000$), and the improvement becomes more gradual as $n_w$ increases, indicating diminishing returns with larger reference panel sizes. 
Our findings highlight the significance of recent initiatives in the genetic community aimed at expanding the reference panel sample size, such as the TOPMed project \citep{taliun2021sequencing}. While some widely used reference panel datasets, like 1000 Genomes, have only included hundreds of subjects \citep{10002015global}, our results underscore the importance of increasing the sample size of reference panels to enhance the performance of reference panel-based estimators.

Figure~\ref{main_fig_2} also examines the prediction accuracy curve as the sample size $n_x = n_w$ increases from $1,000$ to $400,000$. Our results reveal that the sparsity level plays an important role in shaping the curve and determining the relative performance of $\hat{\bbeta}_{\textnormal{L}}(\lambda)$ and $\hat{\bbeta}_{\LW}(\lambda)$. Specifically, as the sparsity level increases, the gap in prediction accuracy between the two estimators also increases.
For very sparse signals (e.g., $m/p=0.001$), the two estimators exhibit very similar performance across the entire range of sample sizes. However, as the signals become denser, the difference between the two estimators becomes more pronounced. Moreover, the sample size at which the two estimators show the largest difference also increases. For example, the largest difference is observed at $n_x \approx 100,000$ when $m/p=0.005$ and increases to $\approx 400,000$ when $m/p=0.05$. 

In summary, our theoretical results uncover the influences of sparsity, heritability, and sample sizes on the relative performance of $\hat{\bbeta}_{\textnormal{L}}(\lambda)$ and $\hat{\bbeta}_{\LW}(\lambda)$. 
With the continuous collection of more data by the genetic community \citep{zhou2022global}, the sample size of the training data is expected to increase. However, using reference panel-based estimators may lead to a reduction in performance compared to using estimators directly generated from the training data.
Such  prediction performance decay suggests that there is a trade-off for the conveniences offered by using a reference panel instead of the original training data.  
Increasing the current reference panel sample size could help mitigate this reduction, although a substantial decay may still persists, particularly for traits with high heritability and polygenicity \citep{boyle2017expanded}.
}

}

\subsection{A case study to compare \texorpdfstring{$\hat{\bbeta}_{\textnormal{L}}(\lambda)$}{TEXT} and 
\texorpdfstring{$\hat{\bbeta}_{\LW}(\lambda)$}{TEXT}}
\label{sec: case study}
Due to the complexity of fixed point equations, finding the optimal tuning parameters for $\hat{\bbeta}_{\textnormal{L}}(\lambda)$ and $\hat{\bbeta}_{\LW}(\lambda)$  can be challenging. 
In our numerical illustrations of Section~\ref{sec3.1}, we explore different values of $\lambda$ individually. The results presented in the figures correspond to the best-performing $\lambda$ values for each estimator. This approach enables a fair comparison between the two estimators, considering that their best tuning parameters may not necessarily be the same. 
Figure~\ref{supp_fig_theory_5} illustrates the existence of best-performing $\lambda$ values for both  $R_{\LW}(\lambda)$ and $A_{\LW}^2(\lambda)$. Furthermore, we find that the best-performing $\lambda$ and associated $R_{\LW}(\lambda)$ decreases as heritability increases, sparsity decreases, as well as sample sizes $n_x$ and $n_w$ increase, and changes in $n_x$ have a greater impact on $R_{\LW}(\lambda)$ compared to changes in $n_w$.



\begin{figure}[!ht]
\includegraphics[page=1,width=0.85\linewidth]{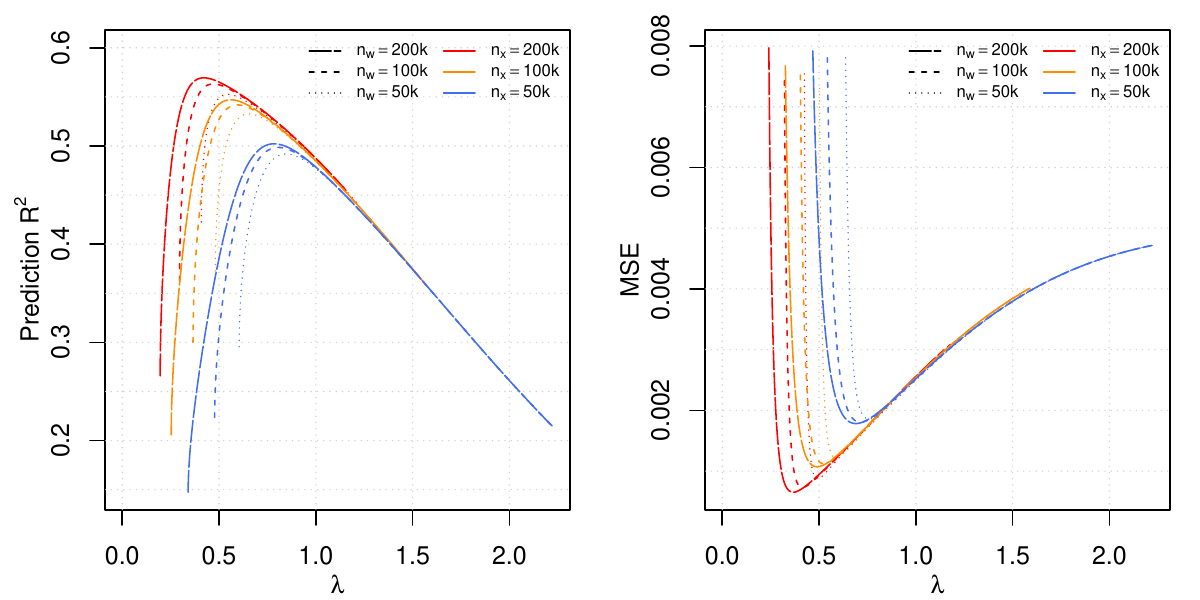}
\centering
\caption{
\textbf{Illustrating the theoretical out-of-sample $R^2$ and MSE of $\hat{\bbeta}_{\LW}(\lambda)$ as $\lambda$ varies.} 
The out-of-sample $R^2$  and MSE are calculated according to Theorem~\ref{thm: i.i.d. lasso mse + R2}. Here we set $\bmSigma = \bI_{p}$, $p~=~$461,488, heritability $h_x^2=h_s^2=0.6$, $n_x~=~$50,000, 100,000, 200,000, $n_w~=~$50,000, 100,000, 200,000, and $m/p~=~$0.005.
Entries of $\bbeta_0$ are i.i.d. random variables following the Bernoulli-Gaussian distribution. 
}
\label{supp_fig_theory_5}
\end{figure}

Based on the observations regarding the existence of a best-performing $\lambda$, we introduce additional assumptions to provide a more rigorous and detailed comparison between $\hat{\bbeta}_{\textnormal{L}}(\lambda)$ and $\hat{\bbeta}_{\LW}(\lambda)$.
We focus on the case $n_w=n_x$ and assume the existence of optimal $\lambda$ for both estimators. 
The following condition summarizes our assumptions, which are primarily used to ensure the analytical tractability of the problem.

\begin{condition}
\label{cond: comparison techinical}
    \begin{enumerate}
        \item[(a)] We assume $n_x=n_w$, and entries of $\bbeta_0$ are i.i.d.  Bernoulli-Gaussian random variables following the density function given in Equation~\eqref{true_beta0_bernoulli_Gaussian}. 
        
        \item[(b)] The  optimal $\lambda$ values exist for  $R_{\LW}(\lambda)$ and $A_{\LW}^2(\lambda)$, as well as $ R_{\textnormal {L}}(\lambda)$ and $A_{\LA}^2(\lambda)$.
        \item[(c)] We use the approximation $e^{-x^2}/(\sqrt{\pi} x) \approx \erfc(x)$, where  
        $\erfc(x)=1 - 2/\sqrt{\pi} \cdot \int_{0}^{x} e^{-t^2} dt$. 
        Furthermore, we assume that the  optimal $\lambda$ value of $R_{\LW}(\lambda)$, denoted as ${\lambda^*_{\LW,\M}}$, 
        satisfies 
         \begin{align*}
            {\lambda^*_{\LW,\M}} \leq \sigma_{\bbeta} +  \frac{1}{\sigma_{\bbeta}} \left(\frac{\tilde{\tau}_*}{1 + \tilde{b}_*} \right)^2 -  2 \gamma_x \kappa {\sigma_{\beta}} \cdot \erfc \left[\frac{\lambda^*_{\LW,\M} (1 + \tilde{b}_*)}{\sqrt{2 \left\{\left({\tilde{\tau}_*} \right)^2 + \sigma_{\beta}^2 (1 + \tilde{b}_*)^2 \right\}}} \right]. 
        \end{align*}
        \item[(d)] 
        The function
        \begin{align} \label{eqn: R^2 vs. Theta_*}
        \frac{\EE \big\{ \eta_{\soft}(\bar{\beta} + \Theta z, \alpha \Theta) \cdot \bar{\beta} \big\} }{\sqrt{\EE \bar\beta^2} \sqrt{\EE \eta^2_{\soft}(\bar{\beta} + \Theta z, \alpha \Theta)}}
        \end{align}
        is strictly decreasing with respect to $\Theta$.
    \end{enumerate}
\end{condition}

\begin{remark}[Remark on Condition \ref{cond: comparison techinical}]
    Condition \ref{cond: comparison techinical} is primarily made due to the lack of closed-form solutions for $L_{1}$ estimators. We would like to emphasize that these assumptions are practical and reasonable. 
    For Condition \ref{cond: comparison techinical} (a), we note that the $R_{\LW}(\lambda)$ and $R_{\textnormal{L}}(\lambda)$ are determined by the solution to the fixed point equation in  Equation \eqref{eqn: i.i.d. l1 state evo}, which does not have a closed-form solution in general due to its connections with the density function of $\bbeta_0$. 
    Thus, to compare $R_{\LW}(\lambda)$ and $R_{\textnormal{L}}(\lambda)$, we make the assumption that entries of $\bbeta_0$ follow the Bernoulli-Gaussian distribution, which allows us to have a closed-form solution to the fixed point equation. 
    Furthermore, as demonstrated in  {Figure~\ref{supp_fig_theory_5}},
    both $R_{\LW}(\lambda)$ and $R_{\textnormal{L}}(\lambda)$ have numerical best-performing tuning parameters in all scenarios. 
    Thus, we assume the existence of the optimal tuning parameters in Condition \ref{cond: comparison techinical} (b).
    In addition, Condition \ref{cond: comparison techinical} (c) is primarily due to technical difficulties. 
   It is known that $e^{-x^2}/\sqrt{\pi} x$ is a tight upper bound on $\erfc(x)$ such that
    \begin{align*}
        \frac{e^{-x^2}}{\sqrt{\pi}} \left(\frac{1}{x} - \frac{1}{2 x^3} \right) \leq \erfc(x)\leq \frac{e^{-x^2}}{\sqrt{\pi} x}\quad \mbox{and} \quad \lim_{x \to \infty}\frac{e^{-x^2}/\sqrt{\pi} x}{\erfc(x)} = 1. 
    \end{align*}
    Moreover, Condition \ref{cond: comparison techinical} (d) suggests that for any fixed $\alpha$, as $\Theta$ decreases to $0$, $\eta_{\soft}(\barbeta + \Theta z, \alpha \Theta)$ converges to $\barbeta$.
    Supplementary Figure \ref{fig: R^2 vs. Theta} displays the pattern of Equation \eqref{eqn: R^2 vs. Theta_*} when entries of $\bbeta_0$ follow the Bernoulli-Gaussian distribution defined in Equation \eqref{true_beta0_bernoulli_Gaussian}, where Equation \eqref{eqn: R^2 vs. Theta_*} is decreasing with respect to $\Theta$.
\end{remark}

The following proposition compares the $\hat{\bbeta}_{\LW}(\lambda)$ and $\hat{\bbeta}_{\textnormal{L}}(\lambda)$ at their respective optimal tuning parameters. 

\begin{proposition} \label{prop: comparison of L_1}
    Under the assumptions of Theorem \ref{thm: i.i.d. lasso mse + R2}, Proposition \ref{prop: i.i.d. non-ref lasso mse + R2}, and Condition \ref{cond: comparison techinical} (a) - (c),
    we have 
    \begin{align*}
        \min_{\lambda \in \RR_+} R_{\LW}(\lambda) > \min_{\lambda \in \RR_+} R_{\textnormal{L}}(\lambda). 
    \end{align*}
    In addition, if Condition \ref{cond: comparison techinical} (d) holds, we have 
    \begin{align*}
        \max_{\lambda \in \RR_+} A^2_{\LW}(\lambda) < \max_{\lambda \in \RR_+} A^2_{\textnormal{L}}(\lambda). 
    \end{align*}
\end{proposition}
Proposition~\ref{prop: comparison of L_1} suggests that $\hat{\bbeta}_{\LW}(\lambda)$ exhibits poorer performance than  $\hat{\bbeta}_{\textnormal{L}}(\lambda)$, even when the reference panel sample size $n_w$ matches the training data sample size $n_x$.
In Section~\ref{sec: L_2 estimator}, we conduct similar comparisons for $L_{2}$ regularized estimators and find consistent results with the $L_1$ estimators, without requiring additional technical  assumptions like those in Condition \ref{cond: comparison techinical}.

{\bxz 
\section{\texorpdfstring{$L_1$}{TEXT} regularized estimators with general \texorpdfstring{$\bmSigma$}{TEXT}  }\label{sec: L_1 estimator}

In this section, we investigate the behavior of $\hat{\bbeta}_{\LW}(\lambda)$ in the presence of a general covariance matrix $\bmSigma$, which captures the underlying {LD} patterns among the genetic variants. Generalizing the AMP results from i.i.d. data to a general $\bmSigma$ is a challenging task, as the soft-threshold operator function is no longer separable. Recent work by \cite{huang2022lasso} has used techniques from \cite{berthier2020state} to handle non-separable functions and develop AMP results for $\hat{\bbeta}_{\textnormal{L}}(\lambda)$. However, these techniques cannot be applied to $\hat{\bbeta}_{\LW}(\lambda)$ due to the additional randomness introduced by the reference panel data $\W$.

To address this challenge, we develop a novel AMP framework to study $\hat{\bbeta}_{\LW}(\lambda)$. Our approach uses matrix AMP techniques \citep{javanmard2013state,feng2022unifying}, while extending them  to handle the non-separable soft-threshold function associated with the general covariance matrix $\bmSigma$. The non-separable matrix AMP framework is presented in Theorem \ref{matrix AMP}. 
We apply these general theoretical results to analyze the asymptotic behavior of $\hat{\bbeta}_{\LW}(\lambda)$ in Theorems \ref{thm: general LASSO main without normalization} and \ref{thm: general lasso mse + R2}. These results encompass the i.i.d. data results presented in Section~\ref{sec: i.i.d. REF-LASSO} as a special case. Moreover, we use the non-separable matrix AMP framework to obtain asymptotic results for $\hat{\bbeta}_{\RW}(\lambda)$ in Theorem~\ref{thm: ridge mse + R2}.

Similar to Equation~(\ref{eta_soft}), we begin by introducing the soft-threshold operator used in matrix AMP. For any $p \times p$ positive definite matrix ${\bSigma}$, we define $\eta_{\theta}({\bm v}): \RR^{p} \mapsto \RR^p$ as 
\begin{equation} \label{eqn: soft-thres}
    \eta_{\theta}({\bm v}) := \arg\min_{{\bm w} \in \RR^p} \left(\frac{1}{2} \bnorm{{\bm w}- {\bm v}}_{\bSigma}^2 + \theta \bnorm{{\bm w}}_1
    \right).
\end{equation}
When $\bmSigma = \bI_p$, $\eta_{\theta}({\bm v})$ is a separable function and we have $\eta_{\theta}({\bm v}) = \eta_{\soft}(v, \theta)$. However, for general $\bmSigma$, $\eta_{\theta}({\bm v})$ becomes non-separable.
We consider the generalized AMP recursion \citep{rangan2011generalized} for $\hat{\bbeta}_{\LW}(\lambda)$, where  
\begin{equation}\label{eqn: general - lasso - AMP without normalization}
\begin{split}
    \sqrt{p} {\bbeta}^{t+1} &= \eta_{\lambda (1+b_{t})} \bigg\{ -  \frac{\sqrt{N}}{n_{w}} {\bSigma}^{-1} \W^{\T}{\rb^{t}} + \frac{\sqrt{p}}{n_{x}} (1+b_t) {\bSigma}^{-1} \X^{\T} \y_{x} + \sqrt{p} {\bbeta}^{t} \bigg\}
    \\ 
  \mbox{and} \quad  \rb^{t} &=  \sqrt{\frac{p}{N}} \W {\bbeta}^{t} + \frac{b_t}{1 + b_{t-1}} \rb^{t-1}. 
\end{split}
\end{equation}
Here $N = n_{w} + n_{x}$ and the initialization $\bbeta^0$, ${\bbeta}^{t+1} \in \RR^{p}$, $\rb^{t} \in \RR^n_w$ and $b_t$ satisfy 
\begin{equation} \label{bt_recursion_1n_normaliation}
\begin{split}
    \frac{b_t}{1+b_{t-1}} = \frac{1}{n_w} \Div \eta_{\lambda(1+b_{t-1})}  \bigg\{ - \frac{\sqrt{N}}{n_{w}} {\bSigma}^{-1} \W^{\T}{\rb^{t-1}} + \frac{\sqrt{p}}{n_{x}} (1+b_{t-1}) {\bSigma}^{-1} \X^{\T} \y_{x} + \sqrt{p} {\bbeta}^{t-1} \bigg\},
\end{split}
\end{equation}
where the divergence $\Div \eta_{\theta}({\bm v})$  is defined as 
\#\label{def:div}
\Div \eta_{\theta}({\bm v}) = \sum_{j=1}^p \partial \eta_{\theta,j}({\bm v})/\partial v_j.
\#
The following section introduces the development of a non-separable matrix AMP framework, which enables the computation of the state evolution as described by Equation \eqref{eqn: general - lasso - AMP without normalization}. This framework addresses the challenges posed by the non-separable soft-threshold function and general $\bmSigma$, facilitating the analysis of reference panel-based estimators.


\subsection{Non-separable matrix AMP} \label{subsec: matrix AMP}

Following \cite{berthier2020state}, we first establish the framework for a symmetric $\GOE$ and then extend the symmetric matrix AMP framework to asymmetric scenarios. 
We introduce the following definitions. 
Let $\Ab$ be a symmetric $\GOE(N')$ for $N' \in \NN$. Consider matrix $\Xb^{t} = \begin{bmatrix}
(\xb^{t}_1)^{\T} & \cdots & (\xb^{t}_{N'})^{\T}
\end{bmatrix}^{\T} \in \RR^{{N'} \times q}$ and function $\cF^t = \begin{bmatrix}
(f^{t}_1)^{\T} \cdots (f^{t}_{N'})^{\T}
\end{bmatrix}^{\T}: \RR^{{N'}  \times q} \mapsto \RR^{N' \times q}$ with fixed $q$ and $(f^{t}_{i}): \RR^{{N'}  \times q} \mapsto \RR^{q}$ for all $i \leq N'$. 
The symmetric matrix AMP is given by 
\begin{equation} \label{eqn: symmetric matrix AMP}
    \Xb^{t+1} = \Ab \mb^{t} - \mb^{t-1} {\Bb^t}^{\T},
\end{equation}
with the initialization $\Xb^0$, $\mb^{t} = \cF^t(\Xb^{t}) \in \RR^{{N'} \times q}$, and 
$\Bb^t = (N')^{-1} \sum_{i=1}^{N'} {\partial f^{t}_{i}}/{\partial \xb^{t}_i} (\Xb^{t}) \in \RR^{q \times q}$. 
Here ${\partial f^{t}_{i}}/{\partial \xb^{t}_i}$ is a  $q \times q$ Jacobian matrix.
Note that, unlike \cite{javanmard2013state}, we do not require $\cF^{t}$ to be a separable function.
For $1 \leq i,j \leq t$, the state evolution $\Kb_{i,j} \in \RR^{q \times q}$ is defined recursively with initialization
$ \Kb_{1,1} = \lim_{{N'} \to \infty} (N')^{-1} {\cF^0}^{\T}(\Xb^0) {\cF^0}(\Xb^0) \in \RR^{q \times q}$.
Let $\Kb^{(t)} = [\Kb_{i,j}]_{1 \leq i,j \leq t} \in \RR^{qt \times qt}$, we define the joint matrix Gaussian distribution of $\Zb^1, \Zb^2, \cdots, \Zb^{t}$ by 
\begin{equation} \label{Zb_i}
    \vecc(\Zb^1, \Zb^2, \cdots, \Zb^{t}) \sim N(0, \Kb^{(t)} \otimes \bI_{N'}).
\end{equation}
The $\Kb_{t+1, s+1}$ for $s \in [t]$ is given by
\begin{equation} \label{matrix state evolution}
\begin{split}
    \Kb_{t+1, s+1} &= \lim_{{N'} \to \infty} \frac{1}{{N'}} \EE \big\{ \cF^t(\Zb^{t})^{\T} \cF^s(\Zb^s) \big\} \in \RR^{q \times q}.
\end{split}
\end{equation}
Furthermore, for any $p \in \NN_{> 0}$, we say a function $\phi: \RR^{p} \mapsto \RR$ is a normalized pseudo-Lipschitz if there exist a constant $L$ such that for any $\xb, \yb \in \RR^{p}$, we have 
\begin{equation} \label{eqn:normalized_pesudo}
    |\phi(\xb) - \phi(\yb)| \leq L \left(1 + \frac{\bnorm{\xb}_2}{\sqrt{p}} + \frac{\bnorm{\yb}_2}{\sqrt{p}} \right) \frac{\bnorm{\xb - \yb}_2}{\sqrt{p}}. 
\end{equation}
We define uniformly normalized pseudo-Lipschitz  functions similar to that in Section \ref{sec1}.
The asymptotic behavior of $\Xb^{t}$ is summarized in the following theorem. Recall $\overset{P}{\approx}$ indicates convergence in probability.

\begin{theorem} \label{matrix AMP}
Assume $\Kb_{1,1}, \Kb_{2,2} \cdots, \Kb_{t+1, t+1}$ are positive semi-definite. Then for any deterministic sequence $\{\phi^{N'}\}_{N' \in \NN}$ of uniformly normalized pseudo-Lipschitz functions, as $N' \to \infty$, we have
\begin{equation*}
    \phi^{N'} (\Xb^0, \Xb^1, \cdots, \Xb^{t}) \overset{P}{\approx} \EE \big\{ \phi^{N'}(\Xb^0, \Zb^1, \cdots, \Zb^{t}) \big\}. 
\end{equation*}
\end{theorem}
Note that when $q = 1$, our results simplify to Theorem 3 in \cite{berthier2020state}.
Additionally, when the function $\cF^t$ is separable, our results align with Theorem 1 in \cite{javanmard2013state}. 
One of our major technical contributions is the extension of matrix AMP to handle non-separable functions for any fixed $q$. 
The primary challenge in our proof lies in the generalization of Stein's lemmas to the matrix case. 
Moreover, we show in Lemma \ref{prob fact 2} that for some matrix $\ub$, we have $\phi(\Ab \ub) \overset{P}{\approx} \phi(\Zb)$ for $\Ab \in \GOE(N')$ and matrix Gaussian distribution $\Zb$.
Our generalized matrix AMP results may serve as important tools for analyzing the asymptotic behavior of various regression problems, including our results for $\hat{\bbeta}_{\LW}(\lambda)$ and $\hat{\bbeta}_{\RW}(\lambda)$, where we have $N' = n_x + n_w + p$  and $q=3$.
}

\subsection{Existence of a unique solution to the fixed point equation}\label{subsec: calibration}

For each $t$, Theorem \ref{matrix AMP} indicates the $\bbeta^{t}$ given in Equation \eqref{eqn: general - lasso - AMP without normalization} is asymptotically determined by $(\tau_{t}^2, b_{t})$. 
As $t \to \infty$, 
we can show that $\bbeta^{t}$ converges to $\hat{\bbeta}_{\LW}(\lambda)$. 
Additionally, $(\tau_{t}^2, b_{t})$ converges to the fixed point $(\tau_{*}^2, b_{*})$ given by Equation \eqref{fixed point equation} below. 
Therefore, $\hat{\bbeta}_{\LW}(\lambda)$ is asymptotically determined by $(\tau_{*}^2, b_{*})$. In this section, we present Propositions \ref{prop 1} and \ref{prop 2} to 
guarantee the existence of $(\tau_{*}^2, b_{*})$, the solution of fixed point equation. These results will be used to prove our results in Theorem \ref{thm: general LASSO main without normalization}. 
First, we define the function $F(\tau, b) = (F_1, F_2)(\tau, b): \RR^2 \mapsto \RR^2$, where
\#
F_1(\tau^2, b) =\ & \lim_{p \to \infty} \EE \left[ \gamma_w \bnorm{\frac{1}{\sqrt{p}} {\eta}_{\lambda(1 + b)} \big\{ \tau {\bSigma}^{-1/2} \Zb + (1 + b) \sqrt{p}{\bbeta_0} \big\}}_{\bSigma}^2 \right] + \gamma_x (1 + b)^2 \lim_{p \to \infty} \frac{\EE \bnorm{{\bbeta_0}}_{\bSigma}^2}{h_{x}^2}, \nn\\
F_2(\tau^2, b) =\ & (1 + b) \cdot \lim_{p \to \infty} \frac{1}{n_w} \EE \Div \eta_{\lambda(1 + b)} \big\{\tau {\bSigma}^{-1/2} \Zb + (1 + b) \sqrt{p} \bbeta_0\big\}.  \label{F_1, F_2}
\#
Our goal is to show that for any $\lambda > 0$, there exists an unique solution $(\tau_*^2, b_*)$ to non-linear equations 
\begin{equation} \label{fixed point equation}
    \tau_*^2 = F_1(\tau_*^2, b_*)\quad \mbox{and} \quad b_* = F_2(\tau_*^2, b_*).
\end{equation}
It is challenging to directly study Equation~\eqref{fixed point equation}. Instead, we introduce a new parameter $\alpha$ and consider the following equivalent equations
\begin{align}\label{eqn: 4.11 without normalization}
    &\tau_*^2 = \lim_{p \to \infty} \EE \left[\gamma_w \bnorm{\frac{1}{\sqrt{p}} {\eta}_{\alpha \tau_*}\big\{\tau_* {\bSigma}^{-1/2} \Zb + (1 + b_*)  \sqrt{p}{\bbeta}_0 \big\} }_{\bSigma}^2  + \gamma_x (1 + b_*)^2  \bnorm{{\bbeta}_0}_{\bSigma}^2/h_{x}^2 \right], \\ \label{eqn: 4.12 without normalization}
    &b_* = (1 + b_*) \cdot \lim_{p \to \infty} \frac{1}{n_w} \EE \Div \eta_{\alpha \tau_*}\big\{\tau_* {\bSigma}^{-1/2} \Zb + (1 + b_*)  \sqrt{p} \bbeta_0\big\},  \\
    \label{eqn: 4.13 without normalization}
    &\alpha \tau_* = \lambda (1 + b_*).
\end{align}

We will develop the results for Equations \eqref{eqn: 4.11 without normalization}-\eqref{eqn: 4.12 without normalization} and Equation \eqref{eqn: 4.13 without normalization} separately. Specifically, 
Proposition \ref{prop 1} below demonstrates that there exists a unique $\alpha_{\min} > 0$ such that Equations \eqref{eqn: 4.11 without normalization}-\eqref{eqn: 4.12 without normalization} have a unique fixed point solution for any $\alpha > \alpha_{\min}$.
The solution $(\tau_*(\alpha), b_*(\alpha))$ depends on $\alpha$. 
Proposition \ref{prop 2} {establishes} the existence of $\alpha_* > \alpha_{\min}$ such that Equation \eqref{eqn: 4.13 without normalization} holds at the same time. 
As a result of Theorem \ref{thm: general LASSO main without normalization}, Corollary \ref{cor: alpha uniqueness} guarantees the uniqueness of $\alpha_*$, which follows from the uniqueness of $\hat{\bbeta}_{\LW}$ in Equation~\eqref{eqn:ref-panel-lasso-est without normalization}.
Together, we will have the conclusion that $\left(\tau_*(\alpha_*), b_*(\alpha_{*})\right)$ is the unique solution to the fixed point equation $(\tau_*^2, b_*) = F(\tau_*^2, b_*)$.

\begin{proposition}\label{prop 1}
Define the function
\begin{equation}
    g(\alpha) = \lim_{p \to \infty} \frac{1}{n_w} \bnorm{\eta_{\alpha}({\bSigma}^{-1/2} \Zb)}_{\bSigma}^2.
\end{equation}
Suppose that $\gamma_w > 1$, then the equation $g(\alpha) = 1$ has a unique solution, denoted by $\alpha_{\min}$. 
For any $\gamma_w \leq 1$ and $\alpha > 0$, or $\gamma_w > 1$ and $\alpha > \alpha_{\min}$, the following equations have a unique solution $(\tau_*, b_*) \in \RR_{+} \times \RR$:
\begin{gather} 
\tau_*^2 = \lim_{p \to \infty} \EE \left[\gamma_w \bnorm{\frac{1}{\sqrt{p}} {\eta}_{\alpha \tau_*}\big\{\tau_* {\bSigma}^{-1/2} \Zb + (1 + b_*)  \sqrt{p}{\bbeta}_0 \big\} }_{\bSigma}^2  + \gamma_x (1 + b_*)^2  \bnorm{{\bbeta}_0}_{\bSigma}^2/h_{x}^2 \right], \label{eq:prop4_fixed_pt_eqns}\\
\frac{1}{1 + b_*} = \left[1 - \lim_{p \to \infty} \frac{1}{n_w} \EE \Div \eta_{\alpha \tau_*} \big\{\tau_* {\bSigma}^{-1/2} \Zb + (1 + b_*)  \sqrt{p} \bbeta_0 \big\}\right]_+. \label{eq:prop4_fixed_pt_eqns_b}
\end{gather}
\end{proposition}

Proposition \ref{prop 1} suggests that there exists a unique solution $(\tau_{*}, b_{*})$ to Equations \eqref{eqn: 4.11 without normalization} - \eqref{eqn: 4.12 without normalization} for $\alpha> \alpha_{\min}$.
Next, we define a function $\alpha \mapsto \lambda(\alpha)$ on $(\alpha_{\min}, \infty)$ by 
\begin{equation} \label{lambda(alpha)}
     \lambda(\alpha) = \alpha {\tau_*} \Bigg[1 - \lim_{p \to \infty} \frac{1}{n_w} \EE \Div \eta_{\alpha {\tau_*}} \big\{{\tau_*} {\bSigma}^{-1/2} \Zb + (1 + b_*) \sqrt{p} \bbeta_0\big\} \Bigg].
\end{equation}
We will demonstrate the existence of the inverse of this function, denoted as $\lambda \mapsto \alpha(\lambda)$. To do this, we define $\alpha: (0, \infty) \mapsto (\alpha_{\min}, \infty)$ such that 
\begin{equation} \label{alpha(lambda)}
    \alpha(\lambda) \in \{a \in (\alpha_{\min}, \infty): \lambda(a) = \lambda \}.
\end{equation}

Proposition \ref{prop 2} below establishes that the set on the right-hand side of Equation~\eqref{alpha(lambda)} is non-empty, which implies that the function $\lambda \mapsto \alpha(\lambda)$ is well-defined. 
That is, for any given $\lambda$, there exists an $\alpha$ that satisfies Equation~\eqref{alpha(lambda)}. 
\begin{proposition} \label{prop 2}
When $\gamma_w > 1$, the function $\alpha \mapsto \lambda(\alpha)$ is continuous on the interval $(\alpha_{\min}, \infty)$, whereas it is continuous on $(0, \infty)$ when $\gamma_w \leq 1$. 
Furthermore, the function $\lambda \to \alpha(\lambda)$ that satisfies Equation \eqref{alpha(lambda)} exists. 
Additionally, for any $\alpha_*$ satisfying $\lambda(\alpha_*) = \lambda$, the ${1}/(1 + b_*)$ defined in Equation~\eqref{eq:prop4_fixed_pt_eqns_b} is positive. 
\end{proposition}

In Corollary \ref{cor: alpha uniqueness}, we will further demonstrate the uniqueness of the $\alpha$ that satisfies Equation~\eqref{alpha(lambda)}.
Therefore, for any $\lambda>0$, Equations \eqref{eqn: 4.11 without normalization} - \eqref{eqn: 4.13 without normalization} have a unique fixed point solution.  
The generalized Stokes' theorem on smoothly changing manifold from \cite{baddeley1977integrals} plays an important  role in the proof of Proposition \ref{prop 1} and \ref{prop 2}.
The techniques used to prove these propositions for non-separable matrix AMP and $\hat{\bbeta}_{\LW}(\lambda)$ differ substantially from those used in previous papers for $\hat{\bbeta}_{\textnormal{L}}(\lambda)$ \citep{bayati2011lasso,huang2022lasso}. For $\hat{\bbeta}_{\textnormal{L}}(\lambda)$, the state evolution is typically characterized by a single recursive process, whereas our Equations \eqref{eqn: 4.11 without normalization}-\eqref{eqn: 4.13 without normalization} rely on both $\tau_*$ and $b_*$. Additionally, for general $\bmSigma$, the soft-threshold operator $\eta_{\theta}$ does not have a closed-form solution. We also address difficulties regarding the differentiability of Equation \eqref{eqn: soft-thres}.

}


\subsection{Numerical illustration of the calibration between \texorpdfstring{$\alpha$}{TEXT} and \texorpdfstring{$\lambda$}{TEXT}}

Propositions \ref{prop 1} and \ref{prop 2} establish a connection between $\alpha$ and $\lambda$. The fixed point $(\tau_*(\alpha), b_*(\alpha))$ depends on $\alpha$ and serves as the unique solution to the fixed point equation $(\tau_*^2, b_*) = F(\tau_*^2, b_*)$. Consequently, since $R_{\LW}(\lambda)$ and $A^2_{\LW}(\lambda)$ are functions of the fixed points, they will also be functions of $\alpha$. 
In this section, we provide numerical illustrations of the calibration between $\alpha$ and $\lambda$ in our non-separable matrix AMP framework, extending the numerical analysis conducted in Section~\ref{sec3.1}. Specifically, we explore the impact of $\alpha$ on $R_{\LW}(\lambda)$ and $A^2_{\LW}(\lambda)$ and investigate the existence of a unique best-performing value of $\alpha$. This analysis complements the observations made in Figure~\ref{supp_fig_theory_5}, which specifically focuses on the effect of $\lambda$. Additionally, we compare the patterns observed in $\hat{\bbeta}_{\LW}(\lambda)$ with those of $\hat{\bbeta}_{\textnormal{L}}(\lambda)$ to gain further insights.

\begin{figure}[!t] 
\includegraphics[page=1,width=0.85\linewidth]{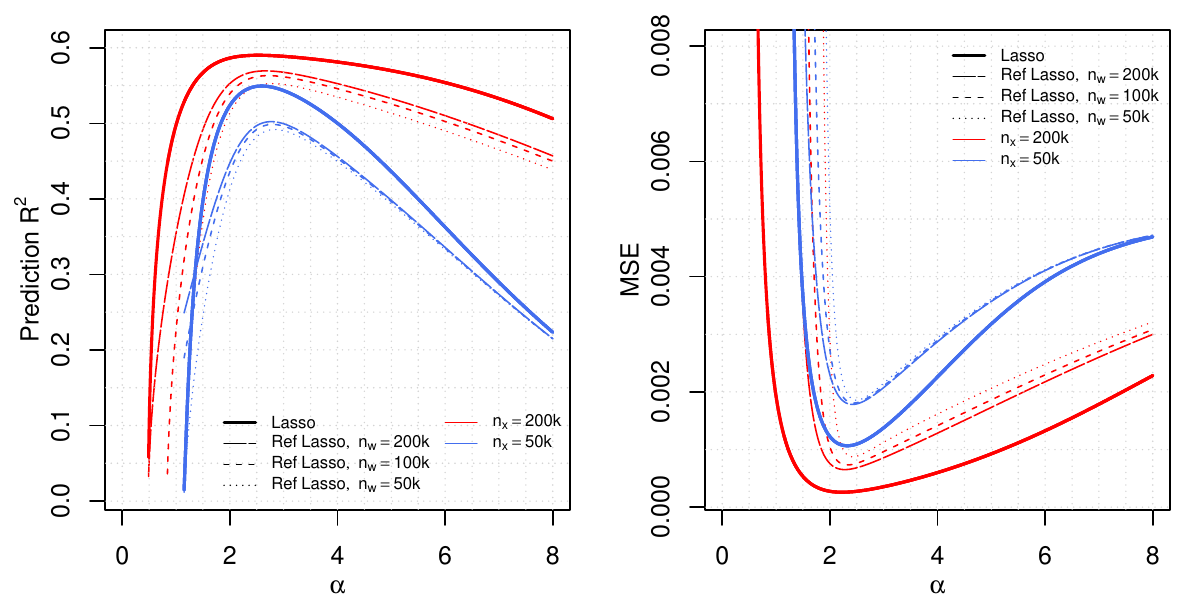}
\centering
\caption{
\textbf{Comparing the theoretical out-of-sample $R^2$  and MSE of $\hat{\bbeta}_{\textnormal{L}}(\lambda)$ and $\hat{\bbeta}_{\LW}(\lambda)$ as $\alpha$ varies.} 
The theoretical results of $\hat{\bbeta}_{\textnormal{L}}(\lambda)$ (`Lasso') and $\hat{\bbeta}_{\LW}(\lambda)$  (`Ref Lasso') are calculated according to Proposition~\ref{prop: i.i.d. non-ref lasso mse + R2} and Theorem~\ref{thm: i.i.d. lasso mse + R2}, respectively. 
Here we set $\bmSigma = \bI_{p}$, 
 $p~=~$461,488, heritability $h_x^2=h_s^2=0.6$, sparsity $m/p=0.005$, $n_x~=~$50,000, 200,000, and $n_w~=~$50,000, 100,000, 200,000. Entries of  $\bbeta_0$ are i.i.d. random variables following the Bernoulli-Gaussian distribution. 
 The relationship between $\alpha$ and $\lambda$ is given in Equation \eqref{alpha(lambda)}.  
}
\label{main_fig_3}
\end{figure}

Figure~\ref{main_fig_3} illustrates the $A^2_{\LW}(\lambda)$, $A^2_{\textnormal{L}}(\lambda)$, $R_{\LW}(\lambda)$, and $R_{\textnormal{L}}(\lambda)$ at different $\alpha$ values under the same setup as in Figure~\ref{supp_fig_theory_5}. 
Each $\lambda$ value in Figure~\ref{supp_fig_theory_5} corresponds to each $\alpha$ value in Figure~\ref{main_fig_3} according to Equation~\eqref{lambda(alpha)}.
Similar to Figure~\ref{supp_fig_theory_5}, 
Figure~\ref{main_fig_3} suggests the existence of best-performing $\alpha$ for both MSE and out-of-sample $R^2$. The respective best-performing $\alpha$ for $R_{\textnormal{L}}(\lambda)$ and $A^2_{\textnormal{L}}(\lambda)$ are close, so are the best-performing $\alpha$ for $R_{\LW}(\lambda)$ and $A^2_{\LW}(\lambda)$. 
Among all estimators of the same $n_x$, for both MSE and out-of-sample $R^2$, the best-performing $\alpha$ of $\hat{\bbeta}_{\textnormal{L}}(\lambda)$ are slightly smaller than those of  $\hat{\bbeta}_{\textnormal{L}}(\lambda)$, and the best-performing $\alpha$ for all reference panel-based estimators of same $n_x$ are numerically close, regardless of $n_w$. 
As $n_x$ increases, the best-performing $\alpha$ of all estimators decreases. 
It is worth mentioning that the relative change in best-performing $\lambda$ in Figure~\ref{supp_fig_theory_5} is larger than the relative change in best-performing $\alpha$ in Figure~\ref{main_fig_3}. That is, the peaks/troughs of $\lambda$ in Figure~\ref{supp_fig_theory_5} are further apart than those of $\alpha$ in Figure~\ref{main_fig_3}.
With $n_x$ fixed, as $n_w$ increases, the best-performing $\alpha$ of reference panel-based $L_1$ estimators decrease, though the magnitude of the decrease is smaller than that for increasing $n_x$. Similarly, the best-performing $\lambda$ also decreases with a magnitude smaller than that for increasing $n_x$ in Figure~\ref{supp_fig_theory_5}.

\begin{figure}[!t] 
\includegraphics[page=1,width=0.9\linewidth]{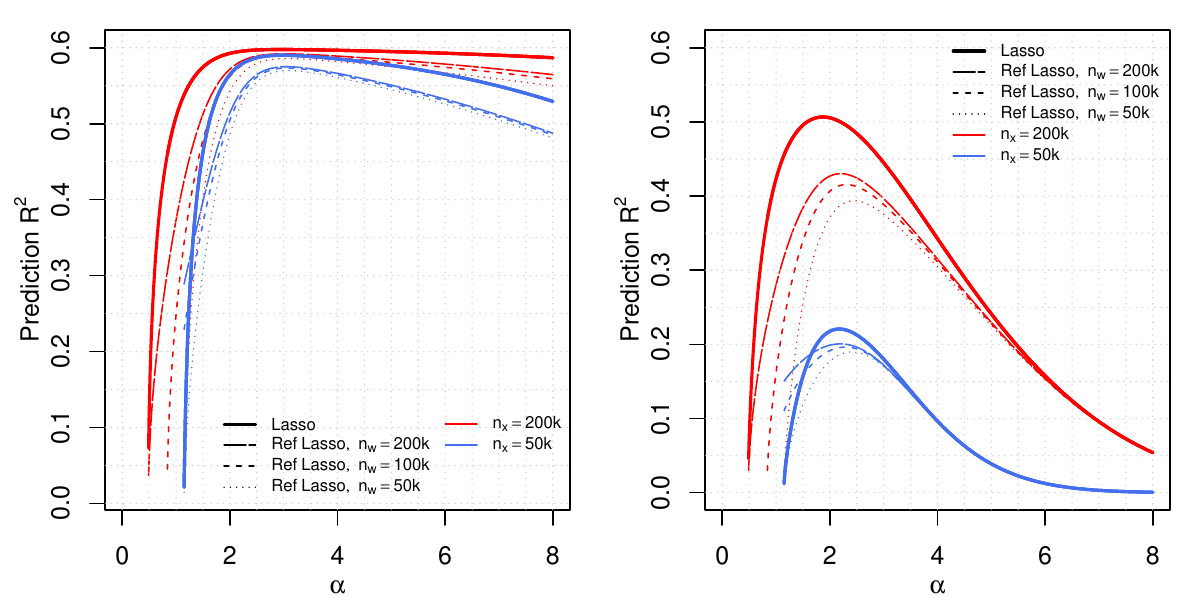}
\centering
\caption{
\textbf{Comparing the theoretical out-of-sample $R^2$ of $\hat{\bbeta}_{\textnormal{L}}(\lambda)$ and $\hat{\bbeta}_{\LW}(\lambda)$ under different sparsity as $\alpha$ varies.} 
The out-of-sample $R^2$  of $\hat{\bbeta}_{\textnormal{L}}(\lambda)$ (`Lasso') and $\hat{\bbeta}_{\LW}(\lambda)$  (`Ref Lasso') is calculated according to Proposition~\ref{prop: i.i.d. non-ref lasso mse + R2} and Theorem~\ref{thm: i.i.d. lasso mse + R2}, respectively. 
Here we set $\bmSigma = \bI_{p}$, 
 $p~=~$461,488, heritability $h_x^2=h_s^2=0.6$, $n_x~=~$50,000, 200,000, and $n_w~=~$50,000, 100,000, 200,000. Entries of  $\bbeta_0$ are i.i.d. random variables following the Bernoulli-Gaussian distribution. 
{\textbf {Left:}} Sparsity $m/p=0.001$. {\textbf {Right:}} Sparsity $m/p=0.05$.
}
\label{main_fig_4}
\end{figure}

We further explore the patterns for $\alpha$ at different sparsity and heritability levels.
{Figure}~\ref{main_fig_4}  shows that  both $A^2_{\LW}(\lambda)$ and $A^2_{\textnormal{L}}(\lambda)$  decreases much more slowly in $\alpha$ under $m/p=0.001$ than under $m/p=0.05$. These results indicate that both $\hat{\bbeta}_{\textnormal{L}}(\lambda)$ and  $\hat{\bbeta}_{\textnormal{L}}(\lambda)$ are more sensitive to $\alpha$ (or $\lambda$)  when signals become denser. 
In summary, our numerical results support the calibration between $\alpha$ and $\lambda$ for non-separable matrix AMP 
and provide insights into the similarities and differences between $\hat{\bbeta}_{\textnormal{L}}(\lambda)$  and $\hat{\bbeta}_{\LW}(\lambda)$.

\subsection{Asymptotic results}\label{subsec: main results}


In this section, we use our non-separable matrix AMP framework to derive the asymptotic risks of $\hat{\bbeta}_{\LW}(\lambda)$. Theorem~\ref{thm: general LASSO main without normalization} provides a characterization of the asymptotic behavior of $\hat{\bbeta}_{\LW}(\lambda)$ for any deterministic sequence of uniformly pseudo-Lipschitz function.

\begin{theorem} \label{thm: general LASSO main without normalization}
Let $\{\bbeta_0, \bepsilon_x,\bmeps_s, \bSigma, \bX,  \Sbb, \bW\}$ be a converging sequence of instances with $\PP\left({\bbeta_0}(p) \neq 0 \right) > 0$. 
Each row of $\X$, $\Sbb$, and $\Wb$ is i.i.d. Gaussian with mean $\bm 0$ and variance ${\bSigma}$.
For any deterministic sequence $\{ \phi^{p}: \RR^{p} \times \RR^p \mapsto \RR \}_{p \in \NN}$, of uniformly pseudo-Lipschitz functions, we have 
\begin{align} \label{eqn: general LASSO main without normalization formula}
    \phi^{p} \big\{ \hat{\bbeta}_{\LW}(\lambda), \bbeta_0 \big\} \overset{P}{\approx} \EE\ \phi^{p} \left[ \frac{1}{\sqrt{p}} \eta_{\lambda (1 + b_*)} \big\{ (1+b_*) \sqrt{p} \bbeta_0 + \tau_* \bSigma^{-1/2} \Zb \big\}, \bbeta_0 \right],
\end{align}
where $\Zb \sim N(0, {\bI_p})$ is independent of $\bbeta_0$. The $\tau_* = \tau_*(\alpha(\lambda))$ and $b_* = b_*(\alpha(\lambda))$ are given in Equations \eqref{eqn: 4.11 without normalization} - \eqref{eqn: 4.13 without normalization}. 
\end{theorem}
In the proof of Theorem~\ref{thm: general LASSO main without normalization}, we use the formulation of the multivariate Ornstein-Uhlenbeck semi-group from \citep{gardiner1985handbook} to rigorously establish Lemma \ref{OU-process}. The lemma guarantees that for any jointly Gaussian vectors $\zb_1$ and $\zb_2$ in $\RR^{p}$, the expectation $\EE f(\zb_{1}) f(\zb_2)$ is increasing with respect to $\EE \zb_1^{\T} \zb_2/p$ for some function $f$. This lemma, which was not included in previous AMP literature \citep{huang2022lasso}, is crucial for our analysis. 
Equation \eqref{eqn: general LASSO main without normalization formula} provides a characterization of $\hat{\bbeta}_{\LW}(\lambda)$  in the high-dimensional regime. 
According to the right-hand side of the equation, after the soft-threshold function $\eta_{\lambda (1 + b_*)}$ introduces sparsity, $\hat{\bbeta}_{\LW}(\lambda)$ is centered around $(1 + b_*) \bbeta_0$ with a variance of $\tau^2_*$. The positivity of $(1 + b_*)$ is ensured by Propositions \ref{prop 1} and \ref{prop 2}.

The model and normalization setups used in Theorem \ref{thm: general LASSO main without normalization} are specified in Conditions \ref{cond1-np-ratio}-\ref{cond-distn-data}. 
Similar setups are widely used in the genetic and RMT literature, such as \cite{zhao2022block}.
These conditions enable us to understand the asymptotic behavior of $\hat{\bbeta}_{\LW}(\lambda)$ in practical genetic applications and its dependence on key parameters of genetic datasets and traits, such as sample size, sparsity, and heritability. 
To establish a connection with previous AMP literature, we use a different normalization setting and present an alternative version of Theorem \ref{thm: general LASSO main without normalization} within the proof. 
Additionally, we demonstrate the equivalence between the two versions.

Our approach for proving Theorem \ref{thm: general LASSO main without normalization} can be extended to other estimators that have been studied in previous AMP literature. For example, we can provide a similar characterization for $\hat{\bbeta}_{\textnormal{L}}(\lambda)$ by introducing $(\bar{\tau}_*, \bar{b}_*)$, which is defined as follows:
\begin{align} \label{tau' equation}
    &\bar{\tau}_*^2 = \gamma_x \EE \| \bbeta_0 \|^2_{\bmSigma} \cdot \frac{1 - h_{x}^2}{h_{x}^2} + \lim_{p \to \infty} \gamma_x \EE \left[\left\| \frac{1}{\sqrt{p}} \eta_{\lambda (1 + \bar{b}_*)} \big\{ \sqrt{p} \bbeta_0 + \bar{\tau}_* \bmSigma^{-1/2} \Zb \big\} - \bbeta_0 \right\|^2_{\bmSigma} \right] \\ \label{eqn: 1/1+barb equation}
    &\mbox{and} \quad (1 + \bar{b}_*)^{-1} = 1 - \lim_{p \to \infty} \frac{1}{n_x} \EE \Div \eta_{\lambda (1 + \bar{b}_*)} \big(\sqrt{p} \bbeta_0 + \bar{\tau}_* \bmSigma^{-1/2} \Zb \big).
\end{align}

These results align well with the Theorem 1 in \cite{huang2022lasso} and Theorem 1.5 in \cite{bayati2011lasso}. For example, 
Theorem 1 of \cite{huang2022lasso} shows that $\hat{\bbeta}_{\textnormal{L}}(\lambda)$ asymptotically concentrates around $\bbeta_0$ with variance $\bar{\tau}^2$ and sparsity introduced by $\eta_{\lambda (1 + \bar{b})}$.  Additionally, \cite{mousavi2018consistent} show that $\bar{b}/(1 + \bar{b}) = \lim_{p \to \infty} \gamma_x \|\hat{\bbeta}_{\LA}(\lambda)\|_0$ when $\bmSigma=\bI_{p}$, which also holds with our Equation \eqref{eqn: 4.12 without normalization}.

As a consequence of Theorem \ref{thm: general LASSO main without normalization}, the function $\lambda \to \alpha(\lambda)$ is uniquely determined. Together with Propositions \ref{prop 1} and \ref{prop 2}, the following corollary ensures the existence and uniqueness of the fixed point solution to Equation \eqref{fixed point equation}.

\begin{corollary} \label{cor: alpha uniqueness}
For any $\lambda$ and $\sigma_{{\bepsilon}_x}^2$ greater than zero, there exists a unique $\alpha > \alpha_{\min}$ when $\gamma_w > 1$ and $\alpha > 0$ when $\gamma_w \leq 1$ such that $\lambda(\alpha) = \lambda$, where the function $\lambda(\alpha)$ is defined in Equation \eqref{lambda(alpha)}. 
Consequently, the function $\lambda \to \alpha(\lambda)$ is continuous and non-decreasing.
\end{corollary}

It is important to note that the uniqueness of $\alpha$ is a direct result of Theorem \ref{thm: general LASSO main without normalization}. Theorem \ref{thm: general LASSO main without normalization} also enables us to calculate the $\MSE$ and out-of-sample $R^2$ for $\hat{\bbeta}_{\LW}(\lambda)$. By selecting the corresponding function $\phi^{p}$, we have the following results for $R_{\LW}(\lambda)$ and $A^2_{\LW}(\lambda)$.

\begin{theorem} \label{thm: general lasso mse + R2}
Let $\{\bbeta_0, \bepsilon_x,\bmeps_s, \bSigma, \bX,  \Sbb, \bW\}$ be a converging sequence of instances with $\PP\left({\bbeta_0}(p) \neq 0 \right) > 0$. 
Each row of $\X$, $\Sbb$, and $\Wb$ is i.i.d. Gaussian with mean $\bm 0$ and variance ${\bSigma}$.
In probability, the $\MSE$ of $\hat{\bbeta}_{\LW}(\lambda)$ is  
\begin{equation*} 
    R_{\LW}(\lambda) = \lim_{p \to \infty} \bnorm{\hat{\bbeta}_{\LW}(\lambda) - \bbeta_0}_\bmSigma^2 = \lim_{p \to \infty} \bnorm{\frac{1}{\sqrt{p}} \eta_{\lambda (1 + b_*)}\big\{ \tau_{*} {\bSigma}^{-1/2} \Zb + (1 + b_*) \sqrt{p} \bbeta_0 \big\} - \bbeta_0}_{\bSigma}^2,
\end{equation*}
and the out-of-sample $R^2$ of $\hat{\bbeta}_{\LW}(\lambda)$ is  
\begin{equation*} 
   A^2_{\LW}(\lambda) = \lim_{p \to \infty} h_{s}^2 \cdot \frac{\EE \left\langle  {\bbeta_0},  \eta_{\theta_*} \big\{ \tau_* {\bSigma}^{-1/2} \Zb + (1 + b_*) \sqrt{p} {\bbeta_0} \big\} \right \rangle_{\bSigma}^2}{\EE \bnorm{\bbeta_0}_{{\bSigma}}^2 \cdot \EE \bnorm{\eta_{\theta_*} \big\{\tau_* {\bSigma}^{-1/2} \Zb + (1 + b_*) \sqrt{p} {\bbeta_0} \big\}}_{\bSigma}^2},
\end{equation*}
where $(\tau_*, b_*)$ satisfy the fixed point equation in Equation~\eqref{fixed point equation}. 
\end{theorem}

{\bxz
\section{\texorpdfstring{$L_2$}{TEXT} regularized estimators}\label{sec: L_2 estimator}
Our non-separable matrix AMP approach developed in the previous section is not limited to $L_{1}$ regularized estimators. 
In this section, we extend our analysis to include $L_{2}$ regularized estimators $\hat{\bbeta}_{\RA}(\lambda)$ and $\hat{\bbeta}_{\RW}(\lambda)$, as defined in Equation~\eqref{eqn:ref-panel-ridge-est without normalization}. These $L_{2}$ regularized estimators have been previously studied in RMT literature \citep{ledoit2011eigenvectors,dobriban2018high,zhao2022block}. While analyzing $L_{1}$ regularized estimators using RMT techniques is challenging, our AMP framework allows us to provide a unified analysis of both reference panel-based $L_{1}$ and $L_{2}$ regularized estimators.

\subsection{\texorpdfstring{$L_2$}{TEXT} regularized estimators with isotropic features}\label{sec: L_2 estimator1}
To facilitate easy comparison and obtain closed-form results, we first focus on the special case $\bmSigma = \bI_{p}$. Theorem \ref{thm: i.i.d. ridge mse + R2} establishes the AMP asymptotic results  of  $\hat{\bbeta}_{\RW}(\lambda)$. 
\begin{theorem}\label{thm: i.i.d. ridge mse + R2}
Let $\{\bbeta_0, \bepsilon_x,\bmeps_s, \bSigma, \bX,  \Sbb, \bW\}$ be a converging sequence of instances with $\PP\left({\bbeta_0}(p) \neq 0 \right) > 0$.
Each row of $\X$, $\Sbb$, and $\Wb$ is i.i.d. Gaussian with mean $\bm 0$ and variance ${\bI_p}$.
Then in probability, the out-of-sample $R^2$ of $\hat{\bbeta}_{\RW}(\lambda)$ is 
\begin{equation} \label{eqn: i.i.d. ref ridge R^2}
    A^2_{\RW}(\lambda) = \frac{h_{s}^4}{h_{s}^2 + \gamma_x} \cdot \Bigg[1 - 4 \gamma_w \Bigg(\frac{1}{ 1 + \lambda + \gamma_w + \sqrt{(1 - \lambda - \gamma_w)^2 + 4\lambda} } \Bigg)^2 \Bigg], 
\end{equation}
which is consistent with Corollary 1 in \cite{zhao2022block}. The optimal out-of-sample $R^2$ is obtained by letting $\lambda \to \infty$, and we have
\begin{align*}
    \lim_{\lambda \to \infty} A^2_{\RW}(\lambda) = \frac{h_{s}^4}{h_{s}^2 + \gamma_x}.
\end{align*}
In addition, let $R_{\lambda, \gamma_w}=\sqrt{4 \lambda +\left(\lambda +\gamma_w-1\right)^2}$, in probability, the MSE of $\hat{\bbeta}_{\RW}(\lambda)$ is given by 
\begin{align*}
    R_{\RW}(\lambda) =\ & \kappa \sigma_{\beta}^2\cdot  \Bigg\{ \frac{\left(\lambda ^2+\lambda  \gamma_w+\lambda  R_{\lambda, \gamma_w}-R_{\lambda, \gamma_w}-\gamma_w+1\right)^2}{\lambda ^2   \left(\lambda + R_{\lambda, \gamma_w}+\gamma_w+1\right)^2}\\
    &+ \frac{\gamma_x \left(\lambda +R_{\lambda, \gamma_w}+\gamma_w-1\right)^2}{h_x^2 \lambda ^2 \left(\lambda   +R_{\lambda, \gamma_w}+\gamma_w+1\right)^2}\\
    &\cdot \frac{{2 h_x^2 \gamma_w}/{\gamma_x}+(\lambda +1) \left(\lambda +R_{\lambda, \gamma_w}+1\right)+\gamma_w \left(2 \lambda +R_{\lambda, \gamma_w}\right)+\gamma_w^2}{\gamma_w \left(2 \lambda +R_{\lambda, \gamma_w}\right)+(\lambda +1) \left(\lambda +R_{\lambda, \gamma_w}+1\right)+\gamma_w^2-2 \gamma_w} \Bigg\}.
\end{align*}
\end{theorem}
In addition, the following proposition provides the AMP asymptotic  results for $\hat{\bbeta}_{\RA}(\lambda)$. 
\begin{proposition} \label{prop: ridge regression mse + R2}
Let $\{\bbeta_0, \bepsilon_x,\bmeps_s, \bSigma, \bX,  \Sbb\}$ be a non-reference panel converging sequence of instances with $\PP\left({\bbeta_0}(p) \neq 0 \right) > 0$. 
Each row of $\Xb$ and $\Sbb$
is i.i.d. Gaussian with mean ${\bm 0}$ and variance $\bI_p$. 
In probability,, the out-of-sample $R^2$ of $\hat{\bbeta}_{\RA}(\lambda)$  is given by
\begin{equation} \label{eqn: i.i.d. ridge R^2}
    A^2_{\textnormal {R}}(\lambda) = \frac{h_{s}^4}{(1 - c)h_{s}^2 + \gamma_x} \cdot \Bigg[1 - 4 \gamma_x \Bigg\{\frac{1}{1 + \lambda + \gamma_x + \sqrt{(1 - \lambda - \gamma_x)^2 + 4\lambda} } \Bigg\}^2 \Bigg]
\end{equation}
with 
\begin{equation} \label{eqn: constant c}
    c = \frac{4\gamma_x}{1 + \lambda + \gamma_x + \sqrt{(1 - \lambda - \gamma_x)^2 + 4\lambda}}.
\end{equation}
When $\lambda_{\RA}^* = \gamma_x \cdot (1 - h_x^2)/h_x^2$, we have the optimal out-of-sample $R^2$  
\begin{align*}
    A^2_{\textnormal {R}}(\lambda^{*}_{\textnormal{R}}) = \frac{h_s^4 }{(1 - R_{h_x,\gamma_x}) h_s^2 + \gamma_x} \cdot \left(1 - 4 \gamma_x R_{h_x,\gamma_x}^2 \right),
\end{align*}
where $R_{h_x,\gamma_x} =h_x^2/[h_x^2 \{\sqrt{\gamma_x^2/h_x^4+(2/h_x^2-4) \gamma_x+1}+1\}+\gamma_x]$. 
In addition, the $\MSE$ of $\hat{\bbeta}_{\RA}(\lambda)$ is given by
\begin{equation*}
\begin{split}
    R_{\textnormal {R}} (\lambda) &= \kappa \sigma_\beta^2\cdot \Bigg\{\frac{(\gamma_x + 1)\lambda + (\gamma_x - 1)^2 + (\gamma_x - 1)\sqrt{(1 - \gamma_x - \lambda )^2 + 4 \lambda}}{2 \gamma_x \sqrt{(1 - \gamma_x - \lambda )^2 + 4 \lambda}} \Bigg\}\\ &+ \kappa\sigma_\beta^2 \frac{1 - h_x^2}{h_x^2} \cdot \Bigg\{\frac{1 + \gamma_x + \lambda - \sqrt{(1 - \gamma_x - \lambda )^2 + 4 \lambda}}{2 \sqrt{(1 - \gamma_x - \lambda )^2 + 4 \lambda}} \Bigg\} .
\end{split}
\end{equation*}
When $\lambda_{\RA, M}^{*} = \gamma_x \cdot (1 - h_x^2)/h_x^2$, we have the optimal $\MSE$
\begin{align*}
    R_{\RA}(\lambda_{\RA, M}^{*}) = \kappa \sigma_\beta^2\cdot \frac{h_x^4 \left\{2 (M_{h_x,\gamma_x}-2) \gamma_x-d+1\right\} -(M_{h_x,\gamma_x}-2) h_x^2 \gamma_x+\gamma_x^2}{2 M_{h_x,\gamma_x} h_x^4 \gamma_x}
\end{align*}
where $M_{h_x,\gamma_x} = \sqrt{{\gamma_x^2}/{h_x^4}+\left({2}/{h_x^2}-4\right) \gamma_x+1}$. 
\end{proposition}

Proposition \ref{prop: ridge regression mse + R2} provides insights into the asymptotic behavior of $\hat{\bbeta}_{\RA}(\lambda)$, which align with the expressions derived in previous RMT studies such as \cite{dobriban2018high} and \cite{zhao2022genetic}. 
From these closed-form results,  it is clear that $\hat{\bbeta}_{\RA}(\lambda)$ has higher prediction accuracy than $\hat{\bbeta}_{\RW}(\lambda)$. 
Specifically, the main difference between $A_{\RW}^2(\lambda)$ and $A_{\textnormal{R}}^2(\lambda)$ lies in the constant $c$ given in Equation \eqref{eqn: constant c}. 
As $c$ is always positive, it follows that $A^2_{\textnormal {R}}(\lambda)$ is always greater than $A^2_{\RW}(\lambda)$. 
Additionally, it is worth noting that $A^2_{\RW}(\lambda)$ is  a monotonically increasing function with respect to $\lambda$, while $A^2_{\textnormal {R}}(\lambda)$ has an optimal $\lambda$ value. 

Proposition \ref{prop: comparison of L_2} provides a formal comparison between $\hat{\bbeta}_{\RA}(\lambda)$ and $\hat{\bbeta}_{\RW}(\lambda)$ when the sample size of the training data and reference panel is the same ($n_{x} = n_{w}$). These results indicate that $\hat{\bbeta}_{\RW}(\lambda)$  performs worse in terms of prediction accuracy compared to $\hat{\bbeta}_{\RA}(\lambda)$. These findings align with the results presented in Proposition~\ref{prop: comparison of L_1} for the comparison between $\hat{\bbeta}_{\LW}(\lambda)$ and $\hat{\bbeta}_{\textnormal{L}}(\lambda)$.

\begin{proposition} \label{prop: comparison of L_2}
    Under the assumptions of Theorem \ref{thm: i.i.d. ridge mse + R2} and Proposition \ref{prop: ridge regression mse + R2}, 
    when $n_{x} = n_{w}$, we have
    \begin{align*}
        \min_{\lambda \in \RR_+} R_{\RW}(\lambda) > \min_{\lambda \in \RR_+} R_{\textnormal{R}}(\lambda) \quad \mbox{and} \quad  \max_{\lambda \in \RR_+} A^2_{\RW}(\lambda) < \max_{\lambda \in \RR_+} A^2_{\RA}(\lambda).
    \end{align*}
\end{proposition}

\subsection{\texorpdfstring{$L_2$}{TEXT} regularized estimators with general \texorpdfstring{$\bmSigma$}{TEXT} }\label{sec: L_2 estimator2}
Next, we establish the AMP results of $\hat{\bbeta}_{\RW}(\lambda)$ for general $\bmSigma$. 
The generalized AMP recursion approximating for $\hat{\bbeta}_{\RW}(\lambda)$ is given by 
\begin{align*}
    \sqrt{p}{\bbeta}^{t+1} &= (\bI_p + \lambda (1 + c_t) \bmSigma^{-1})^{-1} \Bigg\{  -  \frac{\sqrt{N}}{n_{w}} {\bSigma}^{-1} \W^{\T}{\rb^{t}} + \frac{\sqrt{p}}{n_{x}} (1+c_t) {\bSigma}^{-1} \X^{\T} \y_{x} + \sqrt{p} {\bbeta}^{t} \Bigg\} \quad \mbox{and} \\
    \rb^{t} &= \sqrt{\frac{p}{N}} \Wb {\bbeta}^{t} + \frac{c_t}{1 + c_{t-1}} \rb^{t-1},
\end{align*}
where we reiterate $N = n_w + n_x$ and $c_t$ is defined through the recursion
\begin{equation*}
    \frac{c_t}{1+ c_{t-1}} = \frac{1}{n_w} \tr \bigg[ \big\{\bI_p + \lambda (1 + c_t) \bmSigma^{-1} \big\}^{-1} \bigg].
\end{equation*}
Moreover, we define the function $G(\rho, c) = (G_1, G_2)(\rho, c): \RR^{2} \mapsto \RR^{2}$, where
\begin{align} \label{G_1, G_2}
\begin{split}
    G_{1}(\rho, c) =\ & \lim_{p \to \infty} \EE \left[\gamma_w \bnorm{ \frac{1}{\sqrt{p}} \big\{{\bI_p} + \lambda (1 + c) {\bSigma}^{-1} \big\}^{-1} \big\{\rho {\bSigma}^{-1/2} \Zb + (1 + c) \sqrt{p}{\bbeta_0} \big\}}_{\bSigma}^2 \right]\\
    &+ \gamma_x (1 + c)^2 \lim_{p \to \infty} \EE  \bnorm{{\bbeta_0}}_{\bSigma}^2/ h_{x}^2 \quad \mbox{and} \\
    G_{2}(\rho, c) =\ & \frac{1}{n_w} (1 + c) \tr \bigg[ \big\{\bI_p + \lambda (1 + c) \bmSigma^{-1} \big\}^{-1} \bigg].
\end{split}
\end{align}
We denote the state evolution $(\rho_{*}^2, c_*)$ to be the unique solution of 
\begin{equation} \label{ridge state evolution}
\begin{split}
    \rho_{*}^2 = G_1(\rho_*, c_*)\quad \mbox{and} \quad c_{*} = G_{2}(\rho_*, c_*).
\end{split}
\end{equation}
To establish the existence and uniqueness of a solution for Equation~\eqref{ridge state evolution}, we use a similar proof technique as used for Equation~\eqref{F_1, F_2}. Notably, the second line of Equation~\eqref{G_1, G_2} is independent of $\rho_*$. As a result, by directly applying the intermediate value theorem, we can confirm the existence of a unique solution for $c_*$. This observation significantly simplifies our proof regarding the existence and uniqueness of Equation~\eqref{ridge state evolution}. We present the following proposition to summarize our results.


\begin{proposition}
    For any $\lambda \in \RR_+$, the fixed point equation in Equation~\eqref{ridge state evolution} admits a unique solution $(\rho_*, c_*) \in \RR_+ \times \RR_+$. 
\end{proposition}


The fixed point solution of Equation \eqref{ridge state evolution} determines the asymptotic behavior of $\hat{\bbeta}_{\RW}(\lambda)$. 
Theorem~\ref{thm: general ref Ridge main without normalization} provides a characterization of the asymptotic behavior of $\hat{\bbeta}_{\RW}(\lambda)$ for any deterministic sequence of uniformly pseudo-Lipschitz function.

\begin{theorem} \label{thm: general ref Ridge main without normalization}
Let $\{\bbeta_0, \bepsilon_x,\bmeps_s, \bSigma, \bX,  \Sbb, \bW\}$ be a converging sequence of instances with $\PP\left({\bbeta_0}(p) \neq 0 \right) > 0$. 
Each row of $\X$, $\Sbb$, and $\Wb$ is i.i.d. Gaussian with mean $\bm 0$ and variance ${\bSigma}$.
For any deterministic sequence $\{ \phi^{p}: \RR^{p} \times \RR^p \mapsto \RR \}_{p \in \NN}$ of uniformly pseudo-Lipschitz functions, we have 
\begin{align} \label{eqn: general ridge main without normalization formula}
    \phi^{p} \big\{ \hat{\bbeta}_{\RW}(\lambda), \bbeta_0 \big\} \overset{P}{\approx} \EE\ \phi^{p} \left[ \frac{1}{\sqrt{p}} \big\{{\bI_p} + \lambda (1 + c_*) {\bSigma}^{-1} \big\}^{-1} \big\{\rho_{*} {\bSigma}^{-1/2} \Zb + (1 + c_*) \sqrt{p} \bbeta_0 \big\}, \bbeta_0 \right],
\end{align}
where $\Zb \sim N(0, {\bI_p})$ and is independent of $\bbeta_0$. The $\rho_*$ and $c_*$ are given in Equation \eqref{ridge state evolution}. 
\end{theorem}

By selecting the corresponding function $\phi^{p}$, Theorem~\ref{thm: ridge mse + R2} summarizes the $\MSE$ and out-of-sample $R^2$ of $\hat{\bbeta}_{\RW}(\lambda)$. 
\begin{theorem} \label{thm: ridge mse + R2}
Let $\{\bbeta_0, \bepsilon_x,\bmeps_s, \bSigma, \bX,  \Sbb, \bW\}$ be a converging sequence of instances with $\PP\left({\bbeta_0}(p) \neq 0 \right) > 0$. 
Each row of $\X$, $\Sbb$, and $\Wb$ is i.i.d. Gaussian with mean $\bm 0$ and variance ${\bSigma}$.
In probability, we have 
\$
R_{\RW}(\lambda) 
&= \lim_{p \to \infty} \bnorm{\hat{\bbeta}_{\RW}(\lambda) - \bbeta_0}_\Sigma^2 \\
&=\lim_{p \to \infty} \bnorm{ \frac{1}{\sqrt{p}} \big\{{\bI_p} + \lambda (1 + c_*) {\bSigma}^{-1} \big\}^{-1} \big\{\rho_{*} {\bSigma}^{-1/2} \Zb + (1 + c_*) \sqrt{p} \bbeta_0 \big\} - \bbeta_0}_{\Sigma}^2 \quad \textnormal{and} \\ 
A^2_{\RW}(\lambda) &= \lim_{p \to \infty} h_{s}^2 \cdot \frac{ \EE \left\langle {\bbeta}_0, \big\{{\bI_p} + \lambda(1 + c_*) {\bSigma}^{-1} \big\}^{-1} \big\{\rho_{*} {\bSigma}^{-1/2} \Zb + (1 + c_*) \sqrt{p}{\bbeta}_0 \big\} \right\rangle_{\Sigma}^2}{\EE \bnorm{\bbeta_0}_{\bSigma}^2 \cdot \EE \bnorm{ \big\{{\bI_p} + \lambda(1 + c_*) {\bSigma}^{-1} \big\}^{-1} \big\{\rho_{*} {\bSigma}^{-1/2} \Zb + (1 + c_*) \sqrt{p} {\bbeta}_0 \big\}}_\Sigma^2},
\$
where $(\rho_{*}, c_*)$ satisfy the fixed point equation in Equation \eqref{ridge state evolution}. 
\end{theorem}
}

{\bxz
\section{UK Biobank data analysis} \label{sec: numerical experiments}

\subsection{Simulation studies with real genotype data}
In this section, we perform simulations using real genotype data from the UKB study. After downloading the genotype data, we apply the following quality control (QC) steps: 1) exclude subjects with more than 10\% missing genotypes; 2) exclude SNPs with minor allele frequency (MAF) < 0.01; 3) exclude SNPs with larger than 10\% missing genotype rate; 4) exclude SNPs that fail the Hardy-Weinberg equilibrium test at \textit{p}-value $=1\times 10^{-7}$ level. After QCs, 488,371 subjects and 461,488 SNPs remain. Furthermore, we limit our analysis to the 366,355 unrelated individuals of British ancestry from the QC'ed dataset. We set $h_x^2 = h_s^2 = 0.3$ and $0.6$ and randomly select 0.05\%  of the 461,488 SNPs to have nonzero genetic effects, which are independently generated from a Gaussian distribution \citep{yang2011gcta}. 

We examine the following genetic prediction methods: 1) lasso for large-scale SNP data \citep{qian2020fast}; 2) elastic-net for large-scale SNP data \citep{qian2020fast}; and 3) lassosum, a popular reference panel lasso estimator developed by \cite{mak2017polygenic}. Ridge estimator for large-scale SNP data has been implemented by \cite{qian2020fast} but is not performed due to computational resource limits. We experiment different training sample size $n_x$ for all three methods, and various reference panels of different sizes and data sources for reference panel lasso. For reference panel lasso, the training samples consisted of 50,000, 100,000, and 200,000 randomly selected subjects from the 366,355 unrelated British individuals in the UKB study. For lasso and elastic-net, due to computation resource limits, the training sample size is 50,000 and 100,000. Both validation and test sets have 20,000 randomly selected subjects that are independent of training samples. 
The reference panels in the reference panel lasso come from four datasets: the UKB training set, the UKB validation set, the UKB testing set, and the 1000 Genomes reference panel of European ancestry.  The UKB reference panels have various sizes ranging from $25$ to $50,000$. All UKB reference panel subjects are selected at random. The 1000 Genomes reference panel of European ancestry has 503 subjects. 
We perform 100 simulation replications for each choice of heritability and sparsity. 
The hyperparameters for each method are automatically selected by the software. 

\begin{figure}[!ht] 
\includegraphics[page=1,width=0.8\linewidth]{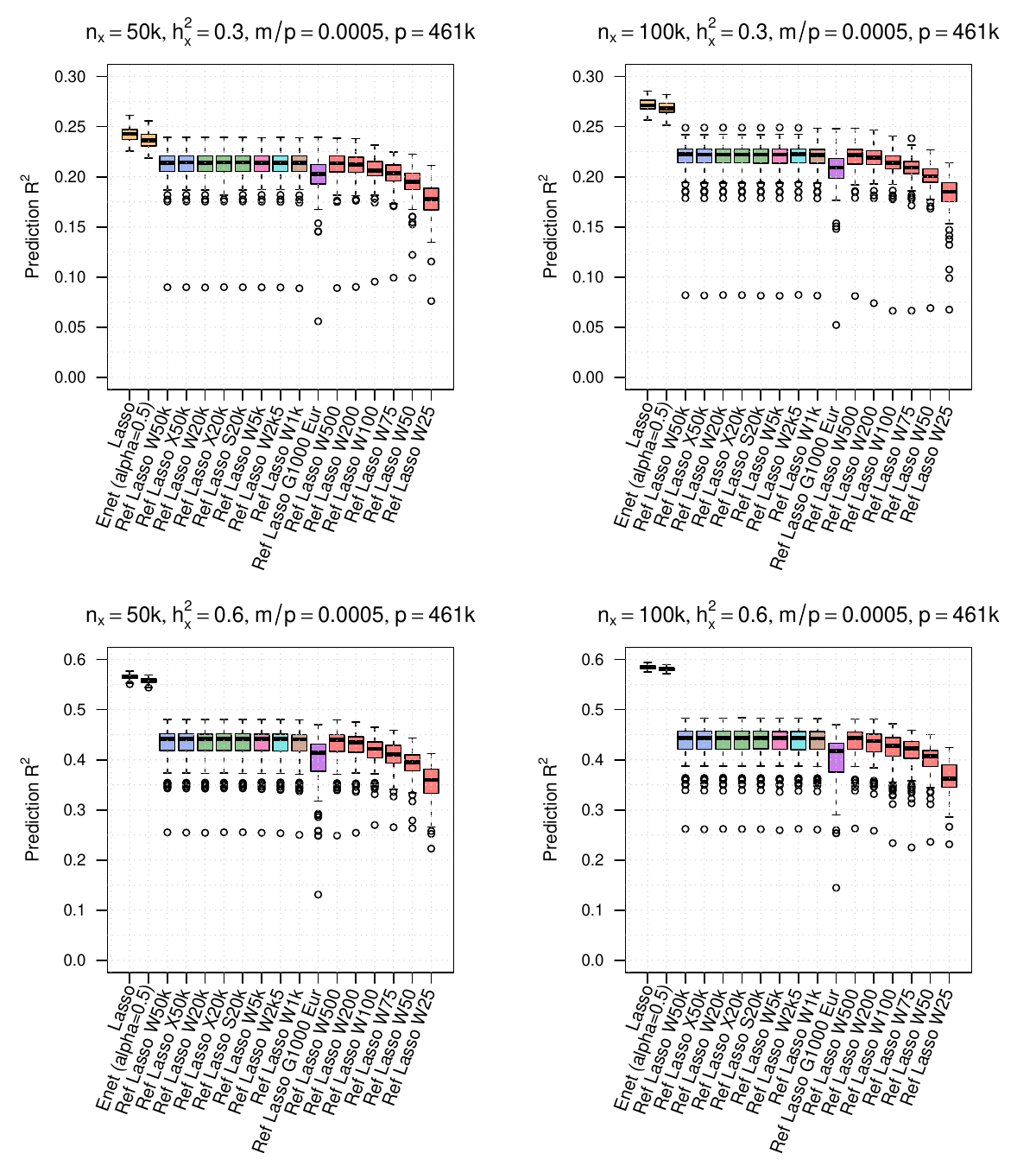}
\centering
\caption{\textbf{Out-of-sample $R^2$ of different estimators in UKB simulation data analysis.}
We set $p=461,488$, $m/p=0.0005$, $n_x=50,000$ or $100,000$, $n_s=20,000$, $h_x^2=h_s^2=0.3$ or $0.6$, and varying reference panel size $n_w$ from $25$ to $50,000$. 
From left to right: Lasso, $L_1$ regularized estimator; Enet, Elastic-net estimator; Ref Lasso W:
reference panel-based $L_1$ estimator, 
the reference panel is 
independent of the training and testing datasets;
Ref Lasso X:
the reference panel is from the training dataset; 
Ref Lasso Z: the reference panel is from the testing dataset. Subjects of all the above reference panels are from the UKB study. 
Ref Lasso G1000 Eur: we use the 1000 Genomes reference panel of European ancestry. 
}
\label{numerical_simulated_mp_0_0005}
\end{figure}

The results are displayed in Figure~\ref{numerical_simulated_mp_0_0005}. 
We have the following observations. 
First, we consistently observe that lasso prediction accuracy outperforms reference panel lasso for all choices of reference panels. This finding aligns with the numerical illustration of our theoretical results presented in Figure~\ref{main_fig_2}.
As we increase the heritability $h_x^2=h_s^2$ from 0.3 to 0.6, the performance gap between lasso and reference panel lasso widens. Similarly, as the training sample size $n_x$ increases from 50,000 to 100,000, the performance gap also increases. 
This suggests that lasso is generally more preferable when the training set is relatively large and genetic signals are strong. However, in cases where the training sample size is limited and the traits are less heritable, such as under $n_x=50,000$ and $h_x^2=h_s^2=0.3$,
reference panel lasso can be a viable alternative with a small performance gap compared to lasso. 
Under the scenario with $h_x^2=h_s^2=0.3$, both lasso and reference panel lasso estimators demonstrate very similar performance. 
Second, we observe that the prediction accuracy of the reference panel lasso decreases as the size of the reference panel sample size $n_w$ decreases to less than 500. 
It aligns with our theoretical findings, as illustrated in Figure~\ref{main_fig_2}.
Regardless of the choice of heritability and training sample size, a small reference panel size adversely affects the performance of the reference panel lasso estimator.


It is worth mentioning that lasso and reference panel lasso have different computational requirements, which is an important consideration in genetic studies. 
In Supplementary Figure~\ref{runtime_ram_simulated_lasso_lassosum_ntrain100k_h06}, we provide a comparison of the runtime and memory usage for lasso and reference panel lasso. 
We consider scenarios where $n_x=~$100,000 or 200,000, $h_x^2=h_s^2=0.6$, and $m/p=0.0005$. 
Generally, the runtime of lasso is longer and requires much more memory than reference panel lasso. Meanwhile, the runtime and memory usage of reference panel lasso increase quadratically with the reference panel size $n_w$. We also observe that the runtime and memory usage of reference panel lasso remains unaffected by the sample size $n_x$, which is expected. This is because the training set $(\Xb, \yb_x)$ influences the reference panel estimator through the summary statistics $\Xb^{\T} \yb_x$, and the runtime of reference panel lasso does not depend on $n_x$ when $\Xb^{\T} \yb_x$ is given. On the other hand, lasso under $n_x=200,000$ is not executed due to the substantial memory requirements. In summary, our simulation results are in line with the theoretical findings, reinforcing their validity in real genotype data. 



\subsection{Retinal imaging data analysis} \label{section:real data analysis}
In this section, we perform a real data analysis using retinal imaging biomarkers from the UKB study. Specifically, we focus on 46 imaging traits obtained from optical coherence tomography images \citep{zhao2023eye}. These traits include retinal thickness across different layers \citep{ko2017associations,patel2016spectral} and vertical cup-to-disc ratio \citep{han2019genome}. A complete list of the 46 traits can be found in 
Supplementary Figure~\ref{numerical_eye_lasso_ref_lasso_ridge_all}.
The UKB retinal imaging dataset comprises 45,148 unrelated British individuals. We randomly divide this dataset into a training set of 36,054 subjects ($n_x=36,054$), a validation set of 4,428 subjects, and a testing set of 4,666 subjects ($n_s=4,666$). 
For the reference panel, we use 
20,000 unrelated UKB British individuals that are independent of the training and testing subjects. We also use an external reference panel from the 1000 Genomes dataset, which represents individuals of European ancestry.
We evaluate the following methods for analysis: 1) lasso for large-scale SNP data \citep{qian2020fast}, 2) ridge for large-scale SNP data \citep{qian2020fast}, and 3) reference panel lasso \citep{mak2017polygenic}. The hyperparameters for these methods are automatically selected by the software used.


%

Figure~\ref{numerical_eye_lasso_ref_lasso_indep_g1000eur_ridge_top10} highlights the top 10 imaging biomarkers with the highest overall prediction accuracy using  different estimators. 
Among these imaging biomarkers, lasso consistently achieves the highest prediction accuracy. The reference panel lasso with the UKB reference panel generally performs slightly better than the reference panel lasso with the 1000 Genomes reference panel, although both reference panel lasso methods show {broadly} similar performance. On the other hand, ridge consistently exhibits the lowest prediction accuracy, likely due to the {relatively low level of sparsity present in the signals of} these retinal imaging biomarkers.
The comparable performance of the two reference panel-based lasso methods suggests that a carefully constructed reference panel, such as the 1000 Genomes reference panel, is indeed effective in practice. It is worth noting that the performance of the 1000 Genomes reference panel on real data in Figure~\ref{numerical_eye_lasso_ref_lasso_indep_g1000eur_ridge_top10} is better than its performance on simulated data in Figure~\ref{numerical_simulated_mp_0_0005}. This difference could be attributed to the fact that the SNPs in the 1000 Genomes reference panel only partially overlap with the 461,488 SNPs in our {QC’ed} dataset. In the simulated data, the causal SNPs are randomly selected from the 461,488 SNPs, which means that the 1000 Genomes reference panel may not include all causal SNPs due to the partial overlap.

\begin{figure}[htbp] 
\includegraphics[page=1,angle=0, width=0.8\linewidth]{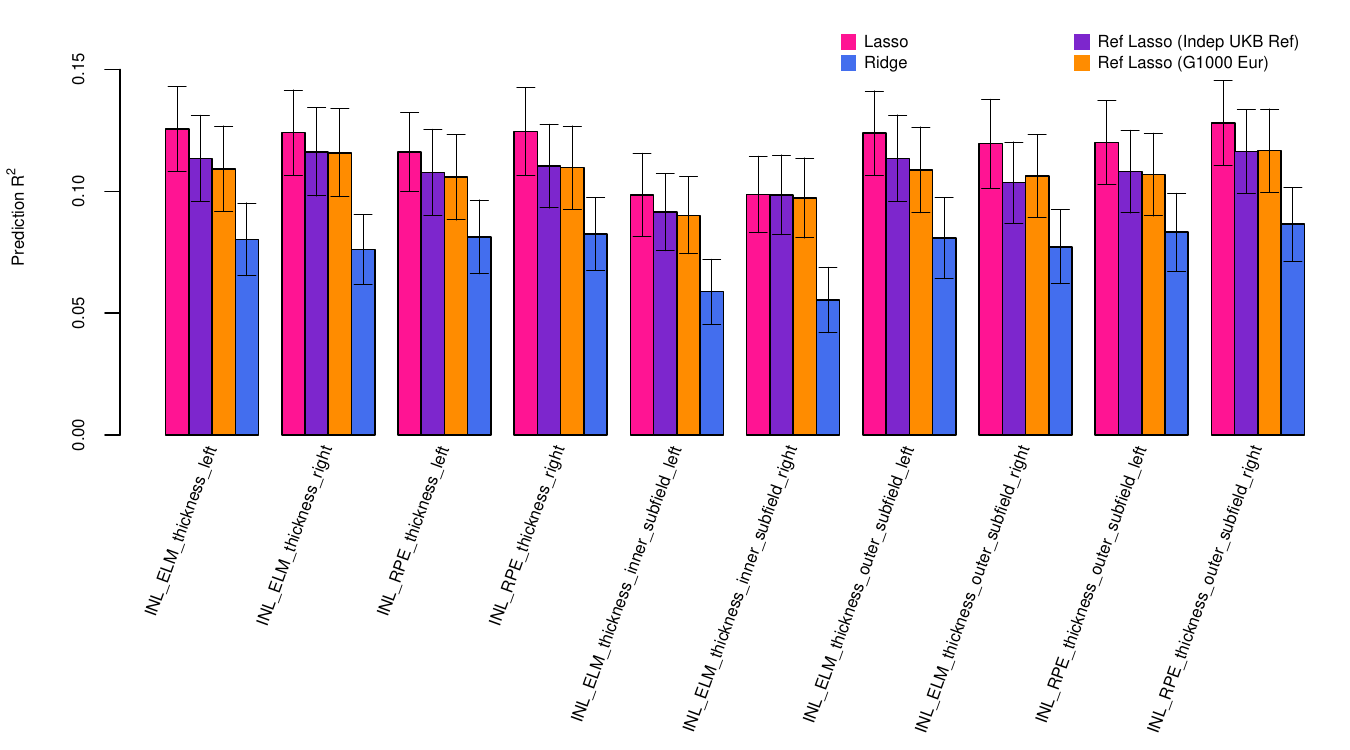}
\centering
\caption{\textbf{Out-of-sample $R^2$ of different estimators in UKB retinal imaging data analysis.}
We show the prediction accuracy for $10$ retinal imaging traits. 
Here $p=461,488$, $n_x=36,054$, $n_w=20,000$, and $n_s=4,666$. 
Different colors indicate the $L_1$ and $L_2$ regularized estimators (Lasso and Ridge), as well as the reference panel-based $L_1$ regularized estimators using UKB data (Indep UKB Ref) and 1000 Genomes (G1000 Eur) as reference panel.  
The standard error of prediction accuracy of each method is calculated using 500 independent bootstrap samples of size $4,666$.
}
\label{numerical_eye_lasso_ref_lasso_indep_g1000eur_ridge_top10}
\end{figure}

Supplementary Figure~\ref{numerical_eye_lasso_ref_lasso_ridge_all} provides a comprehensive overview of the prediction accuracy of different estimators for all 46 imaging biomarkers. Among these biomarkers, 21 of them are left-right pairs, and the prediction accuracy of all methods exhibits high symmetry across these pairs.
Lasso outperforms reference panel lasso for 36 out of the 46 biomarkers. On average, lasso achieves a 5.27\% relative increase in prediction accuracy compared to reference panel lasso. 
When focusing solely on these 36 traits, the average relative increase in prediction accuracy rises to 8.59\%. Notably, the largest gap between lasso and reference panel lasso is observed for the average inner nuclear layer thickness of the left eye, with a gap of 0.0163 in prediction accuracy and a 23.60\% relative increase compared to the reference panel lasso. The corresponding right eye trait also exhibits a relatively large performance gap of 0.0103 and a 13.33\% relative increase compared to the reference panel lasso.

Furthermore, we observe that lasso tends to outperform reference panel lasso when the overall prediction accuracy is relatively high. This is supported by the positive correlation of 0.7411 between the lasso-reference panel lasso performance gap and the lasso prediction accuracy. These findings align with our theoretical results and the analysis of simulated data. For example, in the right panel of Figure~\ref{main_fig_1}, the prediction accuracy gap between traditional and reference panel-based $L_1$ estimators widens as heritability increases. This is further supported by the simulated data analysis shown in Figure~\ref{numerical_simulated_mp_0_0005},
where the prediction accuracy gap between lasso and reference panel lasso estimators widens as heritability $h_x^2=h_s^2$ increases from 0.3 to 0.6. Additionally, when comparing the performance of lasso, reference panel lasso, and ridge estimator across all 46 traits, we find that lasso and reference panel lasso estimators outperform the ridge estimator on 43 out of the 46 traits. Moreover, the prediction accuracy achieved by all three methods is relatively small when the ridge is the best-performing method, with prediction accuracy ranging from 0.0106 to 0.0181.

Overall, our real data analysis provides insights into the prediction accuracy of different estimators for retinal imaging biomarkers in the UKB  study.
These results highlight the better performance of lasso compared to reference panel lasso across a majority of the 46 imaging biomarkers. The substantial performance gaps observed in certain biomarkers indicate the advantages of using lasso for specific traits in the UKB retinal imaging data analysis.



}

\section{Discussion} \label{sec: discussion}

This paper introduces a novel non-separable matrix AMP framework for analyzing reference panel-based regularized estimators in high-dimensional genetic data prediction. Our analysis provides insights into the asymptotic prediction performance of these estimators, considering general data structures represented by $\bmSigma$, and highlights their distinctions from traditional regularized estimators. We model and demonstrate the impact of key factors that influence the performance of reference panel-based estimators, such as model sparsity, heritability, training data sample size, and reference panel sample size. Importantly, we observe similar phenomena for both $L_1$ and $L_2$ regularized estimators. Our theoretical findings align with results obtained from simulation studies using real genotype data and real data analyses using the {UKB} database.

For the special case where $\bmSigma = \bI_p$, we establish the convergence of the $\MSE$ and out-of-sample $R^2$ in Theorems \ref{thm: i.i.d. lasso mse + R2} and \ref{thm: i.i.d. ridge mse + R2} as convergence in probability. These theorems are derived as special cases of Theorems \ref{thm: general lasso mse + R2} and \ref{thm: ridge mse + R2}, which are proved using non-separable matrix AMP techniques for a general $\bmSigma$ and provide convergence in probability results. It is worth noting that by directly following the approach of \cite{javanmard2013state} and replicating the proof provided in Appendix \ref{general - state - proof}, it is possible to replace convergence in probability in Theorems \ref{thm: i.i.d. lasso mse + R2} and \ref{thm: i.i.d. ridge mse + R2} with almost sure convergence for i.i.d. data.

The developed AMP techniques and the comprehensive understanding gained through our analyses contribute to the field of high-dimensional genetic data prediction, shedding light on the strengths and limitations of reference panel-based estimators. Overall, our findings from theoretical, simulated, and real data analyses consistently demonstrate the superiority of traditional regularized estimators over reference panel-based estimators, particularly in scenarios {where high prediction accuracy is observed.} Therefore, our results emphasize the importance of understanding the specific characteristics of the trait under investigation and the real dataset, as well as the desired trade-off between prediction accuracy and computational efficiency when selecting an appropriate estimator for genetic data prediction. Additionally, our study highlights the practical significance of developing larger reference panels. These statistical insights offer valuable guidance to researchers in the field, empowering them to make informed decisions regarding the choice of estimator and the construction of infrastructure for genetic data analysis.

\bibliographystyle{apalike}
\bibliography{ref}

\newpage
\appendix 
\renewcommand{\thetable}{S.\arabic{table}}
\renewcommand{\thefigure}{S.\arabic{figure}}
\renewcommand{\thesection}{S.\arabic{section}}
\renewcommand{\thelemma}{S.\arabic{lemma}}
\renewcommand{\theproposition}{S.\arabic{proposition}}
\renewcommand{\thecondition}{S.\arabic{condition}}

\addcontentsline{toc}{section}{Appendix} 
\part{Appendix} 
\parttoc 

\section{Proofs for Section \ref{sec: case study}}
\label{sec: Proof for sec: case study}

Theorem \ref{thm: i.i.d. lasso mse + R2} can be viewed as a special case of Theorem \ref{thm: general lasso mse + R2}. Therefore, we omit the proof in this section.
We now prove Proposition \ref{prop: comparison of L_1} in Section \ref{sec: case study} under Condition \ref{cond: comparison techinical}. 
Recall that Condition \ref{cond: comparison techinical} (a) assumes that each coordinate of $\bbeta_0$, denoted as $\bbeta_{0,i}$, is an i.i.d. random variable that follows
\$
(1 - \kappa) \delta(0) + \kappa N(0, \sigma_{\bbeta}^2/p).
\$
Therefore, we have
\begin{align} \label{eqn:beta_bar}
\barbeta \sim (1 - \kappa) \delta(0) + \kappa N(0, \sigma_{\bbeta}^2).
\end{align}
We point out that Condition \ref{cond: comparison techinical} is only used in Sections \ref{sec: case study} and \ref{sec: Proof for sec: case study}, while the main results on $R_{\LW}(\lambda)$, $A^2_{\LW}(\lambda)$ assume a generic distribution of $\bbeta_0$.

The proof of Proposition \ref{prop: comparison of L_1} is outlined as follows. 
In  Section \ref{sec: l1 and ref l1 close form}, we compute an equivalent expressions of $R_{\LW}(\lambda)$, $A^2_{\LW}(\lambda)$, $R_{\textnormal{L}}(\lambda)$, and $A^2_{\textnormal{L}}(\lambda)$ based on Theorem \ref{thm: i.i.d. lasso mse + R2} and Proposition \ref{prop: i.i.d. non-ref lasso mse + R2}.
In Section \ref{proof of prop: comparison of L_1}, we compare $R_{\LW}(\lambda), A^2_{\LW}(\lambda)$, $R_{\LA}(\lambda)$, and $ A^2_{\LA}(\lambda)$ utilizing their simplified expressions. These simplified expressions  are also used in the numerical experiments in Sections \ref{sec: numerical experiments} and \ref{sec: Supplementary Figures}.

\subsection{Simplified expressions for risk measures of \texorpdfstring{$L_1$}{TEXT} regularized estimators}
\label{sec: l1 and ref l1 close form}

Section \ref{sec: ref l1 close form} simplifies the expressions of $R_{\LW}(\lambda)$ and $A^2_{\LW}(\lambda)$, while Section \ref{sec: l1 closed form} provides simplified forms for $R_{\textnormal{L}}(\lambda)$ and $A^2_{\textnormal{L}}(\lambda)$. 
The derivations in this section heavily depend on Condition \ref{cond: comparison techinical}. 
These simplified formulations 
serves as a ground for proving Proposition \ref{prop: comparison of L_1} in Section \ref{proof of prop: comparison of L_1}.

\subsubsection{Reference panel-based \texorpdfstring{$L_1$}{TEXT} regularized estimator with isotropic features}
\label{sec: ref l1 close form}
In this section, our objective is to provide simplified expressions of $R_{\LW}(\lambda)$ and $A^2_{\LW}(\lambda)$ when $\bmSigma = \bI_p$. 
The resulting equations are presented in equations~\eqref{eqn: S.2.1.1-8} and \eqref{eqn: S.2.1.1-9}, independent of $\eta_{\textnormal{soft}}$ and expectation. 
Recall that the fixed point equations in \eqref{eqn: i.i.d. l1 state evo} are given by 
\begin{equation*}
\begin{split}
    &\tilde{\tau}^2_* = \gamma_x (1 + \tilde{b}_*)^2 \kappa \sigma_{\bbeta}^2/h_x^2 + \gamma_w \EE \eta_{\soft}^2\big\{ \tilde{\tau}_* z + (1+\tilde{b}_{*})\barbeta, \lambda (1 + \tilde{b}_*) \big\} \quad \mbox{and}\\
    &\frac{\tilde{b}_{*}}{1 + \tilde{b}_{*}} = \gamma_w \EE \eta'_{\soft} \big\{ \tilde{\tau}_* z + (1 + \tilde{b}_*)\barbeta, \lambda (1 + \tilde{b}_*) \big\},
\end{split}
\end{equation*}
where $z \sim N(0,1)$ is independent of $\barbeta$. 
Equivalently, let $\zeta_* = (1 + \Tilde{b}_*)/\Tilde{\tau}_*$, we have
\begin{align} \label{eqn: S.2.1.1-1}
    &1 =  \gamma_x \zeta_*^2 \kappa \sigma_{\bbeta}^2/h_x^2 + \gamma_w \EE \eta_{\soft}^2(z + \zeta_* \barbeta, \lambda \zeta_*)\quad \mbox{and} \\
    \label{eqn: S.2.1.1-2}
    &\frac{\tilde{b}_{*}}{1 + \tilde{b}_{*}} = \gamma_w \EE \eta'_{\soft} (z + \zeta_*\barbeta, \lambda \zeta_*) = \gamma_w \cdot \PP\left(|z + \zeta_*\barbeta| > \lambda \zeta_* \right).
\end{align}
We begin by rewriting \eqref{eqn: S.2.1.1-1} in a form that does not involve the soft-thresholding function $\eta_{\textnormal{soft}}$ and the expectation. 
We can express $\EE_{z, \barbeta}\ \eta_{\soft}^2(z + \zeta_* \barbeta, \lambda \zeta_*)$ as
\begin{align*}
\begin{split}
    &\EE_{z, \barbeta}\ \eta_{\soft}^2(z + \zeta_* \barbeta, \lambda \zeta_*)\\
    =\ &  \EE_{z, \barbeta}\ \eta_{\soft}^2(z + \zeta_* \barbeta, \lambda \zeta_*) \one(z + \zeta_* \barbeta > \lambda \zeta_*) + \EE \eta_{\soft}^2(z + \zeta_* \barbeta, \lambda \zeta_*) \one(z + \zeta_* \barbeta < - \lambda \zeta_*)\\
    =\ & \EE_{z, \barbeta} \left\{( z + \zeta_* \barbeta - \zeta_* \lambda )^2 \one(z + \zeta_* \barbeta > \lambda \zeta_*) + ( z + \zeta_* \barbeta + \zeta_* \lambda )^2 \one(z + \zeta_* \barbeta < - \lambda \zeta_*) \right\}\\
    =\ & (1 - \kappa) \cdot \EE_{z, \barbeta} \left\{ ( z + \zeta_* \barbeta -   \zeta_* \lambda )^2 \one(z + \zeta_* \barbeta > \lambda \zeta_*) + ( z + \zeta_* \barbeta + \zeta_* \lambda )^2 \one(z + \zeta_* \barbeta < - \lambda \zeta_*) \bigg|\ \barbeta = 0 \right\}\\ 
    &+ \kappa \cdot \EE_{z, \barbeta}\ \left\{ ( z + \zeta_* \barbeta - \zeta_* \lambda )^2 \one(z + \zeta_* \barbeta > \lambda \zeta_*) + ( z + \zeta_* \barbeta + \zeta_* \lambda )^2 \one(z + \zeta_* \barbeta < - \lambda \zeta_*) \bigg|\ \barbeta \neq 0 \right\}.
\end{split}
\end{align*}
By Equation \eqref{eqn:beta_bar}, we have
\begin{align*}
    &\EE_{z, \barbeta}\ \eta_{\soft}^2(z + \zeta_* \barbeta, \lambda \zeta_*)\\
    &= (1 - \kappa) \cdot \EE_{z}\ \left\{ ( z - \zeta_* \lambda )^2 \one(z  > \lambda \zeta_*) + ( z + \zeta_* \lambda )^2 \one(z < - \lambda \zeta_*)  \right\}\\ 
    & \qquad + \kappa \cdot \EE_{z, \bar{G}}\ \bigg\{ ( z + \zeta_* \bar{G} - \zeta_* \lambda )^2 \one(z + \zeta_* \bar{G} > \lambda \zeta_*) \\
    &\qquad \qquad+ ( z + \zeta_* \bar{G} + \zeta_* \lambda )^2 \one(z + \zeta_* \bar{G} < - \lambda \zeta_*) \bigg|\ \bar{G} \sim N(0, \sigma_{\bbeta}^2) \bigg\}.
\end{align*}
Let $G = z + \zeta_* \bar{G}$.
When $\bar{G} \sim N(0, \sigma_{\bbeta}^2)$, we have 
\begin{align*}
    G \sim N(0, \zeta_*^2 \cdot \sigma_{\bbeta}^2 + 1).
\end{align*}
Then it follows that 
\begin{align}\label{eqn: S.2.1.1-6}
\begin{split}
    &\EE_{z, \barbeta}\ \eta_{\soft}^2(z + \zeta_* \barbeta, \lambda \zeta_*)\\
    &= \kappa \cdot \EE_{G} \left\{ ( G - \zeta_* \lambda )^2 \one(G > \lambda \zeta_*) + ( G + \zeta_* \lambda )^2 \one(G < - \lambda \zeta_*) \right\}\\ 
    &\qquad + (1 - \kappa) \cdot \EE_{z} \left\{ ( z - \zeta_* \lambda )^2 \one(z  > \lambda \zeta_*) + ( z + \zeta_* \lambda )^2 \one(z < - \lambda \zeta_*)  \right\}
    \\
    &=: \kappa \cdot \Rom{1} + (1 - \kappa) \cdot \Rom{2}.
\end{split}
\end{align}
We will evaluate the two terms $\Rom{1}$ and $\Rom{2}$ separately. First, since both $G$ and $z$ are Gaussian random variables and thus are symmetric, we have 
\begin{align*}
    \Rom{1} =\ &\EE_{G} \left\{ ( G - \zeta_* \lambda )^2 \one(G > \lambda \zeta_*) + ( G + \zeta_* \lambda )^2 \one(G < - \lambda \zeta_*) \right\}\\
    =\ & \EE_{G} \left\{ G^2 \one(|G| > \lambda \zeta_*) + \lambda^2 \zeta_*^2 \one(|G| > \lambda \zeta_*) - 4  \zeta_* \lambda \cdot G \one(G > \lambda \zeta_*) \right\} \\
    =\ & (1 + {\zeta_*^2 \sigma_{\beta}^2} + \zeta_*^2 \lambda^2 )\cdot \left\{1 - \erf \left(\sqrt{\frac{ \zeta_*^2 \lambda^2}{2 + 2 \zeta_*^2 \sigma_{\beta}^2}}\right) \right\} - \sqrt{\frac{2}{\pi}} \zeta_* \lambda \cdot e^{-\frac{ \zeta_*^2 \lambda^2}{2 + 2 \zeta_*^2 \sigma_{\beta}^2}} \sqrt{1 + {\zeta_*^2 \sigma_{\beta}^2}}.
\end{align*}
Similarly, we can write the the second term $\Rom{2}$  as
\begin{align*}
    \Rom{2} =\ &\EE_{z} \left\{ \left( z - \zeta_* \lambda \right)^2 \one(z  > \lambda \zeta_*) + \left( z + \zeta_* \lambda \right)^2 \one(z < - \lambda \zeta_*)  \right\}\\
    =\ & (1 + \zeta_*^2 \lambda^2) \cdot \left\{ 1 - \erf\left(\frac{\zeta_* \lambda}{\sqrt{2}}\right) \right\} - \sqrt{\frac{2}{\pi}} \zeta_* \lambda e^{-\zeta_*^2 \lambda^2/2}.
\end{align*}
Then the fixed point equation defined in Equation~\eqref{eqn: S.2.1.1-1} for $\zeta_* \in \RR_{+}$ can be rewritten as  
\begin{equation} \label{eqn:S.2.1.1-3}
\begin{split}
    1 =\ & \frac{\gamma_x \kappa \zeta_*^2 \sigma_{\beta}^2}{h_x^2} + \gamma_w (1 - \kappa) \cdot \left[(1 + \zeta_*^2 \lambda^2) \cdot \left\{ 1 - \erf\left(\frac{\zeta_* \lambda}{\sqrt{2}}\right) \right\} - \sqrt{\frac{2}{\pi}} \zeta_* \lambda e^{-\zeta_*^2 \lambda^2/2} \right]\\
    &+ \gamma_w \kappa \cdot \Bigg[\left(1 + {\zeta_*^2 \sigma_{\beta}^2} + \zeta_*^2 \lambda^2 \right)\cdot \left\{1 - \erf \left(\sqrt{\frac{\zeta_*^2 \lambda^2}{2 + 2 \zeta_*^2 \sigma_{\beta}^2}}\right) \right\}\\
    &- \sqrt{\frac{2}{\pi}} \zeta_* \lambda \cdot e^{-\frac{\zeta_*^2 \lambda^2}{2 + 2 \zeta_*^2 \sigma_{\beta}^2}} \sqrt{1 + {\zeta_*^2 \sigma_{\beta}^2}} \Bigg].
\end{split}
\end{equation}

We proceed to simplify the second fixed point equation. 
The $\tilde{b}_*$ in equation~\eqref{eqn: S.2.1.1-2} satisfies 
\begin{align*}
    \frac{\tilde{b}_{*}}{1 + \tilde{b}_{*}} =\ & \gamma_w \cdot \PP\left(|z + \zeta_*\barbeta| > \lambda \zeta_* \right)
    \\
    =\ & \gamma_w \kappa \cdot \PP\left(|G| > \lambda \zeta_* \right) + \gamma_w (1 - \kappa) \cdot \PP(|z| > \lambda \zeta_*)
    \\
    =\ & \gamma_w \kappa \cdot \Rom{3} + \gamma_w (1 - \kappa) \cdot \Rom{4},
\end{align*}
Terms $\Rom{3}$ and $\Rom{4}$ are
\begin{align*}
    \Rom{3} = \left\{1 - \erf\left(\sqrt{\frac{\zeta_*^2 \lambda^2}{2 + 2\zeta_*^2 \sigma_{\beta}^2}} \right) \right\}, \quad \Rom{4} = \left\{1 - \erf\left(\sqrt{\frac{\zeta_*^2 \lambda^2}{2}} \right) \right\}. 
\end{align*}
Therefore,
\begin{align} \label{eqn:S.2.1.1-4}
    \frac{\tilde{b}_{*}}{1 + \tilde{b}_{*}} =\ & \gamma_w \kappa \left\{1 - \erf\left(\sqrt{\frac{\zeta_*^2 \lambda^2}{2 + 2\zeta_*^2 \sigma_{\beta}^2}} \right) \right\} + \gamma_w (1 - \kappa) \left\{1 - \erf\left(\sqrt{\frac{\zeta_*^2 \lambda^2}{2}} \right) \right\}.
\end{align}
Using these results, we can obtain the simplified expressions for $R_{\LW}(\lambda)$ and $\A^2_{\LW}(\lambda)$ as
\begin{align*} 
\begin{split}
    R_{\LW}(\lambda) =\ & \EE \bigg[\eta_{\soft}\{(1 + \tilde{b}_*)\bar{\beta} + \tilde{\tau}_{*} z, \lambda (1 + \tilde{b}_*)\} - \bar{\beta} \bigg]^2,
    \\
    A_{\LW}(\lambda) =\ & h_s \cdot \frac{\EE \eta_{\soft}(\zeta_* \bar{\beta} + z, \lambda \zeta_*) \cdot \bar{\beta}}{\sqrt{\EE {\barbeta}^2} \cdot \sqrt{\EE \eta^2_{\soft}(\zeta_* \bar{\beta} + z, \lambda \zeta_*)}}.
\end{split}
\end{align*}
We have
\begin{align*}
    &\EE \big\{ \eta_{\soft}(\zeta_* \bar{\beta} + z, \lambda \zeta_*) \cdot \bar{\beta} \big\}\\
    =\ & \EE \left\{ (\zeta_* \bar{\beta} + z - \lambda \zeta_*) \cdot \bar{\beta} \one(\zeta_* \bar{\beta} + z > \lambda\zeta_*) \right\} + \EE \left\{ (\zeta_* \bar{\beta} + z + \lambda \zeta_*) \cdot \bar{\beta} \one(\zeta_* \bar{\beta} + z < - \lambda\zeta_*) \right\} \\
    =\ & \EE \left\{ (\zeta_* \bar{\beta} - \lambda \zeta_*) \cdot \bar{\beta} \one(z > \lambda\zeta_* - \zeta_* \bar{\beta}) \right\} + \EE \left\{(\zeta_* \bar{\beta} + \lambda \zeta_*) \cdot \bar{\beta} \one(z < - \lambda\zeta_* - \zeta_* \bar{\beta}) \right\}\\
    &+ \EE \left\{z \cdot \bar{\beta} \one(z > \lambda\zeta_* - \zeta_* \bar{\beta}) \right\} + \EE \left\{ z \cdot \bar{\beta}\one (z < - \lambda\zeta_* - \zeta_* \bar{\beta} ) \right\}\\
    =\ & \kappa \EE \left\{ (\zeta_* \bar{G} - \lambda \zeta_*) \cdot \bar{G} \one \left(\bar{G} > \frac{\lambda\zeta_* - z}{\zeta_*} \right) \right\} + \kappa \EE \left\{(\zeta_* \bar{G} + \lambda  \zeta_*) \cdot \bar{G} \one \left(\bar{G} < \frac{- \lambda\zeta_* - z}{\zeta_*} \right) \right\}\\
    &+ \kappa \EE \left\{\frac{1}{\sqrt{2 \pi}} e^{-(\lambda\zeta_* - \zeta_* \bar{G})^2/2} \cdot \bar{G} \right\} - \kappa \EE \left\{\frac{1}{\sqrt{2 \pi}} e^{-(\lambda\zeta_* + \zeta_* \bar{G})^2/2} \cdot \bar{G}\right\},
\end{align*}
where $\bar{G} \sim N(0, \sigma_{\bbeta}^2)$. 
Therefore, we have
\begin{align} \label{eqn: S.2.1.1-7}
\begin{split}
    &\EE \left\{ \eta_{\soft}(\zeta_* \bar{\beta} + z, \lambda \zeta_*) \cdot \bar{\beta} \right\}\\
    =\ & \kappa \EE_z \left\{- \frac{1}{\sqrt{2 \pi}} {\sigma_{\beta}} e^{-\frac{(\zeta_* \lambda - z)^2}{2 \zeta_*^2 \sigma_\beta^2}} z + \frac{\zeta_* \sigma_{\beta}^2}{2} +  \frac{\zeta_* \sigma_\beta^2}{2} \erf \left(\frac{z - \lambda \zeta_*}{\sqrt{2} \zeta_* \sigma_{\beta}} \right) \right\}\\
    &+ \kappa \EE_z \left\{\frac{1}{\sqrt{2 \pi}} \sigma_{\beta} e^{-\frac{(z + \lambda \zeta_*)^2}{2 \zeta_*^2 \sigma_\beta^2}} z + \frac{\zeta_* \sigma_{\beta}^2}{2} - \frac{\zeta_* \sigma_\beta^2}{2} \erf \left(\frac{z + \lambda \zeta_*}{\sqrt{2} \zeta_* \sigma_{\beta}} \right) \right\}\\
    &+ \kappa \EE_{\bar{G}} \left\{ \frac{1}{\sqrt{2 \pi}} e^{-(\lambda\zeta_* - \zeta_* \bar{G})^2/2} \cdot \bar{G} \right\} - \kappa \EE_{\bar{G}} \left\{\frac{1}{\sqrt{2 \pi}} e^{-(\lambda\zeta_* + \zeta_* \bar{G})^2/2} \cdot \bar{G} \right\}.
\end{split}
\end{align}
As before, we define $\EE \left\{ \eta_{\soft}(\zeta_* \bar{\beta} + z, \lambda \zeta_*) \cdot \bar{\beta} \right\} =: \Rom{5} + \Rom{6}$ where
\begin{align*}
    \Rom{5} :=\ & \frac{\zeta_* \sigma_\beta^2}{2} \left\{ \kappa \EE_z  \erf \left(\frac{z - \lambda \zeta_* }{\sqrt{2} \zeta_* \sigma_{\beta}} \right) - \kappa \EE_z \erf \left(\frac{z + \lambda \zeta_*}{\sqrt{2} \zeta_* \sigma_{\beta}} \right) \right\}\\
    \Rom{6} :=\ & \kappa \EE_z \left\{ - \frac{1}{\sqrt{2 \pi}} \sigma_{\beta} e^{-\frac{(\zeta_* \lambda - z)^2}{2 \zeta_*^2 \sigma_\beta^2}} z \right\} + \frac{\zeta_* \sigma_{\beta}^2}{2}  + \kappa \EE_z \left\{ \frac{1}{\sqrt{2 \pi}} \sigma_{\beta} e^{-\frac{(\zeta_* \lambda + z)^2}{2 \zeta_*^2 \sigma_\beta^2}} z \right\} + \frac{\zeta_* \sigma_{\beta}^2}{2} \\ 
    &+ \kappa \EE_{\bar{G}} \left\{ \frac{1}{\sqrt{2 \pi}} e^{-(\lambda\zeta_* - \zeta_* \bar{G})^2/2} \cdot \bar{G} \right\} - \kappa \EE_{\bar{G}} \left\{\frac{1}{\sqrt{2 \pi}} e^{-(\lambda\zeta_* + \zeta_* \bar{G})^2/2} \cdot \bar{G} \right\}
\end{align*}
We will evaluate the summation of $\Rom{5}$ and $\Rom{6}$. 
By Formula 13 in Section 4.3 of \cite{ng1969table}, we have 
\begin{align*}
    \Rom{5} =\ & \frac{\zeta_* \sigma_\beta^2}{2} \left\{ \kappa \EE_z  \erf \left(\frac{z - \lambda \zeta_* }{\sqrt{2} \zeta_* \sigma_{\beta}} \right) - \kappa \EE_z \erf \left(\frac{z + \lambda \zeta_*}{\sqrt{2} \zeta_* \sigma_{\beta}} \right) \right\}\\
    =\ & - \kappa {\zeta_* \sigma_\beta^2} \cdot \erf \left(\frac{\lambda \zeta_*}{\sqrt{2 \zeta_*^2 \cdot {\sigma_{\beta}^2} + 2}}\right).
\end{align*}
Moreover, $\Rom{6}$ is
\begin{align*}
    \Rom{6} =\ &\kappa \EE_z \left\{ - \frac{1}{\sqrt{2 \pi}} \sigma_{\beta} e^{-\frac{(\zeta_* \lambda - z)^2}{2 \zeta_*^2 \sigma_\beta^2}} z \right\} + \frac{\zeta_* \sigma_{\beta}^2}{2}  + \kappa \EE_z \left\{ \frac{1}{\sqrt{2 \pi}} \sigma_{\beta} e^{-\frac{(\zeta_* \lambda + z)^2}{2 \zeta_*^2 \sigma_\beta^2}} z \right\} + \frac{\zeta_* \sigma_{\beta}^2}{2} \\ 
    &+ \kappa \EE_{\bar{G}} \left\{ \frac{1}{\sqrt{2 \pi}} e^{-(\lambda\zeta_* - \zeta_* \bar{G})^2/2} \cdot \bar{G} \right\} - \kappa \EE_{\bar{G}} \left\{\frac{1}{\sqrt{2 \pi}} e^{-(\lambda\zeta_* + \zeta_* \bar{G})^2/2} \cdot \bar{G} \right\}\\
    =\ & \kappa \frac{e^{-\frac{\zeta_*^2 \lambda^2}{2( 1+ \zeta_*^2 \sigma_{\beta}^2)}} \sqrt{\frac{2}{\pi}}  \zeta_*^2 \lambda \sqrt{\zeta_*^2 + \frac{1}{\sigma_{\beta}^2}} \sigma_{\beta}^4}{\sigma_{\beta} (1 + \zeta_*^2 \sigma_{\beta}^2)^2} + \kappa \zeta_* \sigma_{\beta} \left\{ \sigma_{\beta} - \frac{e^{-\frac{\zeta_*^2 \lambda^2}{2(1 + \zeta_*^2 \sigma_{\beta}^2)}} \sqrt{\frac{2}{\pi}} \lambda }{\sqrt{1 + \frac{1}{\sigma_{\beta}^2 \zeta_*^2}} (1 + \zeta_*^2 \sigma_{\beta}^2)} \right\}\\
    =\ & \kappa{\zeta_* \sigma_{\beta}^2}.
\end{align*}
It follows that 
\begin{align} \label{eqn: S.2.1.1-5}
    \EE \left\{ \eta_{\soft}(\zeta_* \bar{\beta} + z, \lambda \zeta_*) \cdot \bar{\beta} \right\}
    = \kappa {\zeta_* \sigma_{\beta}^2} - \kappa {\zeta_* \sigma_\beta^2} \cdot \erf \left( \frac{\lambda \zeta_*}{\sqrt{2 \zeta_*^2 \cdot {\sigma_{\beta}^2} + 2}}\right).
\end{align}
Combining equations~\eqref{eqn: S.2.1.1-6} and \eqref{eqn: S.2.1.1-5}, the simplified expression of $R_{\LW}(\lambda)$ is
\begin{align} \label{eqn: S.2.1.1-8}
\begin{split}
    &R_{\LW}(\lambda)\\
    &=\EE \bigg[\eta_{\soft}\{(1 + \tilde{b}_*)\bar{\beta} + \tilde{\tau}_{*} z, \lambda (1 + \tilde{b}_*)\} - \bar{\beta} \bigg]^2\\
    &= \EE \bigg\{\tilde{\tau}_* \eta_{\soft}(\zeta_* \bar{\beta} + z, \lambda \zeta_*) - \bar{\beta} \bigg\}^2\\
    &= \frac{\Tilde{\tau}_*^2}{\gamma_w} \left( 1 + \gamma_x \zeta_*^2 \frac{\kappa \sigma_{\bbeta}^2}{h_x^2} \right) + {\kappa \sigma_{\bbeta}^2} - 2 \Tilde{\tau}_* \left\{\kappa {\zeta_* \sigma_{\beta}^2} - \kappa {\zeta_* \sigma_\beta^2} \cdot \erf \left( \frac{\lambda \zeta_*}{\sqrt{2 \zeta_*^2 \cdot {\sigma_{\beta}^2} + 2}}\right) \right\}.
\end{split}
\end{align}
Moreover, the simplified expression of $A_{\LW}(\lambda)$ is
\begin{align} \label{eqn: S.2.1.1-9}
\begin{split}
    A_{\LW}(\lambda)
    &= h_s \cdot \frac{\EE \eta_{\soft}(\zeta_* \bar{\beta} + z, \lambda \zeta_*) \cdot \bar{\beta}}{\sqrt{\EE {\barbeta}^2} \cdot \sqrt{\EE \eta^2_{\soft}(\zeta_* \bar{\beta} + z, \lambda \zeta_*)}}\\
    &= h_s \cdot \frac{\kappa {\zeta_* \sigma_{\beta}^2} - \kappa {\zeta_* \sigma_\beta^2} \cdot \erf \left( \frac{\lambda \zeta_*}{\sqrt{2 \zeta_*^2 \cdot {\sigma_{\beta}^2} + 2}}\right)}{\sqrt{{\kappa \sigma_{\bbeta}^2}} \cdot \sqrt{(1  - \gamma_x \zeta_*^2 {\kappa \sigma_{\bbeta}^2}/{h_x^2})/\gamma_w}}\\
   & = h_s \zeta_* \cdot \sqrt{{\kappa \sigma_{\bbeta}^2}} \cdot \frac{1 - \erf \left( \frac{\lambda \zeta_*}{\sqrt{2 \zeta_*^2 \cdot {\sigma_{\beta}^2} + 2}}\right)}{\sqrt{(1 - \gamma_x \zeta_*^2 {\kappa \sigma_{\bbeta}^2}/{h_x^2})/\gamma_w}}.
\end{split}
\end{align}

\subsubsection{The \texorpdfstring{$L_1$}{TEXT} regularized estimator with isotropic features}
\label{sec: l1 closed form}


Similar to the previous section, we first express the simplified expression of the fixed point equations~\eqref{eqn:S.1.1.2-2} and \eqref{eqn:S.1.1.2-3}. 
Subsequently, we present the simplified results for $R_{\LA}(\lambda)$ and $A^2_{\LA}(\lambda)$.
Recall from \eqref{eqn: lasso state evolution}, when $\bmSigma = \bI_p$, the fixed point equations  of the $L_1$ regularized estimator for $(\bar{\tau}_*, \bar{b}_*)$ are:
\begin{align} \label{eqn:S.1.1.2-2}
    &\bar{\tau}_*^2 = \gamma_x \kappa \sigma_{\bbeta}^2 \frac{1 - h_x^2}{h_x^2} + \gamma_x \EE \big[ \eta_{\soft} \{\barbeta + \bar{\tau}_* z, \lambda (1 + \bar{b}_*)\} - \barbeta \big]^2
    \\ \label{eqn:S.1.1.2-3}
    &(1 + \bar{b}_*)^{-1} = 1 - \gamma_x \EE \eta'_{\soft}\{ \barbeta + \bar{\tau}_* z, \lambda (1 + \bar{b}_*)\}.
\end{align}
We aim to simplify the following expressions on $R_{\textnormal{L}} (\lambda)$ and $A_{\textnormal{L}}(\lambda)$: 
\begin{align} \label{eqn:S.1.1.2-4}
    R_{\textnormal{L}} (\lambda) =\ & \EE \big[ \eta_{\soft} \{\barbeta + \bar{\tau}_* z, \lambda (1 + \bar{b}_*)\} - \barbeta \big]^2
    \\ \label{eqn:S.1.1.2-5}
    A_{\textnormal{L}}(\lambda) =\ & h_{s} \cdot \frac{ \EE \left[\barbeta \cdot \eta_{\soft}\{\bar{\tau}_* z + \bar{\beta}, \lambda (1 + \bar{b}_*)\} \right]}{\sqrt{\EE \bar{\beta}^2 \cdot \EE \eta^2_{\soft}\{\bar{\tau}_* z + \bar{\beta}, \lambda (1 + \bar{b}_*)\}}}.
\end{align}
To this end, we begin by computing the equivalent expression for  \eqref{eqn:S.1.1.2-2}:  
\begin{align*}
    &\bar{\tau}_*^2 - \gamma_x \kappa \sigma_{\bbeta}^2 \frac{1 - h_x^2}{h_x^2}\\
    =\ & \gamma_x \EE  \big[ \eta_{\soft} \{\barbeta + \bar{\tau}_* z, \lambda (1 + \bar{b}_*)\} - \barbeta  \big]^2\\
    =\ & \gamma_x \EE \big[ \{\barbeta + \bar{\tau}_* z - \lambda (1 + \bar{b}_*)\} - \barbeta  \big]^2 \one\{\barbeta + \bar{\tau}_* z > \lambda (1 + \bar{b}_*)\}\\ 
    &+ \gamma_x \EE \big[ \{\barbeta + \bar{\tau}_* z + \lambda (1 + \bar{b}_*)\} - \barbeta \big]^2 \one\{\barbeta + \bar{\tau}_* z < - \lambda (1 + \bar{b}_*)\}\\
    =\ & \gamma_x \EE \left[ \big\{ \bar{\tau}_* z - \lambda (1 + \bar{b}_*) \big\}^2 \one\{\barbeta + \bar{\tau}_* z > \lambda (1 + \bar{b}_*)\} \right] + \gamma_x \EE \left[ \big\{ \bar{\tau}_* z + \lambda (1 + \bar{b}_*) \big\}^2 \one\{\barbeta + \bar{\tau}_* z < - \lambda (1 + \bar{b}_*)\} \right]\\
    =\ & \gamma_x \EE \left[ \big\{ \bar{\tau}_* z - \lambda (1 + \bar{b}_*) \big\}^2 \one \left\{z > \frac{\lambda (1 + \bar{b}_*) - \barbeta}{\bar{\tau}_*} \right\} \right] + \gamma_x \EE \left[ \big\{ \bar{\tau}_* z + \lambda (1 + \bar{b}_*) \big\}^2 \one\left\{z < \frac{- \lambda (1 + \bar{b}_*)- \barbeta }{\bar{\tau}_*} \right\} \right].
    \\
    =\ & \gamma_x (1 - \kappa) \EE \left[ \big\{ \bar{\tau}_* z - \lambda (1 + \bar{b}_*) \big\}^2 \one \left\{z > \frac{\lambda (1 + \bar{b}_*)}{\bar{\tau}_*} \right\} \right] + \gamma_x \EE \left[ \big\{ \bar{\tau}_* z + \lambda (1 + \bar{b}_*) \big\}^2 \one\left\{z < \frac{- \lambda (1 + \bar{b}_*) }{\bar{\tau}_*} \right\} \right]\\
    &+ \gamma_x \kappa \EE \left[ \big\{ \bar{\tau}_* z - \lambda (1 + \bar{b}_*) \big\}^2 \one \left\{z > \frac{\lambda (1 + \bar{b}_*) - \bar{G}}{\bar{\tau}_*} \right\} \right] + \gamma_x \EE \left[ \big\{ \bar{\tau}_* z + \lambda (1 + \bar{b}_*) \big\}^2 \one\left\{z < \frac{- \lambda (1 + \bar{b}_*)- \bar{G} }{\bar{\tau}_*} \right\} \right].
\end{align*}
Therefore, $\bar{\tau}_*$ satisfies
\begin{align} \label{eqn:S.1.1.2-6}
\begin{split}
\bar{\tau}_{*}^2 
=& \kappa \gamma_x \sigma_{\beta}^2 \cdot \frac{1-h_x^2}{h_x^2} +  
\kappa \gamma_x \cdot \bar{\tau}_{*}^2 
\left\{ \left({\frac{\lambda (1 + \bar{b}_*)}{\bar{\bar{\tau}_{*}}_*}} \right)^2+1 \right\} \left\{1 - \erf \left(\frac{1}{\sqrt{2}} \sqrt{\frac{\lambda^2 (1 + \bar{b}_*)^2}{\bar{\tau}_{*}^2 + \sigma^2_{\beta}}} \right) \right\} 
\\
&- \kappa \gamma_x \cdot \bar{\tau}_{*}^2  \sqrt{\frac{2}{\pi}} \sqrt{\frac{\lambda^2 (1 + \bar{b}_*)^2}{\bar{\tau}_{*}^2 + \sigma^2_{\beta}}} \left( 1+\frac{\sigma^2_{\beta}}{\bar{\tau}_{*}^2+\sigma^2_{\beta}} \right) \exp\left( -\frac{1}{2} \frac{\lambda^2 (1 + \bar{b}_*)^2}{\bar{\tau}_{*}^2+\sigma^2_{\beta}} \right) 
\\
&+
\kappa \gamma_x \cdot 
\left\{ 
\sigma^2_{\beta} \cdot \erf\left(\frac{1}{\sqrt{2}} \sqrt{\frac{\lambda^2 (1 + \bar{b}_*)^2}{\bar{\tau}_*^2 + \sigma^2_{\beta}}} \right)
- \sqrt{\frac{2}{\pi}} {{\lambda (1 + \bar{b}_*)}} \sigma_{\beta} \left( \frac{\sigma^2_{\beta}}{\bar{\tau}_*^2+\sigma^2_{\beta}} \right)^{3/2} \exp\left( -\frac{1}{2} \frac{\lambda^2 (1 + \bar{b}_*)^2}{\bar{\tau}_*^2+\sigma^2_{\beta}} \right) \right\} 
\\
&+ (1-\kappa) \gamma_x \cdot \tau^2 \left[
\left\{\left(\frac{\lambda (1 + \bar{b}_*)}{\bar{\tau}_*}\right)^2+1 \right\} \left\{ 1-\erf\left({\frac{\lambda (1 + \bar{b}_*)}{\sqrt{2} \bar{\tau}_*}}\right) \right\} - \sqrt{\frac{2}{\pi}} {\frac{\lambda (1 + \bar{b}_*)}{\bar{\tau}_*}} \exp\left\{ - \frac{1}{2} \left(\frac{\lambda (1 + \bar{b}_*)}{\bar{\tau}_*}\right)^2 \right\}
\right].
\end{split}
\end{align}
Let $\bar{G} \sim N(0, \sigma_{\bbeta}^2)$. Then Equation \eqref{eqn:S.1.1.2-3} is equivalent to
\begin{align} \label{eqn:S.1.1.2-7}
\begin{split}
    (1 + \bar{b}_*)^{-1} =& 1 - \gamma_x \EE \eta'_{\soft}\{\barbeta + \bar{\tau}_* z, \lambda (1 + \bar{b}_*)\} \\
    =\ & 1 - \gamma_x \PP \left\{|\barbeta + \bar{\tau}_* z| > \lambda (1 + \bar{b}_*) \right\}\\
    =\ & 1 - \gamma_x (1 - \kappa) \PP \left\{|\bar{\tau}_* z| > \lambda (1 + \bar{b}_*) \right\} - \gamma_x \kappa \PP \left\{ |\bar{G} + \bar{\tau}_* z| > \lambda (1 + \bar{b}_*) \right\}\\
    =\ & 1 - \gamma_x (1 - \kappa) \left[ 1 - \erf\left\{\frac{(1 + \bar{b}_*) \lambda}{\sqrt{2} \bar{\tau}_*} \right\} \right] - \gamma_x \kappa \left[ 1 - \erf\left\{\frac{(1 + \bar{b}_*) \lambda}{\sqrt{2(\bar{\tau}_*^2 + \sigma_{\bbeta}^2)}} \right\} \right].
\end{split}
\end{align}
Therefore, $R_{\textnormal{L}}(\lambda)$ in \eqref{eqn:S.1.1.2-4}  is 
\begin{align*}
    R_{\textnormal{L}} (\lambda)= \bar{\tau}_*^2/\gamma_x - \kappa \sigma_{\beta}^2 \cdot \frac{1-h_x^2}{h_x^2},
\end{align*}
with $\bar{\tau}_*$ satisfying equations \eqref{eqn:S.1.1.2-6} and \eqref{eqn:S.1.1.2-7}. We proceed to simplify  $A^2_{\textnormal{L}}(\lambda)$. 
Let $\Tilde{G} \sim N(0, \bar{\tau}_*^2 + \sigma_{\beta}^2)$.
Then $\EE \eta^2_{\soft}\{\bar{\tau}_* z + \bar{\beta}, \lambda (1 + \bar{b}_*)\} $ in the denominator of Equation \eqref{eqn:S.1.1.2-5} can be written as
\#
    &\EE \eta^2_{\soft}\{\bar{\tau}_* z + \bar{\beta}, \lambda (1 + \bar{b}_*)\} \nn\\
    &= (1 - \kappa) \EE \eta^2_{\soft}\{\bar{\tau}_* z, \lambda (1 + \bar{b}_*)\} + \kappa \EE \eta^2_{\soft}\{\Tilde{G}, \lambda (1 + \bar{b}_*)\}  \nn \\
    &= \kappa \EE \{\Tilde{G} - \lambda (1 + \bar{b}_*)\}^2 \one\{\Tilde{G} > \lambda (1 + \bar{b}_*)\} + \kappa \EE \{\Tilde{G} + \lambda (1 + \bar{b}_*)\}^2 \one\{\Tilde{G} < - \lambda (1 + \bar{b}_*)\}  \nn  \\
    &\qquad + (1 - \kappa) \EE \{\bar{\tau}_* z - \lambda (1 + \bar{b}_*)\}^2 \one\{\bar{\tau}_* z > \lambda (1 + \bar{b}_*)\} \nn \\
    &\qquad + (1 - \kappa) \EE \{\bar{\tau}_* z + \lambda (1 + \bar{b}_*)\}^2 \one\{\bar{\tau}_* z < - \lambda (1 + \bar{b}_*)\} \nn \\
    &= \kappa \left\{ (1 + \bar{b}_*)^2 \lambda^2 + \sigma_{\beta}^2 + \bar{\tau}_*^2 \right\} \cdot \left\{1 - \erf\left(\frac{(1 + \bar{b}_*) \lambda}{\sqrt{2 \sigma_{\beta}^2 + 2 \bar{\tau}_*^2}} \right) \right\} - \kappa (1 + \bar{b}_*) \sqrt{\frac{2}{\pi}} \lambda \sqrt{ \sigma_{\beta}^2 + \bar{\tau}_*^2} e^{-\frac{(1 + \bar{b}_*)^2 \lambda^2}{2 \sigma_{\beta}^2 + 2 \bar{\tau}_*^2}} \nn \\
    &\qquad + (1 - \kappa) \left\{ (1 + \bar{b}_*)^2 \lambda^2 + \bar{\tau}_*^2 \right\} \cdot \left\{1 - \erf\left(\frac{(1 + \bar{b}_*) \lambda}{\sqrt{2}{\bar{\tau}_*}} \right) \right\} - (1 - \kappa) (1 + \bar{b}_*) \sqrt{\frac{2}{\pi}} \lambda {\bar{\tau}_*} e^{-\frac{(1 + \bar{b}_*)^2 \lambda^2}{2 \bar{\tau}_*^2}}.  \label{eqn: S.2.1.2-1}
\#
Combining above pieces together, $A_{\textnormal{L}}(\lambda)$ simplifies to
\begin{align*}
    A_{\textnormal{L}}(\lambda) =\ & h_{s} \cdot \frac{ \EE \left[\barbeta \cdot \eta_{\soft}\{\bar{\tau}_* z + \bar{\beta}, \lambda (1 + \bar{b}_*)\} \right]}{\sqrt{\EE \bar{\beta}^2 \cdot \EE \eta^2_{\soft}\{\bar{\tau}_* z + \bar{\beta}, \lambda (1 + \bar{b}_*)\}}}\\
    =\ & h_{s} \cdot \frac{\EE \bar{\beta}^2 + \EE \eta^2_{\soft}\{\bar{\tau}_* z + \bar{\beta}, \lambda (1 + \bar{b}_*)\} - \EE [\bar{\beta} - \eta_{\soft}\{\bar{\tau}_* z + \bar{\beta}, \lambda (1 + \bar{b}_*)\}]^2}{2 \sqrt{\EE \bar{\beta}^2 \cdot \EE \eta^2_{\soft}\{\bar{\tau}_* z + \bar{\beta}, \lambda (1 + \bar{b}_*)\}}}\\
    =\ & h_{s} \cdot \frac{\kappa \sigma_{\beta}^2 - \bar{\tau}_*^2/\gamma_x  + \kappa \sigma_{\beta}^2 \cdot (1-h_x^2)/h_x^2 + \EE \eta^2_{\soft}\{\bar{\tau}_* z + \bar{\beta}, \lambda (1 + \bar{b}_*)\} }{2 \sqrt{\kappa \sigma_{\beta}^2 \cdot \EE \eta^2_{\soft}\{\bar{\tau}_* z + \bar{\beta}, \lambda (1 + \bar{b}_*)\}}}.
\end{align*}
with $\EE \eta^2_{\soft}\{\bar{\tau}_* z + \bar{\beta}, \lambda (1 + \bar{b}_*)\}$ being given in Equation~\eqref{eqn: S.2.1.2-1}.

\subsection{Proof of Proposition \ref{prop: comparison of L_1}}
\label{proof of prop: comparison of L_1}

In this section, we use ${\lambda^*_{\LW, M}}$ and ${\lambda^*_{\LA, M}}$ to denote the optimal tuning parameters minimizing $R_{\LW}(\lambda)$ and $R_{\LA}(\lambda)$, respectively. 
Similarly, 
we use ${\lambda^*_{\LW, R}}$ and ${\lambda^*_{\LA, R}}$ to denote  the optimal tuning parameters that maximize $A^2_{\LW}(\lambda)$ and $A^2_{\LA}(\lambda)$, respectively. 



\paragraph{Comparing $R_{\LW}(\lambda_{\LW, M}^*)$ and $R_{\textnormal{L}}(\lambda_{\LA, M}^*)$:}

We first compare 
$R_{\LW}(\lambda_{\LW, M}^*)$ 
and 
$R_{\textnormal{L}}(\lambda_{\LA, M}^*)$.
To show $R_{\LW}(\lambda_{\LW, M}^*) > R_{\textnormal{L}}(\lambda_{\LA, M}^*)$, we proceed in two steps:  
\begin{enumerate}
  \item[1] We find an immediate quantity  $\Tilde{R}_{\LW}(\lambda_{\LW, M}^*)$ such that $R_{\LW}(\lambda_{\LW, M}^*) > \Tilde{R}_{\LW}(\lambda_{\LW, M}^*)$,
  \item[2] We show $\Tilde{R}_{\LW}(\lambda_{\LW, M}^*) > R_{\textnormal{L}}(\lambda_{\LA, M}^*)$. 
\end{enumerate}

\paragraph{Step 1}
We first rewrite $R_{\LW}(\lambda_{\LW, M}^*)$ in terms of $\Theta_* = \Theta_*({\lambda})$, where  $\Theta_*$ is the solution to the following fixed point equation:
\# \label{eqn: S.2.1.2-3}
    \Theta_*^2 
    &= \sigma_{\beta}^2 \kappa \gamma_x/h_x^2 + \gamma_w \EE {\BS \eta_{\soft}}^2(\Theta_* z + \barbeta, {\lambda}) \nn \\
    &= \sigma_{\beta}^2 \kappa \gamma_x/h_x^2 + \gamma_w \kappa \bigg[ (\Theta_*^2 + {\lambda}^2 + \sigma_{\beta}^2) \cdot\erfc \left\{\frac{{\lambda}}{\sqrt{2(\Theta_*^2 + \sigma_{\beta}^2)}} \right\}  - e^{- \frac{{\lambda}^2}{2 (\Theta_*^2 + \sigma_{\beta}^2)}} \sqrt{\frac{2}{\pi}} {\lambda} \sqrt{\Theta_*^2 + \sigma_{\beta}^2} \bigg] \nn \\
    &\qquad + \gamma_w (1- \kappa) \bigg\{ (\Theta_*^2 + {\lambda}^2) \cdot\erfc \left(\frac{{\lambda}}{\sqrt{2} \Theta_*} \right)  - e^{- \frac{{\lambda}^2}{2 \Theta_*^2}} \sqrt{\frac{2}{\pi}} {\lambda} \Theta_* \bigg\}.
\#
Note  that Equation \eqref{eqn: S.2.1.2-3} is equivalent to both \eqref{eqn:S.2.1.1-3} and \eqref{eqn: i.i.d. l1 state evo} by substituting $\Theta_{*} = 1/\zeta_* = \tilde{\tau}_*/(1 + \tilde{b}_*)$. 
Similar to equation~\eqref{eqn: S.2.1.1-5}, we obtain 
\begin{align*}
    \EE \left\{ \eta_{\soft}(\bar{\beta} + \Theta_* z, {\lambda^*_{\LW, M}}) \cdot \bar{\beta} \right\}
    = \kappa \sigma_\beta^2 \cdot \erfc \left( \frac{{\lambda^*_{\LW, M}}}{\sqrt{2 {\sigma_{\beta}^2} + 2 \Theta_*^2}}\right).
\end{align*}
Therefore,
$R_{\LW}(\lambda_{\LW, M}^*)$ can be rewritten as
\begin{align*}
    R_{\LW}(\lambda_{\LW, M}^*) = \frac{(1 + \tilde{b}_*)^2}{\gamma_w} (\Theta_*^2  - \gamma_x {\kappa \sigma_{\bbeta}^2}/h_x^2) + {\kappa \sigma_{\bbeta}^2} - 2 (1 + \tilde{b}_*) \kappa {\sigma_{\beta}^2} \cdot\erfc \left\{\frac{{\lambda^*_{\LW, M}}}{\sqrt{2(\Theta_*^2 + \sigma_{\beta}^2)}} \right\}.
\end{align*}
A key observation is that $R_{\LW}(\lambda_{\LW, M}^*)$ is a decreasing function with respect to (\wrt) $b_* \in \RR_{+}$.
To verify this, differentiating $R_{\LW}(\lambda_{\LW, M}^*)$ w.r.t. $\tilde{b}_*$ yields
\begin{align*}
    \frac{\partial R_{\LW}(\lambda_{\LW, M}^*)}{\partial \tilde{b}_*} =\ & 2\tilde{b}_* (\Theta_*^2/\gamma_x - {\kappa \sigma_{\bbeta}^2}/h_x^2) + 2 \Bigg[\Theta_*^2/\gamma_x - {\kappa \sigma_{\bbeta}^2}/h_x^2 - \kappa {\sigma_{\beta}^2} \cdot\erfc \left\{\frac{{\lambda^*_{\LW, M}}}{\sqrt{2(\Theta_*^2 + \sigma_{\beta}^2)}} \right\} \Bigg]\\
    =\ & 2\tilde{b}_* (\Theta_*^2/\gamma_x - {\kappa \sigma_{\bbeta}^2}/h_x^2)\\
    &+ \kappa \Bigg[ (\Theta_*^2 + {\lambda^*_{\LW, M}}^2) \cdot\erfc \left\{\frac{{\lambda^*_{\LW, M}}}{\sqrt{2(\Theta_*^2 + \sigma_{\beta}^2)}} \right\}  - e^{- \frac{{\lambda^*_{\LW, M}}^2}{2 (\Theta_*^2 + \sigma_{\beta}^2)}} \sqrt{\frac{2}{\pi}} {\lambda^*_{\LW, M}} \sqrt{\Theta_*^2 + \sigma_{\beta}^2} \Bigg]\\
    &+ \gamma_w (1- \kappa) \bigg[ (\Theta_*^2 + {\lambda^*_{\LW, M}}^2) \cdot\erfc \left(\frac{{\lambda^*_{\LW, M}}}{\sqrt{2} \Theta_*} \right)  - e^{- \frac{{\lambda^*_{\LW, M}}^2}{2 \Theta_*^2}} \sqrt{\frac{2}{\pi}} {\lambda^*_{\LW, M}} \Theta_* \bigg]
    \\
    =\ & 2\tilde{b}_* (\Theta_*^2/\gamma_x - {\kappa \sigma_{\bbeta}^2}/h_x^2) + \kappa \Theta_*^2 \cdot\erfc \left\{\frac{{\lambda^*_{\LW, M}}}{\sqrt{2(\Theta_*^2 + \sigma_{\beta}^2)}} \right\} \\
    &+ \gamma_w (1- \kappa) \bigg[ (\Theta_*^2 + {\lambda^*_{\LW, M}}^2) \cdot\erfc \left(\frac{{\lambda^*_{\LW, M}}}{\sqrt{2} \Theta_*} \right)  - e^{- \frac{{\lambda^*_{\LW, M}}^2}{2 \Theta_*^2}} \sqrt{\frac{2}{\pi}} {\lambda^*_{\LW, M}} \Theta_* \bigg]> 0,
\end{align*}
where the last inequality follows from Condition \ref{cond: comparison techinical}~(c).
Thus we can obtain a lower bound of $R_{\LW}(\lambda_{\LW, M}^*)$, denoted as $\Tilde{R}_{\LW}(\lambda_{\LW, M}^*)$, by setting $\tilde{b}_* = 0$, and 
\begin{align*}
    \Tilde{R}_{\LW}(\lambda_{\LW, M}^*) := \Theta_*^2/\gamma_x  - \kappa \sigma_{\bbeta}^2 (1 - h_x^2)/{h_x^2} - 2 \kappa {\sigma_{\beta}^2} \cdot\erfc \left\{\frac{{\lambda^*_{\LW, M}}}{\sqrt{2(\Theta_*^2 + \sigma_{\beta}^2)}} \right\} 
    \leq R_{\LW}(\lambda_{\LW, M}^*)
\end{align*}

\paragraph{Step 2}
Define $\tilde\Theta_*^2= \tilde\Theta_*^2(\lambda^*_{\LW, M})$ to be
\begin{align} \label{eqn: S.14.9}
    \tilde\Theta_*^2 :=\ & \Theta_*^2 -  2 \gamma_x \kappa {\sigma_{\beta}^2} \cdot \erfc \left(\frac{{\lambda^*_{\LW, M}}}{\sqrt{2(\Theta_*^2 + \sigma_{\beta}^2)}} \right) < \Theta_*^2.
\end{align}
We  have 
$
    \Tilde{R}_{\LW}(\lambda_{\LW, M}^*) = \tilde\Theta_*^2/\gamma_x - \kappa \sigma_{\bbeta}^2 (1 - h_x^2)/{h_x^2}.
$
Then it suffices to find some $\lambda_{*} \in \RR_{\geq 0}$ such that  
\begin{align} \label{eqn: S.2.1.2-2}
    \Tilde{R}_{\LW}({\lambda^*_{\LW, M}}) {\BS \geq} R_{\textnormal{L}}(\lambda_{*}),
\end{align}
which then gives 
$$
R_{\LW}({\lambda^*_{\LW, M}}) > \Tilde{R}_{\LW}({\lambda^*_{\LW, M}}) {\BS \geq} R_{\textnormal{L}}(\lambda_{*}) {\BS \geq} R_{\textnormal{L}}(\lambda_{\LA, M}^{*}).
$$
Recall that  the state evolution equations for the $L_1$ regularized estimator at $\lambda_{*}$ are
\begin{align*} 
    \bar{\tau}(\lambda_{*})^2 =& \gamma_x \kappa \sigma_{\bbeta}^2 \frac{1 - h_x^2}{h_x^2} + \gamma_x \EE \left[ \eta_{\soft} \{\barbeta + \bar{\tau} z, \lambda_{*} (1 + \bar{b})\} - \barbeta \right]^2
\end{align*}
and 
\begin{align*}
    R_{\textnormal{L}}(\lambda_{*}) = \bar{\tau}(\lambda_{*})^2/\gamma_x - \kappa \sigma_{\bbeta}^2 (1 - h_x^2)/{h_x^2}.
\end{align*}
Therefore, finding a $\lambda_*$ that satisfies \eqref{eqn: S.2.1.2-2} is equivalent to finding a $\lambda_{*}$ that satisfies ${\BS \tilde\Theta_*^2} > \bar{\tau}(\lambda_{*})^2$. 
Using \eqref{eqn: S.14.9}, we have 
\begin{align*}
    \tilde\Theta_*^2/\gamma_x 
    &=  \Theta_*^2/\gamma_x - 2 \EE \left\{ \eta_{\soft}(\bar{\beta} + \Theta_* z, {\lambda^*_{\LW, M}}) \cdot \bar{\beta} \right\}\\
    &= \sigma_{\beta}^2 \kappa/h_x^2 + \EE {\BS \eta_{\soft}}^2(\Theta_* z + \barbeta, {\lambda^*_{\LW, M}}) - 2 \EE \left\{ \eta_{\soft}(\bar{\beta} + \Theta_* z, {\lambda^*_{\LW, M}}) \cdot \bar{\beta} \right\}\\
    &= \kappa\sigma_{\beta}^2\frac{1- h_x^2 }{h_x^2} + \EE \left\{{\BS \eta_{\soft}} (\Theta_* z + \barbeta, {\lambda^*_{\LW, M}}) - \bar{\beta} \right\}^2, 
\end{align*}
where we use the fact that $\barbeta \sim (1 - \kappa) \delta(0) + \kappa N(0, \sigma_{\bbeta}^2)$ in the last equality. 
It can be observed that the function $\EE \left\{{\BS \eta_{\soft}} (\Theta_* z + \barbeta, {\lambda^*_{\LW, M}}) - \bar{\beta} \right\}^2$ is strict monotone with respect to $\Theta_*$. 
To establish this, we consider its derivative \wrt~$\Theta_*$. 
We first write $\EE \left\{{\BS \eta_{\soft}}(\Theta_* z + \barbeta, {\lambda^*_{\LW, M}}) - \bar{\beta} \right\}^2$ as
\begin{align*}
    &\EE \left\{{\BS \eta_{\soft}} (\Theta_* z + \barbeta, {\lambda^*_{\LW, M}}) - \bar{\beta} \right\}^2\\ =\ & \kappa \sigma_{\beta}^2 + \kappa \bigg[(\Theta_*^2 + {\lambda^*_{\LW, M}}^2 + \sigma_{\beta}^2) \cdot\erfc \left\{\frac{{\lambda^*_{\LW, M}}}{\sqrt{2(\Theta_*^2 + \sigma_{\beta}^2)}} \right\}  - e^{- \frac{{\lambda^*_{\LW, M}}^2}{2 (\Theta_*^2 + \sigma_{\beta}^2)}} \sqrt{\frac{2}{\pi}} {\lambda^*_{\LW, M}} \sqrt{\Theta_*^2 + \sigma_{\beta}^2} \bigg]  \\
    &- 2 \kappa \sigma_\beta^2 \cdot \erfc \left( \frac{{\lambda^*_{\LW, M}}}{\sqrt{2 {\sigma_{\beta}^2} + 2 \Theta_*^2}}\right)\\
    &+ (1- \kappa) \bigg\{ (\Theta_*^2 + {\lambda^*_{\LW, M}}^2) \cdot\erfc \left(\frac{{\lambda^*_{\LW, M}}}{\sqrt{2} \Theta_*} \right)  - e^{- \frac{{\lambda^*_{\LW, M}}^2}{2 \Theta_*^2}} \sqrt{\frac{2}{\pi}} {\lambda^*_{\LW, M}} \Theta_* \bigg\}\\
    =\ & \kappa \sigma_{\beta}^2+ (1- \kappa) \bigg\{ (\Theta_*^2 + {\lambda^*_{\LW, M}}^2) \cdot\erfc \left(\frac{{\lambda^*_{\LW, M}}}{\sqrt{2} \Theta_*} \right)  - e^{- \frac{{\lambda^*_{\LW, M}}^2}{2 \Theta_*^2}} \sqrt{\frac{2}{\pi}} {\lambda^*_{\LW, M}} \Theta_* \bigg\}\\
    &+ \kappa \bigg[ (\Theta_*^2 + {\lambda^*_{\LW, M}}^2 - \sigma_{\beta}^2) \cdot\erfc \left\{\frac{{\lambda^*_{\LW, M}}}{\sqrt{2(\Theta_*^2 + \sigma_{\beta}^2)}} \right\}  - e^{- \frac{{\lambda^*_{\LW, M}}^2}{2 (\Theta_*^2 + \sigma_{\beta}^2)}} \sqrt{\frac{2}{\pi}} {\lambda^*_{\LW, M}} \sqrt{\Theta_*^2 + \sigma_{\beta}^2} \bigg]
    \\
    =\ &\kappa \sigma_{\beta}^2+ (1- \kappa) \cdot \Rom{3} + \kappa \cdot \Rom{4}.
\end{align*}
We evaluate the derivatives of $\Rom{3}$ and $\Rom{4}$ \wrt~$\Theta_*$ separately. 
The derivatives are
\begin{align*}
    \frac{\partial \Rom{3}}{\partial \Theta_*} =  2 \Theta_* \cdot \erfc\left(\frac{\lambda^*_{\LW, M}}{\sqrt{2} \Theta_*} \right) > 0, 
\end{align*}
and
\begin{align*}
    \frac{\partial \Rom{4}}{\partial \Theta_*} 
    =\ & 2 \Theta_* \left[\erfc \left\{\frac{{\lambda^*_{\LW, M}}}{\sqrt{2(\Theta_*^2 + \sigma_{\beta}^2)}} \right\} - \frac{e^{- \frac{{\lambda^*_{\LW, M}}^2}{2 (\Theta_*^2 + \sigma_{\beta}^2)}} \sqrt{\frac{2}{\pi}} {\lambda^*_{\LW, M}} \sigma_{\beta}^2}{(\Theta_*^2 + \sigma_{\beta}^2)^{3/2}} \right]\\
    =\ & 2 \Theta_* \left[e^{-\frac{{\lambda^*_{\LW, M}}^2}{2(\Theta_*^2 + \sigma_{\beta}^2)} } \sqrt{\frac{2}{\pi}} \left\{\frac{\sqrt{(\Theta_*^2 + \sigma_{\beta}^2)}}{{\lambda^*_{\LW, M}}} \right\} - \frac{e^{- \frac{{\lambda^*_{\LW, M}}^2}{2 (\Theta_*^2 + \sigma_{\beta}^2)}} \sqrt{\frac{2}{\pi}} {\lambda^*_{\LW, M}} \sigma_{\beta}^2}{(\Theta_*^2 + \sigma_{\beta}^2)^{3/2}} \right] > 0
\end{align*}
where the last inequality follows from Condition \ref{cond: comparison techinical}(c). 
Therefore, $\EE \left\{{\BS \eta_{\soft}}(\Theta_* z + \barbeta, {\lambda^*_{\LW, M}}) - \bar{\beta} \right\}^2$ is strictly increasing with respect to $\Theta_*$. 
This implies that
\begin{align*}
\tilde\Theta_*^2/\gamma_x 
&= \kappa \sigma_{\bbeta}^2 \frac{1 - h_x^2}{h_x^2} + \EE \left\{{\BS \eta_{\soft}}(\Theta_* z + \barbeta, {\lambda^*_{\LW, M}}) - \bar{\beta} \right\}^2  \\
&> \kappa \sigma_{\bbeta}^2 \frac{1 - h_x^2}{h_x^2} + \EE \left\{{\BS \eta_{\soft}}(\tilde\Theta_* z + \barbeta, {\lambda^*_{\LW, M}}) - \bar{\beta} \right\}^2, 
\end{align*}
as $\Theta_*> \Tilde \Theta_*$ indicated by \eqref{eqn: S.14.9}. 
Let $\alpha_{\LW, M}^{*} := {\lambda^*_{\LW, M}} / \tilde\Theta_*$. We have
\begin{align*}
    \tilde\Theta_*^2/\gamma_x >\ & \kappa \sigma_{\bbeta}^2 \frac{1 - h_x^2}{h_x^2} + \EE \left\{{\BS \eta_{\soft}}(\tilde\Theta_* z + \barbeta, \alpha_{\LW, M}^{*} \tilde\Theta_*) - \bar{\beta} \right\}^2 =: L(\tilde\Theta_*^2).
\end{align*}
We now choose $\lambda_{*}$ such that $\lambda_{*} \cdot \{1 + \bar{b}(\lambda_{*}) \} = \alpha_{\LW, M}^{*} \bar{\tau}(\lambda_{*})$, which is possible due to Propositions 1.3 and 1.4 by \cite{bayati2011lasso}. 
Summarizing our findings, we have 
\begin{itemize}
    \item[(a)] $\Tilde{R}_{\LW}(\lambda_{\LW, M}^{*}) = \tilde\Theta_*^2/\gamma_x - \kappa \sigma_{\bbeta}^2 \frac{1 - h_x^2}{h_x^2} > L(\tilde\Theta_*^2) - \kappa \sigma_{\bbeta}^2 \frac{1 - h_x^2}{h_x^2}$. 
    \item[(b)] $R_{\LA}(\lambda_{*}) = \bar{\tau}_{*}^{2}/\gamma_x - \kappa \sigma_{\bbeta}^2 \frac{1 - h_x^2}{h_x^2} = L(\bar{\tau}_{*}^{2}) - \kappa \sigma_{\bbeta}^2 \frac{1 - h_x^2}{h_x^2}$ with $\bar{\tau}_* = \bar{\tau}(\lambda_{*})$. 
    \item[(c)] By \cite[Appendix A.1]{bayati2011lasso}, the function $L(x^2)$ is concave with respect to $x^2$ and  strictly increasing for large enough  $x^2$. Moreover, when $x = 0$, $L(0) > 0$. To better understand the function $L(x)$, we provide an illustrative diagram in Supplementary Figure \ref{illu_prop_comparision_L_1}. 
\end{itemize}
Thus, from (a) - (c) and Figure \eqref{illu_prop_comparision_L_1}, we can conclude that $\tilde{\Theta}_{*}^2 > \bar{\tau}_{*}^{2}$.
This completes the proof of $R_{\LW}({\lambda^*_{\LW, M}}) > R_{\LA}(\lambda^*_{\LA, M})$.

\paragraph{Comparing $A^2_{\LW}(\lambda^*_{\LW, R})$ and $A^2_{\LA}(\lambda^*_{\LA, R})$:}
We now  compare $A^2_{\LW}(\lambda^*_{\LW, R})$ and $A^2_{\LA}(\lambda^*_{\LA, R})$.
With  $\Theta_*$ defined by equation~\eqref{eqn: S.2.1.2-3}, $A^2_{\LW}(\lambda)$ is
\begin{align*} 
    A_{\LW}(\lambda) =\ & \frac{\EE \eta_{\soft}\{(1 + \tilde{b}_*)\bar{\beta} + \tilde{\tau}_* z, \lambda (1 + \tilde{b}_*)\} \cdot \bar{\beta}}{\sqrt{\EE \bar\bbeta^2} \sqrt{\EE \eta^2_{\soft}\{(1 + \tilde{b}_*)\bar{\beta} + \tilde{\tau}_* z, \lambda (1 + \tilde{b}_*)\}}}\\
    =\ & \frac{\EE \eta_{\soft}(\bar{\beta} + \Theta_* z, \lambda) \cdot \bar{\beta}}{\sqrt{\EE \bar\bbeta^2} \sqrt{\EE \eta^2_{\soft}(\bar{\beta} + \Theta_* z, \lambda)}}.
\end{align*}
From Equation \eqref{eqn: 4.13 without normalization} and the relationship between $\alpha$ and $\lambda$, there exists some  
$\alpha_{\LW,R}^{*}$ such that $\lambda_{\LW,R}^{*} = \alpha_{\LW, R}^{*} \tilde{\tau}_*/(1+\tilde{b}_*) = \alpha_{\LW,R}^{*} \Theta_*(\lambda_{\LW,R}^{*})$. 
Therefore, it suffices to show
\begin{align*}
A_{\LW}(\alpha_{\LW,R}^{*}) < A_{\textnormal{L}}(\alpha_{\LW,R}^{*}),
\end{align*}
where $A_{\LW}(\alpha)$ and $A_{\LA}(\alpha)$ are given by
\begin{align*} 
    A_{\LW}(\alpha) = \frac{\EE \eta_{\soft}(\bar{\beta} + \Theta_* z, \alpha \Theta_*) \cdot \bar{\beta}}{\sqrt{\EE \bar\bbeta^2} \sqrt{\EE \eta^2_{\soft}(\bar{\beta} + \Theta_* z, \alpha \Theta_*)}}\quad\mbox{and} \quad A_{\LA}(\alpha) = \frac{\EE \eta_{\soft}(\bar{\beta} + \bar{\tau}_* z, \alpha \bar{\tau}_*) \cdot \bar{\beta}}{\sqrt{\EE \bar\bbeta^2} \sqrt{\EE \eta^2_{\soft}(\bar{\beta} + \bar{\tau}_* z, \alpha \bar{\tau}_*)}}
\end{align*}
respectively. 
Following a similar argument as for comparing MSEs, we define $\bar{\Theta}_{*}$ 
\begin{align*}
    \bar{\Theta}_{*}^{2}/\gamma_x 
    &:= \sigma_{{\bepsilon}_x}^2 + \left\{\eta(\Theta_* z + \barbeta, \alpha_{\LW,R}^{*} \Theta_{*} ) - \bar{\beta} \right\}^2\\
    &> \sigma_{{\bepsilon}_x}^2 + \left\{\eta(\Theta_* z + \barbeta, \alpha_{\LW,R}^{*} \bar{\Theta}_{*} ) - \bar{\beta} \right\}^2.
\end{align*}
By the concavity of function $L$, we are able to obtain $\Theta_* \geq \bar{\tau}_*$ under Conditions \ref{cond: comparison techinical}~(a)-(c).
By Condition \ref{cond: comparison techinical}(d), we conclude
$A^2_{\LW}(\lambda^*_{\LW, R}) < A^2_{\LA}(\lambda^*_{\LA, R})$.

\section{Proofs for Section \ref{subsec: matrix AMP}}
\label{Proof of Theorem matrix AMP}

In this section, we provide the proof of Theorem \ref{matrix AMP}. 
We begin by proving Theorem \ref{matrix AMP} in Section \ref{proof of theorem matrix amp}, where we omitted certain details to avoid redundancy with  \cite{berthier2020state}. 
The proof builds upon findings and facts related to AMP as in Sections \ref{sec: supporting_AMP}, $\GOE$ as in Sections \ref{sec: random matrix 1}, and probability facts as in Section \ref{sec: useful prob facts}. 
Detailed proofs of probability facts are given in Section \ref{sec: proof of useful facts}.

\subsection{Proof of Theorem \ref{matrix AMP}}
\label{proof of theorem matrix amp}
Let $\cF^{t}: \RR^{N' \times q} \mapsto \RR^{N' \times q}$ be a function with fixed $q$. 
Let  $\Ab$ be a symmetric $\GOE(N')$. 
Moreover, for two matrices $\Qb$ and $\Pb$ with the same number of rows, let $[\Qb | \Pb ]$ denote the matrix by concatenating $\Qb$ and $\Pb$ horizontally. 
For any matrix $\Mb \in \RR^{N' \times qt}$, we denote the orthogonal projection onto its range to be  $\Pb_{\Mb}$, with $\Pb_{\Mb}^{\bot} = \bI_{N'} - \Pb_{\Mb}$. Then the long AMP recursion is given by
\begin{align} \label{eqn: longamp}
\begin{split}
    \hb^{t+1} =\ & \Pb^{\bot}_{\Qb^{t-1}} \Ab \Pb^{\bot}_{\Qb^{t-1}} \qb^{t} + \Hb^{t-1} \balpha^{t} \quad \mbox {and}\\
    \qb^{t} =\ & \cF^{t}(\hb^{t}),
\end{split}
\end{align}
where at each step $t$, the matrices $\Qb^{t} \in \RR^{N' \times qt}$, $\Hb^{t-1} \in \RR^{N' \times qt}$, and $\balpha^{t} \in \RR^{qt \times q}$ are defined as: 
\begin{align*}
    &\Qb^{t} = \left[\qb^0 | \qb^1 | \cdots | \qb^{t-1} \right],\\
    &\balpha^{t} = \left({\Qb^{t-1}}^{\T} \Qb^{t-1} \right)^{-1} {\Qb^{t-1}}^{\T} \qb^{t}, \quad \mbox{and} \\
    &\Hb^{t-1} = \left[\hb^1 | \hb^2 | \cdots | \hb^{t} \right].
\end{align*}
The initialization is $\qb^0 = f^0(\xb^0)$ and $\hb^1 = \Ab \qb^0$. 
We first prove a special case of Theorem \ref{matrix AMP} under the following assumptions. 
We say that the long AMP iteration \eqref{eqn: longamp} satisfies the non-degeneracy assumption if
    \begin{itemize}
    \item[(a)] almost surely, for all $t \in \NN$ and all $N' \geq t \times q$, $\Qb^{t-1}$ has full column rank. 

    \item[(b)] for all $t \in \NN$, there exists some constant $c_{t} > 0$ independent of $N'$ such that almost surely, there exists a random $N_0$ such that, for $N' \geq N_0$, we have $\sigma_{\min}(\Qb^{t-1})/\sqrt{N'} \geq c_{t} > 0$.
    \end{itemize}
The following theorem indicates that without loss of generality, we may assume long AMP iteration \eqref{eqn: longamp} satisfies the non-degeneracy assumption. 
\begin{theorem} \label{thm: non-deg-matrix-amp}
    Consider the symmetric matrix AMP iteration \eqref{eqn: symmetric matrix AMP} and long AMP  iteration \eqref{eqn: longamp} that satisfies the non-degeneracy assumption. 
    Assume $\Kb_{1,1}, \Kb_{2,2} \cdots, \Kb_{t+1, t+1}$ defined in Equation~\eqref{matrix state evolution} are positive semi-definite. Then for any deterministic sequence $\{\phi^{N'}\}_{N' \in \NN}$ of uniformly normalized pseudo-Lipschitz functions, as $N' \to \infty$ and fixed $t$, we have, as $N' \to \infty$,
    \begin{equation*}
        \phi^{N'} (\Xb^0, \Xb^1, \cdots, \Xb^{t}) \overset{P}{\approx} \EE \big\{ \phi^{N'}(\Xb^0, \Zb^1, \cdots, \Zb^{t}) \big\}. 
    \end{equation*}
\end{theorem}
Finally, when $\bQ^{t-1}$ does not satisfy the non-degeneracy assumption, we can prove Theorem \ref{matrix AMP} by using Theorem \ref{thm: non-deg-matrix-amp} and the argument presented in Section 5.4 of \cite{berthier2020state}.
The remainder of this section focuses on proving Theorem \ref{thm: non-deg-matrix-amp}.

\subsection{Supporting AMP Lemmas}
\label{sec: supporting_AMP}
Under the non-degeneracy assumption, the following lemma characterizes the asymptotic behavior of the long AMP iteration \eqref{eqn: longamp}. 
\begin{lemma}[Lemma 5 in \cite{berthier2020state}]
\label{prop: longamp lemma 5}
Consider the long AMP iteration \eqref{eqn: longamp} and suppose it satisfies the non-degeneracy assumption. Then for any $t \in \NN$ and any deterministic sequence $\phi_{N'}: (\RR^{N' \times q})^{t+2} \mapsto \RR$ of uniformly normalized pseudo-Lipschitz functions, we have 
\begin{equation*}
    \phi^{N'} (\xb^0, \hb^1, \cdots \hb^{t}) \overset{P}{\approx} \EE \big\{ \phi^{N'}(\xb^0, \Zb^1, \cdots \Zb^{t}) \big\},
\end{equation*}
where the joint matrix Gaussian distribution of $\Zb^1, \Zb^2, \cdots \Zb^{t}$ is given by 
\begin{equation*}
    \vecc(\Zb^1, \Zb^2, \cdots \Zb^{t}) \sim N(0, \Kb^{(t)} \otimes \bI_{N'}).
\end{equation*}
\end{lemma}
The following lemma states that long AMP iteration is a good approximation to the AMP recursion defined in Equation~\eqref{eqn: symmetric matrix AMP}. 
\begin{lemma}[Lemma 6  in \cite{berthier2020state}]
\label{prop: longamp lemma 6}
    Consider the symmetric matrix AMP iteration \eqref{eqn: symmetric matrix AMP} and long AMP  iteration \eqref{eqn: longamp} that satisfies the non-degeneracy assumption. 
    For any fixed $t \in \NN$, as $N' \to \infty$, we have
    \begin{align*}
        \frac{1}{\sqrt{N'}} \left\|\hb^{t} - \xb^{t} \right\|_{F} \overset{P}{\to} 0, \quad \frac{1}{\sqrt{N'}} \left\|\cF^{t}(\hb^{t}) - \cF^{t}(\xb^{t}) \right\|_{F} \overset{P}{\to} 0. 
    \end{align*}
\end{lemma}
Theorem \ref{thm: non-deg-matrix-amp} is an immediate consequence of lemmas above.
The proofs of Lemma \ref{prop: longamp lemma 5} and \ref{prop: longamp lemma 6} closely resemble the proofs of Lemmas 5 and 6 in \cite{berthier2020state}, by replacing Appendix B in \cite{berthier2020state} with Lemmas \ref{prob fact 1} to \ref{Stein Lemma} in Section~\ref{sec: useful prob facts}.

\subsection{Three lemmas about random matrices}
\label{sec: random matrix 1}

We first introduce three lemmas on $\GOE(p)$ as follows.

\begin{lemma}[Lemma 19 in \cite{berthier2020state}]
\label{lemma: S.23}
    Consider a sequence of matrices $\Ab \sim \GOE(p)$ and two sequences (in $p$) of (nonrandom vectors) $\ub, \vb \in \RR^p$ such that $\norm{\ub}_2 = \norm{\vb}_2 = \sqrt{p}$. We have  
    \begin{enumerate}
        \item[(a)] $\frac{1}{p} \langle \vb, \Ab \ub \rangle {\BS \overset{P}{\to}} 0$.
        
        \item[(b)] Let $\Pb \in \RR^{p \times p}$ be a sequence of projection matrices such that there exists a constant $t$ that satisfy for all $p$, $\rank(\Pb) \leq t$. Then $\frac{1}{p} \norm{\Pb \Ab \ub}_2^2 {\BS \overset{P}{\to}} 0$ as $p \to \infty$. 
        
        \item[(c)] $\frac{1}{p} \norm{\Ab \ub}_2^2 {\BS \overset{P}{\to}} 1$. 
        
        \item[(d)] There exist a sequence (in $p$) of random vectors $\zb \sim N(0, \bI_p)$ such that for any deterministic sequence $\phi^p : \RR^p \mapsto \RR$, $n \geq 1$ of uniformly normalized pseudo-Lipschitz function  we have 
        \begin{align*}
            \phi^{p} (\Ab \ub) \overset{P}{\approx} \phi^p(\Zb).
        \end{align*}
    \end{enumerate}
\end{lemma}

Lemma \ref{lemma: S.24} states that for any $\Zb \sim N(0, \bI_p)$ and uniformly normalized pseudo-Lipschitz function $\phi^p$, $\phi^p (\Zb)$ converges in probability to the expectation $\EE \phi^p (\Zb)$.
\begin{lemma}[Lemma 23 in \cite{berthier2020state}]
\label{lemma: S.24}
    For any deterministic sequence (in $p$) $\phi^p: \RR^{p} \mapsto \RR$ of uniformly normalized pseudo-Lipschitz functions, we have $\phi^p (\Zb) \overset{P}{\approx} \EE \phi^p (\Zb)$.
\end{lemma}

Lemma \ref{lemma: S.26} provides an asymptotic result on the largest singular value of the matrix $\Ab$, which 
will be used in the proof of Lemma \ref{prob fact 2}. 

\begin{lemma} [\cite{bai1988necessary}]
\label{lemma: S.26}
    Consider a sequence of matrices $\Ab \sim \GOE(p)$.
    Then, the operator norm of $\Ab$, $\|\Ab\|_{\textnormal{op}} \to 2$ almost surely as $p \to \infty$. 
\end{lemma}

\subsection{Supporting probability lemmas}
\label{sec: useful prob facts}

Lemmas \ref{prob fact 1} to \ref{Stein Lemma} below will be used in Section \ref{proof of theorem matrix amp} to prove Theorem \ref{matrix AMP}. Proofs of these lemmas can be found in Section \ref{sec: proof of useful facts}. 
In this section, we assume that the matrix $\bf A$ is drawn from the $\GOE(p)$ ensemble. We denote non-random vectors as $\mathbf{u}$ and $\mathbf{v}$, and non-random ${p \times q}$ matrices as $\mathbf{U}$ and $\mathbf{V}$, where $q$ remains fixed.
Lemma \ref{prob fact 1} shows that the inner product between ${\bf A} {\ub}$ and ${\bf A} {\vb}$ converges in probability to $\langle  \ub , \vb \rangle$. 
\begin{lemma} \label{prob fact 1}
If $\ub, \vb \in \RR^p$ non-random vectors, then we have 
\begin{equation*}
    \frac{1}{p} \langle  {\bf A} {\ub}, {\bf A} {\vb} \rangle \overset{P}{\approx} \frac{1}{p} \langle  \ub , \vb \rangle.
\end{equation*}
Moreover, if ${\bf U},{\bf V} \in \RR^{p \times q}$, then we have 
\begin{equation*}
    \frac{1}{p} ({\Ab} {\Ub})^{\T} {\Ab} {\Vb} \overset{P}{\approx} \frac{1}{p} \Ub^{\T} \Vb.
\end{equation*}
\end{lemma}

Let $\{\phi^{p}\}_{p \in \NN}$ denote any deterministic sequence of uniformly normalized pseudo-Lipschitz functions.
Lemma \ref{prob fact 2} indicates that $\Ab \Ub$ converges in probability to some matrix Gaussian distribution after applying functions $\{\phi^{p}\}_{p \in \NN}$.
\begin{lemma}\label{prob fact 2}
If $\Ub = (\ub_1, \ub_2, \dots, \ub_q) \in \RR^{p \times q}$, $\bnorm{\Ub}_F^2 = pq$ with the smallest singular value $\sigma_{\min}(\Ub)/\sqrt{pq} > c_0 > 0$, then there exists a sequence (in $p$) of matrix Gaussian distribution $\Zb \in \RR^{p \times q}$  with $\vecc(\Zb) \sim N \left( 0, \frac{1}{p} \Ub^{\T} \Ub \otimes {\bI_p} \right)$ such that
\begin{equation}
    \phi^{p} (\Ab \Ub) \overset{P}{\approx} \phi^{p} (\Zb),
\end{equation}
\end{lemma}

Lemma \ref{prob fact 3} indicates that the inner product between $\Vb$ and $\Ab \Ub$ converges in probability to $0$. 

\begin{lemma}\label{prob fact 3}
If $\Ub, \Vb \in \RR^{p \times q}$, $\bnorm{\Ub}_{\textnormal{F}}^2 = \bnorm{\Vb}_{\textnormal{F}}^2 = p \cdot q$, then we have $\frac{1}{p} \Vb^{\T} \Ab \Ub \overset{P}{\to} 0$.
\end{lemma}

Lemma \ref{prob fact 4} guarantees the convergence of the Frobenius norm of $\Ab \Ub$.

\begin{lemma}\label{prob fact 4}
Let $\Ub = (\ub_1, \ub_2, \cdots \ub_q) \in \RR^{p \times q}$, $\bnorm{\Ub}_F^2 = p \cdot q$, and $\sigma_{\min}(\Ub)/\sqrt{pq} > c_0 > 0$. As $p \to \infty$, we have 
\begin{equation*}
    \frac{1}{pq} \bnorm{\Ab \Ub}_{\textnormal{F}}^2 {\BS \overset{P}{\to}} 1.
\end{equation*}
\end{lemma}

Lemma \ref{prob fact 5} indicates that the Frobenius norm of the projected $\Ab \Ub$ converges in probability to $0$.

\begin{lemma}\label{prob fact 5}
Let $\Ub \in \RR^{p \times q}$ and $\Pb$ be a projection matrix such that exist a constant $t$ that satisfies $rank(\Pb) \leq t$. Then we have 
\begin{equation*}
    \frac{1}{p} \bnorm{\Pb \Ab \Ub}_{\textnormal{F}}^2 {\BS \overset{P}{\to}} 0.
\end{equation*}
\end{lemma}

Lemma \ref{prob fact 6} is a direct generalization of Lemma \ref{lemma: S.24}. 

\begin{lemma}\label{prob fact 6}
Let $\Zb$ be a matrix Gaussian distribution with $\vecc(\Zb) \sim N(0, {\bI_p} \otimes \bI_q)$, then we have 
\begin{equation*}
    \phi^{p}(\Zb) := \phi^{p}(\vecc(\Zb)) \overset{P}{\approx} \EE \phi^{p}(\vecc(\Zb)) = \EE \phi^{p}(\Zb).
\end{equation*}
\end{lemma}

Lemma \ref{Stein Lemma} states the Stein's Lemma for joint matrix Gaussian distribution. 

\begin{lemma}[Stein's Lemma]\label{Stein Lemma}
Consider the joint matrix Gaussian distribution $(\Zb^1, \Zb^2) \in \RR^{p \times 2q}$ with $\vecc(\Zb^1, \Zb^2) \sim N(0, \Kb \otimes {\bI_p})$ and some $\Kb \in \RR^{2q \times 2q}$. 
Define a function $\phi(\mathbf{x})$ as follows:
\begin{align*}
    \phi(\xb) = \begin{bmatrix}
\phi_1 (\xb)\\
\phi_2 (\xb)\\
\cdots \\
\phi_p (\xb)
\end{bmatrix}: \RR^{p \times q} \mapsto \RR^{p \times q},
\end{align*}
where $\phi_i(\xb): \RR^{p \times q} \mapsto \RR^q$ such that $\EE (\Zb^1)^{\T}\phi(\Zb^2)$ and
$\EE \left(\sum_{i=1}^{p} \frac{\partial \phi_i}{\partial x_i}(\Zb^2) \right)$ exist. Then, we have 
\begin{equation} \label{stein formula}
    \EE_{\Zb^1, \Zb^2}\ (\Zb^1)^{\T}\phi(\Zb^2) = \EE_{\Zb^1, \Zb^2}\ \left( \frac{1}{p} (\Zb^1)^{\T} \Zb^2 \right) \cdot \EE_{\Zb^1, \Zb^2}\ \left(\sum_{i=1}^{p} \frac{\partial \phi_i}{\partial x_i}(\Zb^2) \right)^{\T}.
\end{equation}
\end{lemma}

\subsection{Proofs of the supporting  probability lemmas}
\label{sec: proof of useful facts}
In this section, we present the proof of Lemmas \ref{prob fact 1} to \ref{Stein Lemma} in Section \ref{sec: useful prob facts}. It is worth noting that the proof of Lemmas \ref{prob fact 3} to \ref{prob fact 5} is similar to that of  Lemmas \ref{lemma: S.23} and \ref{prob fact 1}. In addition, Lemma \ref{prob fact 6} follows directly from Lemma \ref{lemma: S.24}. Therefore,  we only  provide the detailed proofs for Lemmas \ref{prob fact 1}, \ref{prob fact 2}, and \ref{Stein Lemma}. 

\subsubsection{Proof of Lemma \ref{prob fact 1}}

By Lemma \ref{lemma: S.23}, we have $\frac{1}{p} \bnorm{\Ab \ub}_2^2 \overset{P}{\approx} \frac{1}{p} \bnorm{\ub}_2^2$ and $\frac{1}{p} \bnorm{\Ab \vb}_2^2 \overset{P}{\approx} \frac{1}{p} \bnorm{\vb}_2^2$, as $p \to \infty$. Moreover, 
\begin{equation*}
    \frac{1}{p} \bnorm{\Ab (\ub+\vb)}_2^2 \overset{P}{\approx} \frac{1}{p} \bnorm{(\ub+\vb)}_2^2,
\end{equation*}
which implies that 
\begin{equation*}
    \frac{1}{p} \langle  {\bf A} {\ub} , {\bf A} {\vb} \rangle \overset{P}{\approx} \frac{1}{p} \langle  \ub , \vb \rangle.
\end{equation*}
Therefore, for any $1 \leq i, j \leq q$ and matrices $\Ub, \Vb$, we have 
\begin{equation*}
    \left[\frac{1}{p} ({\Ab} {\Ub})^{\T} {\Ab} {\Vb} \right]_{i,j} \overset{P}{\approx} \left[\frac{1}{p} \langle  \Ub , \Vb \rangle \right]_{i,j}.
\end{equation*}
This completes the proof of Lemma \ref{prob fact 1}.

\subsubsection{Proof of Lemma \ref{prob fact 2}}
\label{sec: Proof of Lemma prob fact 2}

In order to prove Lemma \ref{prob fact 2}, the following claim computes the covariance matrix of $\vecc(\Ab \Ub)$. 
The proof for Claim \ref{claim S.1.1} will be presented later in this section.
\begin{claim} \label{claim S.1.1}
The covariance matrix  of $\vecc(\Ab \Ub)$ is $\frac{1}{p} \Ub^{\T} \Ub \otimes {\bI_p} + \Cb$, where $\Cb = \left[\Cb_{i,j} \right]_{1 \leq i,j \leq q} \in \RR^{pq \times pq}$ with each block $\Cb_{i,j} = \frac{1}{p} \ub_j \ub_i^{\T} \in \RR^{p \times p}$.  
\end{claim}
With Claim \ref{claim S.1.1}, we can prove Lemma \ref{prob fact 2}. 
We first consider a special case that $\frac{1}{p} \Ub^{\T} \Ub = \bI_q$. 
The proof will be extended to the case of a general $\Ub$ later in this section. 
Note that the covariance matrix of $\vecc(\Ab \Ub)$ is then given by $\bI_q \otimes {\bI_p} + \Cb = \bI_{pq} + \Cb =: {\bSigma_{U}}$.
Therefore, we have $\vecc(\Ab \Ub) = \bSigma_{U}^{1/2} \Zb$ with vector $\Zb \sim N(0, \bI_{pq})$. 
Consider the deterministic sequence of uniformly normalized pseudo-Lipschitz function $\phi^{p}: \RR^{pq} \mapsto \RR$ and the expression $|\phi^{p}(\Ab \Ub) - \phi^{p}(\Zb)|$, where $\Ab \Ub$ is identified with their vectorizations for notational convenience. We have
\begin{equation*}
    |\phi^{p}(\Ab \Ub) - \phi^{p}(\Zb)| \leq L \left( 1 + \frac{\bnorm{\Ab \Ub}_{2}}{\sqrt{pq}} + \frac{\bnorm{\Zb}_{2}}{\sqrt{pq}} \right) \frac{\bnorm{\Ab \Ub - \Zb}_{2}}{\sqrt{pq}},
\end{equation*}
By the definition of $\vecc(\Ab \Ub) = \bSigma_{U}^{1/2} \Zb$, the entries of $\Zb$ are i.i.d. Gaussian random variables. 
It follows that $\bnorm{\Zb}_{2}/{\sqrt{pq}} \to 1$ by the law of large number. Moreover, by Lemma \ref{lemma: S.26}, we have 
\begin{align*}
    \frac{\bnorm{\Ab \Ub}_{2}^2}{pq} =\ & \frac{\bnorm{\Ab \ub_1}_{2}^2 + \bnorm{\Ab \ub_2}_{2}^2 + \cdots + \bnorm{\Ab \ub_q}_{2}^2}{pq}\\
    \leq\ & \bnorm{\Ab}_{\textnormal{op}}^2 \frac{\bnorm{\Ub}_{\textnormal{F}}^2}{pq} \leq \frac{2 \bnorm{\Ub}_{\textnormal{F}}^2}{pq} \to 2.
\end{align*}
Note that 
\begin{align*}
    \frac{\bnorm{\Ab\Ub - \Zb}_{2}}{\sqrt{pq}} =\ & \frac{\bnorm{(\bSigma_{U}^{1/2} - \bI_{pq}) \Zb}_{2}}{\sqrt{pq}}\\
    =\ & \frac{\bnorm{(\bSigma_{U}^{1/2} + \bI_{pq})^{-1}({\bSigma_{U}} - \bI_{pq}) \Zb}_{2}}{\sqrt{pq}}\\
    \leq\ & \sigma^{-1}_{\min}(\bSigma_{U}^{1/2} + \bI_{pq}) \frac{\bnorm{({\bSigma_{U}} - \bI_{pq}) \Zb}_{2}}{\sqrt{pq}}.
\end{align*}
The last line follows by the fact  that ${\bSigma_{U}}$ is positive definite and the smallest eigenvalue of $\bSigma_{U}^{1/2} + \bI_{pq}$ is  larger than $1$. It follows  that
\begin{align*}
    \frac{\bnorm{\Ab\Ub - \Zb}_{2}}{\sqrt{pq}} \leq\  \frac{\bnorm{({\bSigma_{U}} - \bI_{pq}) \Zb}_{2}}{\sqrt{pq}} = \frac{\bnorm{\Cb \Zb}_{2}}{\sqrt{pq}}.
\end{align*}
Thus, it suffices to show $\frac{1}{pq} \bnorm{\Cb \Zb}_{2}^2 = \frac{1}{pq} \Zb^{\T} \Cb^{\T} \Cb \Zb \to 0$.
Consider $\Cb^{\T} \Cb$, where 
\begin{equation*}
\begin{split}
    \Cb = \begin{bmatrix}
    \Cb_{11} & \cdots & \Cb_{1q}\\
     & \cdots & \\
     \Cb_{1q} & \cdots & \Cb_{qq}
    \end{bmatrix}
\end{split}
\end{equation*}
and  $\Cb_{ij} = \frac{1}{p} \ub_{j} \ub_{i}^{\T}$. 
For any fixed $i,j \in [q]$, the $(i,j)$ block of $\Cb^{\T} \Cb$ is 
\begin{align*}
    \left[\Cb^{\T} \Cb\right]_{ij} =\ & \Cb_{1i}^{\T} \Cb_{1j} + \Cb_{2i}^{\T} \Cb_{2j} + \cdots + \Cb_{qi}^{\T} \Cb_{qj}\\
    =\ & \frac{1}{p^2} \ub_1 \ub_{i}^{\T} \ub_{j} \ub_{1}^{\T} + \frac{1}{p^2} \ub_2 \ub_{i}^{\T} \ub_{j} \ub_{2}^{\T} + \cdots + \frac{1}{p^2} \ub_q \ub_{i}^{\T} \ub_{j} \ub_{q}^{\T}\\
    =\ & \frac{1}{p^2} \ub_{i}^{\T} \ub_{j} \left(\ub_1 \ub_{1}^{\T} + \ub_2 \ub_{2}^{\T} + \cdots \ub_q \ub_{q}^{\T} \right)\\
    =\ & \frac{1}{p^2} \ub_{i}^{\T} \ub_{j} \cdot \Ub \Ub^{\T}.
\end{align*}
This implies that 
\begin{align*}
    \Cb^{\T} \Cb =\ & \frac{1}{p} \Ub^{\T} \Ub \otimes \frac{1}{p} \Ub \Ub^{\T}\\
    =\ & \bI_{q} \otimes \frac{1}{p} \Ub \Ub^{\T}\\
    =\ & \begin{bmatrix}
    \frac{1}{p} \Ub \Ub^{\T} & 0 & \cdots & 0\\
    0 & \frac{1}{p} \Ub \Ub^{\T} & \cdots & 0\\
    0 & 0 & \cdots & \frac{1}{p} \Ub \Ub^{\T}
    \end{bmatrix} \in \RR^{pq \times pq}.
\end{align*}
Let $\Zb = (\Zb_1^{\T}, \Zb_2^{\T}, \cdots, \Zb_q^{\T})^{\T}$ with $\Zb_{i} \sim N(0, {\bI_p})$, then we have 
\begin{align*}
    \frac{1}{pq} \Zb^{\T} \Cb^{\T} \Cb \Zb 
    =\ & \frac{1}{pq} \Zb_1^{\T} \left( \frac{1}{p} \Ub \Ub^{\T} \right) \Zb_1 + \frac{1}{pq} \Zb_2^{\T} \left( \frac{1}{p} \Ub \Ub^{\T} \right) \Zb_2 + \cdots + \frac{1}{pq} \Zb_q^{\T} \left( \frac{1}{p} \Ub \Ub^{\T} \right) \Zb_q\\
    =\ & \frac{1}{q} \sum_{k=1}^{q} \frac{1}{p^2} \bnorm{\Ub^{\T} \Zb_k}_2^2.
\end{align*}
In addition, $\frac{1}{p} \Ub^{\T} \Ub = {\bI_q}$ implies that $\frac{1}{p} \ub_i^{\T} \ub_i = 1$ for all $i \leq q$. Therefore, for fixed $q$, as $p \to \infty$, we have 
\begin{align*}
    \frac{1}{pq} \bnorm{\Cb \Zb}_{2}^2 = \frac{1}{q} \sum_{k=1}^{q} \frac{1}{p^2} \bnorm{\Ub^{\T} \Zb_k}_2^2 \leq \frac{1}{q} \sum_{k=1}^{q} \left( \frac{1}{p} \sum_{i=1}^{q} |\ub_{i}^{\T} \Zb_k| \right)^2 \overset{P}{\to} 0,
\end{align*}
where the last convergence follows from the fact that $\frac{1}{p} \ub_i^{\T} \Zb_k$ is a centered Gaussian random variable with variance $\bnorm{\ub_i}_2^2/p^2 = \ub_i^{\T} \ub_i/p^2 = 1/p \to 0$. This completes the proof of Lemma \ref{prob fact 2} under the special case that $\frac{1}{p} \Ub^{\T} \Ub = \bI_q$, except for Claims \ref{claim S.1.1}.

We have established the proof of Lemma \ref{prob fact 2} for the special case $\frac{1}{p} \Ub^{\T} \Ub = \bI_q$. 
Now, we aim to extend the proof to general  ${\Ub}$. 
Denote $\hat{\bmSigma}_{U}:= \frac{1}{p} {\Ub}^{\T} {\Ub}$, where $\hat{\bmSigma}_{U}$ is a positive definite matrix since $\sigma_{\min}(\hat{\bmSigma}_{U})/\sqrt{p q} \geq c_0 > 0$.
Let $\tilde{\Ub} = \Ub \hat{\bmSigma}_{U}^{-1/2}$, then $\tilde{\Ub}$ satisfies the following properties: 
\begin{enumerate}
    \item[(a)] $\frac{1}{p} \tilde{\Ub}^{\T} \tilde{\Ub} = \frac{1}{p} \hat{\bmSigma}_{U}^{-1/2} \Ub^{\T} \Ub \hat{\bmSigma}_{U}^{-1/2} = \bI_q$.
    
    \item[(b)] Since $\bnorm{\Ub}^2_{F} = {pq}$, we have $\bnorm{\tilde{\Ub} \hat{\bmSigma}_{U}^{1/2}}^2_{F} = pq$. This implies that $pq \geq \sigma_{\min}(\hat{\bmSigma}_{U}^{1/2}) \bnorm{\tilde{\Ub}}_{F}^2$. 
    That is, we have $\bnorm{\tilde{\Ub}}_{F}^2 \lesssim p q$.
    
    \item[(c)] Since $\frac{1}{p} \tilde{\Ub}^{\T} \tilde{\Ub} = \bI_q$, $\sigma_{\min}(\tilde{\Ub})/\sqrt{pq}$ has an uniform positive lower bound. 
\end{enumerate}
Combining these observations, for any deterministic sequence of uniformly normalized pseudo-Lipschitz function ${\phi^{p}}'$, we have 
\begin{align*}
    {\phi^{p}}'(\Ab \tilde{\Ub}) \overset{P}{\approx} {\phi^{p}}'(\tilde{\Zb}),
\end{align*}
where $\tilde{\Zb}$ is Gaussian matrix satisfying $\vecc(\tilde{\Zb}) \sim N(0, \bI_q \otimes {\bI_p})$. 
Therefore, for any fixed sequence of uniformly normalized pseudo-Lipschitz function ${\phi^{p}}$, let ${\phi^{p}}'(\Xb) = {\phi^{p}}(\Xb \hat{\bmSigma}_{U}^{1/2})$, we have 
\begin{align*}
    {\phi^{p}}(\Ab \Ub) = {\phi^{p}}'(\Ab \Ub \hat{\bmSigma}_{U}^{-1/2}) = {\phi^{p}}'(\Ab \tilde{\Ub}) \overset{P}{\approx} {\phi^{p}}'(\tilde{\Zb}) = {\phi^{p}}(\tilde{\Zb} \hat{\bmSigma}_{U}^{1/2}).
\end{align*}
Therefore, we have ${\phi^{p}}(\Ab \Ub) \overset{P}{\approx} {\phi^{p}}(\Zb)$, for $\Zb = \tilde{\Zb} \hat{\bmSigma}_{U}^{1/2}$ such that  $\vecc(\Zb) \sim N(0, \hat{\bmSigma}_{U} \otimes {\bI_p}) = N(0, \Ub^{\T} \Ub/p \otimes {\bI_p})$. 
This completes the proof of Lemma \ref{prob fact 2} module Claim \ref{claim S.1.1}. We then prove Claim \ref{claim S.1.1}.

\begin{proof}[Proof of Claim \ref{claim S.1.1}]

Let $\Ub = (\ub_1, \ub_2, \dots, \ub_q) \in \RR^{p \times q}$, then we have $\Ab \Ub = (\Ab \ub_1, \Ab \ub_2, \dots, \Ab \ub_q)$ with $\Ab \ub_i \in \RR^{p}$. For any $i_1, j_1 \in [q]$, we compute $\EE (\Ab \ub_{i_1}) (\Ab \ub_{j_1})^{\T} \in \RR^{p \times p}$. First, for $i_2 \neq j_2 \in [p]$, we have 
\begin{align*}
    \left[\EE (\Ab \ub_{i_1}) (\Ab \ub_{j_1})^{\T} \right]_{i_2, j_2} =\ & \EE (\Ab \ub_{i_1})_{i_2} (\Ab \ub_{j_1})_{j_2}\\
    =\ & \sum_{1 \leq k,l \leq p} \EE\ \Ab_{i_2, k} (\ub_{i_1})_{k} \Ab_{j_2, l} (\ub_{j_1})_l\\
    =\ & \frac{1}{p} (\ub_{i_1})_{j_2} (\ub_{j_1})_{i_2},
\end{align*}
where the last equality follows from independence and symmetry within $\GOE(p)$, that is, when $i_2=l$ and $j_2=k$. Second, when $i_2 = j_2 \in [p]$, we have 
\begin{align*}
    \left[\EE (\Ab \ub_{i_1}) (\Ab \ub_{j_1})^{\T} \right]_{i_2, j_2} =\ & \EE (\Ab \ub_{i_1})_{i_2} (\Ab \ub_{j_1})_{j_2}\\
    =\ & \sum_{k,l \leq p} \EE\ \Ab_{i_2, k} (\ub_{i_1})_{k} \Ab_{j_2, l} (\ub_{j_1})_l\\
    =\ & \sum_{k \leq p} \EE\ \Ab_{i_2, k} (\ub_{i_1})_{k} \Ab_{j_2, k} (\ub_{j_1})_k\\
    =\ & \frac{1}{p} \sum_{k=1}^{p} (\ub_{i_1})_{k} (\ub_{j_1})_k + \frac{1}{p} (\ub_{i_1})_{j_2} (\ub_{j_1})_{i_2}.
\end{align*}
Therefore, the covariance of $(\Ab \ub_{i_1})$ and $(\Ab \ub_{j_1})$ is $\frac{1}{p} \ub_{i_1}^{\T} \ub_{j_1} {\bI_p} + \frac{1}{p} \ub_{j_1} \ub_{i_1}^{\T}$. This completes the proof of Claim \ref{claim S.1.1}. 
\end{proof}

\subsubsection{Proof of Lemma \ref{Stein Lemma}}

We first recall the multivariate Stein's Lemma in Lemma 5 of  \cite{javanmard2013state} or Lemma 1 of  \cite{liu1994siegel}. 
Let $\EE_{\Gb^1, \Gb^2}$ denote the expectation with respect to the joint distribution $(\Gb^1, \Gb^2)$. 
Then, for any $\Kb' \in \RR^{2 \times 2}$ and the joint multivariate Gaussian $(\Gb^1, \Gb^2) \sim N(0, \Kb' \otimes {\bI_p})$, we have 
\begin{align*}
\EE_{\Gb^1, \Gb^2}\ \Gb^1 \psi(\Gb^2)^{\T} = \EE_{\Gb^1, \Gb^2}\ \Gb^1 {\Gb^2}^{\T} \cdot \EE_{\Gb^1, \Gb^2}\ \left\{ \frac{\partial \psi}{\partial x}(\Gb^2) \right\}^{\T},
\end{align*}
where $\psi: \RR^{p} \mapsto \RR^{p}$, and $\frac{\partial \psi}{\partial x}(\Gb^2)$ is the Jacobian matrix of $\psi$ with respect to $\Gb^2$. 
Now, consider $\vecc(\Zb^1, \Zb^2) \sim N(0, \Kb \otimes \bI_p)$ with
\begin{align*}
    \Zb^1 &= \{(\Zb^1_{1, *})^{\T}, (\Zb^1_{2, *})^{\T}, \cdots, (\Zb^1_{p, *})^{\T}\}^{\T} =  \left[\Zb^1_{i,j} \right]_{ 1 \leq i \leq p, 1 \leq j \leq q} \quad \mbox{and}\\
    \Zb^2 &= \{(\Zb^2_{1, *})^{\T}, (\Zb^2_{2, *})^{\T}, \cdots, (\Zb^2_{p, *})^{\T}\}^{\T} =  \left[\Zb^2_{i,j} \right]_{ 1 \leq i \leq p, 1 \leq j \leq q},
\end{align*}
where $\Zb^1_{i, *}$ is the $i$-th rows of $\Zb^1$ with $\Zb^{1}_{i,j}, \Zb^{2}_{i,j} \in \RR$. 
The left-hand side of Equation \eqref{stein formula} is
\begin{align} \label{eqn: stein formula S.5}
\begin{split}
    \EE_{\Zb^1, \Zb^2}\ (\Zb^1)^{\T}\phi(\Zb^2) =\ & \EE_{\Zb^1, \Zb^2}\ \{(\Zb^1_{1, *})^{\T}, (\Zb^1_{2, *})^{\T}, \cdots, (\Zb^1_{p, *})^{\T}\} \phi(\Zb^2)\\
    =\ & \EE_{\Zb^1, \Zb^2}\ \{(\Zb^1_{1, *})^{\T}, (\Zb^1_{2, *})^{\T}, \cdots, (\Zb^1_{p, *})^{\T}\} 
    \begin{bmatrix}
    \phi_{1}(\Zb^2) \\
    \phi_{2}(\Zb^2)\\
    \cdots \\
    \phi_{p}(\Zb^2)
    \end{bmatrix}\\ 
    =\ & \EE_{\Zb^1, \Zb^2}\ (\Zb^1_{1, *})^{\T} \phi_{1}(\Zb^2) + (\Zb^1_{2, *})^{\T} \phi_{2}(\Zb^2) + \cdots + (\Zb^1_{p, *})^{\T} \phi_{p}(\Zb^2).
\end{split}
\end{align}
For each $i \in [q]$, we first consider $\EE_{\Zb^1, \Zb^2}\ (\Zb^1_{i, *})^{\T} \phi_{i}(\Zb^2)$. 
Note the joint distribution is given by $\vecc(\Zb^1, \Zb^2) \sim N(0, \Kb \otimes {\bI_p})$, which implies that the rows of $(\Zb^1, \Zb^2)$ are independent. Then we have  
\begin{align*}
    \EE_{\Zb^1, \Zb^2}\ (\Zb^1_{i, *})^{\T} \phi_{i}(\Zb^2) = \EE_{\{\Zb^1_{k,*}, \Zb^2_{l,*}: k, l \neq i\}}\ \left\{ \EE_{\Zb^1_{i,*}, \Zb^2_{i,*}}\ (\Zb^1_{i, *})^{\T} \phi_{i}(\Zb^2) \right\},
\end{align*}
where the equality holds due to independent rows. 

Next, we consider the expectation {$\EE_{\{\Zb^1_{k,*}, \Zb^2_{l,*}: k,l \neq i\}}\ \left\{ \EE_{\Zb^1_{i,*}, \Zb^2_{i,*}}\ (\Zb^1_{i, *})^{\T} \phi_{i}(\Zb^2) \right\}$}. 
Conditional on $\{\Zb^1_{k,*}, \Zb^2_{k,*}: k \neq i\}$, we can view $\phi_{i}: \RR^{p \times q} \mapsto \RR^q$ as a function for $\Zb_{i, *}^2$. That is, $\phi_{i}(\ \cdot\ | \{\Zb^1_{k,*}, \Zb^2_{l,*}: k, l \neq i\})$ can be viewed as a function from $\RR^q$ to $\RR^q$. 
Therefore, we can apply the multivariate Stein's Lemma to obtain 
\begin{align*} 
    \EE_{\Zb^1_{i,*}, \Zb^2_{i,*}}\ (\Zb^1_{i, *})^{\T} \phi_{i}(\Zb^2) = \EE_{\Zb^1_{i,*}, \Zb^2_{i,*}}\ (\Zb^1_{i, *})^{\T} (\Zb^2_{i, *}) \cdot \EE_{\Zb^1_{i,*}, \Zb^2_{i,*}}\ \frac{\partial \phi_i}{\partial x_i} (\Zb^2),
\end{align*}
where $\frac{\partial \phi_i}{\partial x_i} (\Zb^2)$ is the $q \times q$ Jacobian matrix of $\phi_{i}(\ \cdot\ | \{\Zb^1_{k,*}, \Zb^2_{l,*}: k, l \neq i\})$.
Taking expectation with respect to $\{\Zb^1_{k,*}, \Zb^2_{l,*}: k, l \neq i\}$ on both sides, we have
\begin{align} \label{stein formula S.6}
    \EE_{\Zb^1, \Zb^2}\ (\Zb^1_{i, *})^{\T} \phi_{i}(\Zb^2) = \EE_{\Zb^1_{i,*}, \Zb^2_{i,*}}\ (\Zb^1_{i, *})^{\T} (\Zb^2_{i, *}) \cdot \EE_{\Zb^1, \Zb^2}\ \frac{\partial \phi_i}{\partial x_i} (\Zb^2).
\end{align}
Now we claim that, for any $i \in [p]$, we have 
\begin{align} \label{stein formula S.7}
    \EE_{\Zb^1_{i,*}, \Zb^2_{i,*}}\ (\Zb^1_{i, *})^{\T} (\Zb^2_{i, *}) = \EE_{\Zb^1, \Zb^2}\ \left\{ \frac{1}{p} (\Zb^1)^{\T} \Zb^2 \right\} .
\end{align}
To see this, note that $\vecc(\Zb^1, \Zb^2) \sim N(0, \Kb \otimes {\bI_p})$ implies 
\begin{align*}
    \EE\ \begin{bmatrix}
    \Zb^1_{1,1}\\
    \cdots\\
    \Zb^1_{p,1}\\
    \cdots\\
    \Zb^1_{1,q}\\
    \cdots\\
    \Zb^1_{p,q}
    \end{bmatrix} (\Zb^2_{1,1}, \cdots, \Zb^2_{p,1}, \cdots, \Zb^2_{1,q}, \cdots, \Zb^2_{p,q}) = \begin{bmatrix}
    \Kb_{1,q+1} {\bI_p} & \cdots & \Kb_{1,2q} {\bI_p}\\
     & \cdots & \\
    \Kb_{q,q+1} {\bI_p} & \cdots & \Kb_{q, 2q} {\bI_p}
    \end{bmatrix}.
\end{align*}
Therefore, for any $r,s \in [q]^2$, the $(r,s)$ block of left hand side of equation~\eqref{stein formula S.7} is given by
\begin{align} \label{stein formula S.8}
    \left\{\EE_{\Zb^1_{i,*}, \Zb^2_{i,*}}\ (\Zb^1_{i, *})^{\T} (\Zb^2_{i, *})\right\}_{r,s} = \EE_{\Zb^1_{i,r}, \Zb^2_{i,s}}\ (\Zb^1_{i, r}) (\Zb^2_{i, s}) = \Kb_{r,q+s}.
\end{align}
The right-hand side of equation~\eqref{stein formula S.7} is determined by
\begin{align*}
    \left[ \EE_{\Zb^1, \Zb^2}\ \left\{ \frac{1}{p} (\Zb^1)^{\T} \Zb^2 \right\}  \right]_{r,s} =\ &  \frac{1}{p} \left\{ \sum_{j=1}^{p}  \EE_{\Zb^1, \Zb^2}\ (\Zb^1_{j,*})^{\T} \Zb^2_{j,*} \right\}_{r,s}\\=\ &  \frac{1}{p} \left\{ \sum_{j=1}^{p} \EE_{\Zb^1_{j,*}, \Zb^2_{j,*}}\ (\Zb^1_{j,*})^{\T} \Zb^2_{j,*} \right\}_{r,s}\\
    =\ & \Kb_{r,q+s}.
\end{align*}
The last equality follows from the fact that Equation \eqref{stein formula S.8} holds for any $i \in [q]$. 
This proves Equation \eqref{stein formula S.7}. 
Combining with equations \eqref{eqn: stein formula S.5} and \eqref{stein formula S.6} completes the proof of Lemma \ref{Stein Lemma}.

\section{Proofs for Section \ref{subsec: calibration}}
\label{sec: Proof for subsec: calibration}

In this section, we provide proofs for propositions in Section \ref{subsec: calibration}, which relies on Theorem \ref{stokes theorem}.  
Briefly, since both the integrand and the integration domain depend on the parameter $t$, Theorem \ref{thm: S.13.1} enables us to compute the derivative with respect to $t$.

To ensure the our proofs are consistent with existing AMP literature \citep{bayati2011lasso, berthier2020state}, we will use a different normalization setting introduced in Section \ref{change of normalization} in  Sections \ref{sec: Proof for subsec: calibration}-\ref{sec: Proof for sec: L_2 estimator}. 
The proofs of Propositions \ref{prop 1} and \ref{prop 2} are provided in Sections~\ref{proof of prop 1} and \ref{proof of prop 2}, respectively. 
In addition, the proofs of all supporting lemmas are collected in Section \ref{sec: proof of supporting lemmas for section 4.2}.

\subsection{Generalized Stokes theorem}\label{stokes theorem}

We will repeatedly use the generalized Stokes theorem \citep{baddeley1977integrals}.  Let $M$ be a compact oriented $n$-dimensional subset of $\RR^{n}$.
We say $M$ is an $n$-manifold if the boundary of $M$ is defined by $g(\xb) = 0$, where $g(\xb): \RR^{n} \mapsto \RR$ is a smooth function on $\RR^{n}$. 
\begin{definition}
    A family $\{A_t : t \in [a,b]\}$ of $n$-manifold in $\RR^{n}$ is called a smoothly changing $n$-manifold in $\RR^n$ if the graph 
    \begin{align*}
        A = \bigcup_{t \in [a,b]} \left( A_t \times \{t\}\right)
    \end{align*}
    is an $(n+1)$-manifold in $\RR^{n + 1}$.
\end{definition}
The generalized Stokes theorem from Theorem 1 in \cite{baddeley1977integrals} is collected as follows.
\begin{theorem}[Generalized Stokes Theorem] 
\label{thm: S.13.1}
    Let $\{A_t : t \in [a,b]\}$ be a family of smoothly changing $n$-manifold in $\RR^n$. Then for any smooth function $f(\xb, t): \RR^{n + 1} \mapsto \RR$ on $A$, we have 
    \begin{align} \label{eqn: stokes}
        \frac{\partial}{\partial t} \int_{A_t} f  = \int_{A_t} \frac{\partial f}{\partial t} + \int_{\partial A_t} f \frac{\partial g}{\partial t}.
    \end{align} 
\end{theorem}

\subsection{Normalized linear model}
\label{change of normalization}

We define a new normalized linear model as follows.
We will use this normalization setting in Sections \ref{sec: Proof for subsec: calibration}-\ref{sec: Proof for sec: L_2 estimator}. 
A sequence of instances 
$$
\{\bbeta_0(p), {\bepsilon}_x(n_x), {\bepsilon}_s(n_s), {\bSigma}(p), \X(n_x, p), \Sbb(n_s, p),\W(n_w, p)\}_{(p, n_x, n_s, n_w) \in \NN^4},
$$
indexed by $p, n_x, n_s$ and $n_w$, is said to be a {\BS normalized converging sequence} if $\bbeta_0(p) \in \RR^{p}, {\bepsilon}_x(n_x) \in \RR^{n_x}, {\bSigma}(p) \in \RR^{p \times p}, \X(n_x, p) \in \RR^{n_x \times p}, \W(n_w, p) \in \RR^{n_w \times p}$ satisfy Conditions \ref{cond1-np-ratio} - \ref{cond: exchange limit} with Conditions \ref{cond2-mp} and \ref{cond-distn-data} being replaced by Conditions~\ref{cond2*-mp} and \ref{cond-distn-X*} below. 

\begin{condition}
\label{cond2*-mp} 
We assume that $\bbeta_0$ follows a distribution with mean $\bm {0}$ and covariance $\bmSigma_{\bbeta_0}$, where $\bmSigma_{\bbeta_0} \in \R^{p \times p}$ is diagonal and has uniformly bounded eigenvalues.
For $i=m+1, \cdots, p$, $(\bmSigma_{\bbeta})_{ii} = 0$ and for $i=1, \cdots, m$, $(\bmSigma_{\bbeta})_{ii} \neq 0$ . 
Thus, there are $m$ nonzero signals in $\bbeta_0$, and let $m/p \to \kappa \in (0,1]$ as $m, p \to \infty$. 
There exists $\sigma_{\bbeta}^2>0$ such that $\kappa \cdot \sigma_{\bbeta}^2 = \sum_{i=1}^p (\bmSigma_{\bbeta})_{ii}/p$. 
Moreover, as $n_x, p \to \infty$ with $p/n_x \to \gamma_x >0$, the empirical distributions of the entries of $\bbeta_0$ converges weakly to a probability measure $\probP_{\barbeta}$ on $\RR$ with bounded second moment. 
Further, we assume that $p^{-1} \sum_{j=1}^p (\bbeta_{0,j})^2 \to \E_{\probP_{\barbeta}}(\bar{\beta}^2)$. 
\end{condition}
\begin{condition}
\label{cond-distn-X*}
Let $\X = \X_0 \bmSigma^{1/2} \in \R^{n_x \times p}$, $\Sbb = \Sbb_0 \bmSigma^{1/2} \in \R^{n_s \times p}$, and $\W = \W_0 \bmSigma^{1/2} \in \RR^{n_w \times p}$ where $\bmSigma \in \R^{p \times p}$ is  positive definite  and the entries of $\X_0$, $\Z_0$, and $\W_0$ are i.i.d. $N(0,1/n_x)$, $N(0, 1/n_s)$, and $N(0, 1/n_{w})$, respectively. The eigenvalues of $\bmSigma$ are uniformly bounded, that is, there exists absolute constants ${ \BS c, C}$ such that $c \leq \lambda_{\min}(\bmSigma) \leq \lambda_{\max}(\bmSigma) \leq C$. 
\end{condition}

With these two conditions, the $L_1$ estimator defined in~\eqref{eqn:-lasso-est with normalization} can be rewritten as
\begin{align} \label{eqn:-lasso-est}
\begin{split}
\hat{\bbeta}_{\textnormal{L}}(\lambda) = \arg\min_{\bbeta \in \R^p} \frac{1}{2} \bbeta^{\T} \X^{\T} \X \bbeta -  \bbeta^{\T} \X^{\T} \y_x + \lambda \| \bbeta \|_1,
\end{split}
\end{align}
and the reference panel based $L_1$ estimator defined in \eqref{eqn:ref-panel-lasso-est without normalization} becomes
\begin{align} \label{eqn:ref-panel-lasso-est}
\begin{split}
\hat{\bbeta}_{\LW}(\lambda) = \arg\min_{\bbeta \in \R^p} \frac{1}{2} \bbeta^{\T} \W^{\T} \W \bbeta - \bbeta^{\T} \X^{\T} \y_x + \lambda \| \bbeta \|_1.
\end{split}
\end{align}
Moreover, $\hat{\bbeta}_{\RW}(\lambda)$ and $\hat{\bbeta}_{\RA}(\lambda)$ has closed-form representations: 
\begin{equation} \label{eqn:ref-panel-ridge-est with normalization}
\hat{\bbeta}_{\RA}(\lambda) = (\X^{\T} \X + \lambda \bI_{p})^{-1}\X^{\T}\y_x
\quad \mbox{and} \quad
\hat{\bbeta}_{\RW}(\lambda) = (\W^{\T} \W + \lambda \bI_{p})^{-1}\X^{\T}\y_x.
\end{equation}

\subsection{Proof of Proposition \ref{prop 1}} \label{proof of prop 1}

In this section, we prove Proposition \ref{prop 1}. 
We have 
\begin{align}
    &\tau_*^2 =  \lim_{p \to \infty} \EE \left[ \frac{1}{n_w} \bnorm{{\eta}_{\alpha \tau_*}(\tau_* {\bSigma}^{-1/2} \Zb + (1 + b_*){\bbeta})}_{\bSigma}^2 + \frac{1}{n_x} (1 + b_*)^2 \bnorm{{\bbeta}_0}_{\bSigma}^2 \cdot \frac{1}{h_x^2} \right] \\
    \label{1/1+b equation}  
    &\mbox{and} \quad (1 + b_*)^{-1} = \left\{ 1 - \lim_{p \to \infty} \frac{1}{n_w} \EE \Div \eta_{\alpha \tau_*}(\tau_* {\bSigma}^{-1/2} \Zb + (1 + b_*) \bbeta_0) \right\}_+,
\end{align}
which are the same as equations~\eqref{eq:prop4_fixed_pt_eqns} and \eqref{eq:prop4_fixed_pt_eqns_b} but under the normalization described in Section~\ref{change of normalization}.
Let $\zeta_* = (1 + b_*)/\tau_* \geq 0$. We divide both sides of the first equation by $\tau_*^2$, and denote the resulting equation by $f(\zeta_*, \alpha) = 1$, where $f(\zeta, \alpha)$ is defined as 
\begin{equation} \label{eqn: 8.69}
 f(\zeta, \alpha):=   \zeta^2 \cdot \Bigg( \lim_{p \to \infty} \EE \frac{1}{n_x} \bnorm{{\bbeta}_0}_{\bSigma}^2 \cdot \frac{1}{h_x^2} \Bigg) + \lim_{p \to \infty} \EE \left\{ \frac{1}{n_w} \bnorm{{\eta}_{\alpha}({\bSigma}^{-1/2} \Zb + \zeta {\bbeta}_0)}_{\bSigma}^2 \right\}. 
\end{equation}

We have the following two technical lemmas about the function $f(\zeta, \alpha)$. 
Lemma \ref{f-convex} is used in the proof of the uniqueness of $\zeta_*$, and the proof of Lemma \ref{f-convex} is deferred to Section~\ref{Proof of Lemma f-convex}. 
\begin{lemma}\label{f-convex}
For any fixed $\alpha$ and $\gamma_w$, $f(\zeta, \alpha)$ is strictly convex with respect to $\zeta$ on $\zeta \geq 0$.  
\end{lemma}
Lemma \ref{monotone wrt alpha} below is used in the proof of the uniqueness of $\alpha_{\min}$. We defer its  proof  to Section \ref{proof of Lemma monotone wrt alpha}. 
\begin{lemma} \label{monotone wrt alpha}
For any fixed $\zeta \geq 0$, $f(\zeta, \alpha)$ given in Equation \eqref{eqn: 8.69} is a continuous strictly decreasing function with respect to $\alpha$.
In particular, $g(\alpha):= f(0, \alpha)$ is strictly decreasing with respect to $\alpha$. Moreover, when $\gamma_w > 1$, there exists a unique $\alpha_{\min}$ that satisfies 
    \begin{equation*}
        g(\alpha) = \lim_{p \to \infty} \frac{1}{n_w} \EE \bnorm{\eta_{\alpha}({\bSigma}^{-1/2} \Zb)}_{\bSigma}^2 = 1.
    \end{equation*}
\end{lemma}

With these two lemmas, we now proceed to prove Proposition \ref{prop 1}. 
{\BS When $\gamma_w > 1$, the uniqueness of $\alpha_{\min}$ follows from Lemma \ref{monotone wrt alpha}.} 
For any fixed $\alpha > \alpha_{\min}$ that satisfies the condition in Proposition \ref{prop 1}, we now establish the existence of a solution $\zeta_*$ to the equation $f(\zeta_*, \alpha) = 1$.
When $\gamma_w > 1$ and $\alpha > \alpha_{\min}$, since $f(0, \alpha)$ is strictly decreasing in $\alpha$, we have
\begin{equation*}
    f(0,\alpha) = \lim_{p \to \infty} \frac{1}{n_w} \bnorm{\eta_{\alpha}({\bSigma}^{-1/2} \Zb)}_{\bSigma}^2 
    < f(0, \alpha_{\min})
    =1.
\end{equation*}
When $\gamma_{w} \leq 1$ and $\alpha > 0$, define $\hat{y}_{\theta}: \RR^{p} \mapsto \RR^{p}$ by
\begin{equation} \label{hat{y}}
    \hat{y}_{\theta}(\vb) := \arg\min_{\bbeta} \left\{\frac{1}{2} \|\bbeta - \vb\|_2^2 + \theta \bnorm{{\bSigma}^{-1/2} \bbeta}_1 \right\}.
\end{equation}
Note that $\hat{y}_{\theta} (v) = {\bSigma}^{1/2} \eta_{\theta}({\bSigma}^{-1/2}v)$ because
\begin{equation*}
\begin{split}
    \hat{y}_{\theta}(\vb) &= {\bSigma}^{1/2} \arg\min_{\bbeta \in \RR^{p}} \left\{ \frac{1}{2} \bnorm{\vb - {\bSigma}^{1/2}\bbeta}_2^2 + \theta \bnorm{\bbeta}_1 \right\}\\
    &= \arg\min_{\bbeta \in \RR^{p}} \left\{ \frac{1}{2} \bnorm{\vb - \bbeta}_2^2 + \theta \bnorm{{\bSigma}^{-1/2} \bbeta}_1 \right\}.
\end{split}
\end{equation*}
Therefore, we have 
\begin{equation*}
    \bnorm{\eta_{\alpha}({\bSigma}^{-1/2} \Zb)}_{\bSigma}^2 = \bnorm{\hat{y}_{\alpha}(\Zb)}_2^2.
\end{equation*}
{\BS Moreover, Section 2.3 in \cite{10.1561/2400000003} indicates that proximal operator $\hat{y}_{\alpha}(\vb)$ is firm nonexpansive, satisfying
\begin{align} \label{eqn:1-lip}
    \|\hat{y}_{\alpha}(\vb_{1}) - \hat{y}_{\alpha}(\vb_{2})\|_{2}^{2} \leq \left\langle \vb_{1} - \vb_{2}, \hat{y}_{\alpha}(\vb_{1}) - \hat{y}_{\alpha}(\vb_{2}) \right \rangle,
\end{align}
which, by the Cauchy–Schwarz inequality, is therefore $1$-Lipschitz.}  Then we have $\EE \bnorm{\hat{y}_\alpha(\Zb)}_2^2 \leq \EE \bnorm{\Zb}_2^2 = p$, where $\Zb \sim N(0, \bI_p)$. Thus, when $\gamma_{w} \leq 1$ and $\alpha > 0$, we have
\begin{equation} \label{f(0, alpha) < 1}
    f(0,\alpha) = \lim_{p \to \infty} \frac{1}{n_w} \EE \bnorm{\eta_{\alpha}({\bSigma}^{-1/2} \Zb)}_{\bSigma}^2 < 1.
\end{equation}
On the other hand, for arbitrary $\gamma_{w}$, as $\zeta \to \infty$, we have 
\begin{equation*}
    f(\zeta, \alpha) > \zeta^2 \cdot \left( \lim_{p \to \infty} \EE \frac{1}{n_x} \bnorm{{\bbeta}_0}_{\bSigma}^2 \cdot \frac{1}{h_x^2} \right) \to \infty.
\end{equation*}
Therefore, as $\zeta$ becomes larger and approaches infinity, $f(\zeta, \alpha)$ tends to infinity as well. By the intermediate value theorem, this completes the proof of the existence of the solution $\zeta_*$ satisfying $f(\zeta_*, \alpha) = 1$. 
The uniqueness of the solution $\zeta_*$ of equation $f(\zeta_*, \alpha) = 1$ follows from Lemma \ref{f-convex}.

Next, we show the existence and uniqueness of $b_*$. We first observe that
\begin{equation*}
\begin{split}
    \eta_{\alpha \tau_*}\left\{\tau_* {\bSigma}^{-1/2} \Zb + (1 + b_*) \cdot \bbeta_0 \right\} &= \tau_* \cdot \eta_{\alpha}\left\{{\bSigma}^{-1/2} \Zb + (1 + b_*)/\tau_* \cdot \bbeta_0 \right\}.
\end{split}
\end{equation*}
Taking divergence with respect to $\bbeta_0$ on both side, we have 
\begin{equation*}
\begin{split}
    \Div \eta_{\alpha \tau_*}\{\tau_* {\bSigma}^{-1/2} \Zb + (1 + b_*) \bbeta_0\} = \Div \eta_{\alpha}\{{\bSigma}^{-1/2} \Zb + (1 + b_*)/\tau_* \bbeta_0\}.
\end{split}
\end{equation*}
Using the fact that  $(1 + b_*)/\tau_* = \zeta_*$ and plugging the above equation into \eqref{1/1+b equation}, we obtain
\begin{equation*}
\begin{split}
    \frac{1}{1 + b_{*}} 
    &= \left\{ 1 - \lim_{p \to \infty} \frac{1}{n_w} \EE \Div \eta_{\alpha}({\bSigma}^{-1/2} \Zb + \zeta_* \bbeta_0) \right\}_+.
\end{split}
\end{equation*}
Since $\zeta_*$ exists and is unique, it follows from the above equation that $b_*$ is also uniquely well-defined. Moreover, since $\zeta_*$ and $1 + b_*$ are uniquely well-defined, $\tau_* = (1 + b_*)/\zeta_*$ is also uniquely well-defined.

\subsection{Proof of Proposition \ref{prop 2}} \label{proof of prop 2}

We prove Proposition \ref{prop 2} in 3 steps:
\begin{itemize}
    \item[1] We show that $\lambda(\alpha)$ is continuous with respect to $\alpha$;

    \item[2] We show that the function $\lambda \mapsto \alpha(\lambda)$ is well-defined by using the intermediate value theorem;

    \item[3] We show that, for any $\alpha_*$ satisfying $\lambda(\alpha_*) = \lambda$, ${1}/(1 + b_*)$ defined in equation~\eqref{eq:prop4_fixed_pt_eqns_b} is positive. 
\end{itemize}

\paragraph{Step 1} 
Lemma \ref{zeta continuous} shows that $\zeta_*(\alpha)$ is continuous with respect to $\alpha$, whose proof is deferred to Section \ref{proof:zeta continuous}. 
\begin{lemma} \label{zeta continuous}
Let $\zeta_*(\alpha)$ denote the implicit solution of $f(\zeta, \alpha) = 1$. 
If $\gamma_w > 1$, then $\zeta_*(\alpha)$ is continuous with respect to $\alpha > \alpha_{\min}$.
If $\gamma_w \leq 1$, then $\zeta_*(\alpha)$ is continuous with respect to $\alpha > 0$.
\end{lemma}
Recall the definition of $\lambda(\alpha)$ in \eqref{lambda(alpha)}. With the new normalization, we have 
\begin{align*}
     \lambda(\alpha) = \alpha {\tau_*} \Bigg[1 - \lim_{p \to \infty} \frac{1}{n_w} \EE \Div \eta_{\alpha {\tau_*}} \{{\tau_*} {\bSigma}^{-1/2} \Zb + (1 + b_*)\bbeta_0\} \Bigg]. 
\end{align*}
We first show the continuity of $\lambda(\alpha)$ in $\alpha$. Note that by Proposition~\ref{prop 1}, we have
\begin{align*}
    \frac{1}{1+b_*} = \left\{ 1 - \lim_{p \to \infty} \frac{1}{n_w} \EE \Div \eta_{\alpha}({\bSigma}^{-1/2} \Zb + \zeta_* \bbeta_0) \right\}_+.
\end{align*}
In addition, $\eta_{\alpha}({\bSigma}^{-1/2} \Zb + \zeta_* \bbeta_0)$ can be viewed as the $L_1$ regularized solution with $\Xb ={\bSigma}^{-1/2}$, 
$\yb = \Zb+\zeta_* {\bSigma}^{1/2} \bbeta_0$, and regularization parameter $\alpha$. 
By Lemma~\ref{zeta continuous}, $\eta_{\alpha}({\bSigma}^{-1/2} \Zb + \zeta_* \bbeta_0)$ is continuous in $\alpha$ almost surely with respect to $\Zb$. Thus, $b_* = b_*(\alpha)$ is continuous in $\alpha$. Rewrite the above equation as
\begin{align*}
    \lambda(\alpha) 
    = \alpha {\tau_*} \Bigg\{1 - \lim_{p \to \infty} \frac{1}{n_w} \EE \Div \eta_{\alpha} ({\bSigma}^{-1/2} \Zb + \zeta_* \bbeta_0) \Bigg\},
\end{align*}
and note that $\tau_* = (1+b_*)/\zeta_*$ is also continuous in $\alpha$. It follows that $\lambda(\alpha)$ is continuous in $\alpha$.

\paragraph{Step 2} 
We only prove the results in steps 2 and 3 when $\gamma_w > 1$. Steps 2 and 3 in the case of $\gamma_{w} \leq 1$ can be proved similarly and thus are omitted. 
In this step, we only show that the function $\lambda \mapsto \alpha(\lambda)$ is well-defined when $\gamma_w > 1$ by using the intermediate value theorem. 
For $\gamma_w > 1$, we aim to prove the existence of a unique $\alpha > \alpha_{\min}$ such that the following condition holds 
\begin{equation*} 
\begin{split}
    \lambda &= \alpha {\tau_*} \Bigg\{1 - \lim_{p \to \infty} \frac{1}{n_w} \EE \Div \eta_{\alpha {\tau_*}} ({\tau_*} {\bSigma}^{-1/2} \Zb + (1 + b_*)\bbeta_0) \Bigg\}_{+}\\
    {\iff} \frac{\lambda}{\alpha} &= {\tau_*} \Bigg\{1 - \lim_{p \to \infty} \frac{1}{n_w} \EE \Div \eta_{\alpha {\tau_*}} ({\tau_*} {\bSigma}^{-1/2} \Zb + (1 + b_*)\bbeta_0) \Bigg\}_+\\
    {\iff} \frac{\lambda}{\alpha} &=  \left\{ \tau_* - \lim_{p \to \infty} \frac{1}{n_w} \EE \langle {\bSigma}^{1/2} \eta_{\alpha {\tau_*}} ({\tau_*} {\bSigma}^{-1/2} \Zb + (1 + b_*)\bbeta_0), \Zb \rangle \right\}_+,
\end{split}
\end{equation*}
or equivalently,
\begin{equation} \label{eqn: calibration equation}
    \frac{\lambda}{\alpha} =  \left\{ \tau_* - \tau_* \lim_{p \to \infty} \frac{1}{n_w} \EE \langle {\bSigma}^{1/2} \eta_{\alpha} ( {\bSigma}^{-1/2} \Zb + \zeta_* \bbeta_0), \Zb \rangle \right\}_+.
\end{equation}
{Equation~\eqref{eqn: calibration equation}} follows from { \BS Stein's lemma in Lemma \ref{Stein Lemma} with $q = 1$}, 
along with the chain rule of divergence. Recall that the fixed point equations \eqref{eqn: 8.69} and \eqref{1/1+b equation} are 
\begin{equation*} 
\begin{split} 
    &1 = \zeta_*^2 \cdot \Bigg( \lim_{p \to \infty} \EE \frac{1}{n_x} \bnorm{{\bbeta_0}}_{\bSigma}^2 \cdot \frac{1}{h_x^2} \Bigg) + \lim_{p \to \infty} \EE \left[ \frac{1}{n_w} \bnorm{{\eta}_{\alpha}({\bSigma}^{-1/2} \Zb + \zeta_* {\bbeta_0})}_{\bSigma}^2 \right] \quad \mbox{and}\\
    &\frac{1}{1 + b_*} = \left( 1 - \lim_{p \to \infty} \frac{1}{n_w} \EE \Div \eta_{\alpha}( {\bSigma}^{-1/2} \Zb + \zeta_* \bbeta_0) \right)_+.
\end{split}
\end{equation*}
These two equations admit a unique solution pair, as proven in Proposition \ref{prop 1}.
As $\alpha$ approaches infinity, the unique solution $\zeta_*^2$ of Equation $f(\zeta_{*}, \alpha) = 1$ converges to  $1/\left( \lim_{p \to \infty} \EE \frac{1}{n_x} \bnorm{{\bbeta_0}}_{\bSigma}^2 \cdot \frac{1}{h_x^2} \right)$, which is positive and bounded. 
Besides, as $\alpha$ tends to infinity, by Equation \eqref{1/1+b equation}, we have ${1}/(1+b_*) \to 1$. 
Since $\zeta_* = (1+b_*)/\tau_*$, we have $\tau^2_* \to \left( \lim_{p \to \infty} \EE \frac{1}{n_x} \bnorm{{\bbeta_0}}_{\bSigma}^2 \cdot \frac{1}{h_x^2} \right)$, which is also positive and bounded. 
Because both $\zeta_*$ and $\tau_*$ are  positive and bounded, as $\alpha$ tends to infinity, we have 
\begin{align*}
\lim_{p \to \infty} \frac{1}{n_w} \EE \langle {\bSigma}^{1/2} \eta_{\alpha}({\bSigma}^{-1/2} \Zb + \zeta_* \bbeta_0), \Zb \rangle \to 0.
\end{align*}
Consequently, we conclude that the right-hand side of Equation \eqref{eqn: calibration equation} asymptotically equals {\BS $\tau_* \in (0, \infty)$}.
However, when $\alpha \to \infty$, the left-hand side $\lambda/\alpha \to 0$. 

We now consider the scenario when $\alpha$ tends to $\alpha_{\min}$. By Lemma \ref{zeta continuous}, {as $f(0, \alpha_{\min}) = g(\alpha_{\min}) = 1$}, we have $\zeta_*(\alpha) \to 0$ as $\alpha \to \alpha_{\min}$. We claim that
\begin{equation}\label{eqn: 8.80}
\begin{split} 
    &1 - \lim_{p \to \infty} \frac{1}{n_w} \EE \Div \eta_{\alpha \tau_*}\{\tau_* {\bSigma}^{-1/2} \Zb + (1 + b_*) \bbeta_0\}\\ 
    &= 1 - \lim_{p \to \infty} \frac{1}{n_w} \EE \Div \eta_{\alpha}({\bSigma}^{-1/2} \Zb + \zeta_* \bbeta_0)
    \leq 0.
\end{split}
\end{equation}
To this end, recall $\lim_{\alpha \to \alpha_{\min}} g(\alpha) = g(\alpha_{\min}) = 1$,  it suffices to show 
\begin{equation*} 
\begin{split} 
    \lim_{p \to \infty} \frac{1}{n_w} \EE \Div \eta_{\alpha}({\bSigma}^{-1/2} \Zb)
    &= \lim_{\alpha \to \alpha_{\min}} \lim_{p \to \infty} \frac{1}{n_w} \EE \Div \eta_{\alpha}({\bSigma}^{-1/2} \Zb + \zeta_* \bbeta_0) \\
    &\geq \lim_{\alpha \to \alpha_{\min}} \lim_{p \to \infty} \frac{1}{n_w} \EE\ \bnorm{\eta_{\alpha}({\bSigma}^{-1/2} \Zb)}_{{\bSigma}}^2.
    \end{split}
\end{equation*}
Consider the $L_1$ regularized solution
$$
\hat{\bv} := \arg\min_{\bbeta \in \RR^{p}} \left\{ \frac{1}{2} \|\Zb - \bSigma^{1/2} \bbeta\|_{2}^{2} + \alpha \|\bbeta\|_{1} \right\} = \eta_{\alpha}({\bSigma}^{-1/2} \Zb),
$$
and the solution path of the estimator $\hat{\yb} = {\bSigma}^{1/2} \hat{\bv}$ as we decrease $\alpha$. 
At each step, $\hat{\yb}$ moves toward the least square fit determined by the current active set of variables. 
Rather than fitting the model completely, the process pauses as soon as a new variable enters the active set. 
As shown by \cite{radchenko2011improved}, $\hat{\yb}$ moves toward the least square fit determined by the updated active set of variables. Let $\cA$ be the current active set, 
since $\bnorm{\hat{\yb}}_2^2 = \bnorm{{\bSigma^{1/2}_{\cA}}\hat{\bv}_{\cA}}_2^2$ decreases with $\alpha$, $\bnorm{\hat{\yb}}_{2}^2$ is smaller than the full least square fit $\bnorm{{\bSigma^{1/2}_{\cA}}({\bSigma^{1/2}_{\cA}}^{\T}{\bSigma^{1/2}_{\cA}})^{-1}{\bSigma^{1/2}_{\cA}}^{\T} \yb}_2^2$ according to above discussion. 
Therefore, we have 
\$
\EE\ \bnorm{\eta_{\alpha}({\bSigma}^{-1/2} \Zb)}_{{\bSigma}}^2
&= \EE\ \eta_{\alpha}({\bSigma}^{-1/2} \Zb)^{\T} \bSigma \eta_{\alpha}({\bSigma}^{-1/2} \Zb)  \\
&=  \EE\ \bnorm{\hat{\yb}}_2^2 \\
&< \EE\ \bnorm{{\bSigma^{1/2}_{\cA}}({\bSigma^{1/2}_{\cA}}^{\T}{\bSigma^{1/2}_{\cA}})^{-1}{\bSigma^{1/2}_{\cA}}^{\T} \yb}_2^2 \\
&= \EE\ \bnorm{\hat{\bv}}_0,
\$
where the last equality follows from the following observation
\$
\EE\ \bnorm{{\bSigma^{1/2}_{\cA}}({\bSigma^{1/2}_{\cA}}^{\T}{\bSigma^{1/2}_{\cA}})^{-1}{\bSigma^{1/2}_{\cA}}^{\T} \yb}_2^2 
&= \EE\ \bnorm{{\bSigma^{1/2}_{\cA}}({\bSigma^{1/2}_{\cA}}^{\T}{\bSigma^{1/2}_{\cA}})^{-1}{\bSigma^{1/2}_{\cA}}^{\T}\Zb}_2^2 \\
&= \EE\ \bigg[ \EE \left\{ \bnorm{{\bSigma^{1/2}_{\cA}}({\bSigma^{1/2}_{\cA}}^{\T}{\bSigma^{1/2}_{\cA}})^{-1}{\bSigma^{1/2}_{\cA}}^{\T} \yb}_2^2 \bigg| |\cA| \right\} \bigg].
\$
Consider the distribution of $ \bnorm{{\bSigma^{1/2}_{\cA}}({\bSigma^{1/2}_{\cA}}^{\T}{\bSigma^{1/2}_{\cA}})^{-1}{\bSigma^{1/2}_{\cA}}^{\T} \yb}_2^2$ conditioning on $|\cA|$. It is known that the $\ell_2$-norm of the projection of a $p$-dimensional vector with i.i.d. unit-variance and zero mean Gaussian components onto an independent $|\cA|$-dimensional subspace 
follows chi-squared distribution with $|\cA|$ degree of freedom. 
Therefore, we have 
$$\EE\ \left\{\bnorm{{\bSigma^{1/2}_{\cA}}({\bSigma^{1/2}_{\cA}}^{\T}{\bSigma^{1/2}_{\cA}})^{-1}{\bSigma^{1/2}_{\cA}}^{\T} \yb}^2\bigg| |\cA| \right\} = |\cA|,$$ 
which implies that 
$$
\EE \bnorm{{\bSigma^{1/2}_{\cA}}({\bSigma^{1/2}_{\cA}}^{\T}{\bSigma^{1/2}_{\cA}})^{-1}{\bSigma^{1/2}_{\cA}}^{\T} \yb}^2 = \EE |\cA| = \EE \left\| \hat{\bv} \right\|_0.
$$

Next, we show $\EE \Div \eta_{\alpha}({\bSigma}^{-1/2} \Zb) = \EE\ \left\| \hat{\bv} \right\|_0$.
From the KKT condition on $\eta_{\alpha}(v)$, we have
\begin{equation*}
\begin{split}
   & \Sigma(\eta_{\alpha}(v) - v) + \alpha \partial \norm{\eta_{\alpha}(v)}_1 = 0 \quad \mbox{and} \\
   & \eta_{\alpha}(v) - v = \Sigma^{-1} \alpha \partial \norm{\eta_{\alpha}(v)}_1.
\end{split}
\end{equation*}
Let $\cA$ be the active set of $\eta_{\alpha}(v)$. Then 
\$
\eta_{\alpha}(v)_{\cA} - v_{\cA} = \Sigma^{-1}_{\cA} \alpha \partial \norm{\eta_{\alpha}(v)_{\cA}}_1 = \Sigma^{-1}_{\cA} \alpha \sign{\eta_{\alpha}(v)_{\cA}}.
\$
Taking derivative with respect to $v$, then
\$
\Div{\eta_{\alpha}(v)_{\cA}} = \Div{v_{\cA}} = |\cA| = \bnorm{\hat{\bv}}_0.
\$

As $\alpha \to \alpha_{\min}$, we have $\frac{1}{n_w} \EE\ \bnorm{\eta_{\alpha}({\bSigma}^{-1/2} \Zb)}_{{\bSigma}}^2 \to 1$ by definition of $\alpha_{\min}$ and Lemma \ref{zeta continuous}.
Therefore, for $\alpha$ sufficiently close to $\alpha_{\min}$, we have established Equation \eqref{eqn: 8.80}. Moreover, the right-hand side of equation~\eqref{eqn: calibration equation} can be written as
\begin{equation*}
\begin{split}
    &\left[\tau_* - \lim_{p \to \infty} \frac{1}{n_w} \EE \langle {\bSigma}^{1/2} \eta_{\alpha {\tau_*}} \{{\tau_*} {\bSigma}^{-1/2} \Zb + (1 + b_*)\bbeta_0\}, \Zb \rangle \right]_+\\
    &= \tau_* \left\{1 - \lim_{p \to \infty} \frac{1}{n_w} \EE \langle {\bSigma}^{1/2} \eta_{\alpha} ({\bSigma}^{-1/2} \Zb + \zeta_* \bbeta_0), \Zb \rangle \right\}_+. 
\end{split}
\end{equation*}
Since 
\$
1 - \lim_{p \to \infty} \frac{1}{n_w} \EE \langle {\bSigma}^{1/2} \eta_{\alpha} ({\bSigma}^{-1/2} \Zb + \zeta_* \bbeta_0), \Zb \rangle 
\$ 
is negative when $\alpha$ is close enough to $\alpha_{\min}$, the right-hand side of Equation \eqref{eqn: calibration equation} equals $0$ when $\alpha \to \alpha_{\min}$. 
On the other hand, in Equation \eqref{eqn: calibration equation}, as $\alpha$ tends to $\alpha_{\min}$, the left-hand side converges as ${\lambda}/{\alpha} \to {\lambda}/{\alpha_{\min}}$, and the converging limit is a positive and bounded constant. 
That is, as $\alpha$ decreases from $\infty$ to $\alpha_{\min}$, $\lambda/\alpha$ on the left-hand side of equation~\eqref{eqn: calibration equation} increases from 0 to $\lambda/\alpha_{\min}$, and the right-hand side of equation~\eqref{eqn: calibration equation} decreases from $\tau_*^2$ (a positive and bounded quantity) to 0. 
By the intermediate value theorem, there must exist a solution $\alpha_*$ to Equation \eqref{eqn: calibration equation} {such that the left-hand side equals the right-hand side at $\alpha_*$}. This completes the proof of step 2.

\paragraph{Step 3}
Finally, in step 3, at $\alpha = \alpha_*$, we have
\begin{equation*}
    0 < \frac{\lambda}{\alpha_{*} \tau_*} = \left\{ 1 - \lim_{p \to \infty} \frac{1}{n_w} \EE \langle {\bSigma}^{1/2} \eta_{\alpha} ({\bSigma}^{-1/2} \Zb + \zeta_* \bbeta_0), \Zb \rangle \right\}_+ = \frac{1}{1 + b_*}. 
\end{equation*}

Lastly, we point out again that, when $\gamma_w \leq 1$ and $\alpha > 0$, steps 2 and 3, and thus Proposition \ref{prop 2}, can be proven similarly. This completes the proof. 

\subsection{Proofs for supporting lemmas}
\label{sec: proof of supporting lemmas for section 4.2}
In this section, we provide  proofs for the supporting lemmas used in Sections \ref{proof of prop 1} and \ref{proof of prop 2}.

\subsubsection{Proof of Lemma \ref{f-convex}} \label{Proof of Lemma f-convex}

We aim to show the second derivative of $f(\zeta,\alpha)$ with respect to $\zeta$ is positive. Let $\hat{\bbeta}(\Zb) := {\eta}_{\alpha}({\bSigma}^{-1/2} \Zb + \zeta {\bbeta_0})$, and 
let $\cA$ denote the active set of $\hat{\bbeta}(\Zb)$.  By the KKT condition, we have 
\begin{equation} \label{hat{beta}}
\begin{split}
    \hat{\bbeta}_{\cA} =  ({\bSigma}_{\cA, \cA})^{-1} \bigg\{ ({\bSigma}^{1/2}\Zb)_\cA + \zeta ({\bSigma}\bbeta_0)_\cA - \alpha \sign(\hat{\bbeta}_\cA) \bigg\}.
\end{split}
\end{equation}
Therefore, for any fixed active set $\cA$, we have 
\$
    &\lim_{p \to \infty} \frac{1}{n_w} \bnorm{{\eta}_{\alpha}({\bSigma}^{-1/2} \Zb + \zeta {\bbeta_0})}_{\bSigma}^2 \\
    &= \lim_{p \to \infty} \frac{1}{n_w} \EE\  \hat{\bbeta}_{\cA}^{\T} {\bSigma}_{\cA, \cA} \hat{\bbeta}_{\cA} \\ 
    &= \lim_{p \to \infty} \frac{1}{n_w} \EE \Bigg\{ ({\bSigma}^{1/2}\Zb)_\cA + \zeta ({\bSigma}\bbeta_0)_\cA - \alpha \sign(\hat{\bbeta}_\cA) \Bigg\}^{\T} {\bSigma}_{\cA,\cA}^{-1} \Bigg\{ ({\bSigma}^{1/2}\Zb)_\cA + \zeta ({\bSigma}\bbeta_0)_\cA - \alpha \sign(\hat{\bbeta}_\cA) \Bigg\}\\
    &= \lim_{p \to \infty}  \zeta^2 \cdot \frac{1}{n_w} \EE\ \bbeta_0^{\T} {\bSigma}_{\cA,*}^{\T} {\bSigma}_{\cA,\cA}^{-1} {\bSigma}_{\cA,*} \bbeta_0 + 2\zeta \cdot \bbeta_0^{\T} {\bSigma}_{\cA,*}^{\T} {\bSigma}_{\cA,\cA}^{-1} \left\{({\bSigma}^{1/2}\Zb)_\cA - \alpha \sign(\hat{\bbeta}_\cA) \right\}\\ +
    &= \lim_{p \to \infty} \frac{1}{n_w} \EE \Bigg\{ ({\bSigma}^{1/2}\Zb)_\cA - \alpha \sign(\hat{\bbeta}_\cA) \Bigg\}^{\T} {\bSigma}_{\cA,\cA}^{-1} \Bigg\{ ({\bSigma}^{1/2}\Zb)_\cA - \alpha \sign(\hat{\bbeta}_\cA) \Bigg\}.
\$

Now we compute the second derivative  of $f(\zeta,\alpha)$  with respect to $\zeta$ using Theorem \ref{thm: S.13.1}. 
We divide the whole $p$-dimensional space into regions such that the active set of $\hat{\bbeta}(\Zb) = {\eta}_{\alpha}({\bSigma}^{-1/2} \Zb + \zeta {\bbeta_0})$ remains the same in each region and changes by one variable between two neighboring regions that share a common boundary hyperplane. Let $S_i$ and $S_j$ be two neighboring regions that share a common hyperplane. Denote $\cA_i$ and $\cA_j$ to be the active sets in $S_i$ and $S_j$, respectively. 
Assume $\hat{\bbeta}_{\cA_i}[k] > 0$ in $S_i$ and $\hat{\bbeta}_{\cA_j}[k] = 0$ in $S_j$, where $\hat{\bbeta}_{\cA_i}[k]$ and $\hat{\bbeta}_{\cA_j}[k]$ are the $k-$th components of $\hat{\bbeta}_{\cA_i}$ and $\hat{\bbeta}_{\cA_j}$, respectively.  Then, by Equation \eqref{hat{beta}}, the boundary $F_{ij}$ is given by the following equation
\begin{equation*}
    g_{ij}(\Zb,\zeta) := e_{k}^{\T} \cdot \hat{\bbeta}_{\cA_i} =  e_{k}^{\T} \cdot {\bSigma}_{\cA_i, \cA_i}^{-1} \bigg\{ ({\bSigma}^{1/2}\Zb)_{\cA_i} + \zeta ({\bSigma}\bbeta_0)_{\cA_i} - \alpha \sign(\hat{\bbeta}_{\cA_i}) \bigg\} = 0,
\end{equation*}
where $e_k$ represent the $k-$th coordinate vector for $\hat{\bbeta}_{\cA_i}$. 

Let $u(\Zb, \zeta) := \| \eta_{\alpha}(\bSigma^{-1/2} \Zb + \zeta \bbeta_0) \|_{\bSigma}^2$ for $\Zb \in \R^p$ and $\zeta \geq 0$. 
Then for any fixed $\zeta$, let $S_i = S_i(\zeta) := \text{int} \left[ \Zb \in \R^p, \supp\big\{\hat{\bbeta}(\Zb, \zeta) \big\} = \cA_i \right]$ for $\cA_i \subseteq \{1, 2, \cdots, p\}$, 
where $\text{int}[E]$ denotes the interior of a set $E$. 
We have $u(\Zb, \zeta) = f_i(\Zb, \zeta)$ for all $\Zb \in S_i$, where 
\begin{equation} \label{eq:f_i function form}
\begin{split}
f_i(\Zb, \zeta) := \zeta^2 \cdot \bbeta_0^{\T} {\bSigma}_{{\cA_i},*}^{\T} {\bSigma}_{{\cA_i},{\cA_i}}^{-1} {\bSigma}_{{\cA_i},*} \bbeta_0 + 2\zeta \cdot \bbeta_0^{\T} {\bSigma}_{{\cA_i},*}^{\T} {\bSigma}_{{\cA_i},{\cA_i}}^{-1} \left\{({\bSigma}^{1/2}\Zb)_{\cA_i} - \alpha \sign(\hat{\bbeta}_{\cA_i}) \right\} \\
+ \Bigg\{ ({\bSigma}^{1/2}\Zb)_{\cA_i} - \alpha \sign(\hat{\bbeta}_{\cA_i}) \Bigg\}^{\T} {\bSigma}_{{\cA_i},{\cA_i}}^{-1} \Bigg\{ ({\bSigma}^{1/2}\Zb)_{\cA_i} - \alpha \sign(\hat{\bbeta}_{\cA_i}) \Bigg\}
\end{split}
\end{equation}
is the function on $S_i$. Furthermore, let $F_{ij}$ denote the boundary between two neighboring regions $S_i$ and $S_j$, then $F_{ij}$ has measure $0$ in $\R^p$ and is a hyperplane of dimension $p-1$. 
Recall that the  $L_1$ regularized solution $\hat{\bbeta}_{\LW}$ is continuous in $\yb$ (\cite{tibshirani2017sparsity}) and note that $\eta_{\alpha}(\bSigma^{-1/2} \Zb + \zeta \bbeta_0)$ can be viewed as the $L_1$ regularized solution with $\Xb = \bSigma^{1/2}$ and $\yb = \Zb + \zeta \bSigma^{1/2} \bbeta_0$. Therefore, $u$ is continuous and its value satisfies $f_i(\Zb, \zeta) = f_j(\Zb, \zeta)$ for $\Zb \in F_{ij}$. We can rewrite $u(\Zb, \zeta)$ as 
\begin{equation*}
\begin{split}
u(\Zb, \zeta) = \sum_{i=1}^{3^p} f_i(\Zb, \zeta) \cdot \one(\Zb \in S_i) + \sum_{i,j: \text{neighboring } S_i, S_j} f_i(\Zb, \zeta) \cdot \one(\Zb \in F_{ij}),
\end{split}
\end{equation*}
where $3^p$ is the number of regions $\{S_{i}\}_{i}$. 
Then we define $\bar{u}(\Zb, \zeta)$ as follows, which is equal to $u(\Zb, \zeta)$ almost surely, except on $\cup F_{ij}$, a set of measure $0$:
\begin{equation*}
\begin{split}
\bar{u}(\Zb, \zeta) &:= \sum_{i=1}^{3^p} f_i(\Zb, \zeta) \cdot \one(\Zb \in S_i) + 2 \cdot \sum_{i,j: \text{neighboring } S_i, S_j} f_i(\Zb, \zeta) \cdot \one(\Zb \in F_{ij}) \\
&= \sum_{i=1}^{3^p} f_i(\Zb, \zeta) \cdot \one(\Zb \in \text{cl} S_i),
\end{split}
\end{equation*}
where $\text{cl}[E]$ denotes the closure of a set $E$. From Equation \eqref{eqn: 8.69}, we have 
\begin{equation*}
\begin{split}
f(\zeta, \alpha) = \zeta^2 \cdot \Bigg( \lim_{p \to \infty} \EE \frac{1}{n_x} \bnorm{{\bbeta_0}}_{\bSigma}^2 \cdot \frac{1}{h_x^2} \Bigg) + \lim_{p \to \infty} \EE \left\{ \frac{1}{n_w} u(\Zb, \zeta) \right\}
\end{split}
\end{equation*}
and we define a corresponding $\bar{f}(\zeta, \alpha)$ as 
\begin{equation*}
\begin{split}
\bar{f}(\zeta, \alpha) := \zeta^2 \cdot \Bigg( \lim_{p \to \infty} \EE \frac{1}{n_x} \bnorm{{\bbeta_0}}_{\bSigma}^2 \cdot \frac{1}{h_x^2} \Bigg) + \lim_{p \to \infty} \EE \left\{ \frac{1}{n_w} \bar{u}(\Zb, \zeta) \right\}.
\end{split}
\end{equation*}
We will analyze the second order derivative of $\bar{f}(\zeta, \alpha)$ with respect to $\zeta$ and show that the sign of the second order derivative of $\bar{f}(\zeta, \alpha)$ is same as that of ${f}(\zeta, \alpha)$. Leave out the factor $1/n_w$ for now, the expectation of $\bar{u}(\Zb, \zeta)$ can be written as
\begin{equation*}
\begin{split}
\EE \bar{u}(\Zb, \zeta) 
= \int_{\R^p} \bar{u}(\Zb, \zeta) 
= \int_{\R^p} \sum_{i=1}^{3^p} f_i(\Zb, \zeta) \cdot \one(\Zb \in \text{cl} S_i) 
= \sum_{i=1}^{3^p} \int_{\text{cl} S_i} f_i(\Zb, \zeta).
\end{split}
\end{equation*}
To show the positivity of the second order derivative of $f(\zeta, \alpha)$ with respect to $\zeta$, we apply Theorem \ref{thm: S.13.1} to $\lim_p \EE n_w^{-1} \bar{u}(\Zb, \zeta)$. 
Note that $\partial \text{cl}S_i$ is the boundary of $\text{cl}S_i$ and is a disjoint union of all the boundaries $F_{ij}$ (union over $j$ such that $S_i$ and $S_j$ are neighboring). 
Furthermore, if we use $g_{ij}$ above to describe the boundary $F_{ij}$, we have
\begin{equation} \label{eq:baddeley application to u_bar}
\begin{split}
\frac{\partial}{\partial \zeta} \EE \bar{u}(\Zb, \zeta)
&= \sum_{i=1}^{3^p} \frac{\partial}{\partial \zeta} \int_{\text{cl} S_i} f_i(\Zb, \zeta)  \\
&= \sum_{i=1}^{3^p} \int_{S_i} \frac{\partial f_i(\Zb, \zeta)}{\partial \zeta}  + 
\sum_{i=1}^{3^p} \sum_{j: S_i, S_j \text{ neighboring}} \int_{F_{ij}} f_i(\Zb, \zeta) \frac{\partial g_{ij}(\Zb, \zeta)}{\partial \zeta} .
\end{split}
\end{equation}
Moreover, for neighboring $S_i$ and $S_j$, the boundary contribution of $f_i$ on $F_{ij}$ and the boundary contribution of $f_j$ on $F_{ij}$ are negative to each other. 
Thus, for each boundary $F_{ij}$, the boundary contributions in Equation~\eqref{eq:baddeley application to u_bar} cancel out. It follows that 
\begin{equation*}
\begin{split}
\sum_{i=1}^{3^p} \sum_{j: S_i, S_j \text{ neighboring}} \int_{F_{ij}} f_i(\Zb, \zeta) v_{ij}^0  = 0.
\end{split}
\end{equation*}
Therefore, we have
\begin{equation*}
\begin{split}
&\frac{\partial}{\partial \zeta} \EE \bar{u}(\Zb, \zeta)
= \sum_{i=1}^{3^p} \int_{S_i} \frac{\partial f_i(\Zb, \zeta)}{\partial \zeta}  \\
=& \sum_{i=1}^{3^p} \EE \left[ \zeta \cdot \bbeta_0^{\T} {\bSigma}_{{\cA_i},*}^{\T} {\bSigma}_{{\cA_i},{\cA_i}}^{-1} {\bSigma}_{{\cA_i},*} \bbeta_0 + \bbeta_0^{\T} {\bSigma}_{{\cA_i},*}^{\T} {\bSigma}_{{\cA_i},{\cA_i}}^{-1} \left\{({\bSigma}^{1/2}\Zb)_{\cA_i} - \alpha \sign(\hat{\bbeta}_{\cA_i}) \right\} \cdot \one(\Zb \in S_i) \right].
\end{split}
\end{equation*}
Recall that we have 
\begin{equation*}
\begin{split}
\frac{\partial}{\partial \zeta} \EE u(\Zb, \zeta)
= \sum_{i=1}^{3^p} \frac{\partial}{\partial \zeta} \int_{S_i} f_i(\Zb, \zeta)  + \sum_{i,j: \text{neighboring }S_i, S_j} \frac{\partial}{\partial \zeta} \int_{F_{ij}} f_i(\Zb, \zeta) ,
\end{split}
\end{equation*}
and 
\begin{equation*}
\begin{split}
\frac{\partial}{\partial \zeta} \EE \bar{u}(\Zb, \zeta)
&= \sum_{i=1}^{3^p} \frac{\partial}{\partial \zeta} \int_{S_i} f_i(\Zb, \zeta)  + 2 \cdot \sum_{i,j: \text{neighboring }S_i, S_j} \frac{\partial}{\partial \zeta} \int_{F_{ij}} f_i(\Zb, \zeta)  \\
&= \sum_{i=1}^{3^p} \frac{\partial}{\partial \zeta} \int_{S_i} f_i(\Zb, \zeta) . 
\end{split}
\end{equation*}
That is, 
$$
\sum_{i,j: \text{neighboring }S_i, S_j} \frac{\partial}{\partial \zeta} \int_{F_{ij}} f_i(\Zb, \zeta)  = 0. 
$$
Therefore, we have 
\begin{equation*}
\begin{split}
&\frac{\partial}{\partial \zeta} \EE u(\Zb, \zeta)
= \frac{\partial}{\partial \zeta} \EE \bar{u}(\Zb, \zeta) \\
=& \sum_{i=1}^{3^p} \EE \left[ \zeta \cdot \bbeta_0^{\T} {\bSigma}_{{\cA_i},*}^{\T} {\bSigma}_{{\cA_i},{\cA_i}}^{-1} {\bSigma}_{{\cA_i},*} \bbeta_0 + \bbeta_0^{\T} {\bSigma}_{{\cA_i},*}^{\T} {\bSigma}_{{\cA_i},{\cA_i}}^{-1} \left\{({\bSigma}^{1/2}\Zb)_{\cA_i} - \alpha \sign(\hat{\bbeta}_{\cA_i}) \right\} \cdot \one(\Zb \in S_i) \right] \\
=& \sum_{i=1}^{3^p} \EE\ \left[ ({\bSigma} \bbeta_0)_{\cA_i}^{\T} {\bSigma}_{{\cA_i},{\cA_i}}^{-1} \left\{({\bSigma}^{1/2}\Zb)_{\cA_i} + \zeta ({\bSigma} \bbeta_0)_{\cA_i} - \alpha \sign(\hat{\bbeta}_{\cA_i}) \right\} \right] \cdot \one(\Zb \in S_i) \\
=& \sum_{i=1}^{3^p} \EE\ ({\bSigma} \bbeta_0)_{\cA_i}^{\T} \cdot \hat{\bbeta}_{\cA_i} \cdot \one(\Zb \in S_i).
\end{split}
\end{equation*}
Denote $h_i(\Zb, \zeta) = ({\bSigma} \bbeta_0)_{\cA_i}^{\T} \cdot \hat{\bbeta}_{\cA_i}$ and $h_j(\Zb, \zeta) = ({\bSigma} \bbeta_0)_{\cA_j}^{\T} \cdot \hat{\bbeta}_{\cA_j}$. 
Then we claim $h_i(\Zb, \zeta) = h_j(\Zb, \zeta)$ for $\Zb \in F_{ij}$. Note that $\hat{\bbeta}$ is a $L_1$ regularized solution for $\X = \bSigma^{1/2}$, signal $\zeta \bbeta_0$, noise $\Zb$, $\yb = \zeta \bSigma^{1/2} \bbeta + \Zb$, and $L_1$ parameter $\lambda = \alpha$. 
Thus, the fitted value $\X \hat{\bbeta} = \bSigma^{1/2} \hat{\bbeta} = \bSigma^{1/2}_{*,\cA} \hat{\bbeta}_\cA$ is unique for all $\Zb \in \R^p$. Moreover, we have 
\begin{equation*}
\begin{split}
h_i(\Zb, \zeta) &:= ({\bSigma} \bbeta_0)_{\cA_i}^{\T} \hat{\bbeta}_{\cA_i} 
= \{(\bSigma^{1/2})^{\T} \bSigma^{1/2} \bbeta_0\}_{\cA_i}^{\T} \hat{\bbeta}_{\cA_i} \\
&= \{ (\bSigma^{1/2}_{*,\cA_i})^{\T} \bSigma^{1/2} \bbeta_0 \}^{\T} \hat{\bbeta}_{\cA_i} 
= (\bSigma^{1/2} \bbeta_0)^{\T} (\bSigma_{*,\cA_i}^{1/2} \hat{\bbeta}_{\cA_i}).
\end{split}
\end{equation*}
Thus, for any $\Zb \in F_{ij}$, the fitted value is the same. That is, we have $\bSigma_{*,\cA_i}^{1/2} \hat{\bbeta}_{\cA_i} = \bSigma_{*,\cA_j}^{1/2} \hat{\bbeta}_{\cA_j}$. Then we have $h_i(\Zb, \zeta) = h_j(\Zb, \zeta)$. 
Therefore, when we calculate the second order derivative of $\EE u(\Zb, \zeta)$ with respect to $\zeta$, similar to the calculation of the first order derivative of $\EE u(\Zb, \zeta)$ with respect to $\zeta$, the boundary contributions to the second order derivative also cancel out. Thus, we have 
\begin{equation*}
\begin{split}
&\frac{\partial^2}{\partial \zeta^2} \EE u(\Zb, \zeta)
= \sum_{i=1}^{3^p} \EE\ \left( \bbeta_0^{\T} {\bSigma}_{{\cA_i},*}^{\T} {\bSigma}_{{\cA_i},{\cA_i}}^{-1} {\bSigma}_{{\cA_i},*} \bbeta_0 \right) \cdot \one(\Zb \in S_i) 
= \EE\ \left( \bbeta_0^{\T} {\bSigma}_{{\cA},*}^{\T} {\bSigma}_{{\cA},{\cA}}^{-1} {\bSigma}_{{\cA},*} \bbeta_0 \right)
\end{split}
\end{equation*}
and
\begin{equation*}
\begin{split}
\frac{\partial^2}{\partial \zeta^2} f(\zeta, \alpha)
=& \frac{\partial^2}{\partial \zeta^2} \left\{ \zeta^2 \cdot \left(\lim_{p \to \infty} \EE \frac{1}{n_x} \| \bbeta_0 \|_{\bSigma}^2 \cdot \frac{1}{h_x^2} \right) 
+ \EE \lim_{p \to \infty} \frac{1}{n_w} \| u(\Zb, \zeta) \|_{\bSigma}^2 \right\} \\
=& \lim_{p\to \infty} \frac{1}{n_w} \EE \bbeta_0^{\T} {\bSigma}_{\cA,*}^{\T} {\bSigma}_{\cA,\cA}^{-1} {\bSigma}_{\cA,*} \bbeta_0 + \Bigg( \lim_{p \to \infty} \EE \frac{1}{n_x} \bnorm{{\bbeta_0}}_{\bSigma}^2 \cdot \frac{1}{h_x^2} \Bigg) > 0.
\end{split}
\end{equation*}

\subsubsection{Proof of Lemma \ref{monotone wrt alpha}} \label{proof of Lemma monotone wrt alpha}

Recall that, $f(\zeta, \alpha)$ is given by
\begin{equation*}
    f(\zeta, \alpha) = \zeta^2 \cdot \Bigg( \lim_{p \to \infty} \EE \frac{1}{n_x} \bnorm{{\bbeta_0}}_{\bSigma}^2 \cdot \frac{1}{h_x^2} \Bigg) + \lim_{p \to \infty} \EE \left\{ \frac{1}{n_w} \bnorm{{\eta}_{\alpha}({\bSigma}^{-1/2} \Zb + \zeta {\bbeta_0})}_{\bSigma}^2 \right\}.
\end{equation*}
Similar to proof of Lemma \ref{f-convex}, we have  
\begin{equation*}
\begin{split}
    \hat{\bbeta}_{\cA} =  ({\bSigma}_{\cA, \cA})^{-1} \bigg\{ ({\bSigma}^{1/2}\Zb)_\cA + \zeta ({\bSigma}\bbeta_0)_\cA - \alpha \sign(\hat{\bbeta}_\cA) \bigg\}.
\end{split}
\end{equation*}
Therefore, within each active set $\cA$, the derivative of $\hat{\bbeta}_{\cA}$ with respect to $\alpha$ is given by
\begin{equation*}
    {\bSigma}_{\cA, \cA} \frac{\partial \hat{\bbeta}_{\cA}}{\partial \alpha} = - \sign(\hat{\bbeta}_\cA).
\end{equation*}
The derivative of $f(\zeta, \alpha)$ with respect to $\alpha$ is
\begin{equation*}
    \frac{\partial f(\zeta, \alpha)}{\partial \alpha} = \frac{\partial}{\partial \alpha} \lim_{p \to \infty} \EE \left\{ \frac{1}{n_w} \bnorm{{\eta}_{\alpha}({\bSigma}^{-1/2} \Zb + \zeta {\bbeta_0})}_{\bSigma}^2 \right\}.
\end{equation*}
According to the discussion after dquation \eqref{hat{beta}}, $\bnorm{\hat{\bbeta}}_{{\bSigma}}^2$ is continuous across the entire space $\Zb \in \RR^p$. Thus, the contribution of the derivative at the boundary is $0$. Therefore, we have 
\begin{equation*}
\begin{split}
    \frac{\partial}{\partial \alpha} \EE \left\{ \frac{1}{n_w} \bnorm{{\eta}_{\alpha}({\bSigma}^{-1/2} \Zb + \zeta {\bbeta_0})}_{\bSigma}^2 \right\} &= \frac{2}{n_w} \EE\  \hat{\bbeta}^{\T}_{\cA} {\bSigma}_{\cA, \cA} \frac{\partial \hat{\bbeta}_{\cA}}{\partial \alpha} = - \frac{2}{n_w} \hat{\bbeta}^{\T}_{\cA} \sign(\hat{\bbeta}_\cA) < 0.
\end{split}
\end{equation*}
This completes the proof of the first part of Lemma \ref{monotone wrt alpha}. 
{When $\gamma_w > 1$, recall that $g(\alpha) := f(0, \alpha)$ and note that the partial derivative of $f(\zeta, \alpha)$ with respect to $\alpha$ does not depend on $\zeta$. Thus, we also have $dg/d\alpha < 0$ and $g(\alpha)$ is a decreasing function from $\gamma_w$ to $0$ as $\alpha$ increasing from $0$ to $\infty$. }
Therefore, $f(0, \alpha) = 1$ has a unique solution denoted by $\alpha_{\min}$.

\subsubsection{Proof of Lemma \ref{zeta continuous}}\label{proof:zeta continuous}

We aim to demonstrate the continuity of $\zeta_*(\alpha)$ with respect to $\alpha$.
First, we note that $f(\zeta, \alpha)$ is continuously differentiable with respect to $\zeta$, due to the continuity of $\hat{\bbeta}$ and the discussion in Appendix \ref{Proof of Lemma f-convex}.

Second, when $\gamma_w > 1$ and $\alpha > \alpha_{\min}$, we have $f(0, \alpha) < 1$. 
Similarly, when $\gamma_w \leq 1$ and $\alpha > 0$, $f(0,\alpha) < 1$ holds by Equation \eqref{f(0, alpha) < 1}. 
Moreover, Proposition \ref{prop 1} ensures that the solution is unique for each $\alpha$.
Next, by Lemma \ref{f-convex}, we have  $\frac{\partial f}{\partial \zeta}(\zeta, \alpha) > 0$ at $\zeta = \zeta_*(\alpha)$.

Therefore, according to the implicit function theorem, we conclude that the implicit solution $\zeta_*(\alpha)$ is continuously differentiable with respect to $\alpha$. This completes the proof of Lemma \ref{zeta continuous}.


\section{Proofs for Section \ref{subsec: main results}}
\label{Proof of main results}

In this section, we provide proofs for the results in Section \ref{subsec: main results}.
Section \ref{sec: trace lemmas} presents some useful preliminary results regarding the trace of random matrices. 
We outline the proof of Theorem \ref{thm: general LASSO main without normalization} in Section \ref{subsec: Proof ofthm: general LASSO main without normalization}.
Specifically, we introduce Theorem \ref{general LASSO main}, which is shown to be equivalent to Theorem \ref{thm: general LASSO main without normalization}. 
{\BS The detailed proof of Theorem \ref{general LASSO main} is 
given in Section \ref{Proof for Appendix S.5.1}}. 
Sections \ref{Proof of Corollary alpha uniqueness} and \ref{proof of theorem general lasso mse + R2} provide proofs for Corollary \ref{cor: alpha uniqueness} and Theorem \ref{thm: general lasso mse + R2}, respectively.

\subsection{Trace lemmas for random matrices}
\label{sec: trace lemmas}
We require the following two lemmas in order to derive the asymptotic behavior of the out-of-sample $R^2$.
Lemma \ref{lemma: S.21} states that, for any random vector with covariance $\bmSigma_{\alpha}$, the quadratic form converges in probability to a nonrandom constant determined by $\bmSigma_{\alpha}$.
\begin{lemma}[Lemma 5 in \cite{zhao2022estimating}]\label{lemma: S.21}
Consider a random vector $\alpha \in \RR^p$ with mean $\bm{0}$ and covariance $\bmSigma_{\alpha}$, where $\bmSigma_\alpha$ is positive semi-definite.
Let $\Ab \in \RR^{p \times p}$ be a nonrandom positive semi-definite matrix. Further assume $\tr( \Ab \bmSigma_\alpha) \neq 0$. Then, as $p \to \infty$, we have
\begin{align*}
\frac{\alpha^{\T}\Ab\alpha}{\tr(\Ab \bmSigma_{\alpha})} \to 1, \qprob. 
\end{align*}
\end{lemma}

Lemma \ref{lemma: S.22} states that $n_s^{-1} \tr(\Sbb^{\T} \Sbb)$ converges in probability to the trace of the covariance matrix.
\begin{lemma}[Lemma 2.16 in \cite{yao2015sample}]
\label{lemma: S.22}
Assume Conditions \ref{cond1-np-ratio} and \ref{cond-distn-data}. Let $\hat{\bmSigma} = \Sbb^{\T} \Sbb /n_s$. Then, as $p \to \infty$,  we have
\begin{align*}
\frac{\tr(\Sbb^{\T} \Sbb /n_s)}{\tr(\bmSigma)} \to 1, \qprob. 
\end{align*}
\end{lemma}

\subsubsection{Property for proximal operator}
The following lemma indicates that gradient of proximal operator exists almost surely and  with proof can be found in Proof of Lemma A.3 in \cite{celentano2020lasso} and Table 1 in \cite{bellec2019biasing}. 
\begin{lemma} \label{lemma:proximal}
Consider a proximal operator defined in \eqref{hat{y}}: 
\$
\hat{y}_{\BS \theta}(\vb) = {\bSigma}^{1/2} \eta_{\theta}({\bSigma}^{-1/2}\vb).
\$
For any constant $\tau, b$, $\nabla \hat{y}_{\BS \theta}(\tau \Zb + (1 + b) {\bSigma}^{1/2} \bbeta_0)$ exists almost surely with respect to $\Zb \sim N(0, \bI_{p})$ and has close form
\begin{align*}
    \nabla \hat{y}_{\BS \theta}(\tau \Zb + (1 + b) {\bSigma}^{1/2} \bbeta_0) = ({\bSigma}^{1/2})_{*,\cA} ({\bSigma}_{\cA,\cA})^{-1} ({\bSigma}^{1/2})_{*,\cA}^{\T},
\end{align*}
where $\cA$ denotes the active set of $\eta_{\theta_*}(\vb)$.
Therefore, $\Div \hat{y}_{\BS \theta}(\vb) = \tr \left\{ \nabla \hat{y}_{\BS \theta}(\vb) \right\} \geq 0$.
\end{lemma}

\subsection{Proof of Theorem \ref{thm: general LASSO main without normalization}}
\label{subsec: Proof ofthm: general LASSO main without normalization}

In this section, we introduce Theorem \ref{general LASSO main}, which restates Theorem \ref{thm: general LASSO main without normalization} using the normalization  introduced in Section \ref{change of normalization}. 

The proof of Theorem \ref{general LASSO main} is outlined in this section, with the details deferred to Section \ref{Proof for Appendix S.5.1}.
Briefly, Proposition \ref{prop 3} guarantees that the fixed point of Equation \eqref{general - lasso - AMP} is the minimizer of Equation \eqref{eqn:ref-panel-lasso-est}. 
Theorem \ref{general - state} demonstrates the asymptotic behavior of $\mathbf{\bbeta}^{t}$ from Equation \eqref{general - lasso - AMP} at each step $t$. 
Furthermore, Theorem \ref{AMP approx estimator} demonstrates that the recursion $\mathbf{\beta}^{t}$ converges to the estimator $\hat{\bbeta}_{\LW}(\lambda)$ in $\ell_2$ distance, which is proved in Section \ref{proof of Theorem AMP approx estimator}. 
These results establish the asymptotic behavior of $\hat{\bbeta}_{\LW}(\lambda)$. {Recall that the normalized converging sequence is defined in Section \ref{change of normalization}.}

\begin{theorem} \label{general LASSO main}
Let $\{\bbeta_0, \bepsilon_x,\bmeps_s, \bSigma, \bX,  \Sbb, \bW\}$ be a {\BS normalized converging sequence} of instances and $\PP\left({\bbeta_0}(p) \neq 0 \right) > 0$. 
Assume each row of $\X$, $\Sbb$, and $\Wb$ is i.i.d. Gaussian with mean $\bm 0$ and variance ${\bSigma}$.
For any deterministic sequence of uniformly normalized pseudo-Lipschitz functions $\{ \phi^{p}: \RR^{p} \times \RR^p \mapsto \RR \}_{p \in \NN}$, we have 
\begin{align} \label{general LASSO main formula}
    \phi^{p}(\hat{\bbeta}_{\LW}(\lambda), \bbeta_0) \overset{P}{\approx} \EE\ \phi^{p} \left[\eta_{\lambda (1 + b_*)}\{(1+b_*)\bbeta_0 + \tau_* \bSigma^{-1/2} \zb\}, \bbeta_0 \right],
\end{align}
where $\zb \sim N(0, {\bI_p})$ is independent of $\bbeta_0$, $\tau_* = \tau_*(\alpha(\lambda))$ and $b_* = b_*(\alpha(\lambda))$ are the solutions to
\begin{align}\label{eqn: 4.11}
    \tau_*^2 &=  \lim_{p \to \infty} \EE \left[\frac{1}{n_w} \bnorm{{\eta}_{\lambda (1 + b_*)}\{\tau {\bSigma}^{-1/2} \Zb + (1 + b_*){\bbeta}\}}_{\bSigma}^2 + \frac{1}{n_x} (1 + b_*)^2 \bnorm{{\bbeta}_0}_{\bSigma}^2 \cdot \frac{1}{h_x^2} \right], \\
    \label{eqn: 4.12}
     b_* &= (1 + b_*) \cdot \lim_{p \to \infty} \frac{1}{n_w} \EE \Div \eta_{\lambda (1 + b_*)}\{\tau_* {\bSigma}^{-1/2} \Zb + (1 + b_*) \bbeta\}.
\end{align}
\end{theorem}

Given Theorem Theorem \ref{general LASSO main},   we now prove  Theorem \ref{thm: general LASSO main without normalization}. 

\begin{proof}[Proof of Theorem \ref{thm: general LASSO main without normalization}]
    Let $\{\bbeta_0, \bepsilon_x,\bmeps_s, \bSigma, \bX,  \Sbb, \bW\}$ be a {\BS normalized converging sequence}. We define 
    \begin{equation} \label{change normalization}
    \begin{split}
        &\bbeta'_0 = \bbeta_0/\sqrt{p}, \quad \Xb' = \sqrt{n_x} \Xb, \quad \Wb' = \sqrt{n_w} \Wb,\\ 
        &\bepsilon'_x = \bepsilon_x /\sqrt{\gamma_x},\quad \yb'_x = \yb_x/\sqrt{\gamma_x}, \quad \mbox{and} \quad \lambda' = \lambda/\sqrt{p}.
    \end{split}
    \end{equation}
    Then $\{\bbeta'_0, \bepsilon'_x,\bmeps'_s, \bSigma, \bX',  \Sbb', \bW' \}$ is a converging sequence satisfying Conditions \ref{cond1-np-ratio} - \ref{cond: exchange limit}. 
    Moreover, with $\bbeta' = \bbeta/\sqrt{p}$, the linear models \eqref{linear model} and \eqref{eqn:ref-panel-lasso-est without normalization} are equivalent to 
    \begin{align*}
        \yb'_x =& \Xb' \bbeta'_0 + \bepsilon'_x \quad \mbox{and} \\
        \hat{\bbeta}_{\LW} (\lambda) =& \arg\min_{\bbeta' \in \R^p} \frac{1}{2 n_w} {\bbeta'}^{\T} {\W'}^{\T} {\W'} \bbeta' - \frac{1}{n_x} {\bbeta'}^{\T} {\X'}^{\T} \yb_x' + \lambda' \| \bbeta' \|_1.
    \end{align*}
    Therefore, by plugging~\eqref{change normalization} into Theorem \ref{general LASSO main}, 
    we complete the proof of Theorem \ref{thm: general LASSO main without normalization}. 
\end{proof}

\paragraph{Proof sketch for Theorem \ref{general LASSO main}}
We then sketch the proof for Theorem \ref{general LASSO main}. 
The existence and uniqueness of $(\tau_{*}^2, b_{*})$ have been established in Sections \ref{proof of prop 1} and \ref{proof of prop 2}.
We prove Theorem \ref{general LASSO main} using the limiting distribution of the AMP recursion:
\begin{equation}\label{general - lasso - AMP}
\begin{split}
    {\bbeta}^{t+1} &= \eta_{\lambda (1+b_{t})} \bigg\{ -  \sqrt{\frac{N}{n_{w}}} {\bSigma}^{-1} \W^{\T}{\rb^{t}} + (1+b_t) {\bSigma}^{-1} \X^{\T} \y_{x} + {\bbeta}^{t} \bigg\} \quad \mbox{and} 
    \\ 
    \rb^{t} &=  \sqrt{\frac{n_{w}}{N}} \W {\bbeta}^{t} + \frac{b_t}{1 + b_{t-1}} \rb^{t-1},
\end{split}
\end{equation}
where  $b_t$ satisfies the recursion
\begin{equation*}
\begin{split}
    \frac{b_t}{1+b_{t-1}} = \frac{1}{n_w} \Div \eta_{\lambda(1+b_{t-1})}  \bigg\{ - \sqrt{\frac{N}{n_{w}}} {\bSigma}^{-1} \W^{\T}{\rb^{t-1}} + (1+b_{t-1}) {\bSigma}^{-1} \X^{\T} \y_{x} + {\bbeta}^{t-1} \bigg\}.
\end{split}
\end{equation*}
The AMP algorithm generates a sequence of estimates $\bbeta^{t} \in \RR^{p}$ and $\rb^{t} \in \RR^{n_x}$ iteratively as in \eqref{general - lasso - AMP}. 
To simplify the proof, we adopt an {\it oracle} initialization given by
$$
\bbeta^0 = \eta_{\lambda(1 + b_*)}\{\tau_* \bSigma^{-1/2} \Zb + (1 + b_*) \bbeta_0\}
$$ 
with $\tau_*$ and $b_*$ being defined in equations \eqref{eqn: 4.11} - \eqref{eqn: 4.12}. 
According to Theorem \ref{general - state}, $\bbeta^0$ is the {\it equilibrium} point, indicating that $\bbeta^{t}$ has the same asymptotic behavior for any step $t$. 
It is worth noting that the AMP algorithm in \eqref{general - lasso - AMP} is primarily designed for theoretical analysis rather than practical use, as $\bSigma$ is typically unknown.
Therefore, our choice of the initialization $\bbeta^0$ does not present any practical disadvantage.
The following proposition illustrates the relationship between the fixed-point solution of the AMP algorithm and the optimization solution of the reference panel-based $L_1$ regularized estimator defined in Equation \eqref{eqn:ref-panel-lasso-est}.

\begin{proposition} \label{prop 3}
Fixing any $\lambda \geq 0$. 
Any fixed point $\bbeta^{t} = \bbeta_*$ and $\rb^{t} = \rb_*$ of the AMP iteration in \eqref{general - lasso - AMP} is a minimizer of the reference panel-based $L_1$ regularized estimator defined in  \eqref{eqn:ref-panel-lasso-est}.
\end{proposition}
The proof can be found in Section \ref{proof of prop 3}.
Define the sequence $\tau^2_{t+1}$ and $b_t$ by
\begin{align} \label{tau_t and b_t}
\begin{split}
    &\tau_{t+1}^2 =  \lim_{p \to \infty} \EE \left[\frac{1}{n_w} \bnorm{{\eta}_{\lambda (1 + b_{t})}(\tau_{t} {\bSigma}^{-1/2} \Zb + (1 + b_{t}){\bbeta})}_{\bSigma}^2 + \frac{1}{n_x} (1 + b_{t})^2 \bnorm{{\bbeta}_0}_{\bSigma}^2 \cdot \frac{1}{h_x^2} \right]\\
    &\mbox{and} \quad b_{t} = (1 + b_{t-1}) \cdot \lim_{p \to \infty} \frac{1}{n_w} \EE \Div \eta_{\lambda (1 + b_{t-1})}(\tau_{t} {\bSigma}^{-1/2} \Zb + (1 + b_{t-1}) \bbeta).
\end{split}
\end{align}
For a {\BS normalized converging sequence} of instances 
$\{\bbeta_0, \bepsilon_x,\bmeps_s, \bSigma, \bX, \Sbb, \bW\}$, the asymptotic behaviour of the AMP recursion in ~\eqref{general - lasso - AMP} can be characterized using $\tau_{t+1}^2$ and $b_t$ as follows. 
\begin{theorem} \label{general - state}
Let $\{\bbeta_0, \bepsilon_x,\bmeps_s, \bSigma, \bX,  \Sbb, \bW\}$ be a {\BS normalized converging sequence} of instances 
with $\PP\left({\bbeta_0}(p) \neq 0 \right) > 0$. 
Assume each row of $\X$, $\Sbb$, and $\Wb$ is i.i.d. Gaussian with mean $\bm 0$ and variance ${\bSigma}$.
For, any deterministic sequence of uniformly normalized pseudo-Lipschitz functions  $\{ \phi^{p}: \RR^{p} \times \RR^p \mapsto \RR \}_{p \in \NN}$ and fixed $t$, we have, as $p \to \infty$, 
\begin{equation}
\begin{split}
    \phi^{p} \left(\bbeta_0, \bbeta^{t+1}\right) \overset{P}{\approx} \EE\ \phi^{p} \left[\bbeta_0, \eta_{\lambda(1 + b_t)}\{\tau_{t+1} {\bSigma}^{-1/2} \Zb + (1 + b_t) \bbeta_0\} \right],
\end{split}
\end{equation}
where $\Zb \sim N(0, {\bI_p})$ is  independent of $\bbeta_0$. 
\end{theorem}
A stronger version of Theorem \ref{general - state} is stated in Theorem \ref{general - state - stronger}, and the proofs of both Theorem \ref{general - state} and Theorem \ref{general - state - stronger} can be found in Section \ref{general - state - proof}. 
Corollary~\ref{stationary beta0} below is an immediate corollary of Theorem \ref{general - state} above {\BS and we omit the proof here.} 

\begin{corollary} \label{stationary beta0}
Assume the setting of Theorem \ref{general - state}, and further assume that the AMP recursion in \eqref{general - lasso - AMP} has an oracle initialization $\bbeta^0 = \eta_{\lambda(1 + b_*)}\{\tau_* \bSigma + (1 + b_*)\Zb\}$ with $(\tau_*, b_*)$ being given in \eqref{eqn: 4.11} and  \eqref{eqn: 4.12}. Then, we have
\begin{align}
    \tau_t^2 = \tau_*^2\quad \mbox{and} \quad b_t = b_*.
\end{align}
Moreover, for any deterministic sequence of uniformly normalized pseudo-Lipschitz functions $\{ \phi^{p}: \RR^{p} \times \RR^p \mapsto \RR \}_{p \in \NN}$ and fixed $t$, we have, as $p \to \infty$, 
\begin{equation}
\begin{split}
    \phi^{p} \left(\bbeta_0, \bbeta^{t+1}\right) \overset{P}{\approx} \EE\ \phi^{p} \left[\bbeta_0, \eta_{\lambda(1 + b_*)}\{\tau_{*} {\bSigma}^{-1/2} \Zb + (1 + b_*) \bbeta_0\} \right],
\end{split}
\end{equation}
where $\Zb \sim N(0, \bI_p)$ is independent of $\bbeta_0$. 
\end{corollary}

Theorem \ref{AMP approx estimator} below shows the convergence of $\beta^t(\lambda)$ to $\hat\beta_{\LW}(\lambda)$ in $\ell_2$ distance. 
\begin{theorem} \label{AMP approx estimator}
Assume the setting  of Theorem \ref{general LASSO main}. 
Let $\bbeta^{t}(\lambda)$ be the sequence of estimators produced by the AMP algorithm in \eqref{general - lasso - AMP} with an oracle initialization. 
Then we have {\BS almost surely} 
\begin{align}
    \lim_{t \to \infty} \lim_{p \to \infty} \frac{1}{p} \bnorm{\bbeta^{t}(\lambda) - \hat{\bbeta}_{\LW}(\lambda)}^2_2 = 0.
\end{align}
\end{theorem}
The proof of Theorem \ref{AMP approx estimator} is provided in Section \ref{proof of Theorem AMP approx estimator}.
Note that Theorem \ref{AMP approx estimator} requires taking  $p \to \infty$ before taking $t \to \infty$. 
{\BS 
Proposition \ref{prop 3}, Theorem \ref{general - state} and Theorem \ref{AMP approx estimator} together imply Theorem \ref{general LASSO main}. 
\begin{proof}[Proof of Theorem \ref{general LASSO main}]
In the event 
    $$
    \left\{ \lim_{t \to \infty} \lim_{p \to \infty} \frac{1}{p} \bnorm{\bbeta^{t}(\lambda) - \hat{\bbeta}_{\LW}(\lambda)}^2_2 = 0 \right\}, 
    $$
 for any  $\epsilon > 0$, there exists some $T > 0$ such that for all $t \geq T$, 
    $$
    \lim_{p \to \infty} \frac{1}{p} \bnorm{\bbeta^{t}(\lambda) - \hat{\bbeta}_{\LW}(\lambda)}^2_2 \leq \epsilon.
    $$
    Moreover, Theorem \ref{general - state} implies, in probability, 
    \begin{align*}
        \lim_{p \to \infty} \left( \phi^{p} \left(\bbeta_0, \bbeta^{t+1}\right) - \EE\ \phi^{p} \left[\bbeta_0, \eta_{\lambda(1 + b_t)}\{\tau_{t+1} {\bSigma}^{-1/2} \Zb + (1 + b_t) \bbeta_0\} \right]  \right) =0.
    \end{align*}
    By Corollary \ref{stationary beta0} and the oracle initialization, we have that, in probability, 
    \begin{align*}
        \lim_{p \to \infty}   \left( \phi^{p} \left(\bbeta_0, \bbeta^{t+1}\right) - \EE\ \phi^{p} \left[\bbeta_0, \eta_{\lambda(1 + b_*)}\{\tau_{*} {\bSigma}^{-1/2} \Zb + (1 + b_*) \bbeta_0\} \right]  \right) =0.
    \end{align*}
    Therefore, 
    \begin{align*}
        &\lim_{p \to \infty} \left\{ \phi^{p} \left(\bbeta_0, \bbeta^{t+1}\right) - \phi^{p} \left(\bbeta_0, \hat{\bbeta}_{\LW}(\lambda) \right) \right\}
        \\
        &\qquad + \lim_{p \to \infty}  \left( \phi^{p} \left(\bbeta_0, \hat{\bbeta}_{\LW}(\lambda) \right) - \EE\ \phi^{p} \left[\bbeta_0, \eta_{\lambda(1 + b_*)}\{\tau_{*} {\bSigma}^{-1/2} \Zb + (1 + b_*) \bbeta_0\} \right] \right) = 0.
    \end{align*}
Using the properties of uniformly normalized pseudo-Lipschitz functions, for any $\epsilon > 0$, there exists some $T > 0$ such that  for all $t \geq T$,
    $$
    \lim_{p \to \infty} \left| \phi^{p} \left(\bbeta_0, \bbeta^{t+1}\right) - \phi^{p} \left(\bbeta_0, \hat{\bbeta}_{\LW}(\lambda) \right) \right| \leq \epsilon.
    $$
    Therefore, for any  $\epsilon > 0$, there exists some $T > 0$ such that for all $t \geq T$,
    \begin{align*}
        \lim_{p \to \infty} \left| \phi^{p} \left(\bbeta_0, \hat{\bbeta}_{\LW}(\lambda) \right) - \EE\ \phi^{p} \left[\bbeta_0, \eta_{\lambda(1 + b_*)}\{\tau_{*} {\bSigma}^{-1/2} \Zb + (1 + b_*) \bbeta_0\} \right] \right| < \epsilon.
    \end{align*}
    Since the inequality above is independent of $t$, we complete the proof of Theorem \ref{general LASSO main}.
\end{proof}}

\subsection{Proof of Corollary \ref{cor: alpha uniqueness}} \label{Proof of Corollary alpha uniqueness}

According to the existence result in Proposition \ref{prop 2}, we only need to show that for any $\lambda > 0$, there is at most one $\alpha > \alpha_{\min}$ such that $\lambda(\alpha) = \lambda$.
Assume by contradiction that there exist $\alpha_1$ and $\alpha_2$ satisfying $\lambda(\alpha_1) = \lambda(\alpha_2) = \lambda$. Using  Theorem \ref{general LASSO main} with $(\tau_*(\alpha(\lambda)), b(\alpha(\lambda)))$ being given in equations \eqref{eqn: 4.11} - \eqref{eqn: 4.12}, we have
\begin{equation*}
    \phi^{p} (\bbeta_0, \hat{\bbeta}_{\LW}(\lambda) ) \overset{P}{\approx} \EE\ \phi^{p} [\bbeta_0, \eta_{\lambda (1 + b_*)}\{\tau_{*} {\bSigma}^{-1/2} \Zb + (1 + b_*) \bbeta_0\}].
\end{equation*}
Let
\$
\phi^{p} (x,y) := \frac{1}{n_w} \bnorm{y}^2_{\bSigma}~~~\text{and}~~~
\hat{\sigma}^2 :=  \frac{1}{h_x^2} \cdot \lim_{p \to \infty} \frac{1}{n_x} \bnorm{{\bbeta_0}}_{\bSigma}^2 >0.
\$
Asymptotically, we have
\begin{align} \label{eqn: 8.126}
    &\tau_{*}^2 - (1 + b_{*})^2 \hat{\sigma}^2 = \lim_{p \to \infty} \EE \frac{1}{n_w} \bnorm{\hat{\bbeta}_{\LW}(\lambda)}_{{\bSigma}}^2 \quad \mbox{and} \\
    &\frac{1}{1 + b_{*}} = 1 - \tau_{*}^{-1} \lim_{p \to \infty} \EE\  \langle {\bSigma}^{1/2} \hat{\bbeta}_{\LW}(\lambda), \Zb \rangle \label{eqn: 8.127}.
\end{align}
Note that $\hat{\bbeta}_{\LW}(\lambda)$ does not explicitly depend on the choice of $\alpha$.
From equations \eqref{lambda(alpha)} and \eqref{alpha(lambda)}, we have
\begin{equation} \label{eqn: 8.128}
    \lambda = \frac{\alpha \tau_{*}}{1 + b_{*}}.
\end{equation}
Plugging \eqref{eqn: 8.128} into equations \eqref{eqn: 8.126} and \eqref{eqn: 8.127}, we acquire 
\begin{align}\label{eqn: 8.129}
    &\frac{\lambda}{\alpha} = \tau_{*} - \lim_{p \to \infty} \EE\  \langle {\bSigma}^{1/2} \hat{\bbeta}_{\LW}(\lambda), \Zb \rangle \quad \mbox{and}\\ \label{eqn: 8.130}
    &\tau_{*}^2 \left\{1 - \hat{\sigma}^2 \cdot \left(\frac{\alpha}{\lambda} \right)^2 \right\} = \lim_{p \to \infty} \EE  \frac{1}{n_w} \bnorm{\hat{\bbeta}_{\LW}(\lambda)}_{{\bSigma}}^2.
\end{align}
Note that Equation \eqref{eqn: 8.129} implies
\begin{equation} \label{eqn: S.4.6}
    \tau_{*} = \frac{\lambda}{\alpha} +  \lim_{p \to \infty} \EE\  \langle {\bSigma}^{1/2} \hat{\bbeta}_{\LW}(\lambda), \Zb \rangle.
\end{equation}
Plugging above equation into  \eqref{eqn: 8.130}, we obtain 
\begin{equation} \label{eqn: 8.131}
    \left( \frac{\lambda}{\alpha} +  \lim_{p \to \infty} \EE\ \langle {\bSigma}^{1/2} \hat{\bbeta}_{\LW}(\lambda), \Zb \rangle \right)^2 \left\{1 - \hat{\sigma}^2 \cdot \left(\frac{\alpha}{\lambda} \right)^2 \right\} = \lim_{p \to \infty} \EE \frac{1}{n_w} \bnorm{\hat{\bbeta}_{\LW}(\lambda)}_{{\bSigma}}^2.
\end{equation}
To show that there exists a unique $\alpha$ satisfying \eqref{eqn: 8.131}, we need the following claim. 
\begin{claim} \label{claim: 8.131 positive}
We have
\begin{equation*}
    \lim_{p \to \infty} \EE  \langle {\bSigma}^{1/2} \hat{\bbeta}_{\LW}(\lambda), \Zb \rangle =  \lim_{p \to \infty} \EE\ \langle {\bSigma}^{1/2} \eta_{\alpha \tau_{*}}\{\tau_{*} {\bSigma}^{-1/2} \Zb + (1 + b_{*}) \bbeta_0\}, \Zb \rangle \geq 0.
\end{equation*}
\end{claim}
Assume the claim for now. Let $x$ denote $\lambda/\alpha > 0$ and define the function $\ell(x)$ as
\begin{equation*}
    \ell(x) := \left(x +  \lim_{p \to \infty} \EE \langle {\bSigma}^{1/2} \hat{\bbeta}_{\LW}(\lambda), \Zb \rangle \right)^2 \left(1 - \frac{\hat{\sigma}^2}{x^2} \right).
\end{equation*}
The derivative  $\partial \ell/\partial x$ is given by
\begin{equation*}
    2\left(1 - \frac{\hat{\sigma}^2}{x^2} \right) (x + \EE \lim_{p \to \infty} \langle {\bSigma}^{1/2} \hat{\bbeta}_{\LW}(\lambda), \Zb \rangle) + \left( x +  \EE \lim_{p \to \infty} \langle {\bSigma}^{1/2} \hat{\bbeta}_{\LW}(\lambda), \Zb \rangle \right)^2 \cdot \frac{2 \hat{\sigma}^2}{x^3} > 0.
\end{equation*}
When $x \to 0^+$, $\ell(x)$ approaches negative infinity; when $x \to \infty$, $\ell(x)$ approaches positive infinity. Therefore, for  $\lim_{p \to \infty} \EE \norm{\hat{\bbeta}_{\LW}(\lambda)}_{{\bSigma}}^2/n_{w}$, there exists at most one value of $x > 0$ such that  $\ell(x) = \lim_{p \to \infty} \EE \norm{\hat{\bbeta}_{\LW}(\lambda)}_{{\bSigma}}^2/n_{w}$. 
Together with Proposition \ref{prop 2}, we conclude that there exists a unique value of $\lambda/\alpha > 0$ that satisfies Equation \eqref{eqn: 8.131}.
Using \eqref{eqn: 8.128} and \eqref{eqn: S.4.6}, we find that there exists a unique fixed point pair $(\tau_*, b_*)$. This completes the proof of the corollary, except for the Claim~\ref{claim: 8.131 positive}, which we now prove. 

\begin{proof}[Proof of Claim \ref{claim: 8.131 positive}]
The equality in Claim~\ref{claim: 8.131 positive} follows from the definition of $\hat{\bbeta}_{\LW}(\lambda)$ in \eqref{general LASSO main formula}. 
To see that 
\$
\lim_{p \to \infty} \EE\ \langle {\bSigma}^{1/2} \eta_{\alpha \tau_{*}}\{\tau_{*} {\bSigma}^{-1/2} \Zb + (1 + b_{*}) \bbeta_0\}, \Zb \rangle
\$ 
is positive, consider a proximal operator defined in \eqref{hat{y}}: 
\$
\hat{y}_{\BS \theta}(\vb) = {\bSigma}^{1/2} \eta_{\theta}({\bSigma}^{-1/2}\vb).
\$
Lemma \ref{lemma:proximal} implies that 
\begin{equation*}
\begin{split}
    0 \leq\ & \EE \Div \hat{y}_{\BS \theta}\{\tau_{*} \Zb + (1 + b_{*}) {\bSigma}^{1/2} \bbeta_0\}\\ 
    =\ & \EE \tau_{*}^{-1} \langle \hat{y}_{\BS \theta}\{\tau_{*} \Zb + (1 + b_{*}) {\bSigma}^{1/2} \bbeta_0\}, \Zb \rangle \\ 
    =\ & \EE \tau_{*}^{-1} \langle {\bSigma}^{1/2}  \eta_{\theta} \{\tau_{*} {\bSigma}^{-1/2} \Zb + (1 + b_{*}) \bbeta_0\}, \Zb \rangle.
\end{split}
\end{equation*}
Therefore, we have 
\begin{equation*}
    0 \leq \EE\ \langle {\bSigma}^{1/2} \eta_{\theta} \{\tau_{*} {\bSigma}^{-1/2} \Zb + (1 + b_{*}) \bbeta_0\}, \Zb \rangle .
\end{equation*}
This  completes the proof of the claim. 
\end{proof}

\subsection{Proof of Theorem \ref{thm: general lasso mse + R2}}
\label{proof of theorem general lasso mse + R2}
Let $\phi^{p} (x,y) := \norm{x-y}_2^2$ be a sequence of uniformly pseudo-Lipschitz function. Then, using  Theorem \ref{thm: general LASSO main without normalization}, we have
\begin{align*}
    \bnorm{\hat{\bbeta}_{\LW}(\lambda) - \bbeta_0}_\bmSigma^2 = \bnorm{\frac{1}{\sqrt{p}} \eta_{\lambda (1 + b_*)}\{\tau_{*} {\bSigma}^{-1/2} \Zb + (1 + b_*) \sqrt{p} \bbeta_0\} - \bbeta_0}_{\bSigma}^2.
\end{align*}
Similarly, we have
\begin{align*}
    &\bnorm{\hat{\bbeta}_{\LW}(\lambda)}_\bmSigma^2 = \bnorm{\frac{1}{\sqrt{p}} \eta_{\lambda (1 + b_*)}\{\tau_{*} {\bSigma}^{-1/2} \Zb + (1 + b_*) \sqrt{p} \bbeta_0\}}_{\bSigma}^2 \quad \mbox{and} \\
    &\left\langle \bbeta_0, \hat{\bbeta}_{\LW}(\lambda) \right\rangle^2_{\bmSigma} = \left\langle \bbeta_0, \frac{1}{\sqrt{p}} \eta_{\lambda (1 + b_*)}\{\tau_{*} {\bSigma}^{-1/2} \Zb + (1 + b_*) \sqrt{p} \bbeta_0\} - \bbeta_0 \right\rangle^2_{\bmSigma}.
\end{align*}
Recall the definition of  $A(\hat{\bbeta}_\LW)$ from \eqref{R2 def}:
\begin{align*}
    \begin{split}
        A(\hat{\bbeta}_{\LW}(\lambda)) = \frac{\y_s^{\T} \hat{\bmS}_{\Sbb}(\hat{\bbeta}_{\LW}(\lambda))}{ \| \y_s \|_2 \cdot \| \hat{\bmS}_{\Sbb}(\hat{\bbeta}_{\LW}(\lambda)) \|_2 } = \frac{\y_s^{\T} \hat{\bmS}_{\Sbb}(\hat{\bbeta}_{\LW}(\lambda))/n_s}{ \| \y_s \|_2/\sqrt{n_s} \cdot \| \hat{\bmS}_{\Sbb}(\hat{\bbeta}_{\LW}(\lambda))/\sqrt{n_s} \|_2 },
    \end{split}
\end{align*}
where 
\begin{align*}
    \y_s^{\T} \hat{\bmS}_{\Sbb}(\hat{\bbeta}_{\LW}(\lambda))/n_s = \bbeta_0^{\T} \left( \frac{1}{n_s} \Sbb^{\T} \Sbb \right) \hat{\bbeta}_{\LW}(\lambda) + \frac{1}{n_s} \left(\bmeps_s^{\T} \Sbb \hat{\bbeta}_{\LW}(\lambda)\right).
\end{align*}
The following lemma proves the asymptotic convergence of $\bbeta_0^{\T} \left( \frac{1}{n_s} \Sbb^{\T} \Sbb \right) \hat{\bbeta}_{\LW}(\lambda)$. 
\begin{lemma}\label{DE lemma}
    For any independent random vectors $\ub, \vb \in \RR^{p}$, which are independent of $\Sbb$, and with $\norm{u}_2 = O_p(1)$ and $\norm{v}_2 = O_p(1)$, we have 
    \begin{align*}
        \ub^{\T} \left( \frac{1}{n_s} \Sbb^{\T} \Sbb \right) \ub - \ub^{\T} \bmSigma \ub {\BS \overset{P}{\to}} 0.
    \end{align*}
    Moreover, we have   
    \begin{align*}
        \ub^{\T} \left( \frac{1}{n_s} \Sbb^{\T} \Sbb \right) \vb - \ub^{\T} \bmSigma \vb {\BS \overset{P}{\to}} 0.
    \end{align*}
\end{lemma}

\begin{proof}[Proof of Lemma \ref{DE lemma}]
    Lemma \ref{lemma: S.21} generalizes Lemma B.26 in \cite{bai2010spectral}, which allows us to conclude that
    \begin{align*}
        \frac{\ub^{\T} \left( \frac{1}{n_s} \Sbb^{\T} \Sbb \right) \ub}{\tr \left\{ \frac{1}{n_s} \Sbb^{\T} \Sbb \left(\EE \ub \ub^{\T} \right) \right\}} {\BS \overset{P}{\to}} 1.
    \end{align*}
  Let  $\Cb := \EE \ub^{\T} \ub $. By Lemma \ref{lemma: S.22}, we have 
    \begin{align*}
        \frac{\tr \left\{ \frac{1}{n_s} \Sbb^{\T} \Sbb \left(\EE \ub \ub^{\T} \right) \right\} }{\tr \left( \ub^{\T} \bmSigma \ub \right) } = \frac{\tr \left\{ \frac{1}{n_s} (\Sbb \Cb^{1/2})^{\T} (\Sbb \Cb^{1/2}) \right\} }{\tr \left( \ub^{\T} \bmSigma \ub \right) } = \frac{\tr \left\{ \frac{1}{n_s} (\Sbb \Cb^{1/2})^{\T} (\Sbb \Cb^{1/2}) \right\} }{\tr \left( \ub^{\T} \bmSigma \ub \right) } {\BS \overset{P}{\to}} 1.
    \end{align*}
    This completes the proof of the first equation. 
    Using the identity 
    $$
    2 \ub^{\T} \bmSigma \vb = (\ub + \vb)^{\T} \bmSigma (\ub + \vb) - \ub^{\T} \bmSigma \ub - \vb^{\T} \bmSigma \vb,
    $$
     completes the proof of the lemma. 
\end{proof}

In addition, each entry of $\Sbb \hat{\bbeta}_{\LW}(\lambda)$ is i.i.d. {\BS conditional on $\hat{\bbeta}_{\LW}(\lambda)$}. 
Consequently, $\bmeps_s^{\T} \Sbb \hat{\bbeta}_{\LW} $ can be written as the sum of $n_s$ i.i.d. random variables. 
By the law of large number, we conclude that  $\left(\bmeps_s^{\T} \Sbb \hat{\bbeta}_{\LW}(\lambda) \right)/n_s \overset{P}{\to} 0$, and thus 
\begin{align*}
    \y_s^{\T} \hat{\bmS}_{\Sbb}(\hat{\bbeta}_{\LW}(\lambda))/n_s - \bbeta_0^{\T} \bmSigma \hat{\bbeta}_{\LW}(\lambda) = \bbeta_0^{\T} \left( \frac{1}{n_s} \Sbb^{\T} \Sbb \right) \hat{\bbeta}_{\LW}(\lambda) - \bbeta_0^{\T} \bmSigma \hat{\bbeta}_{\LW}(\lambda) + \frac{1}{n_s} \left(\bmeps_s^{\T} \Sbb \hat{\bbeta}_{\LW}(\lambda) \right) \overset{P}{\to} 0.
\end{align*}
Similarly, we have
\begin{align*}
    \| \hat{\bmS}_{\Sbb}(\hat{\bbeta}_{\LW}(\lambda)) \|^2_2/n_s -  \bnorm{\hat{\bbeta}_{\LW}(\lambda)}^2_{\Sigma} \overset{P}{\to} 0\quad \mbox{and} \quad \| \y_s \|^2_2/n_s - \left(\bnorm{\bbeta_0}^2_{\Sigma} + \sigma_{\bmeps_s}^2 \right) \overset{P}{\to} 0.
\end{align*}
Using  continuous mapping theorem, we have
\begin{align} \label{eqn: S.10.2}
    A^2(\hat{\bbeta}_{\LW}(\lambda)) - \frac{\left\langle \bbeta_0, \hat{\bbeta}_{\LW}(\lambda) \right\rangle^2_{\bmSigma}}{\bnorm{\hat{\bbeta}_{\LW}(\lambda)}^2_{\Sigma} \cdot \left(\bnorm{\bbeta_0}^2_{\Sigma} + \sigma_{\bmeps_s}^2 \right)} \overset{P}{\to} 0.
\end{align}
Using a similar to the previous argument, we have 
\begin{align} \label{eqn: S.10.3}
    h_s^2 - \frac{\bnorm{{\bbeta}_0}^2_{\bmSigma}}{\bnorm{{\bbeta}_0}^2_{\bmSigma} + \sigma^2_{\bmeps_s}} \overset{P}{\to} 0.
\end{align}
Combining equations \eqref{eqn: S.10.2} and \eqref{eqn: S.10.3} with Theorem \ref{thm: general LASSO main without normalization}, we complete the proof of Theorem \ref{thm: general lasso mse + R2}. 

\begin{remark}
    An alternative way to understand Lemma \ref{DE lemma} is using the concept of deterministic equivalence; see for example Lemma S.6.9 in \cite{wu2023ensemble}.  It is important to note that $\frac{1}{n_s} \Sbb^{\T} \Sbb$ is deterministically equivalent to $\bmSigma$. Therefore, we have 
    \begin{equation*}
    \lim_{n\to\infty}\left|\tr \left\{\Eb_p( \Sbb^{\T} \Sbb/n_s - \bmSigma)\right\}\right|=0 \qas, 
    \end{equation*}
 for any sequence $\{\Eb_p\}_{p\ge 1} \in \RR^{p}$ such that 
$\lim\sup_n\|\Eb_p\|_{\tr}<\infty$.
\end{remark}

\section{Proofs for Section \ref{subsec: Proof ofthm: general LASSO main without normalization}}\label{Proof for Appendix S.5.1}

In this section, we prove Theorem \ref{general - state}, 
Proposition \ref{prop 3}, and Theorem \ref{AMP approx estimator}. 
In Section \ref{sec: Equivalence l1 and l2}, we first provide some preliminary results that establish the equivalence between the $\ell_1$ and $\ell_2$ norm. Section~\ref{general - state - proof} proves Theorem \ref{general - state}, Section \ref{proof of prop 3} proves Proposition \ref{prop 3}, and Section~\ref{proof of Theorem AMP approx estimator} proves Theorem \ref{AMP approx estimator}. The proofs for all supporting lemmas are provided in Section~\ref{S.6 lemmas proof}.

\subsection{Equivalence of \texorpdfstring{$\ell_{2}$}{TEXT} and \texorpdfstring{$\ell_{1}$}{TEXT} norms in random vector spaces}\label{sec: Equivalence l1 and l2}

Theorem \ref{thm: S.13.2} and the Cauchy-Schwarz inequality establish the equivalence between the $\ell_{1}$ and $\ell_{2}$ norm with high probability. 


\begin{theorem}[\cite{Kain1977DIAMETERSOS}]
\label{thm: S.13.2}
    For any positive number $v$, 
    let $V_{p,v}$ denote a random subspace of dimension $\lfloor p(1 - v) \rfloor$, which is uniformly drawn from $\mathbb{R}^{p}$. 
    There exists a constant $c_v$ that only depends on $v$ such that, for any $p \geq 1$, the following inequality holds with probability at least $1 - 2^{-p}$:
    \begin{align*}
        \forall{\vb}\, \in V_{p,v}: \quad  c_v \bnorm{\vb}_2 \leq \frac{1}{\sqrt{p}} \bnorm{\vb}_1.
    \end{align*}
\end{theorem}
Moreover, for any vector $\vb \in \RR^{p}$, the Cauchy-Schwarz inequality implies that 
\begin{align*}
    \|\vb\|_{1} \leq \sqrt{p} \|\vb\|_{2}.
\end{align*}
This, together with Theorem \ref{thm: S.13.2}, establishes the equivalence between the $\ell_{1}$ and $\ell_{2}$ norm. This equivalence will be used in the proof of Proposition \ref{lemma S.13}.

\subsection{Proof of Theorem \ref{general - state}}
\label{general - state - proof}
In this section,  we state and prove a stronger version of Theorem \ref{general - state}, aka Theorem \ref{general - state - stronger}, and  Theorem \ref{general - state} follows immediately from  Theorem \ref{general - state - stronger}. 
\begin{theorem} \label{general - state - stronger}
Assume the same conditions as in Theorem \ref{general - state}. Let $\{ \phi^{p}: \RR^{p} \times \RR^p \times \RR^p \mapsto \RR \}_{p \in \NN}$ denote any deterministic sequence of uniformly normalized pseudo-Lipschitz functions defined in \eqref{eqn:normalized_pesudo}.
Then, for all $t \geq 0$, we have
\begin{equation}
\begin{split}
    &\phi^{p} \left(\bbeta_0, \bbeta^{t}, \bbeta^{t+1}\right)\\ 
    &\overset{P}{\approx} \EE\ \phi^{p} \left[\bbeta_0,  \eta_{\lambda(1 + b_{t-1})}\{{\bSigma}^{-1/2} \Zb_{t} + (1 + b_{t-1}) \bbeta_0\}, \eta_{\lambda(1 + b_t)}\{{\bSigma}^{-1/2} \Zb_{t+1} + (1 + b_t) \bbeta_0\} \right],
\end{split}
\end{equation}
where $(\Zb_t^\T, \Zb_{t+1}^\T)^\T$ is jointly Gaussian distributed and is independent of $\bbeta_0$.
$(\Zb_t^\T, \Zb_{t+1}^\T)^\T$ has mean $\bm{0}$, covariance 
\begin{align} \label{eqn:Z_t_cov}
\begin{bmatrix}
\tau_t^2 {\bI_p} & \tau^2_{t,t+1} {\bI_p} \\
\tau^2_{t,t+1} {\bI_p} & \tau_{t+1}^2 {\bI_p}
\end{bmatrix}.
\end{align}
The recursion $\tau^2_{t}$ and $\tau_{t,t+1}^2$ for all $t \geq 0$ is determined by~\eqref{tau_t tau_t,t+1}.
\end{theorem}
We prove Theorem \ref{general - state - stronger} in three steps. 
First, we rewrite the recursions in equation~\eqref{general - lasso - AMP} as recursions in~\eqref{eqn: S.6.1 amp}, which only depend on ${\Xb_0}$ and $\Wb_0$ introduced in step 1. 
Next, in order to apply Theorem \ref{matrix AMP} to recursions~\eqref{eqn: S.6.1 amp}, we convert them into the matrix form in \eqref{eqn: standard matrix AMP}. 
Finally, by employing techniques similar to those introduced in \cite{berthier2020state}, we compute the state evolution for recursions in~\eqref{matrix version 2}.

\paragraph{Step 1: Rewrite AMP recursions.} 
Let ${\Xb_0} \in \RR^{n_x \times p}$ be a random matrix with each row being i.i.d. as $N(0, {\bI_p}/n_x)$, and  ${\Wb_0} \in \RR^{n_w \times p}$ be a random matrix with each row  being i.i.d. as $N(0, {\bI_p}/n_w)$. 
Recall that $\Xb \in \RR^{n_x \times p}$ and ${\Wb} \in \RR^{n_w \times p}$ have rows independently  distributed as $N(0, \bSigma/n_x)$ and $N(0, \bSigma/n_w)$, respectively. 
Thus, we have $\Xb = {\Xb_0} {\bSigma}^{1/2}$ and $\Wb = {\Wb_0} {\bSigma}^{1/2} $. 
We consider the following optimization  problem
\begin{equation} \label{i.i.d. ref panel}
\begin{split}
\widehat{\bbeta}_{r}(\lambda) = \arg\min_{\tilde{\bbeta} \in \R^p} \frac{1}{2} \tilde{\bbeta}^{\T} \Wb_0^{\T} \Wb_0 \tilde{\bbeta} - \tilde{\bbeta}^{\T} \X_0^{\T} \y_x + \lambda \|{\bSigma}^{-\frac{1}{2}} \tilde{\bbeta} \|_1.
\end{split}
\end{equation}
It is obvious that $\widehat{\bbeta}_{r}(\lambda) = \bmSigma^{1/2} \hat{\bbeta}_{\LW}(\lambda)$. 
Let $\tilde{\bbeta}^{t} = \bSigma^{1/2} \bbeta^{t}$ and $\tilde{\bbeta} = \bSigma^{1/2} \bbeta_0$. Recall the definition of  $\hat{y}_{\theta}(v)$ from  \eqref{hat{y}}. 
The following claim shows that the AMP recursions in \eqref{general - lasso - AMP} are equivalent to those in \eqref{bt_recursion_1n_normaliation}.

\begin{claim} \label{claim s.6.1} 
Let $N := n_x + n_w$ and $\tilde{\bbeta}^{t+1}_c =  - \sqrt{\frac{N}{n_{w}}} \Wb_0^{\T}{\rb^{t}} + (1+b_t) \Xb_0^{\T}\yb_x + \tilde{\bbeta}^{t}$, then $\tilde{\bbeta}^{t+1}$ and $ \rb^{t}$  satisfy the following recursions
\begin{equation}\label{eqn: S.6.1 amp}
\begin{split}
    \tilde{\bbeta}^{t+1} &= \hat{y}_{\theta_{t}} \Bigg\{ - \sqrt{\frac{N}{n_{w}}} \Wb_0^{\T}{\rb^{t}} + (1+b_t) \Xb_0^{\T}\yb_x + \tilde{\bbeta}^{t} \Bigg\} \quad \mbox{and} 
    \\ 
    \rb^{t} &= \sqrt{\frac{n_{w}}{N}} {\Wb_0} \tilde{\bbeta}^{t} + \frac{{b}_t}{1 + {b}_{t-1}} \rb^{t-1},
\end{split}
\end{equation}
where $\theta_t = \lambda(1+b_t)$ and
\begin{equation}\label{eqn: S.6.1 b_t}
\begin{split}
    \frac{{b}_t}{1+ {b}_{t-1}} = \frac{1}{n_w} \Div \hat{y}_{\theta_{t-1}} (\tilde{\bbeta}^{t}_c).
\end{split}
\end{equation}
\end{claim}
\begin{proof}[Proof of Claim \ref{claim s.6.1}]
The first equation in \eqref{general - lasso - AMP} and the fact that $\hat{y}_{\theta} (v) = \bSigma^{1/2} \eta_{\theta}(\bSigma^{-1/2}v)$ imply that 
\begin{equation*} 
\begin{split}
    \tilde{\bbeta}^{t+1} &= \bSigma^{1/2} {\bbeta}^{t+1}\\ 
    &= \bSigma^{1/2} \eta_{\theta_{t}} \Bigg\{ - \sqrt{\frac{N}{n_{w}}} \bSigma^{-1} \W^{\T}{\rb^{t}} + (1+b_t) \bSigma^{-1} \X^{\T}\yb_x + {\bbeta}^{t} \Bigg\}\\ 
    &= \hat{y}_{\theta_{t}} \Bigg\{ - \sqrt{\frac{N}{n_{w}}} \bSigma^{-1/2} {\W}^{\T}{\rb^{t}} + (1+b_t) \bSigma^{-1/2} {\X}^{\T}\yb_x + {\bSigma}^{1/2} {\bbeta}^{t} \Bigg\}\\
    &= \hat{y}_{\theta_{t}} \Bigg\{ - \sqrt{\frac{N}{n_{w}}} \Wb_0^{\T}{\rb^{t}} + (1+b_t) \Xb_0^{\T}\yb_x + \tilde{\bbeta}^{t} \Bigg\}.
\end{split}
\end{equation*}
For $\rb^{t}$, we have 
\begin{equation*}
\begin{split}
    \rb^{t} &= \W {\bbeta}^{t} + \frac{b_t}{1 + b_{t-1}} \rb^{t-1}= \Wb_0 \tilde{\bbeta}^{t} + \frac{{b}_t}{1 + {b}_{t-1}} \rb^{t-1}.
\end{split}
\end{equation*}
This proves \eqref{eqn: S.6.1 amp}. It remains to prove \eqref{eqn: S.6.1 b_t}. {By~\eqref{bt_recursion_1n_normaliation}, we have }
\begin{equation*}
\begin{split}
    \frac{{b}_t}{1 + {b}_{t-1}} 
    &= \frac{1}{n_w} \Div \eta_{\theta_{t-1}} \left\{- \sqrt{\frac{N}{n_{w}}}  \bSigma^{-1} \Wb^{\T}{\rb^{t-1}} + (1+b_{t-1}) \bSigma^{-1} \Xb^{\T}\yb_x + {\bbeta}^{t-1} \right\}\\
    &= \frac{1}{n_w} \Div \eta_{\theta_{t-1}} \left\{- \sqrt{\frac{N}{n_{w}}} \bSigma^{-1/2} \Wb_0^{\T}{\rb^{t-1}} + (1+b_{t-1}) \bSigma^{-1/2} \Xb_0^{\T}\yb_x + {\bbeta}^{t-1} \right\}.
\end{split}
\end{equation*}
Denote the $\nabla_{\yb} f(\xb, \yb)$ to be the Jacobian with respect to variable $\yb$. 
We have
\begin{align*}
    \frac{{b}_t}{1 + {b}_{t-1}} &= \frac{1}{n_w} \tr \left[ \nabla_{\bbeta^{t-1}} \eta_{\theta_{t-1}} \left\{- \sqrt{\frac{N}{n_{w}}} {\bSigma}^{-1/2} \Wb_0^{\T}{\rb^{t-1}} + (1+b_{t-1}) {\bSigma}^{-1/2} \Xb_0^{\T}\yb_x + {\bbeta}^{t-1}\right\} \right]
    \\
    &\overset{(a)}{=} \frac{1}{n_w}  \tr \left[ \nabla_{\tilde{\bbeta}^{t-1}} {\bSigma}^{1/2} \eta_{\theta_{t-1}} \left\{- \sqrt{\frac{N}{n_{w}}} \bSigma^{-1/2} \Wb_0^{\T}{\rb^{t-1}} + (1+b_{t-1}) {\bSigma}^{-1/2} \Xb_0^{\T}\yb_x + {\bSigma}^{-1/2} \tilde{\bbeta}^{t-1}\right\} \right]
    \\
    &\overset{(b)}{=} \frac{1}{n_w}  \tr \left[ \nabla_{\tilde{\bbeta}^{t-1}} \hat{y}_{\theta_{t-1}} \Bigg\{ - \sqrt{\frac{N}{n_{w}}}  \Wb_0^{\T}{\rb^{t-1}} + (1+b_{t-1}) \Xb_0^{\T}\yb_x + \tilde{\bbeta}^{t-1} \Bigg\} \right]
    \\
    &= \frac{1}{n_w} \Div \hat{y}_{\theta_{t-1}} \Bigg\{ - \sqrt{\frac{N}{n_{w}}} \Wb_0^{\T}{\rb^{t-1}} + (1+b_{t-1}) \Xb_0^{\T}\yb_x + \tilde{\bbeta}^{t-1} \Bigg\} = \frac{1}{n_w} \Div \hat{y}_{\theta_{t-1}} (\tilde{\bbeta}^{t}_c)
\end{align*}
where the equality $(a)$ follows from the chain rule and the equality $(b)$ follows from equation~\eqref{hat{y}}.
This completes the proof of Claim~\ref{claim s.6.1}.
\end{proof}

\paragraph{Step 2: Convert into asymmetric matrix AMP.}
By Claim \ref{claim s.6.1}, to understand the asymptotic behavior of $\bbeta^{t}$, it suffices to understand that of $\tilde{\bbeta}^{t}$. 
In order to apply Theorem \ref{matrix AMP}, we convert the recursions \eqref{eqn: S.6.1 amp} to a matrix form in~\eqref{eqn: standard matrix AMP}. 
Let {$N' := n_x + n_w + p$} and suppose each row of $\Xb_0' = {\Xb_0} \sqrt{n_{x}/{N}}$ and $\Wb_0' = \Wb_0 \sqrt{n_{w}/N}$ is i.i.d. $N(0, \bI_{p}/N)$.
We rewrite the recursions in~\eqref{eqn: S.6.1 amp} in terms of $\tilde{\bbeta}_c^{t+1}$: 
\begin{equation*}
\begin{split}
    \tilde{\bbeta}_c^{t+1} &= \begin{bmatrix}
    \Wb_0'\\
    \Xb_0'\\
    \end{bmatrix}
    ^{\T} \begin{bmatrix}
    -\frac{N}{n_{w}} \rb^{t}\\
    \sqrt{\frac{N}{n_x}}(1+b_t)\yb_x
    \end{bmatrix} +   \hat{y}_{\theta_{t-1}}(\tilde{\bbeta}_c^{t}) \quad \mbox{and} \\ 
    \rb^{t} &= \Wb_0'  \hat{y}_{\theta_{t-1}}(\tilde{\bbeta}_c^{t}) - \frac{b_t}{1 + b_{t-1}}
    (- \rb^{t-1}).
\end{split}
\end{equation*}
Construct a new sequence $\zb^{t} \in \RR^{n_x}$ such that
\begin{equation} \label{Matrix version step 1}
\begin{split}
    \tilde{\bbeta}_c^{t+1} &= \begin{bmatrix}
    \Wb_0'\\
    \Xb_0'\\
    \end{bmatrix}
    ^{\T} \begin{bmatrix}
    - \frac{N}{n_{w}} \rb^{t}\\
    \sqrt{\frac{N}{n_x}}(1+b_t)\yb_x
    \end{bmatrix} + \hat{y}_{\theta_{t-1}}(\tilde{\bbeta}_c^{t})\quad \mbox{and} \\ 
    \begin{bmatrix}
    \rb^{t}\\
    \zb^{t}\\
    \end{bmatrix} &= \begin{bmatrix}
    \Wb_0'\\
    \Xb_0'\\
    \end{bmatrix}  \hat{y}_{\theta_{t-1}}(\tilde{\bbeta}_c^{t})  - \frac{n_{w}}{N} \cdot \frac{b_t}{1 + b_{t-1}} \begin{bmatrix}
    - \frac{N}{n_{w}} \rb^{t-1}\\
    \sqrt{\frac{N}{n_x}}(1+b_{t-1})\yb_x
    \end{bmatrix}.
\end{split}
\end{equation}
This is a special case of the matrix AMP discussed in Theorem \ref{matrix AMP}. To see this, let
\begin{align*}
    \Ab_0 = \begin{bmatrix} 
    (\Wb_0')^{\T} & (\Xb_0')^{\T}
    \end{bmatrix}^{\T} \in \RR^{N \times p}, \quad 
    \tilde{\bbeta} = {\bSigma}^{1/2} \bbeta_0 \in \RR^{p}.
\end{align*}
Then we can write the recursions in~\eqref{Matrix version step 1} in a matrix form as 
\begin{equation*}
\begin{split}
    &\begin{bmatrix}
    \Wb_{0}^{'} \tilde{\bbeta} & \rb^{t} \\
    \Xb_{0}^{'} \tilde{\bbeta} & \zb^{t} \\
    \end{bmatrix} = 
    \begin{bmatrix}
        \Wb_{0}^{'} \\
        \Xb_{0}^{'} \\
    \end{bmatrix}
    \begin{bmatrix}
    \tilde{\bbeta} & \hat{y}_{\theta_{t-1}}(\tilde{\bbeta}_c^{t})
    \end{bmatrix} - \begin{bmatrix}
    - \frac{N}{n_{w}} \rb^{t-1}\\
    \sqrt{\frac{N}{n_x}}(1+b_{t-1})\yb_x
    \end{bmatrix} \begin{bmatrix}
    0 & \frac{1}{N} \Div \hat{y}_{\theta_{t-1}}(\tilde{\bbeta}_c^{t})
    \end{bmatrix} \quad \mbox{and}
    \\
    &\tilde{\bbeta}_c^{t+1} - (1+b_t) \tilde{\bbeta} = 
    \begin{bmatrix}
        (\Wb_{0}^{'})^{\T} & (\Xb_{0}^{'})^{\T} 
    \end{bmatrix}
    \begin{bmatrix}
    - \frac{N}{n_{w}}\rb^{t}\\
    \sqrt{\frac{N}{n_x}}(1+b_{t}) \yb_x
    \end{bmatrix} - \begin{bmatrix}
    \tilde{\bbeta} & \hat{y}_{\theta_{t-1}}(\tilde{\bbeta}_c^{t})
    \end{bmatrix}\begin{bmatrix}
    1+b_t \\
    -1
    \end{bmatrix},
\end{split}
\end{equation*}
or equivalently:
\begin{equation} \label{matrix version 2}
\begin{split}
    &\begin{bmatrix}
    \Wb_{0}^{'} \tilde{\bbeta} & \rb^{t} & 0 \\
    \Xb_{0}^{'} \tilde{\bbeta} & \zb^{t} & 0 \\
    \end{bmatrix} 
    = 
    \Ab_0 \begin{bmatrix}
    \tilde{\bbeta} & \hat{y}_{\theta_{t-1}}(\tilde{\bbeta}_c^{t}) & 0
    \end{bmatrix} - 
    \begin{bmatrix}
    0 & - \frac{N}{n_{w}} \rb^{t-1} & 0\\
    0 & \sqrt{\frac{N}{n_x}} (1 + b_{t-1}) \yb_x & 0
    \end{bmatrix}
    \begin{bmatrix}
    0 & 0 & 0\\
    0 & \frac{1}{N} \Div \hat{y}_{\theta_{t-1}}(\tilde{\bbeta}_c^{t}) & 0\\
    0 & 0 & 0\\
    \end{bmatrix}\quad \mbox{and} \\
    &\begin{bmatrix}
    0 & \tilde{\bbeta}_c^{t+1} - (1+b_t) \tilde{\bbeta} & 0
    \end{bmatrix} = \Ab_0^{\T} 
    \begin{bmatrix}
    0 & - \frac{N}{n_{w}} \rb^{t} & 0\\
    0 & \sqrt{\frac{N}{n_x}} (1 + b_{t}) \yb_x & 0
    \end{bmatrix} - 
    \begin{bmatrix}
    \tilde{\bbeta} & \hat{y}_{\theta_{t-1}}(\tilde{\bbeta}_c^{t}) & 0
    \end{bmatrix} \begin{bmatrix}
    0 & (1+b_t) & 0\\
    0 & - 1 & 0\\
    0 & 0 & 0\\
    \end{bmatrix}.
\end{split}
\end{equation}
Let $q = 3$ and 
\begin{equation*}
\begin{split}
    \Ub^{t} &= \begin{bmatrix}
    \Wb_{0}^{'} \tilde{\bbeta} & \rb^{t} & 0 \\
    \Xb_{0}^{'} \tilde{\bbeta} & \zb^{t} & 0 \\
    \end{bmatrix}  \in \RR^{N \times q},
    \\
    \Vb^{t+1} &= \begin{bmatrix}
    0 & \tilde{\bbeta}_c^{t+1} - (1+b_t) \tilde{\bbeta} & 0
    \end{bmatrix} \in \RR^{p \times q},
    \\
    F^t(\Vb) &= \begin{bmatrix}
    \tilde{\bbeta} & \hat{y}_{\theta_{t-1}}(\Vb_{*, 2} + (1 + b_{t-1})\tilde{\bbeta}) & 0
    \end{bmatrix} \in \RR^{p \times q},
    \quad \mbox{and} \\
    G^t(\Ub) &= \begin{bmatrix}
    0 & - \frac{N}{n_{w}} (\Ub_{1})_{*, 2} & 0\\
    0 & \sqrt{\frac{N}{n_x}}(1+b_t)(\sqrt{\frac{N}{n_x}} (\Ub_{2})_{*, 1} + {\bepsilon}_{x}) & 0\\
    \end{bmatrix} \in \RR^{N \times q},\\
\end{split}
\end{equation*}
where $\Ub = \begin{bmatrix}
    \Ub_{1}^{\T} & \Ub_{2}^{\T}
\end{bmatrix}^{\T}$ with $\Ub_{1} \in \RR^{n_{w} \times q}$ and $\Ub_{2} \in \RR^{n_x \times q}$ and {\BS $\bmeps_{x}$ given by \eqref{linear model}}. 
Then equation~\eqref{matrix version 2} can be written as
\begin{equation} \label{eqn: standard matrix AMP} 
\begin{split}
    \Vb^{t+1} &= \Ab_0^{\T}
    G^t(\Ub^{t}) - F^{t}(\Vb^{t}) \Db_t^{\T} \quad \mbox{and}\\ 
    \Ub^{t} &= \Ab_0 F^t(\Vb^{t}) - G^{t-1}(\Ub^{t-1}) \Cb_{t}^{\T},
\end{split}
\end{equation}
where 
\begin{align*}
    \Db_t = \frac{1}{N} \sum_{i=1}^{N} \frac{\partial (G^{t})_i (\xb)}{\partial \xb_i}\quad \mbox{and} \quad \Cb_t = \frac{1}{N} \sum_{i=1}^{p} \frac{\partial (F^{t})_i (\xb)}{\partial \xb_i}.
\end{align*}


\paragraph{Step 3: Compute the state evolution.} 
The state evolution of the recursion in Equation \eqref{eqn: standard matrix AMP} is computed using Theorem \ref{matrix AMP}. 
Recall \eqref{eqn: symmetric matrix AMP}:  
\$
\Xb^{t+1} = \Ab \cF^{t}(\Xb^{t}) - \cF^{t-1}(\Xb^{t-1}) {\Bb^t}^{\T},
\$
where $\cF^t = \begin{bmatrix}
(f^{t}_1)^{\T} \cdots (f^{t}_{N'})^{\T}
\end{bmatrix}^{\T}$ and $\Bb^t = (N')^{-1} \sum_{i=1}^{N'} {\partial f^{t}_{i}}/{\partial \xb^{t}_i} (\Xb^{t}) \in \RR^{q \times q}$.
Define $\Xb^{t} = \begin{bmatrix}
(\Xb^{t}_1)^{\T} & (\Xb^{t}_2)^{\T}
\end{bmatrix}^{\T} \in \RR^{N' \times q}$ with sub-matrices $\Xb^{t}_1 \in \RR^{N \times q}$ and $\Xb^{t}_2 \in \RR^{p \times q}$ and 
\begin{align} \label{eqn:amp_sym_to_asym}
\begin{split}
    &\Ab = \sqrt{\frac{N}{N'}}
    \begin{bmatrix}
    \Bb_1 & \Ab_0\\
    \Ab_0^{\T} & \Bb_2
    \end{bmatrix}
    \\
    &\cF^{2t}(\Xb^{2t}) = \begin{bmatrix}
    \sqrt{\frac{N'}{N}} G^t(\Xb^{2t}_1)\\
    0\\
    \end{bmatrix} 
    \\
    &\cF^{2t+1}(\Xb^{2t+1}) = \begin{bmatrix}
    0\\
    \sqrt{\frac{N'}{N}} F^{t+1}(\Xb^{2t+1}_2) \\
    \end{bmatrix}.
\end{split}
\end{align}
where $\Bb_1 \in \R^{N \times N}$ is  a GOE with $N(0, 1/N)$ non-diagonal entries,  and $\Bb_2 \in \R^{p \times p}$ is a scaled GOE with  $N(0, 1/N)$ non-diagonal entries.
It is easy to see that 
\begin{align} \label{eqn:rec_sym_to_asym}
\Vb^{t+1} = \Xb^{2t+1}_2, \Ub^{t} = \Xb^{2t}_1.
\end{align}
Therefore, for the deterministic sequence of uniformly normalized pseudo-Lipschitz functions $\{\phi^{N'}\}_{N' \in \NN}$ and $(\Zb^{2t-1}, \Zb^{2t+1})$ defined in \eqref{Zb_i}, $\phi^{N'} (\Xb^0, \Xb^{2t-1}, \Xb^{2t+1}) \approx \phi^{N'} (\Xb^0, \Zb^{2t-1}, \Zb^{2t+1})$ by Theorem \ref{matrix AMP}.   
For any $k \in \NN$, denote $\Zb^{k}$ as
\begin{equation*}
\begin{split}
    \Zb^{k} &= \begin{bmatrix}
    (\Zb^{k}_1)^{\T} & (\Zb^{k}_2)^{\T} &(\Zb^{k}_3)^{\T}
    \end{bmatrix}^{\T}
\end{split}
\end{equation*}
with $\Zb^{k}_1 \in \RR^{n_{w} \times q}$, $\Zb^{k}_2 \in \RR^{n_{x} \times q}$, and $\Zb^{k}_3 \in \RR^{p \times q}$. 
By equation~\eqref{eqn:rec_sym_to_asym}, we have
\begin{align} \label{eqn:state_form_2}
\begin{split}
    &\phi^{p}\big\{\tilde{\bbeta}, \tilde{\bbeta}_c^{t} - (1+b_{t-1}) \tilde{\bbeta}, \tilde{\bbeta}_c^{t+1} - (1+b_t) \tilde{\bbeta} \big\} 
    \\
    =\ & \phi^{p} \big\{\tilde{\bbeta}, (\Vb^{t})_{*,2}, (\Vb^{t+1})_{*,2} \big\}
    \\
    \approx\ & \phi^{p} \big\{\tilde{\bbeta}, (\Zb_{3}^{2t-1})_{*, 2}, (\Zb_{3}^{2t+1})_{*, 2} \big\}
\end{split}
\end{align}
with covariance of $\big\{ (\Zb_{3}^{2t-1})_{*,2}, (\Zb_{3}^{2t+1})_{*,2} \big\}$  given by
\begin{align*}
    \begin{bmatrix}
        (\Kb_{2t-1, 2t-1})_{2,2} \bI_{p} & (\Kb_{2t-1, 2t+1})_{2,2} \bI_{p} \\
        (\Kb_{2t-1, 2t+1})_{2,2} \bI_{p} & (\Kb_{2t+1, 2t+1})_{2,2} \bI_{p} \\
    \end{bmatrix}
\end{align*}
with $\Kb_{t,s}$ defined in \eqref{matrix state evolution}. 
By the definitions of $\cF^{2t}$ and $\cF^{2t+1}$, we have $(\Kb_{2t+1, 2t-1})_{i,j} = 0$ and $(\Kb_{2t+1, 2t+1})_{i,j} = 0$ for any $(2,2) \neq (i,j) \in [3] \times [3]$. 
Define $\tau_{t+1}^2 := (\Kb_{2t+1, 2t+1})_{2,2}$ and $\tau_{t+1,t}^2 := (\Kb_{2t+1, 2t-1})_{2,2}$. Then $\tau_t^2$ and $\tau^{2}_{t+1, t}$ satisfy the following recursions:
\begin{equation} \label{tau_t chuxing}
\begin{split}
    \tau_{t+1}^2 
    &= \lim_{p \to \infty} \frac{1}{N} \EE \left[ \frac{N^2}{n_{w}^2} \bnorm{(\Zb^{2t}_1)_{*,2}}_2^2 + \frac{N}{n_x} (1 + b_t)^2 \bnorm{\sqrt{\frac{N}{n_x}} (\Zb^{2t}_2)_{*,1} + {\bepsilon}_{x}}_2^2 \right]
    \\
    &=\lim_{p \to \infty} \EE \left[\frac{N}{n_w^2} \bnorm{(\Zb^{2t}_1)_{*,2}}_2^2 + \frac{1}{n_x} (1 + b_t)^2 \left( \frac{N}{n_x} \bnorm{(\Zb^{2t}_2)_{*,1}}_2^2 + \bnorm{{\bepsilon}_{x}}_2^2 \right) \right]\quad \mbox{and} 
    \\
    \tau_{t+1,t}^2 
    &= \lim_{p \to \infty} \frac{1}{N} \EE \left[ \frac{N^2}{n_{w}^2} \left \langle (\Zb^{2t}_1)_{*,2}, (\Zb^{2t-2}_1)_{*,2} \right \rangle + \frac{N}{n_x} (1 + b_t)^2 \left \langle \sqrt{\frac{N}{n_x}} (\Zb^{2t}_2)_{*,1} + {\bepsilon}_{x}, \sqrt{\frac{N}{n_x}} (\Zb^{2t-2}_2)_{*,1} + {\bepsilon}_{x} \right \rangle \right]
    \\
    &= \lim_{p \to \infty} \EE \left[ \frac{N}{n_w^2} \left \langle (\Zb^{2t}_1)_{*,2}, (\Zb^{2t-2}_1)_{*,2} \right \rangle + \frac{1}{n_x} (1 + b_t)(1 + b_{t-1}) \left( {\frac{N}{n_x}} \left \langle (\Zb^{2t}_2)_{*,1}, (\Zb^{2t-2}_2)_{*,1} \right \rangle + \bnorm{{\bepsilon}_{x}}_2^2 \right) \right],
\end{split}
\end{equation}
with {\BS $\bmeps_{x}$ defined in \eqref{linear model}}. 
Since $\vecc(\Zb^{2t}) \sim N(0, \Kb_{2t,2t} \otimes \bI_{N'})$ and $\EE {\vecc(\Zb^{2t})} {\vecc(\Zb^{2t-2})}^{\T} = \Kb_{2t,2t-2} \otimes \bI_{N'}$, by \eqref{Zb_i} , we have 
\begin{align*}
    \EE (\Zb^{2t}_1)_{*,2}(\Zb^{2t}_1)_{*,2}^{\T} 
    &= (\Kb_{2t,2t})_{2,2} \bI_{n_w} = \frac{1}{N} \EE \bnorm{\hat{y}_{\theta_{t-1}}((\Zb^{2t-1}_3)_{*,2} + (1 + b_{t-1})\tilde{\bbeta})}_2^2 \bI_{n_w},\\
    \EE (\Zb^{2t}_2)_{*,1}(\Zb^{2t}_2)_{*,1}^{\T} 
    &= (\Kb_{2t,2t})_{1,1} \bI_{n_x} = \frac{1}{N} \EE \bnorm{\tilde{\bbeta}}_2^2 \bI_{n_x},\\
    \EE (\Zb^{2t}_1)_{*,2}(\Zb^{2t-2}_1)_{*,2}^{\T} 
    &= (\Kb_{2t,2t-2})_{2,2} \bI_{n_w} \\
    &= \frac{1}{N} \EE \left \langle \hat{y}_{\theta_{t-1}}((\Zb^{2t-1}_3)_{*,2} + (1 + b_{t-1})\tilde{\bbeta}) ,\hat{y}_{\theta_{t-2}}((\Zb^{2t-3}_3)_{*,2} + (1 + b_{t-2})\tilde{\bbeta}) \right \rangle \bI_{n_w},\\
    \EE (\Zb^{2t}_2)_{*,1}(\Zb^{2t-2}_2)_{*,1}^{\T} 
    &= (\Kb_{2t,2t-2})_{1,1} \bI_{n_x} = \frac{1}{N} \EE \bnorm{\tilde{\bbeta}}_2^2 \bI_{n_x}.
\end{align*}
Therefore, Equation \eqref{tau_t chuxing} implies that
\begin{equation} \label{tau_t chuxing 2}
\begin{split}
    \tau_{t+1}^2 &= \lim_{p \to \infty} (1 + b_t)^2 \left(\frac{1}{n_x} \bnorm{\tilde{\bbeta}}_2^2 \cdot \frac{1}{h_x^2} \right) + \EE \frac{1}{n_w} \bnorm{\hat{y}_{\theta_{t-1}} \big\{\Zb_t + (1 + b_{t-1})\tilde{\bbeta} \big\}}_2^2.
    \\
    \tau_{t+1,t}^2 
    &= \lim_{p \to \infty} (1 + b_t)(1 + b_{t-1}) \left(\frac{1}{n_x} \bnorm{\tilde{\bbeta}}_2^2 \cdot \frac{1}{h_x^2} \right)\\
    &\qquad + \EE \frac{1}{n_w} \left \langle \hat{y}_{\theta_{t-1}} \big\{\Zb_{t} + (1 + b_{t-1})\tilde{\bbeta}\big\} ,\hat{y}_{\theta_{t-2}} \big\{\Zb_{t-1} + (1 + b_{t-2})\tilde{\bbeta}\big\} \right \rangle,
\end{split}
\end{equation}
where $(\Zb_t^\T, \Zb_{t+1}^\T)^\T$ has mean $\bm{0}$, covariance 
\$
\begin{bmatrix}
        (\Kb_{2t-1, 2t-1})_{2,2} \bI_{p} & (\Kb_{2t-1, 2t+1})_{2,2} \bI_{p} \\
        (\Kb_{2t-1, 2t+1})_{2,2} \bI_{p} & (\Kb_{2t+1, 2t+1})_{2,2} \bI_{p} \\
\end{bmatrix} = \begin{bmatrix}
\tau_t^2 {\bI_p} & \tau^2_{t,t+1} {\bI_p} \\
\tau^2_{t,t+1} {\bI_p} & \tau_{t+1}^2 {\bI_p}
\end{bmatrix}.
\$
Alternatively, we can rewrite equation~\eqref{tau_t chuxing 2} as
\begin{equation} \label{tau_t tau_t,t+1}
\begin{split}
    \tau_{t+1}^2 &= \lim_{p \to \infty} \EE \left[ \frac{1}{n_w} \bnorm{{\eta}_{\theta_{t-1}}\big\{\tau_{t} {\bSigma}^{-1/2} Z + (1 + b_{t-1}){\bbeta}_0\big\}}_{\bSigma}^2 + \frac{1}{n_x} (1 + b_t)^2 \bnorm{{\bbeta}_0}_{\bSigma}^2 \cdot \frac{1}{h_x^2} \right] \quad \mbox{and} 
    \\
    \tau_{t+1,t}^2 &= \lim_{p \to \infty} \EE \bigg[\frac{1}{n_x} (1 + b_t)(1 + b_{t-1}) \bnorm{{\bbeta}_0}_{\bSigma}^2  \cdot \frac{1}{h_x^2}\\
    &\qquad\quad + \frac{1}{n_w} \left \langle \eta_{\theta_{t-1}} \big\{ {\bSigma}^{-1/2} \Zb_{t} + (1 + b_{t-1}){\bbeta}_0 \big\} ,\eta_{\theta_{t-2}} \big\{ {\bSigma}^{-1/2} \Zb_{t-1} + (1 + b_{t-2}){\bbeta}_0 \big\} \right \rangle_{\bSigma} \bigg].
\end{split}
\end{equation}
In summary, equation~\eqref{eqn:state_form_2} implies that for any deterministic sequence of uniformly normalized pseudo-Lipschitz functions $\{\psi^{p}\}_{p \in \NN}$, 
\begin{equation*}
\begin{split}
    &\psi^{p} \left[\tilde{\bbeta}, \tilde{\bbeta}_c^{t} - (1+b_{t-1}) \tilde{\bbeta}, \tilde{\bbeta}_c^{t+1} - (1+b_t) \tilde{\bbeta} \right] \overset{P}{\approx} \EE\ \psi^{p} \left(\tilde{\bbeta}, \Zb_t, \Zb_{t+1} \right),
\end{split}
\end{equation*}
with $(\Zb_t^\T, \Zb_{t+1}^\T)^\T$ has mean $\bm{0}$, covariance determined by \eqref{eqn:Z_t_cov} and recursion \eqref{tau_t tau_t,t+1}. 
This implies that 
\begin{equation*}
\begin{split}
    &\psi^{p} \left(\tilde{\bbeta}, \tilde{\bbeta}_c^{t}, \tilde{\bbeta}_c^{t+1} \right) \overset{P}{\approx} \EE\ \psi^{p} \left[\tilde{\bbeta}, \Zb_t + (1+b_{t-1}) \tilde{\bbeta}, \Zb_{t+1} + (1+b_t) \tilde{\bbeta} \right], \\
    &\psi^{p} \left(\tilde{\bbeta}, \tilde{\bbeta}^{t}, \tilde{\bbeta}^{t+1} \right) \overset{P}{\approx} \EE\ \psi^{p} \left[\tilde{\bbeta}, \hat{y}_{\theta_t} \big\{ \Zb_t + (1 + b_{t-1})\tilde{\bbeta} \big\}, \hat{y}_{\theta_t} \big\{ \Zb_{t+1} + (1 + b_t)\tilde{\bbeta} \big\} \right].
\end{split}
\end{equation*}
Let $\psi^{p}(\xb, \yb) = \phi^{p}({\bSigma}^{-1/2}\xb, {\bSigma}^{-1/2}\yb)$. Then, for deterministic sequence of uniformly normalized pseudo-Lipschitz functions $\phi^{p}$, we have 
\begin{equation*}
\begin{split}
    \phi^{p} \left(\bbeta_0, \bbeta^{t}, \bbeta^{t+1} \right) &= \psi^{p} \left({\bSigma}^{1/2}\bbeta_0, {\bSigma}^{1/2}\bbeta^{t},  {\bSigma}^{1/2}\bbeta^{t+1} \right)
    \\
    &= \psi^{p} \left(\tilde{\bbeta}, \tilde{\bbeta}^{t}, \tilde{\bbeta}^{t+1} \right)
    \\
    &\overset{P}{\approx} \EE\ \psi^{p} \left[\tilde{\bbeta}, \hat{y}_{\theta_t} \big\{ \Zb_t + (1 + b_{t-1})\tilde{\bbeta} \big\}, \hat{y}_{\theta_t} \big\{ \Zb_{t+1} + (1 + b_t)\tilde{\bbeta} \big\} \right]
    \\
    &= \EE\ \phi^{p} \left[{\bSigma}^{-1/2} \tilde{\bbeta}, {\bSigma}^{-1/2} \hat{y}_{\theta_t} \big\{ \Zb_t + (1 + b_{t-1})\tilde{\bbeta} \big\}, {\bSigma}^{-1/2} \hat{y}_{\theta_t} \big\{ \Zb_{t+1} + (1 + b_t)\tilde{\bbeta} \big\} \right].
\end{split}
\end{equation*}
It follows that 
\begin{equation*}
\begin{split}
&\phi^{p}\left(\bbeta_0, \bbeta^{t}, \bbeta^{t+1}\right)  \\ 
&\overset{P}{\approx}\EE\ \phi^{p} \left[\bbeta_0,  \eta_{\lambda(1 + b_{t-1})} \big\{ {\bSigma}^{-1/2} \Zb_{t} + (1 + b_{t-1}) \bbeta_0 \big\}, \eta_{\lambda(1 + b_t)} \big\{ {\bSigma}^{-1/2} \Zb_{t+1} + (1 + b_t) \bbeta_0 \big\} \right],
\end{split}
\end{equation*}
where $\Zb_t$ and $\Zb_{t+1}$ are defined in Theorem \ref{general - state - stronger}.

\subsection{Proof of Proposition \ref{prop 3}}
\label{proof of prop 3}


In this section, our goal is to prove that the fixed-point of the recursion in~\eqref{general - lasso - AMP} is the solution to  Equation \eqref{eqn:ref-panel-lasso-est}. Note that the first equation in~\eqref{general - lasso - AMP} implies that
\begin{align*}
    \bSigma \left\{\bbeta_* + \sqrt{\frac{N}{n_{w}}} \bSigma^{-1} \Wb^{\T} \rb_* - (1 + b_*) \bSigma^{-1} \Xb^{\T} \yb_x - \bbeta_* \right\} + \lambda(1 + b_*) \partial \bnorm{\bbeta_*}_1 = 0.
\end{align*}
which simplifies to 
\begin{align*}
    \sqrt{\frac{N}{n_{w}}} \Wb^{\T} \rb_* - (1 + b_*) \Xb^{\T} \yb_x + \lambda(1 + b_*) \partial \bnorm{\bbeta_*}_1 = 0.
\end{align*}
Moreover, the second line of Equation \eqref{general - lasso - AMP} implies that
\begin{align*}
    \left(1 - \frac{b_*}{1 + b_*} \right) \rb_* &= \sqrt{\frac{n_{w}}{N}} \Wb \bbeta_*,
\end{align*}
and thus 
\$
\left(1 - \frac{b_*}{1 + b_*} \right) \Wb^{\T} \rb_* &=  \sqrt{\frac{n_{w}}{N}} \Wb^{\T} \Wb \bbeta_*.
\$
Therefore, we have 
\begin{align*}
    \Wb^{\T} \rb_* &= \sqrt{\frac{n_{w}}{N}} (1 + b_*) \Wb^{\T} \Wb \bbeta_*.
\end{align*}
It then follows that 
\begin{align*}
    &(1 + b_*) \Wb^{\T} \Wb \bbeta_* - (1 + b_*) \Xb^{\T} \yb_x + \lambda(1 + b_*) \partial \bnorm{\bbeta_*}_1 = 0, 
\end{align*}
and thus 
\$
\Wb^{\T} \Wb \bbeta_* - \Xb^{\T} \yb_x + \lambda \partial \bnorm{\bbeta_*}_1 = 0,
\$
which is the solution to equation~\eqref{eqn:ref-panel-lasso-est} for any tuning parameter $\lambda$.

\subsection{Proof of Theorem \ref{AMP approx estimator}}
\label{proof of Theorem AMP approx estimator}

The proof of Theorem \ref{AMP approx estimator} relies on a series of lemmas collected below,  which are inspired by \cite{huang2022lasso} and are proved in Section~\ref{S.6 lemmas proof}. 
The first lemma suggests that the AMP estimates possess Cauchy-type properties in the proportional asymptotic regime considered in this paper. Recall that \as indicates almost surely.

\begin{lemma} \label{lemma S.12}
The estimator $\{\bbeta^{t}, \rb^{t}\}_{t \geq 0}$ of AMP recursion in~\eqref{general - lasso - AMP} satisfies that
\begin{equation} 
    \lim_{t \to \infty} \lim_{p \to \infty} \frac{1}{p} \bnorm{\bbeta^{t} - {\bbeta}^{t-1}}_2^2 = 0\quad \mbox{and} \quad \lim_{t \to \infty} \lim_{p \to \infty} \frac{1}{p} \bnorm{\rb^{t} - \rb^{t-1}}_2^2 = 0 \qas
\end{equation}
\end{lemma}

Moreover, the following lemma shows that the reference panel-based $L_1$ regularized estimator and AMP recursion are all bounded. 
\begin{lemma} \label{lemma S.13}
The estimator $\{\bbeta^{t}, \rb^{t}\}_{t \geq 0}$ of AMP recursion in~\eqref{general - lasso - AMP} and $\hat{\bbeta}_{\LW}(\lambda)$ defined in~\eqref{eqn:ref-panel-lasso-est} satisfy
\begin{align*}
    \frac{1}{\sqrt{p}} \bnorm{\rb^{t-1}}_{2} = O_p(1), \quad \frac{1}{\sqrt{p}} \bnorm{{\bbeta}^{t}}_{2} = O_p(1), \quad 
    \frac{1}{\sqrt{p}} \bnorm{\hat{{\bbeta}}_{\LW}(\lambda)}_{2} = O_p(1) \qas
\end{align*}
\end{lemma}
Let $\sigma_{\min}(\Xb)$ and $\sigma_{\max}(\Xb)$ be the minimum and maximum non-zero singular values of $\Xb$, respectively. Then the following lemma implies that with high probability, $\sigma_{\min}(\Xb)$ and $\sigma_{\min}(\Wb)$ are lower bounded, and $\sigma_{\max}(\Xb)$ and $\sigma_{\max}(\Wb)$ are upper bounded. 
\begin{lemma} \label{lemma S.14}
For every $t \geq 0$, there exists $c_5 > 0$ such that 
\begin{equation*}
    \PP(c_5^{-1} \leq \sigma_{\min}(\Xb) \leq \sigma_{\max}(\Xb) \leq c_5, c_5^{-1} \leq \sigma_{\min}(\Wb) \leq \sigma_{\max}(\Wb) \leq c_5) > 1- 2 \exp(-t^2/2).
\end{equation*}
\end{lemma}
By the KKT condition of the first line in~\eqref{general - lasso - AMP}, let   $\vb^{t} \in \partial \bnorm{\bbeta^{t}}_1$ be the subgradient such that 
\begin{equation} \label{vt KKT}
    \bSigma \{\bbeta^{t} + \bSigma^{-1} \Wb^{\T} \rb^{t-1} - (1 + b_{t-1}) \bSigma^{-1} \Xb^{\T} \yb_x - \bbeta^{t-1}\} + \lambda (1 + b_{t-1}) \vb^{t} = 0.
\end{equation}
Then the next lemma suggests that with high probability, the subgradient $\vb^{t}$ can not have too many coordinates with a magnitude close to $1$. 
\begin{lemma} \label{lemma S.15}
For large enough $t$, there exist $c, C, c_2>0$ such that
\begin{equation}
    \PP \Bigg( \frac{|j \in [p]: |v_j^{t}| \geq 1 - c_2|}{n_w} \geq 1 - \frac{1}{2(1 + b_*)} \Bigg) \leq C \exp(-c n_w),
\end{equation}
where $b_*$ is defined in equations~\eqref{eqn: 4.11} - \eqref{eqn: 4.12} and 
\begin{align*}
    \frac{1}{1 + b_*} = 1 - \frac{1}{n_w} \EE \Div \eta_{\lambda(1 + b_*)}\{\tau_* {\bSigma}^{-1/2} Z + (1 + b_*) \bbeta\}.
\end{align*}
\end{lemma}
Define the minimum singular value of matrix $\Wb$ over a set $S \subset [p]$ by
\begin{align} \label{eqn:kappa-}
    \kappa_{-}(\Wb, S) = \inf\left\{\bnorm{\Wb \gb}_2: \supp(\gb) \subset S, \bnorm{\gb}_2 = 1\right\},
\end{align}
and the $s$ sparse singular value by
\begin{align*}
    \kappa_{-}(\Wb, s) = \min_{|S| \leq s} \kappa_{-}(\Wb, S).
\end{align*}
The next lemma implies that $\kappa_{-}(\Wb, s)$ is lower bounded with high probability. 
\begin{lemma} \label{lemma S.16}
For some $c_4 \geq 0$, there exist constants  $C, c > 0$ such that 
\begin{equation} \label{eqn: lemma S.16}
    \PP\Bigg[\kappa_{-} \left\{\Wb, n_w \left(1 - \frac{1}{4(1 + b_*)} \right) \right\} \leq c_4 \Bigg] \leq  C e^{-c n_w}.
\end{equation}
\end{lemma}


Now we proceed to prove Theorem \ref{AMP approx estimator}. 
The subsequent argument takes place in the intersection of the high probability events described in Lemmas \ref{lemma S.12} - \ref{lemma S.16}.
Let 
\begin{align} \label{eqn: cC def}
    \cC(\vb) := \frac{1}{2} \vb^{\T} \W^{\T} \W \vb - \vb^{\T} \X^{\T} \y_{x} + \lambda \|\vb\|_{1}
\end{align}
and $\zb^{t} = \hat{\bbeta}_{\LW}(\lambda) - \bbeta^{t}$ denote  the difference between the reference panel-based $L_1$ regularized estimator and the AMP estimate at $t$-th iteration, 
then it suffices to show $\lim_{t \to \infty} \lim_{p \to \infty} \|\zb^{t}\|_{2}^{2}/p = 0$. 
To this end, we proceed in the following steps.  Consider $\zb^{t} = \zb^{t}_{\bot} + \zb^{t}_{\|}$ with $\zb^{t}_{\|} \in \ker(\Wb)$ and $\zb^{t}_{\bot} \bot \ker(\Wb)$. 
Using the triangle inequality, 
we have $\|\zb^{t}\|_{2} \leq \|\zb^{t}_{\bot}\|_{2} + \|\zb^{t}_{\|}\|_{2}$. 
We will prove that $\|\zb^{t}_{\bot}\|_{2}$ and $\|\zb^{t}_{\|}\|_{2}$ converge to $0$ separately.

\paragraph{Step 1: $\|\zb^{t}_{\bot}\|_{2}/p$ converges to $0$.}
We have 
\begin{equation*}
\begin{split}
    0 \geq\ & \frac{\cC({\bbeta}^{t} + \zb^{t}) - \cC({\bbeta}^{t})}{p}\\
    =\ & \frac{1}{2p} ({\bbeta}^{t} + \zb^{t})^{\T} \Wb^{\T} \Wb ({\bbeta}^{t} + \zb^{t}) - \frac{1}{p}({\bbeta}^{t} + \zb^{t})^{\T} \Xb^{\T} \yb_x + \frac{\lambda}{p} \bnorm{{\bbeta}^{t} + \zb^{t}}_1\\ 
    &- \frac{1}{2p} ({\bbeta}^{t})^{\T}  \Wb^{\T} \Wb {\bbeta}^{t} + \frac{1}{p} {\bbeta}^{t} \Xb^{\T}\yb_x  - \frac{\lambda}{p} \bnorm{{\bbeta}^{t}}_1.
\end{split}
\end{equation*}
This further implies that
\begin{align} \label{eqn: 8.20}
    0 \geq \frac{1}{2p} \bnorm{\Wb \zb^{t}}_2^2 + \frac{1}{p} \langle  \zb^{t} , \sg\cC({\bbeta}^{t}) \rangle + \frac{\lambda}{p} \left(\bnorm{{\bbeta}^{t} + \zb^{t}}_1 - \bnorm{{\bbeta}^{t}}_1 - \langle  \zb^{t} , \vb^{t} \rangle \right),
\end{align}
where $\sg\cC({\bbeta}^{t}) := \Wb^{\T} \Wb {\bbeta}^{t} - \Xb^{\T} \yb_x + \lambda \vb^{t}$ and $\vb^{t}$ is defined in Equation \eqref{vt KKT}. We first consider the second term of Equation \eqref{eqn: 8.20}. Our goal is to show that $\langle  \zb^{t} , \sg\cC({\bbeta}^{t}) \rangle /p \to 0$. 
We first write  $\sg\cC({\bbeta}^{t})$ as 
\begin{equation*}
\begin{split}
    \sg\cC({\bbeta}^{t}) 
    &= \Wb^{\T} \Wb {\bbeta}^{t} - \Xb^{\T} \yb_x + \lambda \vb^{t}\\
    &= \Wb^{\T} \Bigg( \rb^{t} - \frac{b_t}{1 + b_{t-1}} \rb^{t-1} \Bigg) - \Xb^{\T} \yb_x \\
    &\qquad - \frac{\lambda}{\lambda (1 + b_{t-1})} {\bSigma}\{{\bbeta}^{t} + {\bSigma}^{-1} \Wb^{\T} \rb^{t-1} - (1 + b_{t-1}){\bSigma}^{-1} \Xb^{\T} \yb_x - {\bbeta}^{t-1} \}\\
    &= \Wb^{\T}( \rb^{t} - \rb^{t-1} ) - \Bigg\{- \frac{1}{1 + b_{t-1}} + \frac{\lambda}{\lambda (1 + b_{t-1})} \Bigg\}\Wb^{\T} \rb^{t-1} - \frac{\lambda}{\lambda (1 + b_{t-1})} {\bSigma} ({\bbeta}^{t} - {\bbeta}^{t-1}).
\end{split}
\end{equation*}
The last equality is a consequence of Corollary \ref{stationary beta0} due to the "oracle" initialization in Theorem \ref{AMP approx estimator}.  
Therefore, by Lemmas \ref{lemma S.12} and \ref{lemma S.14}, we have in probability
\begin{equation*}
    \frac{1}{\sqrt{p}}\bnorm{\sg\cC({\bbeta}^{t})}_2 \to 0.
\end{equation*}
By Lemma \ref{lemma S.13}, we have $\norm{{\bbeta}^{t}}_2/\sqrt{p} = O_p(1)$ and $\norm{\hat{{\bbeta}}_{\LW}}_2/\sqrt{p} = O_p(1)$ in probability. It follows that $\bnorm{\zb^{t}}_2/\sqrt{p} = O_p(1)$ in probability. Thus, we have $\langle  \zb^{t} , \sg\cC({\bbeta}^{t}) \rangle/p \to 0$ in probability. Therefore, by \eqref{eqn: 8.20}, for any $\epsilon > 0$ and sufficient large $p$, we have
\begin{equation} \label{eqn: S.7.8}
    \frac{1}{2p} \bnorm{\Wb \zb^{t}}_2^2 + \frac{\lambda}{p} \Bigg(\bnorm{{\bbeta}^{t} + \zb^{t}}_1 - \bnorm{{\bbeta}^{t}}_1 - \langle  \zb^{t} , \vb^{t} \rangle \Bigg) \leq c_1 {\epsilon}.
\end{equation}
Note that both terms on the left-hand side of \eqref{eqn: S.7.8} are non-negative. 
The first term is clearly non-negative. 
Recall that $\vb^{t} \in \partial \bnorm{\bbeta^{t}}_1$. 
{\BS For any set $A \in [p]$, let $\vb^{t}_{A} \in \RR^{|A|}$ denote the vector such that $(\vb^{t}_{A})_{i} = (\vb^{t})_{i}$ for all $i \in A$.} 
Let $S$ be the active set of ${\bbeta}^{t}$.
The second item is also non-negative since 
\begin{equation*}
\begin{split}
    &\bnorm{{\bbeta}^{t} + \zb^{t}}_1 - \bnorm{{\bbeta}^{t}}_1 - \langle  \zb^{t} , \vb^{t} \rangle\\
    = & \bnorm{{\bbeta}_S^{t} + \zb_S^{t}}_1 - \bnorm{{\bbeta}_S^{t}}_1 - \langle  \zb_S^{t} , \sign({\bbeta}^{t}_S) \rangle + \bnorm{\zb^{t}_{\bar{S}}}_1 - \langle  \zb_{\bar{S}}^{t} , \vb_{\bar{S}}^{t} \rangle\\
    = & \left\langle {\bbeta}_S^{t} + \zb_S^{t},  \ \sign({\bbeta}^2_S + \zb^{t}_S) - \sign({\bbeta}^{t}_S)  \right\rangle + \bnorm{\zb^{t}_{\bar{S}}}_1 - \langle  \zb_{\bar{S}}^{t} , \vb_{\bar{S}}^{t} \rangle \geq 0,
\end{split}
\end{equation*}
where $\bar{S} = [p]\setminus S$ is the complement of $S$. The last inequality follows from $$\left\langle {\bbeta}_S^{t} + \zb_S^{t},  \ \sign({\bbeta}^2_S + \zb^{t}_S) - \sign({\bbeta}^{t}_S)  \right\rangle \geq 0,$$  and the absolute value of each coordinate of $\vb_{\bar{S}}^{t} \leq 1$. Therefore, we have
\begin{align}
     &\bnorm{\Wb \zb^{t}}_2^2\leq 2 p c_{1} \epsilon, \label{eqn: S.7.9} 
\end{align}
and thus 
\#
 &\bnorm{\zb^{t}_{\bar{S}}}_1 - \langle  \zb_{\bar{S}}^{t} , \vb_{\bar{S}}^{t} \rangle \leq 2 p c_{1} \epsilon. \label{eqn: S.7.10}
\#
Recall $\zb^{t} = \zb^{t}_{\bot} + \zb^{t}_{\|}$ with $\zb^{t}_{\|} \in \ker(\Wb)$ and $\zb^{t}_{\bot} \bot \ker(\Wb)$, it follows from \eqref{eqn: S.7.9} and Lemma \ref{lemma S.14} that 
\begin{equation} \label{eqn: S.7.11}
\bnorm{\zb^{t}_{\bot}}_2^2 
    \leq 2 p c_5 c_1 \epsilon. 
\end{equation}

\paragraph{Step 2: $\|\zb^{t}_{\|}\|_{2}/p$ converges to $0$.}

We need to obtain an analogous bound for $\zb^{t}_{\|}$. 
Let $\zb^{t}_{\bot, \bar{S}}$ be composed of the coordinates from $\zb^{t}_{\bot}$ corresponding to the set $\bar{S}$. 
We have $\bnorm{\zb^{t}_{\bot, \bar{S}}}_1 \leq \sqrt{p}\bnorm{\zb^{t}_{\bot, \bar{S}}}_2 \leq \sqrt{p} \bnorm{\zb^{t}_{\bot}}_2 \leq p \sqrt{2 c_5 c_{1} \epsilon}$. Then, from \eqref{eqn: S.7.10}, we have
\begin{equation*}
    \bnorm{\zb^{t}_{\|, \bar{S}}}_1 - \langle  \zb_{\|, \bar{S}}^{t} , \vb_{\|, \bar{S}}^{t} \rangle \leq p \left(2 c_{1} \epsilon + \sqrt{2 c_5 c_{1} \epsilon} \right) =: p \zeta_1({\epsilon}),
\end{equation*}
converging to $0$ as ${\epsilon} \to 0$.
Define $S(c_2) = \{j \in \NN: |v^{t}_j| \geq 1 - c_2 \}$,  then $\overline{S(c_2)} \subset \bar{S}$, and it follows that
\begin{equation} \label{eqn: S.7.14}
    \bnorm{\zb^{t}_{\|, \bar{S}}}_1 - \left \langle  \zb_{\|, \bar{S}}^{t} , \vb_{\|, \bar{S}}^{t} \right \rangle \geq \bnorm{\zb^{t}_{\|, \overline{S(c_2)}}}_1 - \left \langle  |\zb_{\|, \overline{S(c_2)}}^{t}| , |\vb_{\|, \overline{S(c_2)}}^{t}| \right \rangle \geq c_2 \bnorm{\zb^{t}_{\|, \overline{S(c_2)}}}_1.
\end{equation}
Therefore,  we have 
\begin{equation*}
    \bnorm{\zb^{t}_{\|, \overline{S(c_2)}}}_1 \leq c_2^{-1} p \zeta_1({\epsilon}).
\end{equation*}
Let $c_3 = \frac{1}{4 \gamma_w (1 + b_*)}$, we have $|S(c_2)| \leq n_w - 2p c_3$ from Lemma \ref{lemma S.15}. Thus, if $|\overline{S(c_2)}| \leq p c_3/2$, we have $p \leq n_w - 3p c_3/2$. In this case, $\ker(\Wb) = \{0\}$ and the proof follows. 

Next, we consider the case $\overline{S(c_2)} \geq p c_3/2$ and partition $\overline{S(c_2)} = \cup_{l=1}^K S_l$, where $p c_3/2 \leq |S_l| \leq p c_3$.  
For any $i \in S_l$ and $j \in S_{l+1}$, we have $|\zb^{t}_{\|, i}| \geq |\zb^{t}_{\|, j}|$. 
In addition, let $\overline{S_{+}} = \cup_{l=2}^K S_l \subset \overline{S(c_2)}$. For any $i \in S_l$, we have $|\zb^{t}_{\|, i}| \leq \bnorm{\zb^{t}_{\|, S_{l-1}}}_1/|S_{l-1}|$. It follows that 
\begin{align} \label{eqn: S.7.15}
\begin{split}
    \left\|\zb^{t}_{\|, \overline{S_{+}}}\right\|_2^2 &= \sum_{l=2}^{K} \bnorm{\zb^{t}_{\|,S_l}}_2^2 \leq \sum_{l=2}^{K} |S_l| \Bigg(\frac{\left\| \zb^{t}_{\|, S_{l-1}}\right\|_1}{|S_{l-1}|} \Bigg)^2\\
    &\leq \frac{4}{pc_3} \sum_{l=2}^{K} \left\|\zb^{t}_{\|, S_{l-1}} \right\|_1^2 \leq \frac{4}{pc_3} \Bigg( \sum_{l=2}^{K} \left\|\zb^{t}_{\|, S_{l-1}}\right\|_1 \Bigg)^2\\
    &\leq \frac{4}{p c_3} \left\|\zb^{t}_{\|, \overline{S(c_2)}}\right\|_1^2 \leq \frac{4 \zeta_1({\epsilon})^2}{c_2^2 c_3} \cdot p =: p \zeta_2({\epsilon}),
\end{split}
\end{align}
where $\zeta_2({\epsilon}) \to 0$ as ${\epsilon} \to 0$.
To conclude the proof, it is sufficient to prove an analogous bound for $\bnorm{\zb^{t}_{\|, {S}_{+}}}_2^2$ with $S_{+} = S(c_2) \cup S_1$. Since $|S_1| \leq p c_3$ and $|S(c_2)| \leq n_w - 2p c_3$, we have $|S_+| \leq n_w - p c_3$. 
By Lemma \ref{lemma S.16}, we have $\sigma_{\min}(\Wb_{S+}) \geq c_4$ with high probability. Since $0 = \Wb \zb^{t}_{\|} = \Wb_{S_+} \zb^{t}_{\|, S_+} + \Wb_{\overline{S_{+}}} \zb^{t}_{\|, \overline{S_{+}}}$, we have
\begin{equation} \label{eqn: S.7.16}
    c_4^2 \bnorm{\zb^{t}_{\|, S_+}}_2^2 \leq  \bnorm{\Wb_{S_+} \zb^{t}_{\|, S_+}}_2^2 = \bnorm{\Wb_{\overline{S_{+}}} \zb^{t}_{\|, \overline{S_{+}}}}_2^2 \leq c_5 \bnorm{\zb^{t}_{\|, \overline{S_{+}}}}_2^2 \leq c_5 p \zeta_2({\epsilon}).
\end{equation}
Combining~\eqref{eqn: S.7.11}, \eqref{eqn: S.7.15}, and \eqref{eqn: S.7.16} finishes the proof of Theorem \ref{AMP approx estimator}.


\subsection{Proofs of supporting lemmas}\label{S.6 lemmas proof}

In this section, we  provide proofs for  Lemmas \ref{lemma S.12} - \ref{lemma S.16}. 
The proof of Lemma \ref{lemma S.12} 
 relies on  the multivariate Ornstein–Uhlenbeck process, which will be introduced first in Section \ref{sec: Preliminary on multivariate OU}.

\subsubsection{Multivariate Ornstein–Uhlenbeck process}
\label{sec: Preliminary on multivariate OU}

We begin by proving the following lemma, which involves an analysis of the multivariate Ornstein-Uhlenbeck process.


\begin{lemma} \label{OU-process}
Let $f:\RR^{p} \mapsto \RR$ be an element in  $L^2(\mu)$,  which consists of all  functions that are  $L^2$ integrable with respect to Gaussian measure $\mu$. Let $(\zb_1^\T, \zb_2^\T)^\T$ be a $2p$-dimensional Gaussian distributed random vector with block covariance $\cov(\zb_1)=\cov(\zb_2)=\bI_p$ and cross covariance $\cov(\zb_1, \zb_2) =x \cdot \bI_p$, then $\EE\ f(\zb_{1}) f(\zb_2)$ is an  increasing function  with respect to $x \in \RR_+$. 
\end{lemma}

To prove Lemma~\ref{OU-process}, we first introduce some preliminary results, most of which are adapted from \cite[Section 4.4.6]{gardiner1985handbook},  \cite[Section 2.7.1]{bakry2014analysis}, or \cite[Section 9.4]{GOIA20161}. We consider the standard multivariate Ornstein–Uhlenbeck process given by the solution to  the following equation
\begin{align} \label{standard OU}
    d\xb(t) = - \bI_p \xb(t) dt + \sqrt{2} \bI_p d\wb(t),
\end{align}
where $\wb(t)$ is an $p$-dimensional standard Wiener process. Throughout this subsection, we assume $\xb(0)$ to be a standard $p$-dimensional Gaussian random variable. 
The unconditional mean and covariance of $\xb(t)$ is given by
\begin{align*}
\EE\ \xb(t) 
&= \exp(-t \bI_p) \EE\ \xb(0)\quad \mbox{and} \\
\EE\ \xb(t) \xb^{\T}(s) 
&= \exp(-t \bI_p) \EE\ \xb(0) \xb(0)^{\T} \exp(-s \bI_p)\\
&\qquad + 2 \int_{0}^{\min\{s,t\}} \exp(-(t-t')\bI_p) \exp(-(s-t')\bI_p) dt'.
\end{align*}
When $\xb(0)$ is a standard $p$-dimensional standard Gaussian random variable, for any $t \geq 0$,  we have 
\begin{align} \label{OU mean var}
    \EE\ \xb(t) = 0_p, \quad \cov(\xb(t)) = \bI_p, \quad \mbox{and} \quad \EE\ \xb(t) \xb(0)^{\T} = e^{-t} I_p.
\end{align}
To analyze the covariance of $f(\xb(t))$ and $f(\xb(0))$, for any function in $f \in L^2(\mu)$ and $t \geq 0$, we define the  Ornstein–Uhlenbeck semi-group $(P_t)_{t \geq 0}$ by 
\begin{align*}
    P_t f(x) = \EE \left( f(\xb_t)|\xb_0 = \xb \right).
\end{align*}
We are interested in the spectral structure of the Ornstein–Uhlenbeck semi-group. Consider a sequence $\{H_k\}_{k \in \NN}$ of Hermite polynomials in $\RR$. 
It is well-known that the sequence of Hermite polynomials forms the orthonormal basis in $L^2(\mu)$.
In $\RR^p$ with $p \geq 1$, we define a sequence of functions $\{H_I\}_{I \in \NN^{p}}$ on $L^2(\mu)$. For $I = (I_1, \cdots I_p) \in \NN^{p}$, let
\begin{align*}
    H_{I} = \prod_{i=1}^{p} H_{k_i}(x_i), \quad (k_1, k_2, \cdots k_p) \in \NN^{p},\quad \mbox{and} \quad (x_1, x_2, \cdots x_p) \in \RR^{p}.
\end{align*}
Here $\{H_I\}_{I \in \NN^p}$ defines a $p$-dimensional orthonormal basis of $L^2(\mu)$. Moreover, $\{H_I\}_{I \in \NN^p}$ satisfies
\begin{align*}
    P_t H_I = \exp\left(- \sum_{i=1}^{p} k_i t\right) H_{I}.
\end{align*}
Therefore, for any function $f \in L^2(\mu)$, there exist $a_I$ and $\lambda_I \in \RR$ such that 
\begin{equation} \label{eigenfunction}
f = \sum_{I \in \NN^p} a_I H_I\quad \mbox{and} \quad  P_t f = \sum_{I \in \NN^p} e^{- \lambda_I t} a_I H_I
\end{equation}
hold. 
With these preliminary results, we prove Lemma \ref{OU-process} below.

\begin{proof}[Proof of Lemma \ref{OU-process}]
Consider the multivariate Ornstein–Uhlenbeck process given in~\eqref{standard OU}
\begin{equation*}
    d\xb(t) = - \bI_p \xb(t) dt + \sqrt{2} \bI_p d\wb(t).
\end{equation*}
Let $(\zb_1^\T, \zb_2^\T)^\T$ be a $2p$-dimensional Gaussian distributed random vector with block covariance $\cov(\zb_1)=\cov(\zb_2)=\bI_p$ and cross covariance $\cov(\zb_1, \zb_2) =x \cdot \bI_p$. 
Consider  $\xb(t)$ at $t = 0$ and $t = t_0 := -\log(x)$.
By Equation \eqref{OU mean var}, we have 
\begin{align*}
    (\xb(0), \xb(t_0)) {=} (\zb_1, \zb_2)
\end{align*}
in distribution.
It follows that 
\begin{equation*}
\begin{split}
    \EE\ f(\zb_1) f(\zb_2) =\ & \EE\ f(\xb(0)) f(\xb(t_0))\\ =\ & \EE_{\xb(0) \sim N(0,\bI_p)} \left\{f(\xb(0))\ \EE \left[f(\xb(t_0))|\xb(0) \right] \right\}\\
    =\ & \EE_{\xb(0) \sim N(0,\bI_p)} \left\{ f(\xb(0))\ P_{t_0} f(\xb(0)) \right\}\\
    =\ & \sum_{I \in \NN^p} e^{- \lambda_I t_0} a_I^2,
\end{split}
\end{equation*}
where the last equality follows from \eqref{eigenfunction} and orthonormality. At $t_0 = \log(1/x)$, we have
\begin{equation*}
    \EE\ f(\zb_1) f(\zb_2) =\sum_{I \in \NN^p} x^{\lambda_I} a_I^2.
\end{equation*}
Since $\lambda_I$ is positive,  $\EE\ f(\zb_1) f(\zb_2)$ is increasing with respect to $x$. This completes the proof of Lemma \ref{OU-process}.
\end{proof}

\subsubsection{Proof of Lemma \ref{lemma S.12}}
In this section, we present the proof of Lemma \ref{lemma S.12}. 
This lemma directly extends Lemma C1 in \cite{bayati2011lasso} to the multivariate case.  
Our proof is motivated by the proof of Lemma 2 by \cite{huang2022lasso}. 
While \cite{huang2022lasso} omitted  some details in the proof of  Lemma 2 in their paper, we find these details highly  nontrivial,  and thus provide the comprehensive proof in this section.

\label{Proof of Lemma S.12}
Considering the "oracle" initialization and recalling Corollary \ref{stationary beta0} and Theorem \ref{general - state - stronger}, we have
\begin{equation} \label{eqn: S.7.2.23}
\begin{split}
    \phi^{p} ({\bbeta}_0, {\bbeta}^{t}, {\bbeta}^{t+1}) \overset{P}{\approx}\ \EE\ \phi^{p} \left[{\bbeta}_0, \eta_{\theta_{*}}\{ {\bSigma}^{-1/2} \Zb_t + (1 + b_{*}){\bbeta}_0\}, \eta_{\theta_{*}}\{ {\bSigma}^{-1/2} \Zb_{t+1} + (1 + b_{*}){\bbeta}_0\} \right],
\end{split}
\end{equation}
where $(\Zb_t^\T,\Zb_{t+1}^\T)^\T$ is  a  joint Gaussian random vector satisfying $\EE \Zb_t \Zb_t^{\T} = \tau_t^2 \bI_p = \tau_*^2 \bI_p$, $\EE \Zb_{t+1} \Zb_{t+1}^{\T} = \tau^2_{t+1} \bI_p = \tau_*^2 \bI_p$, and $\EE \Zb_t \Zb_{t+1}^{\T} = \tau^2_{t+1,t} \bI_p$. 
Define the sequence of $\{y_t\}_{t \geq 0}$ to be 
\begin{align*}
    y_t =\ & \lim_{p \to \infty} \frac{1}{n_w} \bnorm{{\bbeta}^{t} - {\bbeta}^{t-1}}_{\bSigma}^2\\ 
    =\ & \lim_{p \to \infty} \frac{1}{n_w} \EE\  \bnorm{\eta_{\theta_{*}}\big\{ {\bSigma}^{-1/2} \Zb_t + (1 + b_{*}){\bbeta}_0 \big\} - \eta_{\theta_{*}} \big\{ {\bSigma}^{-1/2} \Zb_{t-1} + (1 + b_{*}){\bbeta}_0 \big\} }_{\bSigma}^2.
\end{align*}
Combining equations \eqref{eqn: S.7.2.23} and \eqref{tau_t tau_t,t+1}, we have 
\begin{align} \label{eqn: S.7.2.24}
    y_t = 2 \tau_*^2 - 2 \tau_{t+1,t}^2.
\end{align}
We will show that $y_t \to 0$ in probability  which in turn yields $\tau_{t+1, t} \to \tau_*$  in probability  using \eqref{eqn: S.7.2.24}. 
The following representation holds in terms of the two independent random vectors. 
Let $\Gb_1$ and $\Gb_2$ be independent standard Gaussian distributed random variables. Then we have 
\begin{align} \label{eqn: S.7.2.25}
    \left(\sqrt{\tau_*^2 - \frac{y_{t-1}}{4}} \Gb_2 + \sqrt{\frac{y_{t-1}}{4}} \Gb_1, \sqrt{\tau_*^2 - \frac{y_{t-1}}{4}} \Gb_2 - \sqrt{\frac{y_{t-1}}{4}} \Gb_1 \right) \overset{P}{=} (\Zb_{t}, \Zb_{t-1}).
\end{align}
Using the definition of $\hat{y}_{\theta}(\vb)$ from~\eqref{hat{y}},  Equation \eqref{eqn: S.7.2.25} indicates that
\begin{equation}
\begin{split}
    y_t =\ & \lim_{p \to \infty} \frac{1}{n_w} \EE\  \bigg\| \eta_{\theta_{*}} \bigg\{{\bSigma}^{-1/2} \sqrt{\tau_*^2 - \frac{y_{t-1}}{4}} \Gb_2 + {\bSigma}^{-1/2} \sqrt{\frac{y_{t-1}}{4}} \Gb_1 + (1 + b_{*}){\bbeta}_0\bigg\}\\ 
    &- \eta_{\theta_{*}}\bigg\{{\bSigma}^{-1/2} \sqrt{\tau_*^2 - \frac{y_{t-1}}{4}} \Gb_2 - {\bSigma}^{-1/2} \sqrt{\frac{y_{t-1}}{4}} \Gb_1 + (1 + b_{*}){\bbeta}_0\bigg\} \bigg\|_{\bSigma}^2\\
    =\ & \lim_{p \to \infty} \frac{1}{n_w} \EE\  \bigg\|\hat{y}_{\theta_{*}}\bigg\{\sqrt{\tau_*^2 - \frac{y_{t-1}}{4}} \Gb_2 + \sqrt{\frac{y_{t-1}}{4}} \Gb_1 + (1 + b_{*}){\bSigma}^{1/2}{\bbeta}_0\bigg\}\\ 
    &- \hat{y}_{\theta_{*}}\bigg\{ \sqrt{\tau_*^2 - \frac{y_{t-1}}{4}} \Gb_2 - \sqrt{\frac{y_{t-1}}{4}} \Gb_1 + (1 + b_{*}){\bSigma}^{1/2}{\bbeta}_0\bigg\} \bigg\|_2^2 =: R(y_{t-1}).
\end{split}
\end{equation}
We consider $y_t$ as a function of $y_{t-1}$, denoted by $y_t = R(y_{t-1})$.  The following lemma calculates the derivative of $R$ with respect to $y_{t-1}$. 
\begin{lemma}\label{R'(y_{t-1})}
The derivative of $R(y_{t-1})$ with respect to $y_{t-1}$ is given by
\begin{equation*}
\begin{split}
    R'(y_{t-1}) =\ & \lim_{p \to \infty} \frac{1}{n_w} \EE\Bigg[ \tr \left({\bSigma}^{-1/2} \{\nabla \eta_{\theta_{*}}((1 + b_*) {\bbeta}_0 + {\bSigma}^{-1/2}\Zb_t)\}^{\T} {\bSigma} \nabla \eta_{\theta_{*}}((1 + b_*) {\bbeta}_0 + {\bSigma}^{-1/2}\Zb_{t-1}){\bSigma}^{-1/2} \right) \Bigg]\\
    =\ & \lim_{p \to \infty} \frac{1}{n_w} \EE\ \tr \left[ \nabla \hat{y}_{\theta_{*}}\{ \Zb_{t-1} + (1 + b_{*}){\bSigma}^{1/2}{\bbeta}_0\} \nabla \hat{y}_{\theta_{*}}\{\Zb_t + (1 + b_*) {\bSigma}^{1/2} {\bbeta}_0\}^{\T}  \right].
\end{split}
\end{equation*}
\end{lemma}
The proof of Lemma~\ref{R'(y_{t-1})} is provided at the end of this section. 
Assuming Lemma~\ref{R'(y_{t-1})} for now, we proceed to show  that $R'(y_{t-1})$ is a decreasing function. By Lemma \ref{lemma:proximal}, we have 
\begin{equation} \label{eqn: S.7.2.27}
    \nabla \hat{y}_{\theta_{*}}\{ \Zb + (1 + b_{*}){\bSigma}^{1/2}{\bbeta}_0\} = ({\bSigma}^{1/2})_{*,\cA} ({\bSigma}_{\cA,\cA})^{-1} ({\bSigma}^{1/2})_{*,\cA}^{\T},
\end{equation}
where $\cA$ denotes the active set of $\eta_{\theta_*}\{{\bSigma}^{-1/2} \Zb + (1 + b_{*}){\bbeta}_0\}$. 
Let $f_{i,j}(\Zb): \RR^{p} \mapsto \RR$ be the $(i,j)$ entry of $\nabla \hat{y}_{\theta_{*}}\{ \Zb + (1 + b_{*}){\bSigma}^{1/2}{\bbeta}_0\}$, Lemma \ref{OU-process} shows that $\lim_{p \to \infty} \EE\ f_{i,j}(\Zb_{t-1}) f_{i,j}(\Zb_{t})$ is increasing with respect to $\tau^2_{t+1, t} = \tau_*^2 - y_{t}/2$. 
Therefore, by Lemma \ref{R'(y_{t-1})}, $R'(y_{t-1})$ decreases when $y_{t-1}$ increases.

Next, our goal is to show $G'(0) < 1$. 
At $y_{t-1} = 0$, we have $\Zb_t = \Zb_{t-1} =: \Zb$. By Lemma \ref{R'(y_{t-1})}, we have 
\begin{equation*}
    R'(0) = \lim_{p \to \infty} \frac{1}{n_w} \EE\ \tr \left[\nabla \hat{y}_{\theta_{*}}\{ \tau_* \Zb + (1 + b_{*}){\bSigma}^{1/2}{\bbeta}_0\} \nabla \hat{y}_{\theta_{*}}\{\tau_* \Zb + {\bSigma}^{1/2} (1 + b_*) {\bbeta}_0\}^{\T}  \right].
\end{equation*}
The close form of $\nabla \hat{y}$ is given in Equation~\eqref{eqn: S.7.2.27}.  It follows that 
\begin{equation*}
\begin{split}
    R'(0) =\ & \lim_{p \to \infty} \frac{1}{n_w} \EE\ \tr \left\{ ({\bSigma}^{1/2})_{*,\cA} ({\bSigma}_{\cA,\cA})^{-1} ({\bSigma}^{1/2})_{\cA,*} ({\bSigma}^{1/2})_{*,\cA} ({\bSigma}_{\cA,\cA})^{-1} ({\bSigma}^{1/2})_{\cA,*} \right\} \\
    =\ & \lim_{p \to \infty} \frac{1}{n_w} \EE\ \tr \left\{ ({\bSigma}^{1/2})_{*,\cA} ({\bSigma}_{\cA,\cA})^{-1} {\bSigma}_{\cA,\cA} ({\bSigma}_{\cA,\cA})^{-1} ({\bSigma}^{1/2})_{\cA,*} \right\}\\
    =\ & \lim_{p \to \infty} \frac{1}{n_w} \EE\ \tr \left\{ ({\bSigma}^{1/2})_{*,\cA} ({\bSigma}_{\cA,\cA})^{-1} ({\bSigma}^{1/2})_{\cA,*} \right\}\\
    =\ & \lim_{p \to \infty} \frac{1}{n_w} \EE\ \tr \left[ \nabla \hat{y}_{\theta_{*}}\{ \tau_* \Zb + (1 + b_{*}){\bSigma}^{1/2}{\bbeta}_0\} \right].
\end{split}
\end{equation*}
Note that 
\begin{equation*}
    \EE\ \tr \left[ \nabla \hat{y}_{\theta_{*}}\{ \tau_* \Zb + (1 + b_{*}){\bSigma}^{1/2}{\bbeta}_0\} \right] = \EE\ \Div \hat{y}_{\theta_{*}}\{ \tau_* \Zb + (1 + b_{*}){\bSigma}^{1/2}{\bbeta}_0\}.
\end{equation*}
By Stein's lemma, we have
\begin{equation*}
\begin{split}
    & \EE \Div \hat{y}_{\theta_{*}}\{\tau_* \Zb + (1 + b_{*}) {\bSigma}^{1/2} {\bbeta}_0\}\\ 
    =\ & \EE \tau_*^{-1} \langle \hat{y}_{\theta_{*}}\{\tau_* \Zb + (1 + b_{*}) {\bSigma}^{1/2} {\bbeta}_0\}, \Zb \rangle \\ 
    =\ & \EE \tau_*^{-1} \langle {\bSigma}^{1/2}  \eta_{\theta_*} \{\tau_* {\bSigma}^{-1/2} Z + (1 + b_{*}) {\bbeta}_0\}, \Zb \rangle \\
    =\ & \EE \Div \eta_{\theta_*} \{\tau_* {\bSigma}^{-1/2} \Zb + (1 + b_{*}) {\bbeta}_0\}.
\end{split}
\end{equation*}
By equations \eqref{lambda(alpha)} and \eqref{alpha(lambda)}, $\tau_*$ and $ b_*$ satisfy 
\begin{equation*}
     0 < \lambda = \alpha {\tau_*} \Bigg[1 - \lim_{p \to \infty} \frac{1}{n_w} \EE \Div \eta_{\alpha {\tau_*}} \{{\tau_*} {\bSigma}^{-1/2} \Zb + (1 + b_*){\bbeta}\} \Bigg],
\end{equation*}
and thus
\begin{equation*}
    1 > \lim_{p \to \infty} \frac{1}{n_w} \EE \Div \eta_{\alpha {\tau_*}} \{{\tau_*} {\bSigma}^{-1/2} \Zb + (1 + b_*){\bbeta}\} = R'(0).
\end{equation*}
Moreover, since $R(0) = 0$, we have $R(y_t) \leq R'(0) \cdot y_{t-1} \leq  \cdots \leq R'(0)^{t} \cdot R(y_1)$. It follows that $R(y_t) \to 0$ as $t \to \infty$. 
That is, for any $y_0 > 0$, the iteration procedure $y_t = R(y_{t-1})$ converges and $y_t \to 0$. 
This completes the proof of Lemma \ref{lemma S.12}. 

\begin{proof}[Proof of Lemma \ref{R'(y_{t-1})}]
Our proof is based on a variant of Stein's lemma, aka Lemma 17 of \cite{berthier2020state} or Lemma \ref{Stein Lemma} with $q = 1$. Note that 
\begin{equation*}
\begin{split}
    R'(y_{t-1}) = \lim_{p \to \infty} \frac{2}{p\delta} \EE & \bigg[\eta_{\theta_{*}}\bigg\{(1 + b_*) {\bbeta}_0 + {\bSigma}^{-1/2}\Zb_t\bigg\} - \eta_{\theta_{*}}\bigg\{(1 + b_*) {\bbeta}_0 + {\bSigma}^{-1/2}\Zb_{t-1} \bigg\}\bigg]^{\T} {\bSigma}\\ & \bigg[\nabla\eta_{\theta_{*}}\bigg\{(1 + b_*) {\bbeta}_0 +  {\bSigma}^{-1/2}\Zb_t\bigg\} {\bSigma}^{-1/2} \Zb_t' - \nabla\eta_{\theta_{*}}\bigg\{(1 + b_*) {\bbeta}_0 +  {\bSigma}^{-1/2}\Zb_{t-1}\bigg\} {\bSigma}^{-1/2} \Zb_{t-1}'\bigg]\\ 
    = \lim_{p \to \infty} \frac{2}{p\delta} \EE & \bigg[\eta_{\theta_{*}}\bigg\{(1 + b_*) {\bbeta}_0 + {\bSigma}^{-1/2}\Zb_t\bigg\} - \eta_{\theta_{*}}\bigg\{(1 + b_*) {\bbeta}_0 + {\bSigma}^{-1/2}\Zb_{t-1}\bigg\} \bigg]^{\T} {\bSigma}\\ & \bigg[\nabla\eta_{\theta_{*}}\bigg\{(1 + b_*) {\bbeta}_0 +  {\bSigma}^{-1/2}\Zb_t\bigg\} {\bSigma}^{-1/2} \bigg(-\frac{\Gb_1}{8\sqrt{\tau^2_* - y_{t-1}/4}} + \frac{\Gb_2}{8\sqrt{y_{t-1}/4}}\bigg)\\ &- \nabla\eta_{\theta_{*}}\bigg\{(1 + b_*) {\bbeta}_0 +  {\bSigma}^{-1/2}\Zb_{t-1}\bigg\} {\bSigma}^{-1/2} \bigg(-\frac{\Gb_1}{8\sqrt{\tau^2_* - y_{t-1}/4}} - \frac{\Gb_2}{8\sqrt{y_{t-1}/4}}\bigg)\bigg]
    \\
    &= \Rom{1} + \Rom{2} + \Rom{3} + \Rom{4},
\end{split}
\end{equation*}
where $\Rom{1}$, $\Rom{2}$, $\Rom{3}$, $\Rom{4}$ are
\begin{align*}
    \Rom{1} =\ & \EE \eta^{\T}_{\theta_{*}}\bigg\{(1 + b_*) {\bbeta}_0 + {\bSigma}^{-1/2}\Zb_t\bigg\} {\bSigma} \bigg[\nabla\eta_{\theta_{*}}\bigg\{(1 + b_*) {\bbeta}_0 +  {\bSigma}^{-1/2}\Zb_t\bigg\} {\bSigma}^{-1/2} \bigg(-\frac{\Gb_1}{8\sqrt{\tau^2_* - y_{t-1}/4}} + \frac{\Gb_2}{8\sqrt{y_{t-1}/4}}\bigg)\bigg]
    \\
    \Rom{2} =\ & \EE \eta^{\T}_{\theta_{*}}\bigg\{(1 + b_*) {\bbeta}_0 + {\bSigma}^{-1/2}\Zb_{t-1}\bigg\} {\bSigma}\bigg[\nabla\eta_{\theta_{*}}\bigg\{(1 + b_*) {\bbeta}_0 +  {\bSigma}^{-1/2}\Zb_{t-1}\bigg\} {\bSigma}^{-1/2} \bigg(-\frac{\Gb_1}{8\sqrt{\tau^2_* - y_{t-1}/4}} - \frac{\Gb_2}{8\sqrt{y_{t-1}/4}}\bigg)\bigg]
    \\
    \Rom{3} =\ & \EE \eta_{\theta_{*}}^{\T} \bigg\{(1 + b_*) {\bbeta}_0 + {\bSigma}^{-1/2}\Zb_{t}\bigg\}{\bSigma}\bigg[\nabla\eta_{\theta_{*}}\bigg\{(1 + b_*) {\bbeta}_0 +  {\bSigma}^{-1/2}\Zb_{t-1}\bigg\} {\bSigma}^{-1/2} \bigg(\frac{\Gb_1}{8\sqrt{\tau^2_* - y_{t-1}/4}} + \frac{\Gb_2}{8\sqrt{y_{t-1}/4}}\bigg)\bigg]
    \\
    \Rom{4} =\ & \EE \eta_{\theta_{*}}^{\T} \bigg\{(1 + b_*) {\bbeta}_0 + {\bSigma}^{-1/2}\Zb_{t-1}\bigg\}{\bSigma}\bigg[\nabla\eta_{\theta_{*}}\bigg\{(1 + b_*) {\bbeta}_0 +  {\bSigma}^{-1/2}\Zb_{t}\bigg\} {\bSigma}^{-1/2} \bigg(\frac{\Gb_1}{8\sqrt{\tau^2_* - y_{t-1}/4}} - \frac{\Gb_2}{8\sqrt{y_{t-1}/4}}\bigg)\bigg]
\end{align*}
Now we consider each term separately. First, we have  
\begin{equation*}
\begin{split}
\Rom{1} &= \EE \eta^{\T}_{\theta_{*}}\bigg\{(1 + b_*) {\bbeta}_0 + {\bSigma}^{-1/2}\bigg(\sqrt{\tau_{*}^2 - \frac{y_{t-1}}{4}}\Gb_1 + \sqrt{\frac{y_{t-1}}{4}}\Gb_2\bigg)\bigg\} {\bSigma} \nabla\eta_{\theta_{*}}\bigg\{(1 + b_*) {\bbeta}_0 +  {\bSigma}^{-1/2}\Zb_t\bigg\}\\
    &\quad {\bSigma}^{-1/2} \bigg(-\frac{\Gb_1}{8\sqrt{\tau^2_* - y_{t-1}/4}} + \frac{\Gb_2}{8\sqrt{y_{t-1}/4}}\bigg)\\ 
    &= - \EE \eta^{\T}_{\theta_{*}}\bigg\{(1 + b_*) {\bbeta}_0 + {\bSigma}^{-1/2}\bigg(\sqrt{\tau_{*}^2 - \frac{y_{t-1}}{4}}\Gb_1 + \sqrt{\frac{y_{t-1}}{4}}\Gb_2\bigg)\bigg\} {\bSigma} \nabla\eta_{\theta_{*}}\bigg\{(1 + b_*) {\bbeta}_0 +  {\bSigma}^{-1/2}\Zb_t\bigg\} {\bSigma}^{-1/2} \frac{\Gb_1}{8\sqrt{\tau^2_* - y_{t-1}/4}}\\ 
    &\quad + \EE \eta^{\T}_{\theta_{*}}\bigg\{(1 + b_*) {\bbeta}_0 + {\bSigma}^{-1/2} \bigg( \sqrt{\tau_{*}^2 - \frac{y_{t-1}}{4}}\Gb_1 + \sqrt{\frac{y_{t-1}}{4}}\Gb_2\bigg)\bigg\} {\bSigma} \nabla\eta_{\theta_{*}}\bigg\{(1 + b_*) {\bbeta}_0 +  {\bSigma}^{-1/2}\Zb_t\bigg\} {\bSigma}^{-1/2} \frac{\Gb_2}{8\sqrt{y_{t-1}/4}}\\
    &= - \frac{1}{8} \EE \tr\Bigg[ {\bSigma}^{-1/2} \nabla \eta^{\T}_{\theta_{*}}\bigg\{(1 + b_*) {\bbeta}_0 + {\bSigma}^{-1/2}\Zb_t\bigg\} {\bSigma} \nabla\eta_{\theta_{*}}\bigg\{(1 + b_*) {\bbeta}_0 +  {\bSigma}^{-1/2}\Zb_t\bigg\} {\bSigma}^{-1/2} \Bigg]\\
    &\quad + \frac{1}{8} \EE \tr\Bigg[ {\bSigma}^{-1/2} \nabla \eta^{\T}_{\theta_{*}}\bigg\{(1 + b_*) {\bbeta}_0 + {\bSigma}^{-1/2}\Zb_t\bigg\} {\bSigma} \nabla\eta_{\theta_{*}}\bigg\{(1 + b_*) {\bbeta}_0 +  {\bSigma}^{-1/2}\Zb_t\bigg\} {\bSigma}^{-1/2} \Bigg] \\
    & = 0,
\end{split}
\end{equation*}
where the last line follows from Stein's lemma.

Similarly, we have 
\begin{equation*}
    \Rom{2} = 0.
\end{equation*}
Moreover, using Stein's lemma again, we obtain
\begin{equation*}
\begin{split}
    \Rom{3} =\ & \frac{1}{4} \EE \tr\Bigg[ {\bSigma}^{-1/2} \nabla \eta^{\T}_{\theta_{*}}\bigg\{(1 + b_*) {\bbeta}_0 + {\bSigma}^{-1/2}\Zb_t\bigg\} {\bSigma} \nabla\eta_{\theta_{*}}\bigg\{(1 + b_*) {\bbeta}_0 +  {\bSigma}^{-1/2}\Zb_{t-1}\bigg\} {\bSigma}^{-1/2} \Bigg]
    \\
    \Rom{4} =\ & \frac{1}{4} \EE \tr\Bigg[ {\bSigma}^{-1/2} \nabla \eta^{\T}_{\theta_{*}}\bigg\{(1 + b_*) {\bbeta}_0 + {\bSigma}^{-1/2}\Zb_{t-1}\bigg\} {\bSigma} \nabla\eta_{\theta_{*}}\bigg\{(1 + b_*) {\bbeta}_0 +  {\bSigma}^{-1/2}\Zb_{t}\bigg\} {\bSigma}^{-1/2} \Bigg].
\end{split}
\end{equation*}
Together, we conclude that 
\begin{equation*}
\begin{split}
    R'(y_{t-1}) = \lim_{p \to \infty} \frac{1}{p\delta} \EE\Bigg[ \tr\bigg\{{\bSigma}^{-1} \nabla \eta_{\theta_{*}}\bigg((1 + b_*) {\bbeta}_0 + {\bSigma}^{-1/2}\Zb_t\bigg)^{\T} {\bSigma} \nabla \eta_{\theta_{*}}\bigg((1 + b_*) {\bbeta}_0 + {\bSigma}^{-1/2}\Zb_{t-1}\bigg)\bigg\} \Bigg].
\end{split}
\end{equation*}
This completes the proof of the first equality. 
It is worth noting that ${\bSigma}^{1/2} \eta_{\theta}({\bSigma}^{-1/2}v) = \hat{y}_{\theta}(v)$ as shown in \eqref{hat{y}}. 
It follows that  
\begin{equation*}
    {\bSigma}^{1/2} \nabla \eta({\bSigma}^{-1/2} v) {\bSigma}^{-1/2} = \nabla \hat{y}(v).
\end{equation*}
Combining with properties of the trace function, we conclude the proof of Lemma \ref{R'(y_{t-1})}.
\end{proof}

\subsubsection{Proof of Lemma \ref{lemma S.13}}

From the state evolution in Corollary \ref{stationary beta0}, we have $\frac{1}{\sqrt{p}} \bnorm{{\bbeta}^{t}} = O_p(1)$ and $\frac{1}{\sqrt{p}} \bnorm{\rb^{t}} = O_p(1)$.
To show $\left\| \hat{{\bbeta}}_{\LW}(\lambda) \right\|_{2}/\sqrt{p} = O_p(1)$, we introduce the function $\tilde{\cC}({\vb}): \RR^{p} \mapsto \RR$ defined as
\begin{equation} \label{eqn: S.7.3.28}
    \tilde{\cC}(\vb) = \vb^{\T} \Wb^{\T} \Wb \vb - 2 \vb^{\T} \Xb^{\T} \yb_x + \frac{\bnorm{\Xb^{\T} \yb_x}^2_2}{\sigma^2_{\min}(\Wb)} + 2 \lambda \| \vb \|_1.
\end{equation}
The following lemma summarizes the properties of $\tilde{\cC}(\vb)$.
\begin{lemma} \label{C' observation}
The followings hold. 
   \begin{enumerate}
  \item The  $\tilde{\cC}(\vb)$ has the same minimizer $\hat{\bbeta}_{\LW}(\lambda)$ as $\cC(\vb)$ from \eqref{eqn: cC def}. 
  \item The $\tilde{\cC}$ is always non-negative. 
  \item The $p^{-1}{\tilde{\cC}(\hat{{\bbeta}}_{\LW})}$ has an upper bound. 
  \end{enumerate}
\end{lemma}
\begin{proof}[Proof of Lemma \ref{C' observation}]
    First, we have 
    \$
    \tilde{\cC} = 2 \left( \cC + \frac{\bnorm{\Xb^{\T} \yb_{x}}^2_2}{2 \sigma^2_{\min}(\Wb)} \right). 
    \$
    It follows that 
    \begin{align} \label{eqn: S.7.3.29}
        \hat{\bbeta}_{\LW}(\lambda) = \arg\min_{\vb} \cC(\vb) = \arg\min_{\vb} \tilde{\cC}(\vb). 
    \end{align}
    Moreover, we have 
    \begin{equation*}
    \begin{split}
        &\vb^{\T} \W^{\T} \W \vb - 2 \vb^{\T} \X^{\T} \y_x + \frac{\bnorm{\Xb^{\T}\yb_x}^2_2}{\sigma^2_{\min}(\Wb)}\\
        \geq &\  \bnorm{\Wb \vb}^2_2 - 2 \bnorm{\vb}_2 \bnorm{\Xb^{\T} \yb_x}_2+ \frac{\bnorm{\Xb^{\T} \yb_x}^2_2}{\sigma^2_{\min}(\Wb)}\\
        \geq &\ \sigma_{\min}^2(\Wb) \bnorm{\vb}^2_2 - 2 \bnorm{\vb}_2 \bnorm{\Xb^{\T} \yb_x}_2+ \frac{\bnorm{\Xb^{\T} \yb_x}^2_2}{\sigma^2_{\min}(\Wb)}\\
        \geq &\ \left\{ \sigma_{\min}(\Wb)  \bnorm{\vb}_2 -  \frac{\bnorm{\Xb^{\T} \yb_x}_2}{\sigma_{\min}(\Wb)} \right\}^2 \geq 0.
    \end{split}
    \end{equation*}
    From Equation \eqref{eqn: S.7.3.29},  we have 
    \begin{equation}
    \begin{split}
        \frac{1}{p} {\tilde{\cC}(\hat{{\bbeta}}_{\LW}(\lambda))} \leq \frac{1}{p} {\tilde{\cC}(0)} = \frac{\bnorm{\Xb^{\T} \yb_x}^2_2}{p \cdot \sigma^2_{\min}(\Wb)} \leq \frac{2 \bnorm{\Xb^{\T} \Xb {\bbeta}}^2_2 + 2 \bnorm{\Xb {\bepsilon}_{x}}^2_2}{p \cdot \sigma^2_{\min}(\Wb)}.
    \end{split}
    \end{equation}
    According to Proposition \ref{eqn: S.7.14}, we have $\sigma_{\min}(\Wb) \geq c_5^{-1}$ and $\sigma_{\max}(\Xb) \leq c_5$ with high probability. Therefore, as $p \to \infty$, there exists some $B > 0$ depending on ${\bbeta_0}$ and $\sigma^2_{{\bepsilon}_{x}}$ such that $B$ is the upper bound of $\frac{1}{p} {\tilde{\cC}(\hat{{\bbeta}})}$.
\end{proof}

We consider the decomposition of $\hat{{\bbeta}}_{\LW}(\lambda)$ as $\hat{{\bbeta}}_{\LW}(\lambda) = \hat{{\bbeta}}_{\|} + \hat{{\bbeta}}_{\bot}$ where $\hat{{\bbeta}}_{\|} \in \ker(\Wb)$ and $\hat{{\bbeta}}_\bot \in \ker(\Wb)^{\bot}$. 
When $\gamma_w > 1$, since $\hat{{\bbeta}}_{\|}$ belongs to the random subspace $\ker(\Wb)$ with dimension $p - n_w = n_{w}(\gamma_w - 1)$, Theorem \ref{thm: S.13.2} implies that there exists a positive constant $c_1 > 0$ such that 
\begin{equation} \label{eqn: S.7.3.31}
\frac{1}{p} \bnorm{\hat{{\bbeta}}_{\LW}(\lambda)}^2_2 
\leq \frac{1}{p} \bnorm{\hat{{\bbeta}}_{\|}}^2_2 + \frac{1}{p} \bnorm{\hat{{\bbeta}}_{\bot}}^2_2
\leq c_1 \left(\frac{\bnorm{\hat{{\bbeta}}_{\|}}_1}{p} \right)^2 + \frac{1}{p} \bnorm{\hat{{\bbeta}}_{\bot}}^2_2.
\end{equation}
Therefore, using the  triangle inequality and Cauchy-Schwarz inequality, we obtain 
\begin{equation} \label{eqn: S.7.3.32}
\begin{split}
    \frac{1}{p} \bnorm{\hat{{\bbeta}}_{\LW}(\lambda)}^2_2 &\leq 2c_1 \left(\frac{\bnorm{\hat{{\bbeta}}_{\LW}(\lambda)}_1}{p} \right)^2 + 2c_1 \left(\frac{\bnorm{\hat{{\bbeta}}_{\bot}}_1}{p} \right)^2 + \frac{1}{p} \bnorm{\hat{{\bbeta}}_{\bot}}^2_2\\
    &\leq 2c_1 \left(\frac{\bnorm{\hat{{\bbeta}}_{\LW}(\lambda)}_1}{p} \right)^2 + \frac{2c_1 + 1}{p} \bnorm{\hat{{\bbeta}}_{\bot}}^2_2.
\end{split}
\end{equation}
Note that we have $\ker(\Wb) = \{0\}$ when $\gamma_w \leq 1$. Therefore, equations \eqref{eqn: S.7.3.31} and \eqref{eqn: S.7.3.32} also hold. 
By Equation \eqref{eqn: S.7.3.28}, we have $\bnorm{\hat{{\bbeta}}_{\LW}(\lambda)}_1 \leq \frac{1}{2 \lambda} \tilde{\cC}(\hat{{\bbeta}}_{\LW}(\lambda))$. 
Furthermore, Proposition \ref{lemma S.14} implies the existence of a constant $c_2$, such that
\begin{equation*} 
\begin{split}
    \frac{1}{p} \bnorm{\hat{{\bbeta}}_{\LW}(\lambda)}^2_2 &\leq c_3 \left(\frac{{\tilde{\cC}(\hat{{\bbeta}}_{\LW}(\lambda))}}{p} \right)^2 + \frac{c_2}{p} \bnorm{\Wb \hat{{\bbeta}}_{\bot}}^2_2\\
    &\overset{(a)}{\leq} c_3 \left(\frac{{\tilde{\cC}(\hat{{\bbeta}}_{\LW}(\lambda))}}{p} \right)^2 +  \frac{2 c_2}{p} \Bigg(\bnorm{\W \hat{{\bbeta}}_{\LW}(\lambda)}_2^2 - 2 \hat{{\bbeta}}_{\LW}^{\T} \X^{\T} \y_x + \frac{\bnorm{\Xb^{\T} \yb_x}^2_2}{\sigma^2_{\min}(\Wb)} \Bigg) + \frac{2 c_2}{p} \cdot \frac{\bnorm{\Xb^{\T} \yb_x}^2_2}{\sigma^2_{\min}(\Wb)}\\
    &\leq  c_3 \left(\frac{{\tilde{\cC}(\hat{{\bbeta}}_{\LW}(\lambda))}}{p} \right)^2 +  \frac{2 c_2}{p} \tilde{\cC}(\hat{{\bbeta}}_{\LW}(\lambda)) + \frac{2 c_2}{p} \cdot \frac{\bnorm{\Xb^{\T} \yb_x}^2_2}{\sigma^2_{\min}(\Wb)}\\
    &\leq c_3 \left(\frac{{\tilde{\cC}(\hat{{\bbeta}}_{\LW}(\lambda))}}{p} \right)^2 +  \frac{2 c_2}{p} \tilde{\cC}(\hat{{\bbeta}}_{\LW}(\lambda)) + {4 c_2} \cdot \frac{ \bnorm{\Xb^{\T} \Xb {\bbeta}_0}^2_2 +  \bnorm{\Xb {\bepsilon}_{x}}^2_2}{p \cdot \sigma^2_{\min}(\Wb)} \\
    &< \infty,
\end{split}
\end{equation*}
where the inequality $(a)$ holds because
\begin{equation*}
\begin{split}
    &\frac{2 c_2}{p} \Bigg(\bnorm{\W \hat{{\bbeta}}_{\LW}(\lambda)}_2^2 - 2 \hat{{\bbeta}}_{\LW}(\lambda)^{\T} \X^{\T} \y_x + \frac{\bnorm{\Xb^{\T} \yb_x}^2_2}{\sigma^2_{\min}(\Wb)} \Bigg) + \frac{2 c_2}{p} \cdot \frac{\bnorm{\Xb^{\T} \yb_x}^2_2}{\sigma^2_{\min}(\Wb)}\\
    \geq &\  \frac{c_2}{p} \bnorm{\W \hat{{\bbeta}}_{\LW}(\lambda)}_2^2 + \frac{c_2}{p} \bnorm{\W \hat{{\bbeta}}_{\LW}(\lambda)}_2^2 - \frac{4 c_2}{p} \hat{{\bbeta}}_{\LW}(\lambda)^{\T} \X^{\T} \y_x + \frac{4 c_2}{p} \frac{\bnorm{\Xb^{\T} \yb_x}^2_2}{\sigma^2_{\min}(\Wb)}\\
    \geq &\  \frac{c_2}{p} \bnorm{\W \hat{{\bbeta}}_{\LW}(\lambda)}_2^2 + \frac{c_2}{p} \cdot \Bigg( \bnorm{\W \hat{{\bbeta}}_{\LW}(\lambda)}_2^2 - 4 \hat{{\bbeta}}_{\LW}(\lambda)^{\T} \X^{\T} \y_x + 4 \frac{\bnorm{\Xb^{\T} \yb_x}^2_2}{\sigma^2_{\min}(\Wb)} \Bigg)\\
    \geq &\  \frac{c_2}{p} \bnorm{\W \hat{{\bbeta}}_{\LW}(\lambda)}_2^2 + \frac{c_2}{p} \cdot \Bigg( \sigma_{\min}^2(\Wb) \bnorm{\hat{{\bbeta}}_{\LW}(\lambda)}_2^2 - 4 \bnorm{\hat{{\bbeta}}_{\LW}(\lambda)}_2 \bnorm{\X^{\T} \y_x}_2 + 4 \frac{\bnorm{\Xb^{\T} \yb_x}^2_2}{\sigma^2_{\min}(\Wb)} \Bigg)\\
    \geq &\  \frac{c_2}{p} \bnorm{\W \hat{{\bbeta}}_{\LW}(\lambda)}_2^2 + \frac{c_2}{p} \cdot \Bigg( \sigma_{\min}(\Wb) \bnorm{\hat{{\bbeta}}_{\LW}(\lambda)}_2 - 2 \frac{\bnorm{\Xb^{\T} \yb_x}_2}{\sigma_{\min}(\Wb)} \Bigg)^2 \geq \frac{c_2}{p} \bnorm{\W \hat{{\bbeta}}_{\LW}(\lambda)}_2^2.
\end{split}
\end{equation*}
This completes the proof of Lemma \ref{lemma S.13}.

\subsubsection{Proof of Lemma \ref{lemma S.14}}
Consider the matrices $\Xb  = {\Xb_0} \bSigma^{1/2}$ and $\Wb = {\Wb_0} \bSigma^{1/2}$, where entries of ${\Xb_0}$ and ${\Wb_0}$ are i.i.d. random variables from  $N(0,1/n_x)$ and $N(0, 1/n_w)$, respectively. 
By Corollary 5.35 in \cite{vershynin_2012}, we have 
\begin{align*}
    \PP\left(\frac{1}{\sqrt{\gamma_x}} - 1 - t \leq \sigma_{\min}({\Xb_0}) \leq \frac{1}{\sqrt{\gamma_x}} + 1 + t \right) \geq 1 - 2 \exp(-t^2/2)
\end{align*}
and 
\begin{align*}
    \PP\left(\frac{1}{\sqrt{\gamma_w}} - 1 - t \leq \sigma_{\min}({\Wb_0}) \leq \frac{1}{\sqrt{\gamma_w}} + 1 + t \right) \geq 1 - 2 \exp(-t^2/2).
\end{align*}
Note  that we have 
\begin{align*}
    &\sigma_{\min}(\Xb) \geq \sigma_{\min}({\Xb_0})\sigma_{\min}(\bmSigma^{1/2}), \quad \sigma_{\max}(\Xb) \leq \sigma_{\max} ({\Xb_0})\sigma_{\max}(\bmSigma^{1/2})\\
    &\sigma_{\min}(\Wb) \geq \sigma_{\min}({\Wb_0})\sigma_{\min}(\bmSigma^{1/2}), \quad \mbox{and}\quad \sigma_{\max}(\Wb) \leq \sigma_{\max} ({\Wb_0})\sigma_{\max}(\bmSigma^{1/2}).
\end{align*}
It follows that, for any $t \geq 0$, there exists $c_5 > 0$ such that 
\begin{equation*}
    \PP(c_5^{-1} \leq \sigma_{\min}(\Xb) \leq \sigma_{\max}(\Xb) \leq c_5, c_5^{-1} \leq \sigma_{\min}(\Wb) \leq \sigma_{\max}(\Wb) \leq c_5) > 1- 2 \exp(-t^2/2).
\end{equation*}

\subsubsection{Proof of Lemma \ref{lemma S.15}}    
Let $S(c_2) = \{j \in \NN: |\vb^{t}_j| \geq 1 - c_2 \}$. By Equation \eqref{vt KKT}, we have 
\$
&\frac{1}{p} |S(c_2)| \\
         &= \frac{1}{p} \sum_{i=1}^p \one \Bigg[\frac{1}{\theta_{t-1}} \Bigg|{\bmSigma} \left\{{\bbeta}^{t} + {\bmSigma}^{-1}\Wb^{\T}\rb^{t-1} - (1 + b_{t-1}){\bmSigma}^{-1}\Xb^{\T}\yb_x - {\bbeta}^{t-1} \right\}\Bigg|_i \geq 1-c_2 \Bigg]\\
        &\overset{P}{\approx} \frac{1}{p} \sum_{i=1}^p \one \Bigg[\frac{1}{\theta_{t-1}} \Bigg|{\bmSigma} \left[ \eta_{\theta_{t-1}}\{\tau_t {\bmSigma}^{-1/2}\Zb + (1 + b_{t-1}){\bbeta_0}\} -\{\tau_t {\bmSigma}^{-1/2}\Zb + (1 + b_{t-1}){\bbeta_0}\} \right]\Bigg|_i \geq 1-c_2 \Bigg]\\
        &= \frac{1}{p} \sum_{i=1}^p \one \Bigg[\frac{1}{\theta_{t-1}} \Bigg|{\bmSigma} \Bigg[ \tau_t {\bmSigma}^{-1/2}\Zb + (1 + b_{t-1}){\bbeta_0} - \eta_{\theta_{t-1}} \left\{\tau_t {\bmSigma}^{-1/2}\Zb + (1 + b_{t-1}){\bbeta_0} \right\} \Bigg]\Bigg|_i \geq 1-c_2 \Bigg].
\$
    With an oracle initialization, we have
    \begin{equation}\label{eqn: S.7.5.33}
    \begin{split}
        &\frac{1}{p} |S(c_2)|\\ 
        &\overset{P}{\approx} \frac{1}{p} \sum_{i=1}^p \one \Bigg[\frac{1}{\theta_{*}} \Bigg|{\bmSigma} \Bigg[ \tau_* {\bmSigma}^{-1/2}\Zb + (1 + b_{*}){\bbeta_0} - \eta_{\theta_{*}} \left\{\tau_* {\bmSigma}^{-1/2}\Zb + (1 + b_{*}){\bbeta_0} \right\} \Bigg]\Bigg|_i \geq 1-c_2 \Bigg].
    \end{split}
    \end{equation}
    Let $\lambda_{\min}$ and $\lambda_{\max}$ be the smallest and largest singular values of $\bmSigma$, respectively. Let $\bar{{\bmSigma}} = {\bmSigma}/\lambda_{\min}$, $\bar{\tau}_* = \tau_*/\lambda^{1/2}_{\min}$, and $\bar{\theta}_* = \theta_*/\lambda_{\min}$.  Then we have 
    \begin{equation} \label{eqn: S.7.5.34}
    \begin{split}
        \hat{{\bbeta}} &:= \eta_{\theta_*} \left\{\tau_* {\bmSigma}^{-1/2}\Zb + (1 + b_*){\bbeta_0} \right\}\\
        &= \arg\min_{\vb \in \RR^p} \Bigg[\frac{1}{2}\bnorm{{\bmSigma}^{1/2} \{\vb - (1 + b_*) {\bbeta_0}\} -\tau_{t}\Zb }_2^2 + \theta_* \bnorm{\vb}_1
        \Bigg]\\
        &= \arg\min_{\vb \in \RR^p} \Bigg[\frac{1}{2}\bnorm{\bar{{\bmSigma}}^{1/2} \{\vb - (1 + b_*) {\bbeta_0}\} - \bar{\tau}_{*}\Zb }_2^2 + \bar{\theta}_* \bnorm{\vb}_1 \Bigg].
    \end{split}
    \end{equation}
    The KKT condition of this optimization problem gives 
    \begin{equation*}
        \bar{{\bmSigma}}^{1/2} \bigg[\bar{{\bmSigma}}^{1/2} \{\hat{{\bbeta}} - (1 + b_*) {\bbeta_0}\} - \bar{\tau}_{*}\Zb  \bigg] + \bar{\theta}_* \partial \bnorm{\hat{{\bbeta}}}_1 = 0.
    \end{equation*}
    Let $\Tilde{y} = \hat{{\bbeta}} - \bar{{\bmSigma}}^{1/2} \bigg[\bar{{\bmSigma}}^{1/2} \{\hat{{\bbeta}} - (1 + b_*) {\bbeta_0}\} - \bar{\tau}_{t}\Zb  \bigg]$.  We have
    \begin{equation*}
        \hat{{\bbeta}} = \eta_{\soft}(\Tilde{y}; \bar{\theta}_*),
    \end{equation*}
    where $\eta_{\soft}(x,a)$ is defined in  \eqref{eta_soft} and is applied coordinate-wise. 
    From  \eqref{eqn: S.7.5.33} above, we have
    \begin{equation} \label{eqn: S.7.5.35}
        \vb^{t} = \frac{1}{\bar{\theta}_*} \left\{\Tilde{y} - \eta_{\soft}(\Tilde{y}; \bar{\theta}_*) \right\}.
    \end{equation}
    Let $f(\bar{\tau}_{t}\Zb) = (I_p - \bar{{\bmSigma}}^{-1}) \bar{{\bmSigma}}^{1/2} \{(1 + b_*){\bbeta_0} - \hat{{\bbeta}}\}$. Then we have
    \begin{equation*}
    \begin{split}
        \Tilde{y} &= {\bbeta_0} + \bar{{\bmSigma}}^{1/2} \bigg[\bar{\tau}_{t}\Zb + (I_p - \bar{{\bmSigma}}^{-1}) \bar{{\bmSigma}}^{1/2} \{(1 + b_*){\bbeta_0} - \hat{{\bbeta}}\} \bigg]\\
        &= {\bbeta_0} + \bar{{\bmSigma}}^{1/2} \left\{\bar{\tau}_{*}\Zb + f(\bar{\tau}_{*}\Zb) \right\}.
    \end{split}
    \end{equation*}
    Let $\sigma_j$ be the $j^{th}$ row of $\bar{{\bmSigma}}^{1/2}$ and $\sigma_j^{\T} \Zb = x$, then we have $x \sim N(0, \bnorm{\sigma_j}_2^2)$. Let $P_j^{\bot}$ be the projection operator onto the orthogonal complement of the span of $\sigma_j$, then we have 
    \begin{equation} \label{eqn: S.7.5.36}
    \begin{split}
        \Tilde{y}_j &= {\bbeta_{0,j}} + \bar{\tau}_* \sigma_j^{\T} \Zb + \sigma_j^{\T} f\left(\bar{\tau}_{t} \langle \sigma_j,\Zb \rangle \sigma_j/\norm{\sigma_j}_2^2 + \bar{\tau}_* P^{\bot}_j \Zb \right)\\
        &= {\bbeta_{0,j}} + \bar{\tau}_* x + \sigma_j^{\T} f\left(\bar{\tau}_{t} x \sigma_j/\norm{\sigma_j}_2^2 + \bar{\tau}_* P^{\bot}_j \Zb \right) =: h_j(x).
    \end{split}
    \end{equation}
    By Equation \eqref{eqn: S.7.5.34} and the proximal operator property described in Equation \eqref{eqn:1-lip}, $\bar{{\bmSigma}}^{1/2} \{(1 + b_*){\bbeta_0} - \hat{{\bbeta}}\}$ is 1-Lipschitz in $\bar{\tau}_{t}\Zb$. This implies that $f(\bar{\tau}_* \Zb)$ is $(1 - \kappa_{\mathrm{cond}}^{-1})$-Lipschitz in $\bar{\tau}_* \Zb$ and $\bar{\tau}_*(1 - \kappa_{\mathrm{cond}}^{-1})/\norm{\sigma_j}_2$-Lipschitz in $x$, where $\kappa_{\mathrm{cond}} = \lambda_{\max}/\lambda_{\min}$.
    For any $x_1, x_2 \in \RR$, we have 
\begin{equation} \label{eqn: S.7.5.37}
    \begin{split}
        &|h_j(x_1) - h_j(x_2)|\\ 
        \geq& \bar{\tau}_* |x_1 - x_2| - \Bigg|\sigma_j^{\T} \Bigg\{ f\Bigg(\bar{\tau}_{t} x_1 \frac{\sigma_j}{\norm{\sigma_j}_2^2} + \bar{\tau}_* P^{\bot}_j \Zb \Bigg) - f\Bigg(\bar{\tau}_{t} x_2 \frac{\sigma_j}{\norm{\sigma_j}_2^2} + \bar{\tau}_* P^{\bot}_j \Zb \Bigg) \Bigg\} \Bigg|\\
        \geq&  \bar{\tau}_* |x_1 - x_2| - \bar{\tau}_*(1 - \kappa_{\mathrm{cond}}^{-1}) |x_1 - x_2|\\
        =& \bar{\tau}_* \kappa_{\mathrm{cond}}^{-1} |x_1 - x_2|.
    \end{split}
\end{equation}
    By Equation \eqref{eqn: S.7.5.35} and the definition of $S(c_2)$, we have 
    \begin{equation*}
        S(c_2) = \{j \in [p]: |\Tilde{y}_j| \geq \bar{\theta}_*(1 - c_2) \}.
    \end{equation*}
    It then follows that 
    \begin{equation} \label{eqn: S.7.5.38}
        \frac{|S(c_2)|}{n_w} = \frac{|j \in [p]: |\Tilde{y}_j| \geq \bar{\theta}_*|}{n_w} + \frac{|j \in [p]: 1 - |\Tilde{y}_j|/\bar{\theta}_* \in (0, c_2] |}{n_w}.
    \end{equation}
    Consider the function 
    \begin{equation*}
        g(\Tilde{y}, c_2) := \frac{1}{n_w} \sum_{i=1}^p g'(\Tilde{y}_j, c_2),
    \end{equation*}
    where $g'(\Tilde{y}, c_2) = \min \Bigg\{1, \Bigg(\frac{|\Tilde{y}|}{\bar{\theta_*}c_2} - \frac{1}{c_2} + 2 \Bigg)_+ \Bigg\}$. Since
\begin{equation*}
\begin{split}
     |g(\Tilde{y}_1, c_2) - g(\Tilde{y}_2, c_2)| &\leq \frac{1}{n_w} \sum_{i=1}^p |g'(\Tilde{y}_{1, j}, c_2) - g'(\Tilde{y}_{2, j}, c_2)|\\
     &\leq \frac{1}{n_w} \sum_{i=1}^p \frac{1}{\bar{\theta}_* c_2} |\Tilde{y}_{1, j} - \Tilde{y}_{2, j}|\\
     &\leq \frac{\sqrt{p}}{n_w \bar{\theta}_* c_2} \bnorm{\Tilde{y}_1 - \Tilde{y}_2}_2,
\end{split}
\end{equation*}
$g(\Tilde{y}, c_2)$ is $\frac{\sqrt{p}}{n_w \bar{\theta}_* c_2}$-Lipschitz in $\Tilde{y}$. Note that for all $\Tilde{y}$, we have 
\begin{equation}
    \frac{|S(c_2)|}{n_w} \leq g(\Tilde{y}, c_2) \leq \frac{|S(2 c_2)|}{n_w}.
\end{equation}
Moreover, by equations \eqref{eqn: S.7.5.35} and \eqref{eqn: S.7.5.38}, we have 
\begin{equation*}
\begin{split}
    \EE g(\Tilde{y}, c_2) &\leq \EE \Bigg(\frac{\norm{\hat{{\bbeta}}}_0}{n_w} \Bigg) + \EE \Bigg(\frac{|j \in [p]: 1 - |\Tilde{y}_j|/\bar{\theta}_* \in (0, 2 c_2] |}{n_w} \Bigg)\\
    &\leq \frac{b_*}{1 + b_*} + 2 \sup_{a} \EE_x \Bigg\{ \frac{1}{n_w} \sum_{j=1}^p \one \left(a \leq \frac{h_j(x)}{\theta_*} \leq  a + 2 c_2 \right) \Bigg\}\\
    &\leq 1 - \frac{1}{1 + b_*} + 2 \frac{1}{n_w} \sum_{j=1}^p \sup_{a} \EE_x \Bigg\{ \one \left(a \leq \frac{h_j(x)}{\theta_*} \leq  a + 2 c_2 \right) \Bigg\}\\ 
    &\leq 1 - \frac{1}{1 + b_*} + 2 \frac{1}{n_w} \sum_{j=1}^p \sup_{a} \EE_x \Bigg\{ \one \left(\frac{a \kappa_{cond}}{\bar{\tau}_{t}} \leq \frac{x}{\bar{\theta}_*} \leq  \frac{(a + 2 c_2)\kappa_{cond}}{\bar{\tau}_{t}} \right) \Bigg\}\\
    &\leq 1 - \frac{1}{1 + b_*} + \frac{4 c_2 \kappa_{cond} \bar{\theta}_* \gamma_w}{\sqrt{2\pi} \bar{\tau}_{t}}.
\end{split}
\end{equation*}
The fourth inequality above follows from the following argument: 
Since $|h_j(x_1) - h_j(x_2)| \geq \bar{\tau}_* \kappa_{\mathrm{cond}}^{-1} |x_1 - x_2|$, all $x$ that satisfy $a \leq h_j(x)/\theta_* \leq  a + 2 c_2$ must be contained in an interval of length less than $2c \cdot \kappa_{\mathrm{cond}}/\bar{\tau}$. 
Otherwise, if there exists two $x_1$ and $x_2$ satisfying the condition, then $|h(x_1) - h(x_2)|/\bar{\theta}_* > 2c$, which leads to a contradiction.
By Equation~\eqref{eqn: S.7.5.36}, $\Tilde{y}$ is $2 \kappa^{1/2}_{\mathrm{cond}} \bar{\tau}_*$-Lipschitz in $\Zb$. Therefore, $g(\Tilde{y}, c_2)$ is $\frac{2\sqrt{p} \kappa_{\mathrm{cond}}^{1/2} \bar{\tau}_*}{n \bar{\theta}_* c_2}$-Lipschitz in $\Zb$. 
By the Gaussian Lipschitz inequality, Theorem 5.2.2 in \cite{vershynin2018high}, we have
\begin{equation*}
\begin{split}
    &\PP \Bigg(\frac{|S(c_2)|}{n_w} \geq 1 - \frac{1}{1 + b_*} + \frac{4 c_2 \kappa_{\mathrm{cond}} \bar{\theta}_* \gamma_w}{\sqrt{2\pi} \bar{\tau}_{t}} \Bigg)\\
    \leq &\ \PP \Bigg(g(\Tilde{y}, c_2) \geq 1 - \frac{1}{1 + b_*} + \frac{4 c_2 \kappa_{\mathrm{cond}} \bar{\theta}_* \gamma_w}{\sqrt{2\pi} \bar{\tau}_{t}} \Bigg)\\
    \leq &\ \PP \Bigg(g(\Tilde{y}, c_2) \geq \EE g(\Tilde{y}, c_2) + {\bepsilon} \Bigg)\\
    \leq &\ \exp \left(-\frac{n_w \theta^2_* c^2_2}{8 \gamma_w \kappa_{\mathrm{cond}} \bar{\tau}^2_*} {\bepsilon}^2 \right).
\end{split}
\end{equation*}
By choosing a suitable $c_2$, we conclude that there exist $C$ and $c_1 > 0$ such that
\begin{equation*}
    \PP \Bigg(\frac{|S(c_2)|}{n_w} \geq 1 - \frac{1}{2(1 + b_*)} \Bigg) \leq C \exp(-n_w c_1).
\end{equation*}

\subsubsection{Proof of Lemma \ref{lemma S.16}}

Define 
$$
k := \left[n_w \left\{ 1 - \frac{1}{4(1 + b_*)} \right\} \right]
$$
and we may assume $k < p$. Otherwise, we have $\kappa_{-}(\Wb, k) = \kappa_{-}(\Wb, p)$ and can thus  conclude~\eqref{eqn: lemma S.16} by Lemma \ref{lemma S.14}. 
By the definition of $\kappa_{-}$ in~\eqref{eqn:kappa-}, we have $\kappa_{-}(\Xb, S') \geq \kappa_{-}(\Xb, S)$ for any $S' \subset S$. 
Therefore, we have $\kappa_{-}(\Wb, k) = \min_{|S| = k} \kappa_{-} (\Wb, S)$. 
A union bound allows us to conclude that 
\begin{align*}
    \PP\Bigg[\kappa_{-} \left\{\Wb, k \right\} \leq t \Bigg] \leq \sum_{|S| = k} \PP\left[\kappa_{-}(\Wb, S) \leq t\right]. 
\end{align*}
Denote $\Wb_{S}$ to be the $n_{w} \times p$ matrix such that the $i$-th column is the same as the $i$-th column of $\Wb$ for $i \in S$, and $j$-th column equal to $\bm{0}$ for $j \in \bar{S} = [p] \setminus S$.
With these definitions, $\kappa_{-}(\Wb, S) = \sigma_{\min}(\Wb_{S})$.
Moreover, $\Wb_{S} = \Wb_{0, S} \bmSigma_{S,S}^{1/2}$ where $\Wb_{0, S}$ has i.i.d. entries following $N(0, 1/n_{w})$. 
Thus, we have
\begin{align*}
    \sigma_{\min}(\Wb_{S}) \geq \sigma_{\min}(\Wb_{0, S}) \cdot \sigma_{\min}(\bmSigma_{S,S}^{1/2}). 
\end{align*}
Due to i.i.d. properties, $\PP\left[\kappa_{-}(\Wb, S) \leq t\right]$ is independent of the choice of $S$. Therefore, 
\begin{align*}
    \PP\Bigg[\kappa_{-} \left\{\Wb, k \right\} \leq t \Bigg] \leq \binom{p}{k} \cdot \PP\Bigg[\sigma_{\min}(\Wb_{0, S}) \leq t \Bigg]. 
\end{align*}
Let $f_{\min}(k,n_{w}, \lambda)$ denote the probability density function for the smallest singular value $\sigma_{\min}(\Wb_{0, S})$. 
By  \cite[Proposition 5.2]{edelman1988eigenvalues}, $f_{\min}(k,n_{w}, \lambda)$ satisfies the inequality that $f_{\min}(k,n_{w}, \lambda) \leq g_{\min}(k,n_{w}, \lambda)$, where $g_{\min}(k,n_{w}, \lambda)$ is defined as
\begin{align*}
    g_{\min}(k,n_{w}, \lambda)  & := \frac{\Gamma\big\{ (n_{w} + 1)/2 \big\}}{\Gamma\big\{ k/2 \big\} \cdot \Gamma\big\{ (n_{w} - k + 1)/2 \big\} \cdot \Gamma\big\{ (n_{w} - k + 2)/2 \big\}}\\
    &\qquad \cdot \left(\frac{\pi}{2 n_{w} \lambda} \right)^{1/2} \left(\frac{n_{w} \lambda}{2} \right)^{(n_{w} - k)/2} \exp(- n_w \lambda /2).
\end{align*}
Fist note that $g_{\min}(k,n_{w}, \lambda)$ is strictly increasing in $\lambda \in [0, (n_{w} - k - 1)/n_{w}]$.
Lemma 2.9 in \cite{blanchard2011compressed} indicates that as $k/n_{w} \to \rho \in (0, 1]$ and $n \to \infty$, 
\begin{align*}
    g_{\min}(k,n_{w},\lambda) \to Q(n_{w},\lambda) \exp \big\{ n_{w} \phi_{\min}(\lambda,\rho) \big\}, 
\end{align*}
where $Q(\lambda, \rho)$ is a polynomial in $n_{w}$ and $\lambda$.
Moreover,we define 
\begin{align*}
    &\phi_{\min}(\lambda,\rho) = H(\rho) + \frac{1}{2} \left[(1 - \rho) \log(\lambda) + 1 - \rho + \rho \log(\rho) - \lambda \right],\\ 
    & H(\rho) = \rho \log(1/\rho) + (1 - \rho) \log(1/(1 - \rho)). 
\end{align*}
Therefore, for any $t/\sigma_{\min}(\bmSigma_{S,S}^{1/2}) \leq 1 - \rho$, we have
\begin{align*}
    \PP\Bigg[\sigma_{\min}(\Wb_{0, S}) \leq t/\sigma_{\min}(\bmSigma_{S,S}^{1/2}) \Bigg] =\ & \int_{0}^{t/\sigma_{\min}(\bmSigma_{S,S}^{1/2})} f_{\min}(k,n_{w}, \lambda) d \lambda\\
    \leq\ & \int_{0}^{t/\sigma_{\min}(\bmSigma_{S,S}^{1/2})} g_{\min}(k,n_{w}, \lambda) d \lambda\\
    \leq\ & t/\sigma_{\min}(\bmSigma_{S,S}^{1/2}) \cdot g_{\min}\big\{k,n_{w}, t/\sigma_{\min}(\bmSigma_{S,S}^{1/2}) \big\}\\
    =\ & P \big\{n_{w}, t/\sigma_{\min}(\bmSigma_{S,S}^{1/2}) \big\} \exp \left[ n_{w} \phi_{\min} \big\{t/\sigma_{\min}(\bmSigma_{S,S}^{1/2}) ,\rho\big\} \right],
\end{align*}
where $P \big\{n_{w}, t/\sigma_{\min}(\bmSigma_{S,S}^{1/2}) \big\}$ is a polynomial in both $\big\{n_{w}, t/\sigma_{\min}(\bmSigma_{S,S}^{1/2}) \big\}$. 
Moreover,  Binet's second expression for $\log\Gamma(z)$ function  in \cite[Section 12.3]{whittaker_watson_1996} allows us to obtain
\begin{align*}
    \frac{1}{n_{w}} \binom{p}{k} \to H(\rho/\gamma_{w}) \cdot \gamma_{w}.
\end{align*}
We conclude that 
\begin{align*}
    \PP \Bigg[\kappa_{-} \left\{\Wb, k \right\} \leq t \Bigg] \leq P \big\{n_{w}, t/\sigma_{\min}(\bmSigma_{S,S}^{1/2}) \big\} \exp \left[ n_{w} \left(H(\rho/\gamma_{w})\cdot \gamma_{w} + \phi_{\min} \big\{t/\sigma_{\min}(\bmSigma_{S,S}^{1/2}) ,\rho\big\} \right) \right]. 
\end{align*}
{\BS Therefore, 
\begin{align*}
    &\PP \Bigg( \kappa_{-} \left[\Wb, n_w \left\{ 1 - \frac{1}{4(1 + b_*)} \right\} \right] \leq t \Bigg) 
    \\
    &\qquad \leq P \big\{n_{w}, t/\sigma_{\min}(\bmSigma_{S,S}^{1/2}) \big\} \exp \left[ n_{w} \left(H(\rho/\gamma_{w})\cdot \gamma_{w} + \phi_{\min} \big\{t/\sigma_{\min}(\bmSigma_{S,S}^{1/2}) ,\rho\big\} \right) \right]. 
\end{align*}
with $\rho = 1 - \frac{1}{4(1 + b_*)}$.
For any $t$ satisfying 
\begin{align}\label{eqn:c_4_cond}
    0 \leq t <\ & \sigma_{\min}(\bmSigma^{1/2}) \cdot \exp\left\{ - 1 -  8(\gamma_{w} + 2) \log 2 \cdot (1 + b_*) \right\}
\end{align}
Let $c, C$ satisfy
\begin{align}\label{eqn:c_cond}
    \log \big\{ t/\sigma_{\min}(\bmSigma^{1/2}) \big\} \leq -1 -  8\big\{(\gamma_{w} + 2) \log 2 + c\big\} \cdot (1 + b_*)
\end{align}
and asymptotically, $P \big\{n_{w}, t/\sigma_{\min}(\bmSigma_{S,S}^{1/2}) \big\}$ is upper bounded by a constant $C$. 
We have 
$$
\PP\Bigg[\kappa_{-} \left\{\Wb, k \right\} \leq t \Bigg] \leq C \exp(-c n_{w}).
$$
To see this, noticing $H(\rho/\gamma_{w}) \leq \log 2$ by Jensen's inequality, we have
\begin{align*}
    &H(\rho/\gamma_{w})\cdot \gamma_{w} + \phi_{\min} \big\{t/\sigma_{\min}(\bmSigma_{S,S}^{1/2}) ,\rho\big\}
    \\
    \leq\ & \gamma_w \cdot \log 2 + \frac{1}{2} \left[(1 - \rho) \log(t/\sigma_{\min}(\bmSigma_{S,S}^{1/2})) + 1 - \rho + \rho \log(\rho) - t/\sigma_{\min}(\bmSigma_{S,S}^{1/2}) \right]
    \\
    \leq\ & \gamma_w \cdot \log 2 + \frac{1}{2} + \frac{1}{2} (1 - \rho) \log(t/\sigma_{\min}(\bmSigma_{S,S}^{1/2})
    \\
    \leq\ & \gamma_w \cdot \log 2 + \frac{1}{2} + \frac{1}{8 (1 + b_{*})} \log(t/\sigma_{\min}(\bmSigma^{1/2})
    \\
    \overset{(a)}{\leq}\ & \gamma_w \cdot \log 2 + \frac{1}{2} - \frac{1}{8(1 + b_*)} - (\gamma_{w} + 2) \log 2 - c \overset{(b)}{\leq} - c,
\end{align*}
where inequality $(a)$ holds due to \eqref{eqn:c_cond} and inequality $(a)$ holds due to the fact $1/2 < 2 \log 2$.
Thus, for any $t$ satisfying \eqref{eqn:c_4_cond}, there exists some constant $c > 0$ such that
\begin{align*}
    H(\rho/\gamma_{w}) \cdot \gamma_{w} + \phi_{\min} \big\{t/\sigma_{\min}(\bmSigma_{S,S}^{1/2}) ,\rho\big\} \leq -c.
\end{align*}
We finally conclude that there exist constants $C, c > 0$ such that
\begin{align*}
    \PP\Bigg[\kappa_{-} \left\{\Wb, n_w \left(1 - \frac{1}{4(1 + b_*)} \right) \right\} \leq c_4 \Bigg] \leq  C e^{-c n_w}.
\end{align*}
for any fixed $c_4 < \sigma_{\min}(\bmSigma^{1/2}) \cdot \exp\left\{ - 1 -  8(\gamma_{w} + 2) \log 2 \cdot (1 + b_*) \right\}$.
}

\section{Proofs for Section \ref{sec: L_2 estimator}}
\label{sec: Proof for sec: L_2 estimator}

In this section, we provide proofs of the results in Section \ref{sec: L_2 estimator}. 
Specifically, Section \ref{sec: Proof of Theorem i.i.d. ridge mse + R2} computes $R_{\RW}(\lambda)$ and $A^2_{\RW}(\lambda)$ with isotropic features. 
Moreover, we show that these AMP results match the RMT results in 
the existing literature.
In Section \ref{Proof of prop: ridge regression mse + R2}, we 
prove Proposition \ref{prop: ridge regression mse + R2}. 
In Section \ref{proof of comparison of L_2}, we compare $R_{\RW}(\lambda), A^2_{\RW}(\lambda)$, $R_{\RA}(\lambda)$,  and $A^2_{\RA}(\lambda)$ to prove Proposition \ref{prop: comparison of L_2}.

\subsection{Proof of Theorem \ref{thm: i.i.d. ridge mse + R2}}
\label{sec: Proof of Theorem i.i.d. ridge mse + R2}
We compute the closed-form expression for  $A^2_{\RW}(\lambda)$ and show that it gives an equivalent result as in \cite[Corollary 1]{zhao2022block} when $\bmSigma = \bI_{p}$.  \cite{zhao2022block} adopted RMT to prove Corollary 1. 
Recall that ${\rho_{*}^2}$ is the solution to the fixed point equation in Equation \eqref{ridge state evolution}:
\begin{equation} \label{eqn: S.11.1}
    \rho_{*}^2 = \left( \EE \Bigg[ \gamma_x (1 + c_*)^2 \bar{\beta}^2/h_x^2 + \gamma_w \Bigg\{\frac{\rho_{*} z + (1 + c_*)\bar{\beta}}{1 + \lambda (1 + c_*)} \Bigg\} \Bigg]^2 \right).
\end{equation}
By Equation \eqref{eqn: S.10.3}, we have
\begin{align*}
    h_{s}^2 - \frac{\kappa \sigma_{\bbeta}^2}{\kappa \sigma_{\bbeta}^2 + \sigma_{\bmeps_s}^2} \overset{P}{\to} 0. 
\end{align*}
Therefore, Equation \eqref{eqn: S.11.1} implies that 
\begin{equation*} 
\begin{split}
    &\rho_{*}^2 = \gamma_x (1 + c_*)^2 \kappa \sigma_{\bbeta}^2/h_{x}^{2} + \frac{\gamma_w \rho_{*}^2}{\{1 + \lambda (1 + c_*)\}^2}  + \gamma_w \Bigg\{\frac{1 + c_*}{1 + \lambda (1 + c_*)} \Bigg\}^2 \kappa \sigma_{\bbeta}^2,\\
    &\Bigg\{1 - \frac{\gamma_w}{\{1 + \lambda (1 + c_*)\}^2} \Bigg\} \rho_{*}^2 = \gamma_x (1 + c_*)^2 \kappa \sigma_{\bbeta}^2/h_{x}^{2} + \gamma_w \Bigg\{\frac{1 + c_*}{1 + \lambda (1 + c_*)} \Bigg\}^2 \kappa \sigma_{\bbeta}^2,\\
    &\Bigg\{ 1 - \frac{\gamma_w}{\{1 + \lambda (1 + c_*)\}^2} \Bigg \} \frac{\rho_{*}^2}{(1 + c_*)^2} = \gamma_x \kappa \sigma_{\bbeta}^2/h_{x}^{2} + \gamma_w \Bigg\{\frac{1}{1 + \lambda (1 + c_*)} \Bigg\}^2 \kappa \sigma_{\bbeta}^2.
\end{split}
\end{equation*}
Therefore, we have the following simplified fixed point equation, which  allows us to compute the close-form expression for $A^2_{\RW}(\lambda)$ and $R_{\RW}(\lambda)$: 
\begin{align} \label{ref ridge alpha}
    \frac{\rho_{*}^2}{(1 + c_*)^2} = \Bigg[1 - \frac{\gamma_w}{\{1 + \lambda (1 + c_*)\}^2} \Bigg]^{-1} \Bigg[\gamma_x \kappa \sigma_{\bbeta}^2/h_{x}^{2} + \gamma_w \Bigg\{\frac{1}{1 + \lambda (1 + c_*)} \Bigg\}^2 \kappa \sigma_{\bbeta}^2 \Bigg].
\end{align}
For isotropic features, Theorem \ref{thm: ridge mse + R2} implies that 
\begin{align*}
    A^2_{\RW}(\lambda) =\ & h_{s}^2 \cdot \frac{ \EE \left[ \barbeta  \{1 + \lambda(1 + c_*)\}^{-1}\{\rho_{*} z + (1 + c_*) {\barbeta}\} \right]^2 }{\EE \barbeta^2 \cdot \EE \{1 + \lambda(1 + c_*)\}^{-2}\{\rho_{*} z + (1 + c_*) {\barbeta}\}^2}.
\end{align*}
The numerator of $A^2_{\RW}(\lambda)$ is 
\begin{equation} \label{eqn: S.11.3}
\begin{split}
    \left( \EE \left[ \barbeta  \{1 + \lambda(1 + c_*)\}^{-1}\{\rho_{*} z + (1 + c_*) {\barbeta}\} \right] \right)^2 = \Bigg\{\frac{1 + c_*}{1 + \lambda (1 + c_*)} \Bigg\}^2 \left(\kappa \sigma_{\bbeta}^2 \right)^2,
\end{split}
\end{equation}
and the denominator of $A^2_{\RW}(\lambda)$ is 
\begin{equation} \label{eqn: S.11.4}
\begin{split}
    &\EE \barbeta^2 \cdot \EE \{1 + \lambda(1 + c_*)\}^{-2}\{\rho_{*} z + (1 + c_*) {\barbeta}\}^2
    \\ 
    =\ & \kappa \sigma_{\bbeta}^2 \cdot \left[ \Bigg\{\frac{\rho_{*}}{1 + \lambda (1 + c_*)} \Bigg\}^2 + \Bigg\{\frac{  (1 + c_*)}{1 + \lambda (1 + c_*)} \Bigg\}^2 \kappa \sigma_{\bbeta}^2 \right].
\end{split}
\end{equation}
Equations \eqref{eqn: S.11.3} and \eqref{eqn: S.11.4} imply that 
\begin{equation*}
\begin{split}
     A^2_{\RW}(\lambda) &= h_s^2 \frac{ (1 + c_*)^2 \kappa \sigma_{\bbeta}^2}{\rho_{*}^2 + (1 + c_*)^2 \kappa \sigma_{\bbeta}^2} = h_s^2 \frac{\kappa \sigma_{\bbeta}^2}{(\frac{\rho_{*}}{1 + c_*})^2 + \kappa \sigma_{\bbeta}^2}
     \\
     &\overset{(a)}{=} \frac{h_s^2 \kappa \sigma_{\bbeta}^2}{\Bigg[1 - \frac{\gamma_w}{\{1 + \lambda (1 + c_*)\}^2} \Bigg]^{-1} \Bigg[\gamma_x \kappa \sigma_{\bbeta}^2/h_{x}^{2} + \gamma_w \Bigg\{\frac{1}{1 + \lambda (1 + c_*)} \Bigg\}^2 \kappa \sigma_{\bbeta}^2 \Bigg] + \kappa \sigma_{\bbeta}^2}
     \\
     &= \frac{h_s^2}{\Bigg[1 - \frac{\gamma_w}{\{1 + \lambda (1 + c_*)\}^2} \Bigg]^{-1} \Bigg[\gamma_x/h_{x}^{2}  + \gamma_w \Bigg\{\frac{1}{1 + \lambda (1 + c_*)} \Bigg\}^2  \Bigg] + 1 },
\end{split}
\end{equation*}
where the equality $(a)$ follows from Equation \eqref{ref ridge alpha}. 
It then follows that
\begin{equation} \label{eqn: S.11.5}
\begin{split}
    A^2_{\RW}(\lambda) &= \frac{h_s^2}{\Bigg[1 - \frac{\gamma_w}{\{1 + \lambda (1 + c_*)\}^2} \Bigg]^{-1} \Bigg[\gamma_x/h_{x}^{2}  + \gamma_w \Bigg\{\frac{1}{1 + \lambda (1 + c_*)} \Bigg\}^2  \Bigg] + 1 }\\
    &=\frac{h_s^2 \Bigg[1 - \frac{\gamma_w}{\{1 + \lambda (1 + c_*)\}^2} \Bigg]}{\Bigg[\gamma_x \frac{1}{h_s^2} + \gamma_w \Bigg\{\frac{1}{1 + \lambda (1 + c_*)} \Bigg\}^2  \Bigg] + \Bigg[1 - \frac{\gamma_w}{\{1 + \lambda (1 + c_*)\}^2} \Bigg] }\\
    &= \frac{h_s^2 \Bigg[1 - \frac{\gamma_w}{\{1 + \lambda (1 + c_*)\}^2} \Bigg]}{\frac{\gamma_x}{h_s^2} + 1}\\
    &= \frac{h_s^4}{\gamma_x + h_s^2}  \Bigg[1 - \frac{\gamma_w}{\{1 + \lambda (1 + c_*)\}^2} \Bigg].
\end{split}
\end{equation}
When $\bmSigma = \bI_p$, the $c_*$ given by Equation \eqref{ridge state evolution} admits a closed-form expression
\begin{align} \label{eqn: S.11.6}
    c_* = \frac{-(1+\lambda - \gamma_w) + \sqrt{(1 - \lambda - \gamma_w)^2 + 4\lambda}}{2\lambda}.
\end{align}
Combining this with Equation \eqref{eqn: S.11.5}, we have
\begin{equation} \label{eqn: S.11.7}
    A^2_{\RW}(\lambda) = \frac{h_s^4}{\gamma_x + h_s^2} \Bigg[1 - 4 \gamma_w \Bigg\{\frac{1}{ 1 + \lambda + \gamma_w + \sqrt{(1 - \lambda - \gamma_w)^2 + 4\lambda} } \Bigg\}^2 \Bigg].
\end{equation}
This completes the proof of the first part.
Next, we show that Equation \eqref{eqn: S.11.7} is equivalent to the result  reported in the RMT literature. 
Specifically, recalling $A^2_{\RW}(\lambda)$ from Corollary 1 of \cite{zhao2022block}, we have 
\begin{equation} \label{eqn: S.11.8}
    A^2_{\RW}(\lambda) = \frac{h_{s}^4}{h_{s}^2 + \gamma_x} \cdot \frac{1}{1 - b_w'} + o(1)
\end{equation}
with
\begin{equation*}
\begin{split}
        b_w &= \frac{- (-1 + \lambda + \gamma_w) +\sqrt{(1 - \lambda - \gamma_w)^2 + 4 \lambda}}{2} \quad\mbox{and} \\
        b_w' &= - \frac{\gamma_w b_w}{\gamma_w \lambda + (b_w + \lambda)^2}.
\end{split}
\end{equation*}
Our goal is to show that Equation \eqref{eqn: S.11.7} from AMP and Equation \eqref{eqn: S.11.8} from RMT are equivalent. Note that 
\begin{equation*}
\begin{split}
    \frac{1}{1 - b_w'} &= \frac{1}{1 + \frac{\gamma_w b_w}{\gamma_w \lambda + (b_w + \lambda)^2}}\\
    &= \frac{\gamma_w \lambda + (b_w + \lambda)^2}{\gamma_w \lambda + (b_w + \lambda)^2 + \gamma_w b_w}\\
    &= 1 - \frac{\gamma_w b_w}{(b_w + \lambda + \gamma_w) (b_w + \lambda) }\\
    &= 1 - 4 \gamma_w \Bigg\{\frac{1}{ 1 + \lambda + \gamma_w + \sqrt{(1 - \lambda - \gamma_w)^2 + 4\lambda} } \Bigg\}^2,
\end{split}
\end{equation*}
where the last equality holds because
\begin{equation*}
\begin{split}
    \frac{b_w}{b_w + \lambda} &= \Bigg(1 + \frac{\lambda}{b_w} \Bigg)^{-1}\\
    &= \Bigg\{1 + \frac{2\lambda}{- (-1 + \lambda + \gamma_w) +\sqrt{(1 - \lambda - \gamma_w)^2 + 4 \lambda}} \Bigg\}^{-1}\\
    &= \Bigg\{1 + \frac{2\lambda (1 - \lambda - \gamma_w) - \sqrt{(1 - \lambda - \gamma_w)^2 + 4 \lambda})}{- 4 \lambda} \Bigg\}^{-1}\\
    &= \Bigg\{1 - \frac{(1 - \lambda - \gamma_w) - \sqrt{(1 - \lambda - \gamma_w)^2 + 4 \lambda}}{2} \Bigg\}^{-1}\\
    &= \Bigg\{\frac{(1 + \lambda + \gamma_w) + \sqrt{(1 - \lambda - \gamma_w)^2 + 4 \lambda}}{2} \Bigg\}^{-1}.
\end{split}
\end{equation*}
Therefore, our results in Theorem \ref{thm: i.i.d. ridge mse + R2} are consistent with \cite[Corollary 1]{zhao2022block}. 
In addition, by Theorem \ref{thm: ridge mse + R2}, $R_{\RW}(\lambda)$ is  
\begin{equation}
\begin{split}
    R_{\RW} (\lambda) =\ & \EE \Bigg\{\frac{\rho_{*} z  + (1 + c_*)\bar{\beta}}{1 + \lambda (1 + c_*)} -  \bar{\beta} \Bigg\}^2\\
    =\ &  \EE \Bigg[\frac{\rho_{*}}{1 + \lambda (1 + c_*)} z + \Bigg\{\frac{  (1 + c_*)}{1 + \lambda (1 + c_*)} - 1 \Bigg\} \bar{\beta} \Bigg]^2\\
    =\ & \left\{ \frac{\rho_{*}}{1 + \lambda (1 + c_*)} \right\}^2 + \Bigg\{\frac{  (1 + c_*)}{1 + \lambda (1 + c_*)} - 1 \Bigg\}^2 \kappa \sigma_{\beta}^2,
\end{split}
\end{equation}
where $(\rho_{*}, c_*)$ are defined by the equations \eqref{eqn: S.11.1} and \eqref{eqn: S.11.6}. 
Moreover, by Equation \eqref{ref ridge alpha}, we have
\begin{align*}
    R_{\RW}(\lambda) 
    &= \kappa \sigma_{\beta}^2 \Bigg[ \frac{\left\{\lambda ^2+\lambda  \gamma_w+\lambda  \sqrt{4   \lambda +\left(\lambda +\gamma_w-1\right)^2}-\sqrt{4 \lambda +\left(\lambda +\gamma_w-1\right)^2}-\gamma_w+1\right\}^2}{\lambda ^2   \left\{\lambda +\sqrt{4 \lambda +\left(\lambda +\gamma_w-1\right)^2}+\gamma_w+1\right\}^2}\\
    &\qquad + \frac{\gamma_x \left\{\lambda +\sqrt{4 \lambda +\left(\lambda +\gamma_w-1\right)^2}+\gamma_w-1\right\}^2}{h_x^2 \lambda ^2 \left\{\lambda   +\sqrt{4 \lambda +\left(\lambda +\gamma_w-1\right)^2}+\gamma_w+1\right\}^2}\\
    &\qquad \cdot \frac{\frac{2 h_x^2 \gamma_w}{\gamma_x}+(\lambda +1) \left\{\lambda +\sqrt{4 \lambda +\left(\lambda +\gamma_w-1\right)^2}+1\right\}+\gamma_w \left\{2 \lambda +\sqrt{4 \lambda +\left(\lambda +\gamma_w-1\right)^2}\right\}+\gamma_w^2}{\gamma_w \left\{2 \lambda +\sqrt{4 \lambda +\left(\lambda +\gamma_w-1\right)^2}\right\}+(\lambda +1) \left\{\lambda +\sqrt{4 \lambda +\left(\lambda +\gamma_w-1\right)^2}+1\right\}+\gamma_w^2-2 \gamma_w} \Bigg].
\end{align*}

\subsection{Proof of Proposition \ref{prop: ridge regression mse + R2}} \label{Proof of prop: ridge regression mse + R2}

Corollary 1 in \cite{zhao2022block} uses RMT techniques to obtain the out-of-sample $R^2$ of the regular $L_2$ regularized estimator, which is 
\begin{equation} \label{eqn: S.12.0}
\begin{split}
    A^2_{\textnormal {R}}(\lambda) &= \frac{h_s^4 b_r^2}{(b_r^2 - \lambda^2 \bar{b}_r) h_s^2 + (b_r  + \lambda \bar{b}_r) \gamma_{x} (1 - h_s^2)}\\
    &= \frac{h_s^4}{\frac{b_r^2 - \lambda^2 \bar{b}_r}{b_r  + \lambda \bar{b}_r} h_s^2 + \gamma_{x} (1 - h_s^2)} \cdot \frac{b_r^2}{b_r  + \lambda \bar{b}_r}\\
    &= \frac{h_s^4}{\left(\frac{b_r^2 - \lambda^2 \bar{b}_r}{b_r  + \lambda \bar{b}_r} - \gamma_{x} \right)h_s^2 + \gamma_{x}} \cdot \frac{b_r^2}{b_r  + \lambda \bar{b}_r},
\end{split}
\end{equation}
where
\begin{equation} \label{eqn: S.12.1}
\begin{split}
    b_r &= \frac{(1 - \gamma_{x} - \lambda) + \sqrt{(1 - \gamma_{x} - \lambda)^2 + 4\lambda}}{2}\quad \mbox{and} \\
    \bar{b}_r &= - \frac{\gamma_{x} b_r}{\gamma_{x} \lambda + (b_r + \lambda)^2}.
\end{split}
\end{equation}
Denote $\Rom{1}$ and $\Rom{2}$ as 
\begin{align*}
    \Rom{1} = \frac{b_r^2}{b_r  + \lambda \bar{b}_r}, \quad
    \Rom{2} = \frac{b_r^2 - \lambda^2 \bar{b}_r}{b_r  + \lambda \bar{b}_r} - \gamma_{x}.
\end{align*}
First, we compute $\Rom{1}$. Equation~\eqref{eqn: S.12.1} implies that 
\begin{align} \label{eqn: S.12.2}
    (b_r + \lambda)^2 &= \Bigg\{\frac{(1 - \gamma_{x} + \lambda) + \sqrt{(1 - \gamma_{x} - \lambda)^2 + 4\lambda}}{2} \Bigg\}^2.
\end{align}
By equations \eqref{eqn: S.12.1} - \eqref{eqn: S.12.2}, we have 
\begin{equation*}
\begin{split}
    \Rom{1} 
    &= \frac{b_r}{1 - \frac{\gamma_{x} \lambda}{\gamma_{x} \lambda + (b_r + \lambda)^2}}  =b_r \cdot \frac{\gamma_{x} \lambda + (b_r + \lambda)^2}{(b_r + \lambda)^2} = b_r \cdot \Bigg\{ 1 + \frac{\gamma_{x} \lambda}{(b_r + \lambda)^2} \Bigg\}\\
    &= \frac{(1 - \gamma_{x} - \lambda) + \sqrt{(1 - \gamma_{x} - \lambda)^2 + 4\lambda}}{2} \cdot \Bigg[ 1 +4 \gamma_{x} \lambda \Bigg\{\frac{1}{(1 - \gamma_{x} + \lambda) + \sqrt{(1 - \gamma_{x} - \lambda)^2 + 4\lambda}} \Bigg\}^2 \Bigg]\\
    &= \frac{(1 - \gamma_{x} - \lambda) + \sqrt{(1 - \gamma_{x} - \lambda)^2 + 4\lambda}}{2} \cdot \Bigg[ 1 + \frac{\{(1 - \gamma_{x} + \lambda) - \sqrt{(1 - \gamma_{x} - \lambda)^2 + 4\lambda}\}^2}{4 \lambda \gamma_{x}}  \Bigg].
\end{split}
\end{equation*}
It follows that 
\begin{align*}
     \Rom{1} 
     &= \frac{(1 - \gamma_{x} - \lambda) + \sqrt{(1 - \gamma_{x} - \lambda)^2 + 4\lambda}}{2}\\ 
     &\quad \cdot \Bigg\{ \frac{4 \lambda \gamma_{x} + (1 - \gamma_{x} + \lambda)^2 + (1 - \gamma_{x} - \lambda)^2 + 4\lambda - 2 (1 - \gamma_{x} + \lambda) \sqrt{(1 - \gamma_{x} - \lambda)^2 + 4\lambda}}{4 \lambda \gamma_{x}}\Bigg\}\\
    &= \frac{(1 - \gamma_{x} - \lambda) + \sqrt{(1 - \gamma_{x} - \lambda)^2 + 4\lambda}}{2}\\ 
    &\quad\cdot \Bigg\{ \frac{(1 - \gamma_{x} - \lambda)^2 + 4\lambda - (1 - \gamma_{x} + \lambda) \sqrt{(1 - \gamma_{x} - \lambda)^2 + 4\lambda}}{2 \lambda \gamma_{x}}\Bigg\}\\
    &= \frac{-2 \lambda \{(1 - \gamma_{x} - \lambda)^2 + 4\lambda\} + 2 \lambda(1 + \lambda + \gamma_{x}) \sqrt{(1 - \gamma_{x} - \lambda)^2 + 4\lambda} }{4\lambda \gamma_{x}}\\
    &= \frac{- \{(1 - \gamma_{x} - \lambda)^2 + 4\lambda\} + (1 + \lambda + \gamma_{x}) \sqrt{(1 - \gamma_{x} - \lambda)^2 + 4\lambda} }{2 \gamma_{x}}.
\end{align*}
On the other hand, note that 
\begin{equation*}
\begin{split}
    &1 - 4 \gamma_{x} \Bigg\{\frac{1}{ 1 + \lambda + \gamma_{x} + \sqrt{(1 - \lambda - \gamma_{x})^2 + 4\lambda} } \Bigg\}^2\\ 
    &= 1 - 4 \gamma_{x} \Bigg\{\frac{ 1 + \lambda + \gamma_{x} - \sqrt{(1 - \lambda - \gamma_{x})^2 + 4\lambda}}{4 \gamma_{x}} \Bigg\}^2\\
    &= \frac{- \{(1 - \gamma_{x} - \lambda)^2 + 4\lambda\} + (1 + \lambda + \gamma_{x}) \sqrt{(1 - \gamma_{x} - \lambda)^2 + 4\lambda} }{2 \gamma_{x}} = \frac{b_r^2}{b_r  + \lambda \bar{b}_r}.
\end{split}
\end{equation*}
Thus, we have
\begin{equation} \label{eqn: S.12.4}
    \Rom{1} = 1 - 4 \gamma_{x} \Bigg\{\frac{1}{ 1 + \lambda + \gamma_{x} + \sqrt{(1 - \lambda - \gamma_{x})^2 + 4\lambda} } \Bigg\}^2.
\end{equation}
Next, we compute $\Rom{2}$, 
\begin{align*}
    \Rom{2} &= \left\{ b_r^2 + \frac{\gamma_{x} \lambda^2 b_r}{\gamma_{x} \lambda + (b_r + \lambda)^2} \right\} \left\{ \frac{1}{b_r} \cdot \frac{\gamma_{x} \lambda + (b_r + \lambda)^2}{(b_r + \lambda)^2} \right\} - \gamma_{x}\\
    &= \left\{ b_r + \frac{\gamma_{x} \lambda^2}{\gamma_{x} \lambda + (b_r + \lambda)^2} \right\} \left\{ \frac{\gamma_{x} \lambda + (b_r + \lambda)^2}{(b_r + \lambda)^2} \right\} - \gamma_{x}\\
    &= b_r \cdot \left\{ \frac{\gamma_{x} \lambda + (b_r + \lambda)^2}{(b_r + \lambda)^2} \right\} + \frac{\gamma_{x} \lambda^2}{(b_r + \lambda)^2} - \gamma_{x}\\
    &= (b_r + \lambda) \cdot \left\{ \frac{\gamma_{x} \lambda + (b_r + \lambda)^2}{(b_r + \lambda)^2} \right\} -\lambda - \gamma_{x}\\
    &= (b_r + \lambda) \cdot \left\{1 + \frac{\gamma_{x} \lambda}{(b_r + \lambda)^2} \right\} - \lambda - \gamma_{x}. 
\end{align*}
By Equation \eqref{eqn: S.12.2}, we have 
\begin{align*}
    \Rom{2} &= - \lambda - \gamma_{x} + \Bigg[ \frac{(1 - \gamma_{x} + \lambda) + \sqrt{(1 - \gamma_{x} - \lambda)^2 + 4\lambda}}{2}\\ &\cdot  \frac{(1 - \gamma_{x} - \lambda)^2 + 4\lambda - (1 - \gamma_{x} + \lambda) \sqrt{(1 - \gamma_{x} - \lambda)^2 + 4\lambda}}{2 \lambda \gamma_{x}}\Bigg]\\
    &= - \lambda - \gamma_{x} +  \frac{(1 - \gamma_{x} - \lambda)^2 + 4\lambda - (1 - \gamma_{x} + \lambda)^2}{4 \lambda \gamma_{x}} \cdot \sqrt{(1 - \gamma_{x} - \lambda)^2 + 4\lambda}\\
    &= - \lambda - \gamma_{x} + \sqrt{(1 - \gamma_{x} - \lambda)^2 + 4\lambda}\\
    &= 1 - \frac{4\gamma_{x}}{1 + \lambda + \gamma_{x} + \sqrt{(1 - \lambda - \gamma_{x})^2 + 4\lambda}}.
\end{align*}
Combining this with equations \eqref{eqn: S.12.0} and \eqref{eqn: S.12.4} completes the proof of Proposition \ref{prop: ridge regression mse + R2}.

\subsection{Proof of Proposition \ref{prop: comparison of L_2}}
\label{proof of comparison of L_2}

When $\gamma_x = \gamma_w$, 
by equations~\eqref{eqn: i.i.d. ref ridge R^2} and \eqref{eqn: i.i.d. ridge R^2}, we have
\begin{align*}
        \max_{\lambda \in \RR_+} A^2_{\RW}(\lambda) < \max_{\lambda \in \RR_+} A^2_{\RA}(\lambda).
\end{align*}
To show $R_{\RW}(\lambda_{\RW, M}^*) > R_{\textnormal{R}}(\lambda_{\RA, M}^{*})$, 
we repeat the steps in Section \ref{proof of prop: comparison of L_1}. Specifically, 
we consider the following lower bound of $R_{\RW}(\lambda)$ as:
\begin{align*}
    R_{\RW}(\lambda) 
 &> \kappa \sigma_{\beta}^2 \Bigg[\frac{\left\{\lambda ^2+\lambda  \gamma_w+\lambda  \sqrt{4   \lambda +\left(\lambda +\gamma_w-1\right)^2}-\sqrt{4 \lambda +\left(\lambda +\gamma_w-1\right)^2}-\gamma_w+1\right\}^2}{\lambda ^2   \left\{\lambda +\sqrt{4 \lambda +\left(\lambda +\gamma_w-1\right)^2}+\gamma_w+1\right\}^2}\\
    &\qquad + \frac{\gamma_x \left\{\lambda +\sqrt{4 \lambda +\left(\lambda +\gamma_w-1\right)^2}+\gamma_w-1\right\}^2}{h_x^2 \lambda ^2 \left\{\lambda   +\sqrt{4 \lambda +\left(\lambda +\gamma_w-1\right)^2}+\gamma_w+1\right\}^2} \Bigg]\\
    &=: \Tilde{R}_{\RW}(\lambda). 
\end{align*}
Setting the derivative with respect to $\lambda$ to $0$, aka $\partial_{\lambda} \Tilde{R}_{\RW}(\lambda)=0$, we obtain the optimal $\lambda^{**}$ that minimizes the $\Tilde{R}_{\RW}(\lambda)$ as
\begin{align*}
    \lambda^{**} = \frac{\gamma_x \left(h_x^4+h_x^2 \gamma_x+h_x^2+\gamma_x\right)}{h_x^2 \left(h_x^2 \gamma_x+h_x^2+\gamma_x\right)}.
\end{align*}
Plugging the above  into the $\Tilde{R}_{\RW}(\lambda)$, we obtain
\begin{align*}
    \Tilde{R}_{\RW}(\lambda^{**}) = \frac{\kappa \sigma_{\beta}^2  \gamma_x}{h_x^2+\gamma_x}.
\end{align*}
Since
\begin{align*}
    R_{\RW}(\lambda_{\RW, M}^*) = \min_{\lambda} R_{\RW}(\lambda) > \min_{\lambda} \Tilde{R}_{\RW}(\lambda) = \Tilde{R}_{\RW}(\lambda^{**}),
\end{align*}
it suffices to show that $\Tilde{R}_{\RW}(\lambda^{**}) >  R_{\textnormal {R}}(\lambda_{\RA, M}^{*})$. Plugging $\lambda_{\RA, M}^{*} = \left(1-h_x^2\right) \gamma_x/h_x^2$ into $R_{\textnormal {R}}(\lambda_{\RA, M}^{*})$, we obtain 
\begin{align*}
    R_{\textnormal {R}}(\lambda_{\RA, M}^{*}) =&\frac{\left(1-h_x^2\right) \kappa \sigma_{\beta}^2 \left[\frac{\left(1-h_x^2\right) \gamma_x}{h_x^2}-\sqrt{\left\{-\frac{\left(1-h_x^2\right) \gamma_x}{h_x^2}-\gamma_x+1\right\}^2+\frac{4 \left(1-h_x^2\right) \gamma_x}{h_x^2}}+\gamma_x+1\right]}{2 h_x^2 \sqrt{\left\{-\frac{\left(1-h_x^2\right) \gamma_x}{h_x^2}-\gamma_x+1\right\}^2+\frac{4 \left(1-h_x^2\right) \gamma_x}{h_x^2}}}\\
    &+\frac{\kappa \sigma_{\beta}^2 \left[\left(\gamma_x-1\right) \sqrt{\left\{-\frac{\left(1-h_x^2\right) \gamma_x}{h_x^2}-\gamma_x+1\right\}^2+\frac{4 \left(1-h_x^2\right) \gamma_x}{h_x^2}}+\frac{\left(1-h_x^2\right) \gamma_x \left(\gamma_x+1\right)}{h_x^2}+\left(\gamma_x-1\right)^2\right]}{2 \gamma_x \sqrt{\left\{-\frac{\left(1-h_x^2\right) \gamma_x}{h_x^2}-\gamma_x+1\right\}^2+\frac{4 \left(1-h_x^2\right) \gamma_x}{h_x^2}}}\\
    =\ & \frac{\kappa \sigma_{\beta}^2  \left[h_x^4+h_x^4 \left\{-\sqrt{\frac{\gamma_x^2}{h_x^4}+\left(\frac{2}{h_x^2}-4\right) \gamma_x+1}\right\}+\left(2 h_x^2-1\right) h_x^2 \gamma_x \left\{\sqrt{\frac{\gamma_x^2}{h_x^4}+\left(\frac{2}{h_x^2}-4\right) \gamma_x+1}-2\right\}+\gamma_x^2\right]}{2 h_x^4 \gamma_x   \sqrt{\frac{\gamma_x^2}{h_x^4}+\left(\frac{2}{h_x^2}-4\right) \gamma_x+1}}.
\end{align*}
Therefore, $\Tilde{R}_{\RW}(\lambda^{**}) -  R_{\textnormal {R}}(\lambda_{\RA, M}^{*})$ is 
\begin{align*}
    &\frac{\kappa \sigma_{\beta}^2  \gamma_x}{h_x^2+\gamma_x} - \frac{\kappa \sigma_{\beta}^2  \left[h_x^4+h_x^4 \left\{-\sqrt{\frac{\gamma_x^2}{h_x^4}+\left(\frac{2}{h_x^2}-4\right) \gamma_x+1}\right\}+\left(2 h_x^2-1\right) h_x^2 \gamma_x \left\{\sqrt{\frac{\gamma_x^2}{h_x^4}+\left(\frac{2}{h_x^2}-4\right) \gamma_x+1}-2\right\}+\gamma_x^2\right]}{2 h_x^4 \gamma_x   \sqrt{\frac{\gamma_x^2}{h_x^4}+\left(\frac{2}{h_x^2}-4\right) \gamma_x+1}}\\
    =\ & \frac{\kappa \sigma_{\beta}^2 \left[h_x^4+h_x^4 \left\{-\sqrt{\frac{\gamma_x^2}{h_x^4}+\left(\frac{2}{h_x^2}-4\right) \gamma_x+1}\right\}-h_x^2 \gamma_x \left\{2 h_x^2+\sqrt{\frac{\gamma_x^2}{h_x^4}+\left(\frac{2}{h_x^2}-4\right) \gamma_x+1}-2\right\}+\gamma_x^2 \right]}{2 h_x^2 \gamma_x \left(h_x^2+\gamma_x\right)}.
\end{align*}
To show $\Tilde{R}_{\RW}(\lambda^{**}) -  R_{\textnormal {R}}(\lambda_{\RA, M}^{*}) > 0$, it suffices to show 
\begin{align*}
    h_x^4+h_x^4 \left\{-\sqrt{\frac{\gamma_x^2}{h_x^4}+\left(\frac{2}{h_x^2}-4\right) \gamma_x+1}\right\}-h_x^2 \gamma_x \left\{2 h_x^2+\sqrt{\frac{\gamma_x^2}{h_x^4}+\left(\frac{2}{h_x^2}-4\right) \gamma_x+1}-2\right\}+\gamma_x^2 > 0.
\end{align*}
To this end, we have
\begin{align*}
    &h_x^4+h_x^4 \left\{-\sqrt{\frac{\gamma_x^2}{h_x^4}+\left(\frac{2}{h_x^2}-4\right) \gamma_x+1}\right\}-h_x^2 \gamma_x \left\{2 h_x^2+\sqrt{\frac{\gamma_x^2}{h_x^4}+\left(\frac{2}{h_x^2}-4\right) \gamma_x+1}-2\right\}+\gamma_x^2\\
    =\ & -2 h_x^4 \gamma_x+h_x^4+2 h_x^2 \gamma_x-h_x^2 \sqrt{h_x^4+\left(2 h_x^2-4 h_x^4\right) \gamma_x+\gamma_x^2}-\gamma_x \sqrt{h_x^4+\left(2 h_x^2-4 h_x^4\right)
   \gamma_x+\gamma_x^2}+\gamma_x^2\\
   =\ & -2 h_x^4 \gamma_x+\left(h_x^2+\gamma_x\right)^2-\left(h_x^2+\gamma_x\right) \sqrt{\left(h_x^2+\gamma_x\right)^2-4 h_x^4 \gamma_x}\\
   =\ & \frac{4 h_x^4 \gamma_x \left(h_x^2+\gamma_x\right)}{h_x^2+\sqrt{\left(h_x^2+\gamma_x\right)^2-4 h_x^4 \gamma_x}+\gamma_x}-2 h_x^4 \gamma_x\\
   =\ & 2 h_x^4 \gamma_x \Bigg\{\frac{2 \left(h_x^2+\gamma_x\right)}{h_x^2+\sqrt{\left(h_x^2+\gamma_x\right)^2-4 h_x^4 \gamma_x}+\gamma_x} - 1 \Bigg\}\\
   =\ & 2 h_x^4 \gamma_x \Bigg\{\frac{h_x^2+\gamma_x - \sqrt{\left(h_x^2+\gamma_x\right)^2-4 h_x^4 \gamma_x}}{h_x^2+\sqrt{\left(h_x^2 + \gamma_x +\gamma_x\right)^2-4 h_x^4 \gamma_x}} \Bigg\} > 0
\end{align*}
This completes the proof of $\Tilde{R}_{\RW}(\lambda^{**}) -  R_{\textnormal {R}}(\lambda_{\RA, M}^{*}) > 0$ and thus the proof of Proposition \ref{prop: comparison of L_2}.

{
\section{Supplementary figures} \label{sec: Supplementary Figures}

\subsection{UK Biobank simulation studies}
\begin{figure}[H]
\includegraphics[page=1,width=0.9\linewidth]{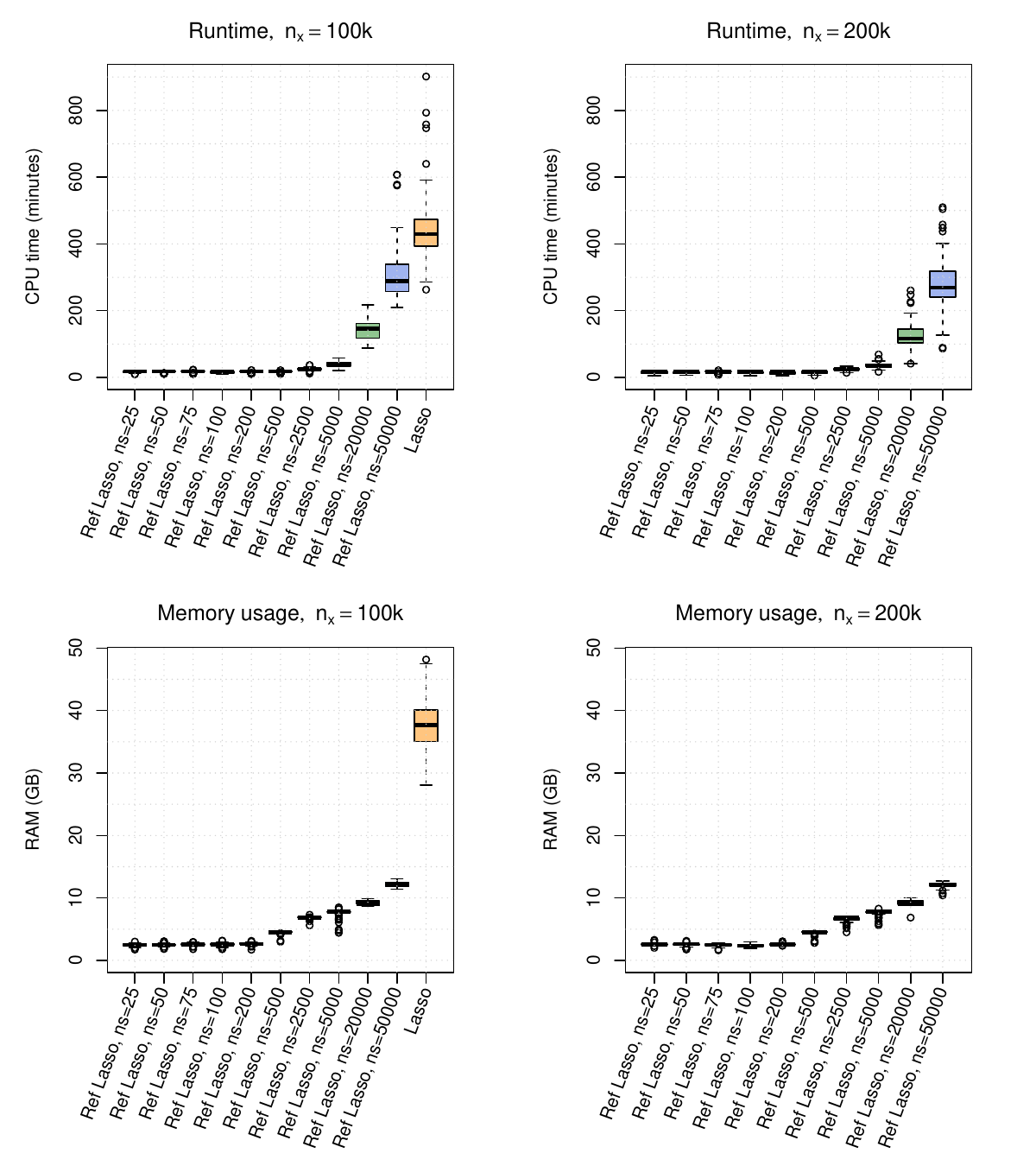}
\centering
\caption{\textbf{Runtime and memory usage of different estimators in UKB simulation studies.} We show the runtime in minutes and the memory usage in GB of $L_1$ regularized estimator (`Lasso') and reference panel-based $L_1$ estimator (`Ref Lasso').
We set $p=461,488$, $n_s=20,000$, $h_x^2=h_s^2=0.6$, $m/p=0.0005$, and varying reference panel size $n_w$ from $25$ to $50,000$. The reference panel is from the UKB study and is independent of the training and testing datasets.  
\textbf{Top left:} Runtime, $n_x=100,000$. \textbf{Top right:} Runtime, $n_x=200,000$. \textbf{Bottom left:} Memory usage, $n_x=100,000$. \textbf{Bottom right:} Memory usage, $n_x=200,000$. 
Lasso under $n_x=200,000$ is not run due to computational resource limits.
}
\label{runtime_ram_simulated_lasso_lassosum_ntrain100k_h06}
\end{figure}

\subsection{UK Biobank retinal imaging data analysis}

\begin{figure}[!ht]
\includegraphics[angle=270, width=0.9\textwidth]{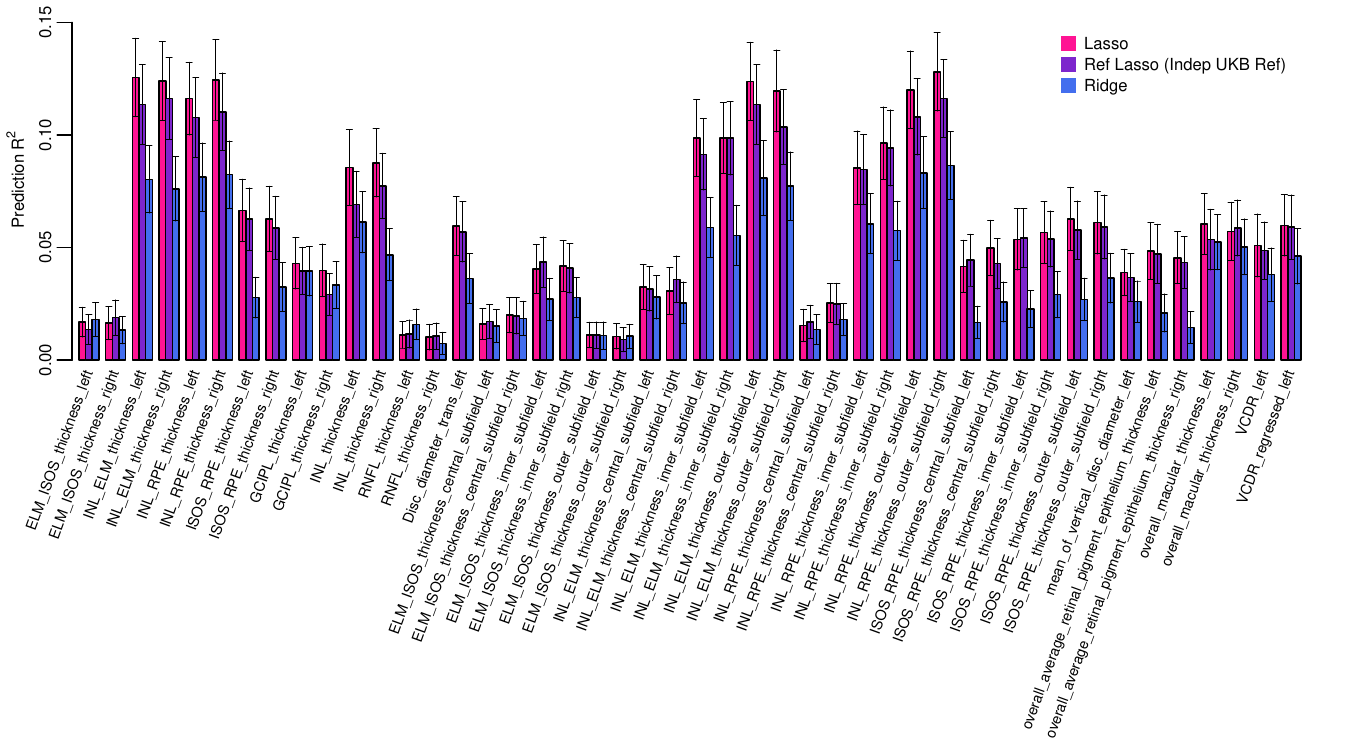}
\centering
\end{figure}
\clearpage
\captionof{figure}{\textbf{Out-of-sample $R^2$ of different estimators in UKB retinal imaging data analysis.}
We show the prediction accuracy for all $46$ retinal imaging traits. 
Here $p=461,488$, $n_x=36,054$, $n_w=20,000$, and $n_s=4,666$. 
Different colors indicate the $L_1$ and $L_2$ regularized estimators (Lasso and Ridge), as well as the reference panel-based $L_1$ regularized estimator with the UKB data being as reference panel (Indep UKB Ref).  
The standard error of prediction accuracy of each method is calculated using 500 independent bootstrap samples of size $4,666$.
} 
\label{numerical_eye_lasso_ref_lasso_ridge_all}
}

\subsection{Numerical illustration of theoretical results}

\begin{figure}[H]
\includegraphics[page=1,width=0.4\linewidth]{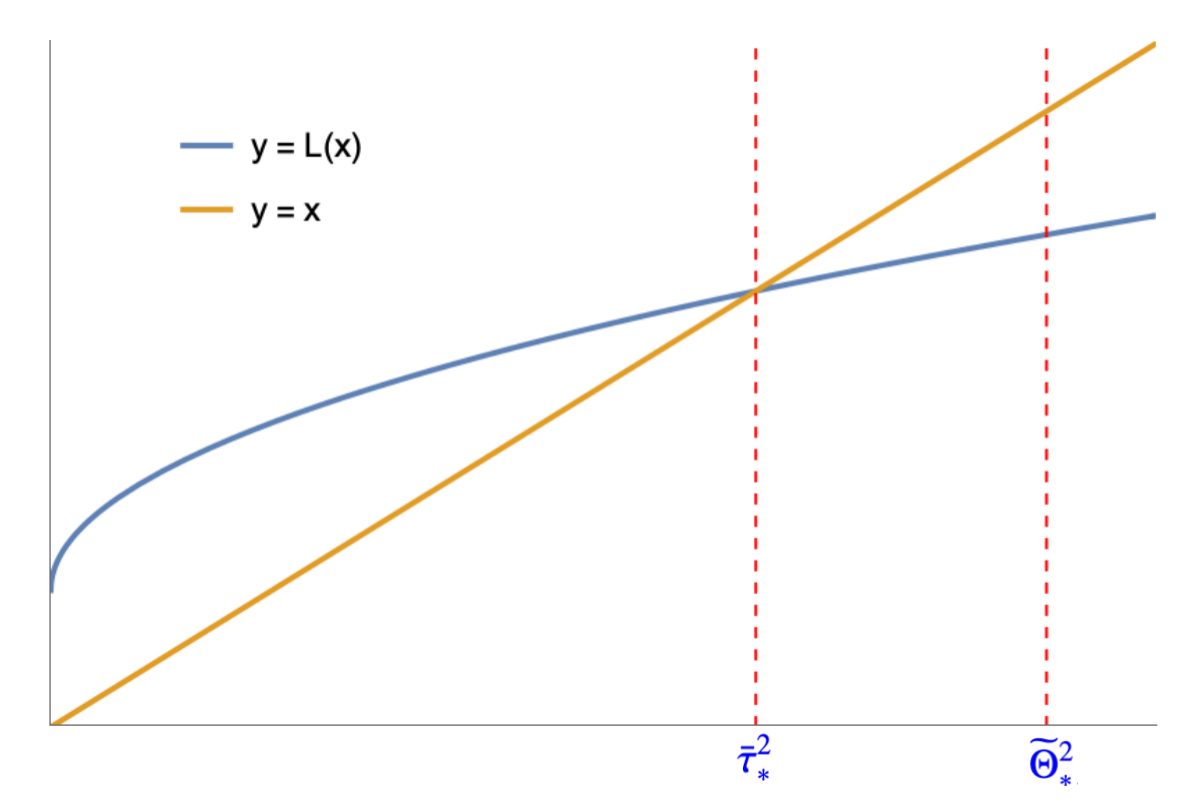}
\centering
\caption{
\textbf{Illustration of equation $y = L(x^2)$ and $y = x$.}
We present a graph of the equations to summarize our finding (a) - (c) in comparing $R_{\LW}(\lambda_{\LW, M}^*)$ and $R_{\textnormal{L}}(\lambda_{\LA, M}^*)$ in Section \ref{proof of prop: comparison of L_1}. 
}
\label{illu_prop_comparision_L_1}
\end{figure}

\begin{figure}[H]
\includegraphics[page=1,width=0.9\linewidth]
{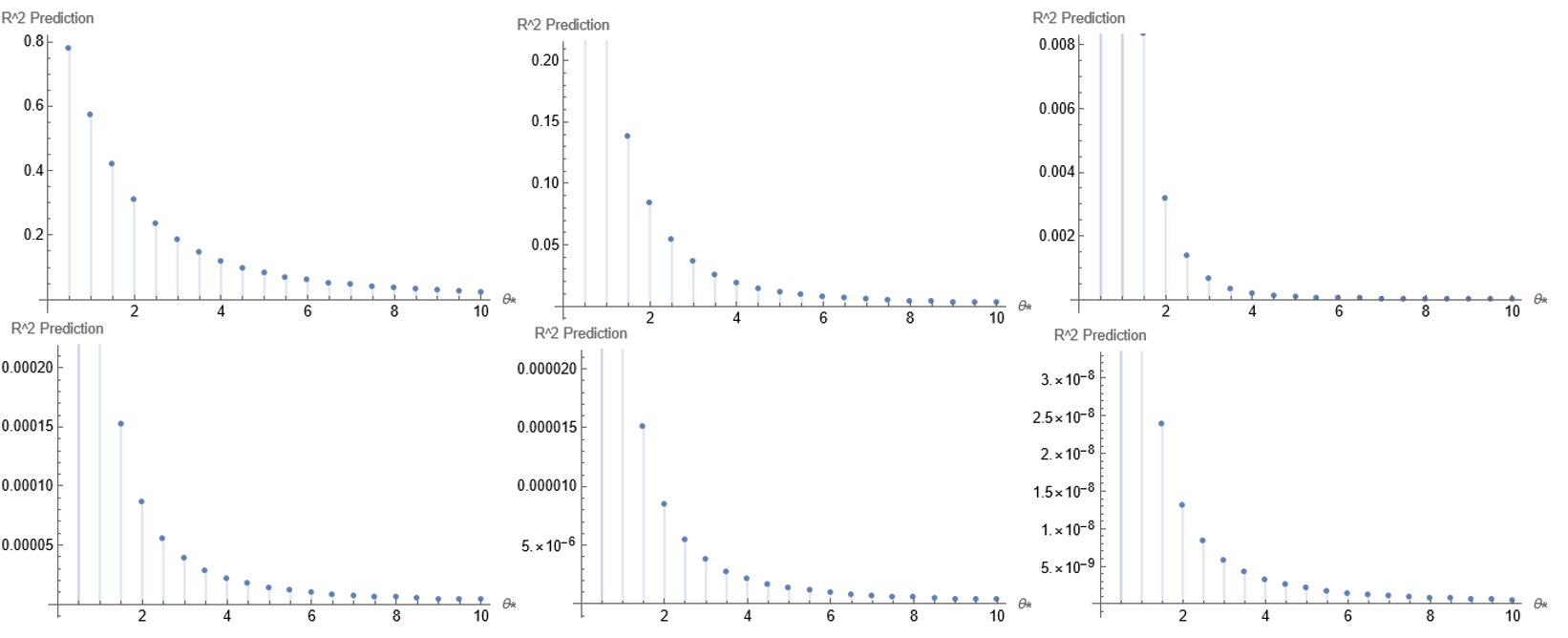}
\centering
\caption{
\textbf{Illustrating the pattern of $A_{\LW}(\alpha)$ versus $\Theta_*$.}
We present a graph of the Equation \eqref{eqn: R^2 vs. Theta_*}, which plots $A^2_{\LW}(\lambda)$ against $\Theta_*$ for different values of $\bar\bbeta$ and $\alpha$. We consider $\bar\bbeta$ drawn from the Bernoulli-Gaussian distribution $(1-\kappa) \delta(0) + \kappa N(0, 1)$, where $\kappa \in \{0.05, 0.95\}$ and $\alpha \in \{0.05, 1.00, 3.00\}$. }
\label{fig: R^2 vs. Theta}
\end{figure}

\end{document}